\documentclass[11pt,twoside,openright,usenatbib]{report}

\newcommand{\fecha}{Marzo 2025}

\long\def\symbolfootnote[#1]#2{\begingroup%
  \def\thefootnote{\fnsymbol{footnote}}\footnote[#1]{#2}\endgroup}

\usepackage{natbib}
\usepackage{epsfig}
\usepackage{graphicx}
\usepackage{lscape}
\usepackage{psfrag}
\usepackage{aas_macros}
\usepackage{amsmath}
\usepackage{amssymb}
\usepackage{url}
\usepackage{supertabular}
\usepackage{textcomp}
\usepackage{mathpazo} 		
\usepackage{tocloft}	
\usepackage{rotating}
\usepackage{hyperref}
\usepackage{amsmath,amssymb}
\usepackage{dcolumn}
\usepackage{bm}
\usepackage{xcolor}
\usepackage{graphicx}
\usepackage{commath}
\usepackage{float}
\usepackage{hyperref}
\usepackage{esint}
\usepackage{booktabs}
\usepackage{commath}
\usepackage{subfig}
\usepackage{rotating}
\newcommand{\RNum}[1]{\uppercase\expandafter{\romannumeral #1\relax}}

\usepackage{sectsty}
\allsectionsfont{\sffamily}

\usepackage{fancyhdr}
\fancypagestyle{plain} {
  \fancyhf{} 
  \fancyfoot[RO,LE]{\thepage}

 }

\makeatletter
\renewcommand{\section}{\@startsection
  {section}                          
  {1}                                
  {\z@}                              
  {-3.75ex \@plus -1ex \@minus -.2ex} 
  {1.5ex}                            
  {\bf \LARGE \sffamily}  	     
}
\renewcommand\subsection{\@startsection
  {subsection}
  {2}
  {\z@}
  {-3.5ex\@plus -1ex \@minus -.2ex}
  {1.3ex}
  {\bf \Large \sffamily}
}
\renewcommand\subsubsection{\@startsection
  {subsubsection}
  {3}
  {\z@}
  {-3.5ex\@plus -1ex \@minus -.2ex}
  {1.0ex}
  {\bf \large \sffamily}
} \makeatother

\makeatletter
\def\@makechapterhead#1{
  \vspace*{70\p@}
  {\parindent \z@ \raggedright \normalfont
    \flushright \Huge \sc \@chapapp\space  \sc \thechapter

    \par\nobreak
         \vskip 60\p@

     \Huge \normalfont \bf \sffamily   
     #1\par\nobreak
    \vskip 40\p@
  }
}
\makeatother

\makeatletter
\long\def\@makecaption#1#2{%
   \vskip 10\p@
   \setbox\@tempboxa\hbox{\small{\bf \sffamily #1:} #2}%
   \ifdim \wd\@tempboxa >\hsize
    {\small{\bf #1:} #2\par}
     \else
       \hbox to\hsize{\hfil\box\@tempboxa\hfil}%
   \fi}
\makeatother

\newcommand{\clearchapterpage}{\newpage{\thispagestyle{empty}\cleardoublepage}}

\begin{document}

\begin{titlepage}

\sffamily

\vspace{10mm}

\begin{centering}

{\huge \bfseries Cosmology without the cosmological principle\\
\vspace{3mm}
{\huge \bfseries A study of the large-scale structure effects in the background universe}}\\
\vspace{2mm}

\noindent\rule{0.9\textwidth}{0.025truein}

\Large

\vspace{15mm}

{\sc Erick C. Pastén }

\large
\vspace{5mm}

Supervisor: Dr. Victor H. Cárdenas (UV) \\
Internal reviewer: Dr. Graeme Candlish (UV) \\
External reviewer: Dr. Christos Tsagas (AUTH)

\vspace{10mm}
\begin{figure}[h!]
\begin{center}
\includegraphics[width=2cm]{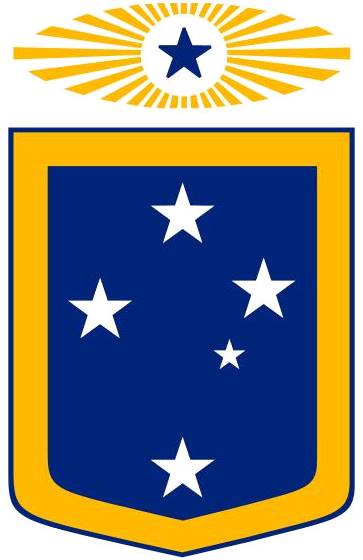}
\end{center}
\end{figure}

\large

\vspace{10mm}
Tesis para optar al grado de Doctor en Astrof\'isica

Instituto de F\'isica y Astronom\'ia

Facultad de Ciencias

Universidad de Valpara\'iso

\vspace{10mm} \noindent\rule{1.5truein}{0.02truein}

{\fecha \\ Valpara\'iso, Chile.}

\end{centering}

\vfill

\end{titlepage}
\normalsize

\clearchapterpage

\pagenumbering{Roman}
\setcounter{page}{1}

\fontsize{11}{13}\selectfont

\begin{center}
\phantom{.}
\vspace{7cm}

To all my homies

\vspace{2cm}
\end{center}

\clearchapterpage

\begin{center}
\phantom{.}
\vspace{7cm}

This thesis is solely my own composition, \\
except where specifically indicated in the text.\\
\vspace{0.5cm}
Total or partial reproduction, for scientific or academic purposes, \\
is authorised including a bibliographic reference to this documment.


\vspace{4cm}

\begin{figure}[h]
\centering
\includegraphics[scale=1]{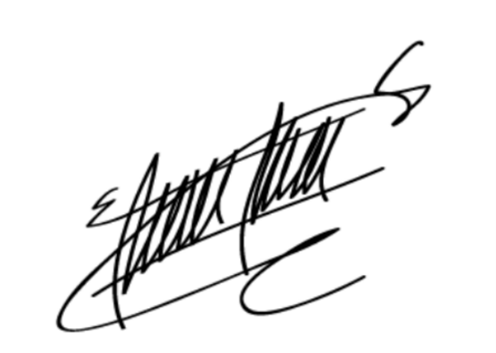}
\end{figure}

Erick Sebastián Contreras Pastén \\
\fecha\\
Valpara\'iso, Chile.
\end{center}

\clearchapterpage

\chapter*{Acknowledgements}

EP acknowledge the support through a graduate scholarship ANID-Subdireccion de Capital Humano/Doctorado Nacional/2021-2121082.

EP acknowledge the support through a mobility benefit of the University of Valparaiso, which allows the author to travel to work in the Aristotle University of Thessaloniki directly with the professor Christos Tsagas.
\clearchapterpage

\chapter*{Abstract}

Recent studies strongly suggest that the $\Lambda$CDM model may no longer serve as the definitive standard model in cosmology, given the increasing tensions between different cosmological observables. Some researchers have described this situation as a "crisis in cosmology." The $\Lambda$CDM model relies on two fundamental assumptions: the isotropy of the universe (supported by observational evidence) and the Copernican principle (a philosophical postulate). Together, these assumptions lead to the well-known Cosmological Principle, with the homogeneity of the universe emerging as a direct consequence. In this thesis, I review the cosmological consequences of relaxing some of these assumptions, exploring inhomogeneous and anisotropic universe models, as well as tilted cosmologies, which consider observers in motion relative to the Hubble flow. We examine methods to study these effects using existing cosmological data. First, we explore a variety of models that describe inhomogeneous and anisotropic universes, including different metric theories, perturbative analyses, averaging effects, tilted scenarios, and cosmographic approaches. We then apply this theoretical framework to analyze SNIA data and the local peculiar velocity field, aiming to constrain key parameters.

On large scales, baryonic matter in the universe tends to cluster into structures known as the cosmic web, which exhibits fractal behavior. The impact of this inhomogeneous large-scale structure on cosmology is an active area of research, spanning from its description through various metrics derived from Einstein's field equations to more complex challenges, such as the back-reaction problem related to the averaging of general relativity. In this work, we investigate the local fractal structure of the universe using luminosity distance relations derived from fractal-like matter distributions in LTB inhomogeneous models, analyzed against SNIA data. Our findings suggest that while a fractal distribution of matter cannot fully account for the cosmic acceleration, it remains a valuable tool for studying the fractality of the local universe, provided the background cosmology is known. It is important to note, however, that this remains a simplified model, and the inclusion of a true fractal distribution of matter in the field equations is an ongoing research challenge involving complex, non-differentiable mathematics.

The effects of general relativity on the evolution of the local large-scale structure remain poorly understood. Some studies indicate that the growth of density contrasts and peculiar velocities accelerates over time when using the covariant formalism of general relativity, as compared to the standard Newtonian or semi-Newtonian approaches. Additionally, the back-reaction problem in cosmology addresses the challenge of averaging cosmological quantities such as matter density, a task complicated by the non-linearity and complexity of the governing equations. Tilted cosmologies, which apply the covariant formalism to study the universe from the perspective of observers moving with peculiar velocities relative to the Hubble expansion, offer significant insights. Theoretical analyses suggest that the deceleration parameter measured by such observers could be negative, even if the background universe is not accelerating, provided the observers are located within a contracting bulk flow extending over large scales. In this work, we take a step toward observing these effects by extracting the volume scalar of the local peculiar velocity field via velocity reconstructions, confirming that the local peculiar velocity field is indeed contracting. Recent blind analyses of SNIA data and independent velocity field reconstructions support this conclusion, further highlighting the importance of studying tilted cosmologies and the relativistic effects on large-scale structure evolution. In this context, we use the covariant formulation to develop a connection between the enhanced growth rate of peculiar velocities in relativistic physics and the semi-Newtonian approach. This effect is strongly linked to the neglect of additional gravitational sources, such as the flux of energy, in Newtonian gravity.

From an observational standpoint, we adopt a phenomenological approach, performing a statistical analysis on the most recent SNIA compilation, Pantheon+, using the cosmographic luminosity distance. We extract the deceleration parameter for different portions of the universe by dividing the data into redshift bins and hemispheres. Our analysis reveals that the deceleration parameters are not consistent across the binned groups, which contradicts the assumption of a homogeneous universe. Statistical tests indicate a trend of increasing deceleration parameters, a result consistent with recent studies. It is also worth noting that very nearby supernovae (at $z \leq 0.008$) appear to be unsuitable for cosmological analysis, as they are contaminated by the velocity dispersion of the local peculiar velocity field, yielding values that do not align with background cosmological parameters. Various corrections using different models have been attempted to improve this data, but they have not yielded the desired results, while other studies opt to analyze the raw SNIA redshift data directly.

In conclusion, the possibility of explaining the apparent accelerated expansion of the universe without invoking dark energy remains a compelling alternative. Models such as tilted cosmologies continue to hold potential, supported by current observational data that challenge the cosmological principle. If dark energy and the acceleration of the universe are indeed real, a deeper understanding of the local structure in cosmological observations is crucial for constraining the physical parameters involved. This highlights the importance of further research into the fractal structure of matter, the physical properties of the bulk flow we inhabit, and the still poorly understood general relativistic effects on the evolution of the large-scale structure. The thesis presents several avenues for future research that could expand our scientific understanding in these areas.

\clearchapterpage

\normalsize


\tableofcontents
\clearchapterpage

\pagenumbering{arabic}
\setcounter{page}{1}

\chapter{Introduction}

Cosmic acceleration of the universe is one of the most important astrophysical discoveries in the last years (\citealp{Riess_1998}, \citealp{Perlmutter99}). The use of the FLRW metric in the modeling of the evolution of the universe has been extremely successful to match observations such as CMB (\citealp{PLANCK20}), SNIA luminosity (\citealp{Brout_2022}) and BAO (\citealp{desicollaboration2024desi2024vicosmological}). This model when compared to data, gives a negative value of the deceleration parameter $q_0$ meaning that the universe is expanding with a positive acceleration. In the General Relativity context, the acceleration of the universe only can be explained adding a \textbf{cosmological constant} $\Lambda$ in the Einstein equations, which has 2 main physical interpretations. If we take the trace of the Einstein equation for a vacuum space-time, we can obtain:

\begin{eqnarray}
    R=4\Lambda
\end{eqnarray}

Where $R$ is the \textit{Ricci-Scalar} representing the 4-curvature of the space-time. In this context $\Lambda$ can be viewed as an intrinsic parameter of the universe. On the other side, $\Lambda$ can also be related with an unknown form of energy $\rho_\Lambda$ called \textit{Dark Energy} (DE henceforth) as:

\begin{eqnarray}
    \frac{\Lambda}{8\pi}=\rho_\Lambda
\end{eqnarray}

Is interesting that this last interpretation is the most widely used in modern cosmology. The reasons behind this preference of the scientific community could be strongly related to the great advance of particle physics in the last years, as an exotic energy density is more appealing to experimental physics because it could be, theoretically, directly measured. 

\section{Challenges for the standard cosmology model}

The negative value of $q_0$ holds as long the cosmological principle and the FLRW metric is assumed in the modeling theory of the universe. Although DE is accepted as the standard explanation for the observations in cosmology, many theoretical and observational issues remains. The fully isotropic and homogeneous description of our universe has becoming strongly questioned in the recent years as long as new anomalies are observed in differents sources of data. In the following we summarize the most important theoretical and observational challenges that standard cosmology faces (see a recent review in \citealp{Peebles_2022}).

\subsection{Theoretical challenges for standard cosmology}

$\Lambda$CDM model theoretical problems comes from the assumption that astronomical observations, corrections and statistical analysis are quite precise. The results then require new physics and new philosophical ideas. Those issues are not new, and the most important could be resumed in the following points:

\begin{enumerate}
    \item \textbf{The Horizon problem:} The scale of the particle horizon at the decoupling epoch in $\Lambda$CDM model is in disagreement with the large observed homogeneity in CMB at quite large scales (\citealp{Lakhal_2019}). The impossibility for the information to travel during the age of the universe from one side to another leading to a similar evolution in different parts of the universe requires new physics to be explained. The standard hypothesis in $\Lambda$CDM is \textit{inflation} (\citealp{Guth_1981}), another unknown form of energy commonly described by a slow-rolling scalar field that has accelerated rapidly the expansion of the universe in very early times. As dark energy, no one knows what exactly inflation field is physically as it cannot be observed from the observational data: inflation presumably occurs before the decoupling of photons, neutrinos and even gravitational waves.
    
    Although others not less strange solutions to this problem has been proposed (see e.g \citealp{Hollands_2002}, \citealp{BRANDENBERGER_2011}), inflationary theory has been questioned as a scientific theory, since that the large amount of inflation possible scenarios could produce different universes which has led to a heated debate (see \citealp{Penrose_1988mg} for a review of inflationary problems). To the date, inflation is a necessary ingredient in the $\Lambda$CDM cosmology. Therefore, inflation is part of $\Lambda$CDM and the horizon problem is "resolved" in this model.
    \item \textbf{The Nature of Dark Energy:} In $\Lambda$CDM, DE is responsible roughly of the 70 \% of the composition of the universe, which produce the cosmic acceleration in $\Lambda$CDM cosmology (see e.g. \citealp{Weinberg_1988} for a review). In principle DE has been attributed to properties of the empty space as quantum field theory predicts on the particle physics standard model. However, the theoretical and observed values disagree by many orders of magnitudes (\citealp{Sol__2013}). This is an important problem to fundamental physics, as to the date it is unknown what is the true nature of DE. 
    
    \item \textbf{The Cold Dark Matter:} A 27\% of the composition of the universe is attributed to a form of matter that we cannot see, different from the baryonic matter that makes up the visible universe. This is modeled as a cold non-interacting fluid is distributed in a similar way that common matter is. Cosmology is not the only evidence for the existence of CDM, as it is required to explain the complex dynamics on the majority of galaxies and clusters (\citealp{Freese_2009}). However, CDM has never been observed directly and the true nature of it still is unknown. Some of the possible explanations consider rare new particles (\citealp{Feng_2010}), modifications of the gravitational theory (\citealp{milgrom2001mondapedagogicalreview}, \citealp{MARTENS2020237}), a mix of both of them (\citealp{Hossenfelder_2020}, \citealp{Mistele_2023}), or even the existence of multiple microscopically primordial black holes that fills the entire universe (\citealp{BIRD2023101231}). Overall, $\Lambda$CDM ignore all the internal dynamics that dark matter could have and the effects of the true nature of dark matter in the dynamics of the universe is still unknown. It is depressing and exciting at the same time that we cannot observe almost the 97\% of the $\Lambda$CDM universe.
    
    \item \textbf{The Flatness Problem:} The universe is presumably flat according to cosmological observations. In $\Lambda$CDM universe, this requires that the value of the initial curvature be extremely fine tuned as little primordial departures from the critical value would lead to large departures of the curvature of the universe (\citealp{dicke1970gravitation}). The common accepted solution for this problem is also inflationary fields. Although being beyond the scope of this thesis, is also worth to mention that some scientists consider that flatness is not really a cosmological problem (see e.g. \citealp{Helbig_2021}). 

    \item \textbf{The Coincidence Problem and the back-reaction:} Dark energy seems to dominate the evolution of the universe at the same epoch that non-linear structure has been formed. This is called the \textit{coincidence problem} \citealp{Velten_2014} and stimulate the research in the possibility of the large-scale structure being the source of cosmic acceleration due to a \textbf{back-reaction} effect \citealp{Schander_2021}. The back-reaction effect has been a controversial topic in cosmological literature due to the different scientific positions about it. Somo researchers think that the contribution is negligible (see e.g \citealp{greenwald2011}) while others believe that is causing indeed the cosmic acceleration (\citealp{Buchert_2000}). Up to now, the debate still continues (\citealp{verweg2024averagingcosmicstructurecosmological}) and the coincidence problem has been relegated to an anecdotal fact in standard cosmology. 

    \item \textbf{The validity of General Relativity:} General relativity is an incomplete theory for gravity since is not compatible with the quantum effects \citealp{Schulz:2014gqa}. It is possible that some other theoretical problems such as the flatness, the horizon or the missing relics are just a product of our lack of knowledge of the evolution of the universe on Planck time scales, where it is believed that the quantum nature of gravity takes on importance. There is an uncountable quantity of modifications to general relativity promising to be the \textit{true theory of gravity}. However, even general relativity effects are not well understood as the numerical computations to perform a full relativistic large-scale simulation are way too expensive. The standard approach is then to use a perturbative analysis over the average values of physical quantities. Whether or not these perturbative analyzes perfectly describe the effects of general relativity is an open debate to this day (\citealp{verweg2024averagingcosmicstructurecosmological}).
    
\end{enumerate}

\subsection{Observational problems of standard cosmology}

On the other side, $\Lambda$CDM model  has several observational problems caused by tensions between the values of different parameters.. The true nature of those tensions is still in debate. Is interesting to note that these discordances are increasing as the quality and the quantity of observational data is improving. The most important observational inconsistencies (to the date) are:

\begin{enumerate}
   \item \textbf{The cosmic discordances} (see \citealp{Abdalla:2022yfr} for a review of cosmological tensions):
    One of the most famous tensions in cosmology is the \textbf{Hubble tension}, which is a significant difference between the rate of local expansion $H_0$ measured in local light sources and the value that CMB predicts (see e.g. \citealp{DiValentino_2021}). However, there are other discordances like the \textbf{$\sigma8$ tension} between CMB (\citealp{PLANCK20}) and low redshift probes as SZ cluster analyses (\citealp{DES2022}). In a interesting research (\citealp{DiValentino21}), it is showed that the rigorous statistical analysis of some cosmic discordances, allowing variation on some extension parameters like neutrino total mass or the state equation of DE, rules out $\Lambda$CDM at $99\%$ CL. This is a blow to $\Lambda$CDM and is a strong motivation to work in new theoretical frameworks. Recent analysis of the local peculiar velocity field seems to worsen the Hubble tension (\citealp{Giani_2024}), while other studies claim that the cosmological parameters are not constant across the universe and evolves with redshift (see e.g \citealp{Perivolaropoulos:2023iqj}, \citealp{Colgain2022}). 


    \item \textbf{SNIA anisotropy:} It has been claimed in different studies (\citealp{Colin2011,Yang_2014,Javanmardi_2015,Sun_2018,Tang_2023}) that SNIA data shows a dipole feature that is not expected in a $\Lambda CDM$ scenario, meanwhile other studies deny the existence of this anomaly (see e.g. \citealp{Deng18b}). If the detection of this dipole is significant, it is possible that its origin is in the hard-to-model peculiar velocity field of very near supernovae which increase the difficulty of source redshift corrections. However, as other dipoles in the sky has been detected in the recent years, a cosmological explanation have to be still considered.  
    
    \item \textbf{The CMB anomalies:} At large angular scales, many strange features have been observed in the CMB by WMAP and Planck (\citealp{PLANCK20}). Some of those features include the lack of variance and correlation on the largest angular scales (\citealp{Chiocchetta_2021}), alignment of the lowest multipole moments with one another and with the motion and geometry of the solar system; the so-called \textit{axis of evil} (\citealp{Oliveira2004}, \citealp{Land2007}) and a large cold spot in the Southern hemisphere (\citealp{Vielva_2004},\citealp{Cruz_2006}). Those features are still unexplained and could potentially be evidence against the statistical isotropy of the universe. 
    
    \item \textbf{The strange multiple dipoles in the sky: }The detection of many mysterious dipoles in the distribution of distant objects in the sky that doesn't match in magnitude with the CMB puts in doubt the veracity of CMB dipole being kinematic in its origin. Some of those dipoles can be found in QSO polarization (\citealp{Pelgrims_2015}), Large Quasar Groups (\citealp{Secrest_2021}, \citealp{Friday_2022}), Galaxy Spin directions, X-ray galaxy clusters (\citealp{Migkas2018}) or even in SNIA (\citealp{Sah2024}). Even without SNIA dipole, the nature of those alignments is mysterious. This suggest the possibility of a CMB not being a true cosmological rest frame and dipoles being from cosmological origin, challenging the isotropic assumption of the universe. If the CMB dipole is an intrinsic cosmological dipole or is being kinematical originated is an active research topic, which analyse other relativistic effects such as aberration of light (see e.g. \citealp{Roldan_2016}).
    
    \item \textbf{Super-voids and other superstructures:} Large voids and superstructures have been detected in the last years with different observational procedures (\citealp{Horvath2015},\citealp{Balazs2015}, \citealp{Mao_2017}). Some of them has been claimed to be in tension with $\Lambda$CDM cosmology. Particularly, the KBC void is a super-void that presumably we live in it, extending over $\sim 300 Mpc$ (\citealp{Keenan13}). This void has been extremely important to cosmology, as some cosmological inhomogeneous models conclude that the existence of a big super-void in our surroundings can mimic the negative deceleration parameter (\citealp{Iguchi2002}, \citealp{Yoo2008}) in $\Lambda$CDM or at least explain the Hubble tension (\citealp{Mazurenko2023}). Even if KBC cannot explain the negative deceleration parameter, it has been claimed that the dimensions of this super-void can falsify $\Lambda$CDM (\citealp{Haslbauer_2020}). However, some studies has put in doubt the existence of KBC (\citealp{Kenworthy19}) or is utility to explain cosmological observations (\citealp{Hoscheit18},\citealp{Camarena22},\citealp{huterer2023enoughlocalvoidsolve}). Also, recent studies suggest that we live in a contracting peculiar velocity field which apparently contradicts the existence of a large void (\citealp{Giani_2024},\citealp{Sorrenti:2024ugq}).

    \item \textbf{Early Galaxies:} The last observations of JWST (\citealp{Labb__2023}, \citealp{Donnan_2022},\citealp{Bouwens_2023}) suggest that early galaxies evolution differs of what was expected in $\Lambda$CDM (\citealp{Boylan_Kolchin_2023}). There are more galaxies, more massive and more luminous than the predictions from standard cosmology at that epoch. The mechanism by which these galaxies are produced is still unknown.
    
    \item \textbf{The peculiar velocities: } The effect of the relative peculiar velocities and the inhomogeneties in the cosmological redshift is an extensive topic in astronomical literature. Usually, the geocentric observed redshift $z_{geo}$ is corrected automatically to the heliocentric $z_{hel}$ via telescope software. Then, in order to apply the FLRW metric and $\Lambda$CDM framework, one has to convert all the observations to a co-moving frame of reference with the Hubble expansion. Usually, CMB is interpreted as this rest frame, leading to the rest frame $z_{cmb}$ redshift. If our peculiar velocity needs to be considered to perform cosmological analysis, also should be considered the peculiar velocity of the light source. This introduce another correction to the redshift which leads finally to a cosmological usable redshift $z_{HD}$ (\citealp{Davis11},\citealp{Peterson_2022}). Usually we don't have exact information of the galaxy source velocity, giving another possible lack of corrections or even overcorrections if we share a bulk flow with the light sources. Those corrections has been discussed and improved in (\citealp{Carr_2022}) using a divergence-less (and therefore irrotational) Newtonian velocity field reconstruction over the last Pantheon + compilation. Also it is worth to mention the existence of recent modelings of the local peculiar velocity field using the last Cosmic Flow-4 released data (\citealp{Curtois_2023},  \citealp{Giani_2024}). As the reconstruction process are Newtonian and the direct astronomical redshift surveys just gives the radial velocity (and assumes the cosmological principle), the effect of the vorticity term of the local peculiar velocity field in cosmology is still not considered. It is worth to mention the possibility that relativistic effects of our peculiar motion are completely neglected when corrections are performed as stated previously, being Tilted Cosmology scenario an interesting new theoretical framework to study deeply.
\end{enumerate}

Are those problems arising from a purely experimental/precision origin or are they caveats in the cosmological model? Focusing in the last possibility, the aim of this thesis is to study what happens if one of the 3 most important ingredients of the standard cosmology, the cosmological principle, is relaxed in different aspects.
In the following section the cosmological principle and all of its features studied in this thesis will be defined.

\section{The Cosmological Principle and the FLRW universe}

The cosmological principle can be defined simply as the assumption of \textbf{global homogeneity and isotropy of the universe}. This can be stated also as \textit{the universe is similar in every place, and looks similar in every direction}. From a geometrical point of view, the cosmological principle can be introduced as the \textbf{Friedmann-Robertson-Walker-Lemaitre metric}:

\begin{eqnarray}
    &ds^2& = -dt^2+d\mathbf{x}^2 \\
    &=&-dt^2 + a^2(t) \left[ \frac{dr^2}{1- k r^2} + r^2(d\theta^2 + sin^2\theta d\phi^2)\right],
\end{eqnarray}

\noindent which is the maximally symmetric geometry of a space-time in general relativity. In tensor notation:

\begin{eqnarray}
    g_{\mu\nu}=\begin{pmatrix}
        -1 & 0 & 0 & 0 \\
         0 & \frac{a(t)^2}{1-kr^2} & 0 & 0 \\
         0 & 0 & a(t)^2 r^2 & 0 \\
         0 & 0 & 0 & a(t)^2 \sin^2\theta    \end{pmatrix}
\end{eqnarray}

With the cosmological principle, it is possible to state that every parameter of interest evolves similarly in every place of the universe. Therefore, we can parameterize these evolutions with the so-called \textbf{cosmic time} parameter $t$. The evolution of the universe in a global homogeneous and isotropic model is then fully included in the cosmic time dependent scale factor $a(t)$. According to this picture, is the space itself which is changing over time.
As this metric is diagonal, according to the Einstein equation, the Energy-Momentum tensor $T_{\mu\nu}$ also should be diagonal. The maximally symmetric EM tensor can be writted as:

\begin{eqnarray}
    T_{\mu\nu}=\begin{pmatrix}
        \rho & 0 & 0 & 0 \\
        0 & p & 0 & 0 \\
        0 & 0 & p & 0 \\
        0 & 0 & 0 & p 
    \end{pmatrix}
\end{eqnarray}

Where $\rho$ is the energy density and $p$ is the pressure associated to each form of energy. The Einstein equation for a FLRW universe became the famous \textbf{Friedmann equation}, the \textbf{acceleration equation} and the \textbf{fluid equation}:

\begin{eqnarray}
     H(t)^2=\frac{8\pi G}{3}\rho-\frac{kc^2}{a^2}+\frac{\Lambda c^2}{3} \\
    q(t) = \frac{4\pi G}{3H}\left (\rho +\frac{3p}{c^2}\right )-\frac{\Lambda c^2}{3H}\\
    \dot{\rho}+3H\left (\rho +\frac{p}{c^2}\right )=0
\end{eqnarray}

Where $H(t)=\dot{a}/a$ is the \textbf{Hubble parameter} and $q(t)=-\Ddot{a}/{a}$ is the \textbf{deceleration parameter}. Those equations together with an state equation relating energy density and pressure in the form of $\rho=w p$, define completely the evolution of the scale factor $a(t)$.

Another way to define an homogeneous and isotropic universe, is to think in \textbf{families of observers} on the covariant formulation of cosmology (see \citealp{Tsagas2008} for a review). This approach is needed to understand the implications of tilted-cosmology models which will be discussed in this thesis. According to covariant cosmology, we can consider the observers co-moving with the Hubble expansion identified as a 4-velocity field $u^{a}$. The dynamics of the universe is then viewed as a evolution of this 4-velocity field. We introduce the 4-gradient of the velocity as the 4-covariant derivative of $u^{a}$:

\begin{equation}
\nabla_bu_a= \frac{1}{3}\,\Theta h_{ab}+ \sigma_{ab}+ \omega_{ab}- A_au_b\,.  \label{Nbua1}
\end{equation}

$\Theta$ is the volume expansion/contraction scalar (when positive/negative respectively), $\sigma_{ab}$ is the shear, $\omega_{ab}$ is the vorticity and $A_a$ is the 4-acceleration. $h_{ab}=g_{ab}+u_au_b$ is defined as the \textbf{projector tensor} which act as the metric tensor in a 3D spatial section. We can note that $\Theta$ is related with the Hubble parameter $H$ as both monitors the expansion of the universe. According to the cosmological principle, $\sigma_{ab}$ and $\omega_{ab}$ should vanish and we can write:

\begin{equation}
H=\frac{\Theta}{3} \label{Htheta}
\end{equation}

as the Hubble parameter. In the same way, the deceleration parameter can be defined as:

\begin{equation}
q=-(1+\frac{3\dot{\Theta}}{\Theta^2}) \label{qtheta}
\end{equation}

The evolution of the universe in the covariant formulation of cosmology is dictated by \textbf{Raychaudhuri equations} (\citealp{Raychaudhuri55}, see \autoref{ss1+3CD} for a deep study of covariant cosmology). Equations \ref{Htheta} and \ref{qtheta} provide a natural way to connect covariant cosmology with theoretical cosmological parameters. However, how those parameters should be interpreted in observational cosmology is a new interesting topic that we will discuss in the following chapters.

\subsection{Philosophy of the cosmological principle}

As it is, it seems like the cosmological principle is an \textit{a priori} assumption of our universe. However, it is strongly supported by 2 statements, one of them being observational and the other purely philosophical:

\begin{enumerate}
    \item \textbf{According to CMB, the universe looks highly isotropic from our perspective}: The CMB is commonly interpreted as the remaining radiation that escaped from the thermal fluid in the epoch of photon decoupling. In this era, the thermal equilibrium between the components of the universe require that this radiation should have features of a black body. The CMB observed today has indeed a black body spectra with a temperature around $\sim 2,7 K$ and has a strong dipole component. This dipole is interpreted as our \textbf{peculiar motion} of $\sim 370 \frac{km}{s}$ w.r.t the Hubble Flow and, when this dipole is removed, the CMB looks incredible isotropic in every direction. The isotropy of the CMB is the main observational evidence that support the cosmological principle.
    \item \textbf{There is no privileged observers in the universe:} The Copernican principle (or best known in science philosophy as \textit{mediocrity Copernican principle}) states that we are not special observers in the universe. Therefore, the universe looks almost equal for other observers as well. This is a completely philosophical assumption and has no observational support, as we cannot travel large distance to observe how does the universe looks like from different places. The non-observational origin of this assumption put cosmology in one of the scientific disciplines influenced the most by philosophy.
\end{enumerate}

\begin{figure}[!ht]
\centering
\includegraphics[scale=0.7]{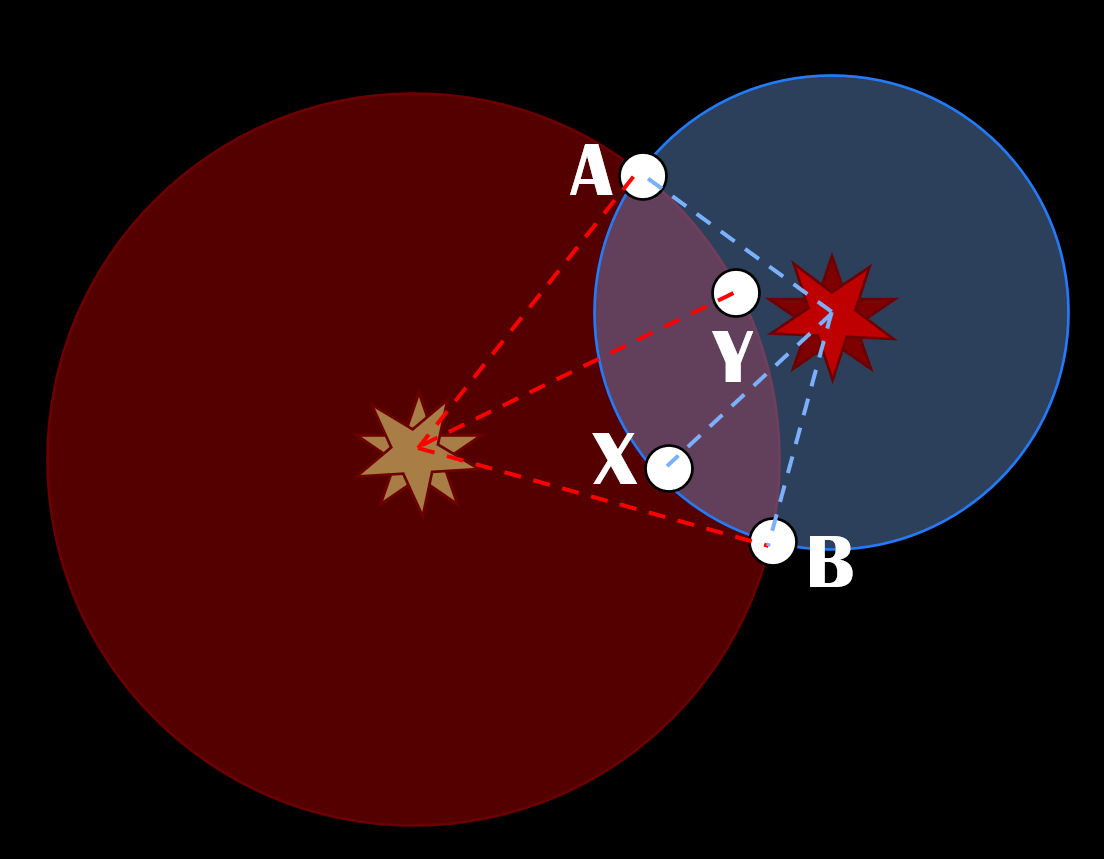}
\caption{The isotropy observed in the universe and the assumption of the Copernican principle, leads to homogeneity. If an observer in the yellow star (the sun) sees the universe isotropic and measure a parameter $\theta$ in $A$, then the parameter should be the same in $B$ and $Y$ ($\theta(A)=\theta(B)=\theta(Y)$). If the Copernican principle holds, then an observer in the red star (Betelgeuse) should measure the same values of the parameter $\theta$ in $A$, $B$ and $X$ ($\theta(A)=\theta(B)=\theta(X)$). Then, homogeneity is derived from this 2 assumptions as $\theta(X)=\theta(Y)$.}
\label{proofcosmo}
\end{figure}

Homogeneity cannot be directly measured as its depend on physical distances and therefore, on the cosmology used. However, the two previous statement leads together in a logic exercise to the homogeneity of the universe (see \ref{proofcosmo}). As the Copernican principle is a purely philosophical statement and is a foundation of the standard model, we can fairly say that Cosmology is the \textit{most philosophical of all the sciences} (see \citealp{pittphilsci9062}). However, the Copernican principle is being questioned in different aspects over the last year. First of all, we are in fact \textit{non-typical} observers as we move in a particular direction with a particular velocity w.r.t. the expansion of the universe which is well observed in the CMB dipole. Also, our local structure is spatially distributed in a particular way and moving differently of other places in the universe in a so called \textit{local bulk-flow} (\citealp{Tully_2023}). Commonly, cosmologists \textit{correct} this "non-typicality" applying cosmological perturbation analysis on the data, which results in redshift modifications (\citealp{Davis11}). The detection of strange alignments in the sky that doesn't match with the CMB dipole and some interesting theoretical results applying covariant cosmological formalism to a "tilted" observer w.r.t. to the Hubble expansion, motivates the further research of the peculiar velocity effect in cosmology. The possibility that corrections in the data are masking some important physical effects must be considered. 

\subsection{Statistical cosmological principle}

It is evident that the universe is far from being perfectly isotropic and homogeneous. If we look in the sky we can see different structures in different directions that are not consistent with a perfect isotropy universe. However, is possible to define \textbf{scales} in which the cosmological principle holds. This approach to understand the cosmological principle is called \textbf{statistical cosmological principle} (\citealp{Neyman2018}) and is the version that is used commonly in astrophysics. If we average a magnitude over a scale in one place (or direction) in the universe, that value should be almost equal to the average of the magnitude at the same scale in any other place (or direction of the universe), then we can say that the cosmological principle is valid at scales greater than the averaged scale. In the $\Lambda$CDM context, the scale of homogeneity goes from $\sim 100-200 \frac{Mpc}{h}$ to $260 \frac{Mpc}{h}$ (\citealp{Yadav2005},\citealp{Melia2023}). Based on those studies, we adopt an homogeneity scale of $\sim 300 \frac{Mpc}{h}$. 

\subsubsection{The Isotropy scale and the peculiar velocities}

\begin{figure}[ht!]
    \centering
    \includegraphics[width=13cm]{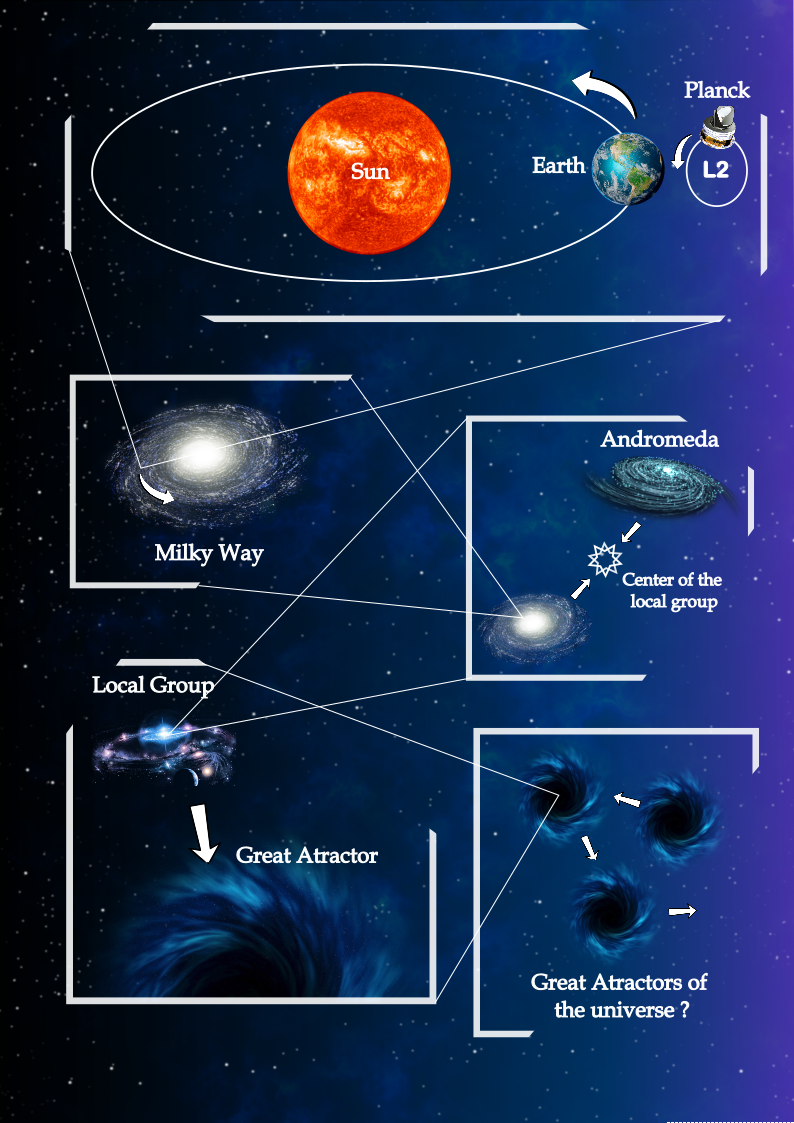}
    \caption{The CMB dipole measured by Planck Satellite could be understood as the addition of many velocity components. Those include the rotation of the satellite around the L2 Lagrange point, the movement of the L2 together with the Earth around the Sun, the movement of the Sun around de Milky Way, the motion of the Milky Way to the center of the Local Group, and finally the fall of the Local Group towards the Great Attractor. It is possible that this dynamics is repeated at larger scales with the existence of a uncountable amount of Great Attractors. (Diagram inspired by 
    George Smoot Nobel lecture, December 8th, 2006).}
    \label{Peculiarcomponents}
\end{figure}

The CMB gives the major constrain over the anisotropy scale of the universe. This anisotropy degree is obtained after subtracting the dipole component of the CMB attributed to our movement w.r.t the Hubble Flow, which is composed by different velocities sources in a complex dynamics (see \ref{Peculiarcomponents}). The kinematical origin for the CMB dipole is one of the main conclusions of the cosmological principle and the standard cosmology model, being the last estimations for our velocity w.r.t to the Hubble flow frame of reference around $\sim 370 km/s$ in the direction of Crater constellation (\citealp{PLANCK20}). However, the interpretation of the CMB dipole being kinematically originated has been challenged by the detection of strange dipoles that doesn't match either in direction or in magnitude with the CMB direction.

If the CMB dipole is true kinematically originated, all observed data has to be corrected to the CMB frame in order to be cosmologically usable. These corrections are commonly performed over the redshift, going from an heliocentric measured $z_{hel}$ to a cosmological $z_{cmb}$ redshift. There is an ongoing debate discussing if this correction is well supported by physics. Some theoretical result suggest that the cosmological effect of our movement w.r.t. CMB cannot only be attributed to corrections in the redshift of the data, having an impact in the estimation of cosmological parameters via relativistic effects. 

More over, these corrections are extended considering the peculiar velocity of the source via peculiar velocity structure modeling (\citealp{Carr_2022}), leading to a presumably cleaner $z_{HD}$ redshift. The final used formulae for the luminosity distance is then:

\begin{equation}
    \bar{D}_L(z_{HD},z_{\odot},z_{sn})=(1+z_{sn})^2 (1+z_{\odot})D_L(z_{HD}),
\end{equation}

Where $z_\odot$ is the peculiar movement of the sun w.r.t to the CMB frame and $z_{sn}$ is the peculiar movement of each supernovae. Two factors of $(1+z_{sn})$ are present in the luminosity distance
correction, due to relativistic beaming contributing as well as Doppler effect. Those correction could potentially improve the usability of very near cosmological data, which is strongly affected by the velocity dispersion of the local large-scale structure. However, the modeling of the peculiar velocity field is commonly made using Newtonian physics, neglecting general relativity effects and therefore using an irrotational velocity field. Even when the corrections over the data are performed, the estimated values for cosmological parameters using very near SNIA completely disagrees with the values extracted using SNIA that are far away (see e.g. \ref{chap:reds}). Notably, to compute cosmological parameters the very near supernovae are completely erased from the analysis to get rid of peculiar velocity effects (\citealp{Brout_2022,Riess_2022}). 

Recently, another correction to the SNIA data using the Cosmicflows-4 velocity survey has been performed modeling the local peculiar velocity field as an ellipsoid (\citealp{Giani_2024}) showing a notably anisotropy in its dynamics. Is interesting to note that these corrections worse the Hubble tension when this anisotropic model is used, as the local peculiar velocity field has been found to be a contracting bulk flow in average. Most notably, the bulk flow has found to be expanding if a spherical model is used, highlighting the importance to study precisely the dynamics of the peculiar velocity field.

\subsubsection{The Homogeneity scale and the fractal large scale structure}

Astronomical objects are not uniformly distributed over across the space, forming nodes, filaments, sheets
and voids (\citealp{EINASTO2001355},\citealp{Springel_2006}). This large-scale structure (LSS) has been called the \textit{Cosmic Web}. The LSS of the universe can be studied observationally by 2 sources of data. The first are the \textit{CMB experiments} whose goal is to understand the initial conditions for the formation of structures. The other one are the \textit{Galaxy surveys} which study the final state of our universe and try to understand its dynamic using statistical methods such as the \textit{2-point correlation function} $\xi(r)$, both in redshift and physical space. The latter is defines as a function of the density contrast $\delta(r)$ as:

\begin{eqnarray}
    \xi(\mathbf{r}_1,\mathbf{r}_2)=\langle\delta(\mathbf{r}_1)\delta(\mathbf{r}_2)\rangle
\end{eqnarray}

If we assume homogeneity and isotropy, then:

\begin{eqnarray}
    \xi(\mathbf{r}_1,\mathbf{r}_2)=\xi(\vert \mathbf{r}_1 - \mathbf{r}_2 \vert)
\end{eqnarray}

Ultimately, the 2-point correlation function is defined as a measure of the excess probability $dP$ to find a galaxy in a volume $dV$ at a distance $r$ of other galaxy, when comparing with a random homogeneous distribution (\citealp{Coil_2013}) . If $n$ is the mean number density of galaxies, then:

\begin{eqnarray}
    dP=n(1+\xi(r))dV
\end{eqnarray}

The power spectrum and the fractal dimension (see chapter \ref{chap:fractal}) of the LSS can be derived from the 2-point correlation function, providing a powerful tool for observational astronomy.  Those observations suggest that the LSS fractality has a cutoff that represent the transition to homogeneity, going from $\sim 100-200 \frac{Mpc}{h}$ to $260 \frac{Mpc}{h}$ (\citealp{Melia2023}). Based on those studies, we can adopt an homogeneity scale of $\sim 300 \frac{Mpc}{h}$. 

The effect of the LSS in cosmology is an interesting topic which is well motivated by the \textit{coincidence problem}: it seems that dark energy begins to dominate the evolution of the universe at the same epoch that non-linear structures form. This topic is well known as the \textbf{back-reaction problem} or \textbf{averaging problem}, since it comes from the difficulty to develop an auxiliary cosmological model with representative matter fields that average the cosmic matter distribution sufficiently well in order to provide relevant results (\citealp{verweg2024averagingcosmicstructurecosmological}). The averaging in cosmology is needed as the numerical and analytical calculations in general relativity using the full complex dynamics of the LSS is extremely demanding. While some authors claim that the back-reaction effect is negligible (see e.g \citealp{greenwald2011} ), others claim that the large-scale structure could be causing the accelerated expansion of the universe (\citealp{Brandenberger2000},\citealp{Buchert_2000}). Even if the back-reaction contribution to the cosmological dynamics is negligible, still there are inhomogeneous "local" cosmological models such as \textit{LTB} (see chapter \ref{chap:models}) that could have a large impact on cosmology just allowing for local inhomogeneities such as a void, an underdensity or a fractal distribution of matter. The latter scenario will be deeply studied in chapter \ref{chap:fractal}.

The aim of this thesis is to explore the different aspects of the cosmological principle, the different scenarios that relax some of its assumptions and study deeply how cosmological parameters could be affected in those different scenarios. This document is a recompilation of the work made in my PhD journey which included both theoretical and observational work. The thesis is structured as follow: in chapter \ref{chap:models} we explore the most important inhomogeneous and anisotropic cosmological models present in literature as well as different frameworks to study deviations from the cosmological principle; in chapter \ref{chap:fractal} we use and inhomogeneous model to explore how a fractal matter distribution can be included in cosmological analysis using an inhomogeneous cosmological metric; in chapter \ref{chap:tilt} we explore the effects of \textit{tilted cosmology} models which brings to the table a new potential explanation for the negative decceleration parameter observed, providing novel ways to connect this theory with observations; we expand the covariant treatment used to develop tilted cosmology to the evolution of peculiar velocity fields in \ref{pec}, increasing the needed to understand relativistic effects in cosmology and the LSS evolution; finally in chapter \ref{chap:reds} we explore the last compilation of SNIA data Pantheon+ with binned analysis revealing possible anomalies in cosmological parameters which seems to contradict the cosmological principle. We discuss and summarize the main results of the whole work in \ref{conc}. Also, the reader can find three appendix at the end of the thesis: the appendix \ref{VFR} develops the theoretical understanding of peculiar velocity field reconstructions and other peculiar velocity evolution features, \ref{s1+3CF} presents a summary of the covariant 3+1 formalism of cosmology to provide a better understanding of tilted cosmologies and in \ref{Marg} I provide deep information about the marginalization of cosmological parameters used in some sections of this work.
\clearchapterpage

%
%
\chapter{Inhomogeneous and anisotropic cosmology}
\label{chap:models}

In the literature there are plenty of approaches to study universes that does not follow the cosmological principle. Some of them works directly with different metric tensors which are solutions of the Einstein equations (such as LTB or Bianchi models), while others deals with a perturbative analysis of general relativity equations, using the covariant formalism of cosmology (like the back-reaction effect or tilted cosmologies). Other methods study the deviations from an isotropic and/or homogeneous universe in an agnostic way, not assuming general relativity or the cosmological principle and allowing for just kinematical cosmological parameters. Those latter approaches are commonly referred as \textbf{cosmographic analysis}. In this chapter I present a brief review of those models and tools used to study inhomogeneous and anisotropic cosmologies.

\section{Inhomogeneous and/or anisotropic metrics}
\subsection{The LTB model}

\citealp{Lemaitre33}, \citealp{Tolman34} and \citealp{Bondi47} were the first to studied isotropic and radial inhomogeneous universes, known as LTB models. Here we restrict ourselves to the study of pressureless matter universes. Assuming for generality the existence of a cosmological constant with equation of state $ \rho_\Lambda=-p_\Lambda$, the energy-momentum tensor is given by:
\begin{equation}\label{eq2}
    T_\mu^\nu=-\rho_M(r,t)\delta^\mu_0\delta^0_\nu-\rho_\Lambda\delta^{\mu}_\nu,
\end{equation}
in which the matter density $\rho_M$ varies in time and radially in space, meanwhile $\rho_\Lambda$ remains fixed. The metric for a LTB universe can be written as:
\begin{equation}\label{LTBmetric}
ds^2 = -dt^2 + \frac{R'^2}{1+2E(r)}dr^2 + R^2d\Omega^2,
\end{equation}
where $R=R(r,t)$ is a generalization of the scale factor and $E(r)$ is a function associated with the total energy. Here a prime indicates a derivative respect to $r$ and a dot a derivative respect to time $t$. From Einstein's equation we get the evolution equation:
\begin{equation}\label{eq4}
    \frac{\dot{R}^2}{R^2}=\frac{2GM(r)}{R^3}+\frac{8\pi G \rho_\Lambda}{3} +\frac{2E}{R^2},
\end{equation}
where $M(r)$ is an integration constant (in time). 

Notice that by writing $R(r,t)=a(t)r$ where $a(t)$ is the scale factor and $2E(r)=-Kr^2$ with $K$ being the curvature constant factor, the metric reduces to the FLRW metric and (\ref{eq4}) reduces to the usual Friedmann equation.

The $M(r)$ function can be interpreted as the mass contained in a sphere of radius $R=ar$ in the FLRW limit. The matter density is defined through:
\begin{eqnarray}\label{eq5}
\frac{M'}{R'R^2} = 4\pi \rho _M.
\end{eqnarray}
So the mass function $M(r)$ is defined as the integral of $\rho_M$ in the LTB space as:
\begin{equation}\label{eq6}
    M(r)=\int_0^r\rho _M (4 \pi R'R^2 dr)
\end{equation}
Defining as usual the local Hubble rate by:
\begin{equation}\label{eq7}
    H(r,t)=\frac{\dot{R}(r,t)}{R(r,t)},
\end{equation}
we can write (\ref{eq4}) as:
\begin{equation}\label{eq8}
    H^2= \frac{2GM(r)}{R^3}+\frac{8\pi G \rho_\Lambda}{3} +\frac{2E}{R^2}
\end{equation}
By defining the local matter density parameter $\Omega_m$ through:
\begin{equation}\label{eq9}
    2GM(r)=H_0^2(r)\Omega_m(r)R_0^3(r),
\end{equation}
where $H_0(r)=H(r,t_0)$ is the present Hubble local rate, $R_0=R(r,t_0)$ being an initial condition that can be gauged, and for the curvature term $\Omega_k$ through
\begin{equation}\label{eq10}
    2E(r)=H_0^2(r)\Omega_k(r)R_0^2(r),
\end{equation}
then, in absens of dark energy, the local Friedman equation takes the form:
\begin{equation}\label{eq11}
H^2(r,t)=H_0^2(r)( \Omega_m(r) A^3 + (1-\Omega_m(r))A^2),
\end{equation}
where $A(r,t)=R_0/R$. This means that we can choose any $R_0(r)$ function and the equations remains the same. This freedom is usually fixed by taking $R_0(r)=r$ (\citealp{Enqvist07}). The Friedman equation (\ref{eq11}) defines completely the relation between the functions $H_0(r)$, $\Omega_m(r)$, $R(r,t)$ and $H(r,t)$.


\subsubsection{Constraints in the LTB model}

In LTB model there is a dynamical relation between 3 features of the universe. These are:

\begin{enumerate}
    \item The Evolution of the Universe, represented in the scale factor $R(r,t)$.
    \item The Energy distribution included in the mass-curvature profile $\Omega_m(r),\Omega_k(r)$ and dark energy.
    \item The Big-Bang spatial structure associated with the $H(r,t)$ Hubble function or with the $t_{BT}(r)$ bang time function as we will see.
\end{enumerate}

One of the most used constraint to LTB models is based on the time that has passed from the big bang until now (\citealp{GarciaBellido08}), the so called `bang time' $t_{BT}(r)$ defined by:
\begin{equation}\label{eq12}
\begin{split}
    H_0(r)(t-&t_{BT}(r))=\int_0^{R(r,t)/R_0(t)}\frac{dx}{\sqrt{\Omega_m/x+\Omega_k}} \\
    &=\frac{R(r,t)}{R_0(r)\sqrt{\Omega_k(r)}}\sqrt{1+\frac{R_0(r)\Omega_m(r)}{R(r,t)\Omega_k(r)}} \\
    &-\frac{\Omega_m(r)}{\sqrt{\Omega_k(r)^3}}\sinh^{-1}{\sqrt\frac{\Omega_m(r)R(r,t)}{\Omega_k(r)R_0(r)}}.
\end{split}
\end{equation}
This function means that not all locations in the universe were created at the same time. If we set $t=t_0$ we can obtain the bang time function $t_{BT}(r)$:
\begin{equation}\label{eq13}
\begin{split}
    &H_0(r)(t_0-t_{BT}(r))=H_0(r)t_{u}(r)=\int_0^{1}\frac{dx}{\sqrt{\Omega_m/x+\Omega_k}} \\
    &=\frac{1}{\sqrt{\Omega_k(r)}}\sqrt{1+\frac{\Omega_m(r)}{\Omega_k(r)}} 
    -\frac{\Omega_m(r)}{\sqrt{\Omega_k(r)^3}}\sinh^{-1}{\sqrt\frac{\Omega_m(r)}{\Omega_k(r)}}, 
\end{split}
\end{equation}
where we have defined $t_{u}(r)$ as the local age of the universe. It is interesting to note that we need to fix two functions here, chosen among $\Omega_m(r)$, $H_0(r)$ and $t_{BT}(r)$, and also fix a gauge (usually $R_0(r,t)=r$) to define completely the scale factor function $R(r,t)$. We obtain a similar result if we use instead the functions $M(r)$ and $E(r)$:
\begin{equation}\label{eq14}
\begin{split}
    \sqrt{2}(t-&t_{BT}(r))=\int_0^{R(r,t)}\frac{dx}{\sqrt{2GM(r)/x+2E(r)}} \\
    &=\frac{R(r,t)}{\sqrt{E}}\sqrt{1+\frac{GM(r)}{R(r,t)E(r)}} \\
    &-\frac{GM(r)}{\sqrt{E(r)^3}}\sinh^{-1}{\sqrt\frac{GM(r)R(r,t)}{E(r)}} 
\end{split}
\end{equation}
In this case, it may seems we need to set three functions, $t_{BT}(r)$, $M(r)$ and $E(r)$, to define $R(r,t)$, but in fact one of them is a pure gauge. Usually $M(r) \propto r^3$ is assumed, so we just need to define two functions as we will see.
Other constraints used in literature are:

\begin{enumerate}
    \item Setting a Mass Profile ignoring Dark Energy, then considering the bang-time function as completely homogeneous $t_{BT}(r)=0$. This approach is commonly used to study the effect of a local inhomogeneity in a EdS background (see e.g. \cite{GarciaBellido08}). There are several mass profiles used in literature that describes a local inhomogeneity in which we can be located, considering parameters that converge to a FLRW universe very outside of the void. The most well-known mass profile in literature is the GBH mass profile used to model a local-void scenario (\citealp{GarciaBellido08}). This profile is defined as:

    \begin{eqnarray}
    \Omega_m(r)=(\Omega_{in} - \Omega_{out})\frac{1-\tanh(\frac{r-r_0}{\Delta r})}{1+\tanh(\frac{r_0}{2\Delta r})}+\Omega_{out}
    \end{eqnarray}

    Where the parameters involved are $\Omega_{in}=\Omega_m(0)$ as the fraction of mass in the center, $\Omega_{out}=\Omega_m(\infty)$ as the fraction of mass very far away from the void, $r_0$ as the size of the void and $\Delta r$ the scale of the transition to homogeneity. When $\Omega_{out}$ is fixed to 1 and $t_{BT}(r)=0$, the local hubble rate can be writed as:

    \begin{eqnarray}
    H_0(r)=H_0 \left (\frac{1}{\Omega_k(r)}-\frac{\Omega_m(r)}{\sqrt{\Omega_k(r)^3}} 
    \sinh^{-1}{\sqrt{\frac{\Omega_m(r)}{\Omega_k(r)}}} \right )
    \end{eqnarray}

    \item Set a Mass Profile in a flat universe with or without Dark Energy, considering the bang-time as completely homogeneous $t_{BT}(r)=0$. A similar approach was used in \citealp{Kenworthy19} to study the effect of a local inhomogeneity in the $H_0$ tension. 
    \item Modify the bang-time function as needed and study a flat non-dark energy universe. As inhomogeneous bang-time function is something undesirable in cosmology, those approaches are to the date used as \textit{toy models} (see \citealp{Isidro16})
\end{enumerate}

Another way to define a LTB model is thinking in a density contrast profile $\delta(r,t)$ defined as:

\begin{equation}\label{eq15}
\delta(r,t)=\frac{\rho(r,t)-\rho_0^\infty}{\rho_0^\infty},
\end{equation}
where $\rho_0^\infty=\rho(\infty,t_0)$ is the mean density being constant very far outside the local inhomogeneity at the present time. This function is useful because this allows us to describe the mass profile using the data directly from the observational local surveys. We have for $t=t_0$:
\begin{equation}\label{eq16}
\rho_0(r)=\rho_0^\infty(\delta_0(r)+1)
\end{equation}
Although the functions $\Omega_m(r)$, $M(r)$ and $E(r)$ have a clear meaning, it is difficult to relate them directly to the observed quantity $\delta_0(r)$. So, instead of using the complexities of (\ref{eq13}), we can approximate a solution by assuming the background universe being described by a FLRW metric and define an integrated density function by:
\begin{equation}\label{eq17}
\rho_0^H(r) \propto \frac{M(r)}{r^3} \propto (\delta^H(r)+1),
\end{equation}
which is exactly true for an homogeneous universe. Using the gauge $R_0(r)=r$ we can write
\begin{equation}\label{eq18}
\delta^H(r)+1 \propto \frac{M(r)}{r^3}  \propto \Omega_m(r)H_0(r)^2,
\end{equation}
where $\delta^H(r)$ is the integrated density contrast. Then, by constraining $t_{BT}(r)$ and $\delta^H(r)$ we can define completely $R(r,t)$. It is important to note that this is not exactly the integrated density contrast as it uses a background FLRW universe, but it is a useful approximation that allows us to define the universe in terms of density parameters. 

\subsubsection{Local void Cosmology}

There are two main reasons why a local void scenario has been subject of great interest in inhomogeneous cosmology over the last years. First, it has been claimed that a local void can explain the Hubble Tension, as the Hubble parameter locally could lead to a greater values compared to the background universe due to the presence of the underdensity. However this possibility has been ruled out in multiple researches, as the modification of the Hubble constant are not large enough (\citealp{Kenworthy19}, \citealp{Camarena22}, \citealp{huterer2023enoughlocalvoidsolve}). On the other hand, according to some authors a local void could potentially explain the negative deceleration parameter observed in cosmological data (see e.g \citealp{Celerier99}). However the void to explain the observations should be greater than those presumably detected (\citealp{GarciaBellido08}) and even the explanation could suffer different theoretical problems (\citealp{Vanderveld06}). Moreover, recent studies suggest that the local peculiar velocity field is contracting rather than expanding, not being compatible with a void scenario. It could be possible to reconcile both, a void and a contracting bulk flow if we consider a void of greater scales in which there is a over-density locating near the center. We discuss this possibility in the chapter \ref{conc}.

\subsection{Bianchi Universes}

Bianchi models (\citealp{Bianchi89}) are the opposite of LTB models as they describe fully homogeneous but anisotropic universes. There are nine different Bianchi universes with different properties, which are classified depending on the isometries of the space (see \citealp{Wainwright_Ellis_1997} for an extensive review of several Bianchi models). For example, the minimal symmetry breaking of the isotropy could be written as:

\begin{eqnarray}
    ds^2 = -dt^2+a(t)^2(dx^2+dy^2)+b(t)^2dz^2\\
\end{eqnarray}

Which corresponds to a flat universe with different scale factor along one of the Cartesian axis. To quantify the possible isometries in a space, is useful to describe the \textit{Lie algebras} of that space. Bianchi discovered 7 different algebras ($\RNum{1}, \RNum{2},\RNum{3},\RNum{4},\RNum{5},\RNum{8},\RNum{9}$) and also two other families of infinities Lie algebras described by a continuous parameter $h$. The algebra $\RNum{6}_h$ with $h$ being a real number except $0$ or $1$ and $\RNum{7}_h$ algebra with $h$ positive or $0$ (\citealp{valent2023axialbianchiixlemaitrehubble}).

The simplest case of a Bianchi universe is a flat anisotropic metric a la FLRW but with different scales factor for each Cartesian coordinate:

\begin{eqnarray}
    ds^2 = -dt^2+a^2dx^2+b^2dy^2+c^2dz^2\\
\end{eqnarray}

with $a,b$ and $c$ being time functions. This is called a \textbf{Bianchi universe type I} and has found different applications in cosmology (\citealp{Fleury_2015},\citealp{schücker2024axialbianchiimeets}). Any of the Bianchi universe can be represented in a Cartesian coordinate form as is described in table \ref{Table:Bianchis}. 

\begin{table}[!ht]
\centering
\begin{tabular}{cc}
 Type &Spatial Metric \\
\hline
$\RNum{1}$ & $a^2dx^2+b^2dy^2+c^2dz^2$  \\
$\RNum{2}$ & $a^2(dx+ydz)^2+b^2dy^2+c^2dz^2$  \\
$\RNum{3}$ &$a^2e^{-2z}dx^2+b^2dy^2+c^2dz^2$  \\
$\RNum{4}$ &$e^{-2z}[a^2(dx-zdy)^2+b^2dy^2]+c^2dz^2$  \\
$\RNum{5}$ & $e^{-2z}[a^2dx^2+b^2dy^2]+c^2dz^2$  \\
$\RNum{6}_h$ &$a^2e^{-2z}dx^2+b^2e^{-2hz}dy^2+c^2dz^2$ \\
$\RNum{7}_h$ & $e^{-2hz}[a^2(\cos{z}dx-\sin{z}dy)^2+b^2(\sin{z}dx+\cos{z}dy)^2]+c^2dz^2$  \\
$\RNum{8}$ & $a^2(\cos{z} dx - \cosh{x} \sin{z} dy)^2+b^2(\sin z dx + \cosh x \cos z dy)^2+c^2( dz- \sinh x dy )^2$  \\
$\RNum{9}$ & $a^2(\cos{z} dx - \cos{x} \sin{z} dy)^2+b^2(\sin z dx + \cos x \cos z dy)^2+c^2( dz+ \sin x dy )^2$  \\
\end{tabular}
\caption{\label{Table:Bianchis}
The spatial metrics in Cartesian coordinates of the 9 Bianchi universe. Note that $\RNum{6}$ and $\RNum{6}_1$ are identical to and $\RNum{3}$ and $\RNum{5}$ respectively. }
\end{table}

It can be shown that universe $ \RNum{5}, \RNum{7}_h$ and $ \RNum{9}$ can be reduced to FLRW, having potential applications in astrophysics (see e.g \citealp{valent2023axialbianchiixlemaitrehubble} for a review of the $\RNum{9}$ universe).

\subsection{Kantowski-Sachs universe}

Another case of an anisotropic but homogeneous universe is the Kantowski-Sachs metric (\citealp{KantowskiSachs}) which could be viewed as a flat universe which different scale factors for the angular and the radial part:

\begin{eqnarray}
    ds^2&=&-dt^2 + a dr^2 + b(d\theta^2 + sin^2\theta d\phi^2)
\end{eqnarray}

This universe is not included in Bianchi classifications because Bianchi isometries include 3-dimensional simply transitive subgroups, which Kantowski-Sachs lack off (\citealp{valent2023axialbianchiixlemaitrehubble}). This universe to the date has no astrophysical applications (\citealp{Bolejko_2011}) as doesn't contain a FLRW universe in any case.

\subsection{Szekeres universe}

A fully generalization of LTB models is the Szekeres universe (\citealp{Szekeres:1974ct},\citealp{Bolejko_2011}), which is an inhomogeneous and anisotropic model for a dust universe with no symmetries. The metric is:

\begin{eqnarray}
    ds^2&=&-dt^2 + e^{2\alpha} dr^2 + e^{2\beta}(dx^2+dy^2)
\end{eqnarray}

Where $\alpha$ and $\beta$ are functions of $r,x,y$ and $t$. The metric is useful for describe late epochs of the universe when pressure-less matter is dominated. There are two families of Szkeres Metric, which are generalizations of other metrics previously seen.  The first family is a generalization of Friedmann and Kantowsky-Sach universe, which is not astrophysical applicable. However, the second familly is astrophysical useful and the functions of the metric can be written as:

\begin{eqnarray*}
    e^\beta&=&\Phi(t,r)e^{\nu(r,x,y)}\\
    e^{\alpha}&=&h(r)\Phi(t,r)\partial_r\beta = h(r)(\partial_r \Phi+\Phi\partial_r\nu)\\
    \nu&=&-\ln{[A(r)(x^2+y^2)+2B_1(r)x+2B_2(r)y+C(r)]}
\end{eqnarray*}

Here, $\Phi(r,t)$ is solution of the generalized Friedmann equation:

\begin{eqnarray}
    \left ( \frac{\partial \Phi}{\partial t} \right )^2=-k(r)+\frac{M(r)}{\Phi}+\frac{1}{3}\Lambda\Phi^2
\end{eqnarray}

Where $\Lambda$ is the cosmological constant. All the other $r$-dependent functions are arbitrary, satisfying the relation:

\begin{eqnarray*}
    g=4(AC-B_1^2-B_2^2)=\frac{1}{h^2}+k^2
\end{eqnarray*}

Some properties of this model are:

\begin{enumerate}
    \item The function $g$ determines the geometry of the 2-surfaces with constant $r$ and $t$. If $g>0$, $g=0$ or $g<0$, we have a spherical, planar or hyperbolic geometry respectively.  
    \item $k$ determines the evolution of the universe when $\Lambda=0$. $k>0$ implies that the universe is expanding and recollapsing, while $k<0$ describes an ever-expanding or ever-collapsing universe. $k=0$ is the intermediate case corresponding to Friedmann flat universe.
    \item As $\frac{1}{h^2}$ have to be positive, for a spherical geometry the three evolutions are allowed. However, For a flat geometry $k$ cannot be positive and for a hyperbolic geometry $k$ should be strictly negative. The first case is the only one that has found cosmological applications (\citealp{Krasinski12},\citealp{Bolejko_2011},\citealp{bolejko2006cosmologicalapplicationsszekeresmodel}) and is called \textit{quasi-spherical Szkeres universe}. The quasi-spherical model could be understood as a generalization of $LTB$ model, where concentric spheres are no longer concentric. In this way, the mass distribution of the Szkeres metric could be understood as composition of a monopole and a dipole mass distribution . 
\end{enumerate}

\subsection{G\"odel universe}

The concept of absolute time in cosmology relies in the cosmological principle. It is assumed that, orthogonal to the world lines of matter there exist a one-parametric system everywhere (the so called cosmic time $t$). The possibility to derive a metric where the concept of absolute time doesn't exist in a dust universe was addressed by Kurt G\"odel in the eighties (\citealp{Godel}). It is shown that the non-existence of this parametric system is equivalent to a rotating universe. If $y$ is the axis of rotation, the G\"odel metric is:

\begin{eqnarray*}
    a^2[-(dt^2+e^{x}dy)^2+dx^2+dz^2+\frac{e^{2x}}{2}dy^2]
\end{eqnarray*}

Where $a$ is a real number. This parameter is related with the angular velocity of the surrounding dust grain around the $y$ axis. The G\"odel metric shows different weird features (such as time-travel and violation of causality) that reduces this universe to an anecdote rather than a potential cosmological model of the universe. Some interesting visualizations of the G\"odel universe have been developed in \citealp{Buser_2013}. Vorticity seems to be the key factor behind the peculiarities of the Godel solution . In fact, the faster the model rotates the larger the acausal domain, while the slower the rotation the smaller the region where causality is violated (e.g.~see \citealp{Barrow_2004} for the covariant description and the stability analysis of the Godel universe).

\section{The Back-reaction problem}

The relation between dark energy and the non-linear dynamics of superstructures is particularly interesting as it has been shown that dark energy impacts strongly the evolution of the large scale-structure. Notably, dark energy seems to dominate the composition of the universe at the same time that non-linear structures form, which is commonly referred as the \textit{coincidence problem}. This could implies that LSS has important effects on the dynamics of the universe at large scales. The effect of the large-scale structure in the dynamics of the expansion of the universe is called the \textit{back-reaction}. A simple example of a back-reaction is the effect of a local void/over-density in the measurement of cosmological parameters. As objects inside this local inhomogeneity moves with a positive/negative radial peculiar velocity respectively, the true kinematical parameters of the expansion of the universe could be over/underestimated when using local data sources. The goal is then to understand which effects produces much more complex large-scale structures with non-linear dynamics in the measurement of cosmological parameters. Some studies claim that the back-reaction effect is completely negligible, while others consider back-reaction as a possible explanation for the negative deceleration parameter. As we will see, the difference of those conclusions comes from different methods used, as there is no scientific consensus to the \textit{averaging problem} in cosmology.

\subsection{Perturbative approach}

One way to study the back-reaction is using a perturbative approach over the metric tensor and the energy-momentum tensor:

\begin{eqnarray*}
    g_{\mu\nu}=\bar{g}_{\mu\nu}+\delta g_{\mu\nu} \\
    T_{\mu\nu}=\bar{T}_{\mu\nu}+\delta T_{\mu\nu}
\end{eqnarray*}

    Where $g_{\mu\nu}$ and $\bar{g}_{\mu\nu}$ are the perturbed and the background metric tensor respectively, while $\delta g_{\mu\nu}$ is the perturbation assumed to be small at linear regime. In the same sense,  $T_{\mu\nu}$, $\bar{T}_{\mu\nu}$ and $\delta T_{\mu\nu}$ are the perturbed, background and the perturbation of the energy-momentum tensor. $\delta g_{\mu\nu}$ and $\delta T_{\mu\nu}$ being symmetric 4x4 tensors have 10 independent degrees of freedom, which can be decomposed in a \textit{scalar}, \textit{vectorial} and \textit{tensorial} part (\citealp{lesgourgues2013tasilecturescosmologicalperturbations}). At the linear regime, the first has 4 degrees of freedom corresponding to the perturbations in the Newtonian potential produced by density perturbations denoted by $\psi$, distortions of the scale factor produced by variations in the isotropic pressure denoted by $\phi$, effects of irrotational fluxes of energy described by a potential and an irrotational part of the shear degree. The vectorial or \textit{gravitomagnetic} part has also 4 degrees of freedom wich are related with vorticity, but in cosmology, vorticity decays rapidly with time and the vectorial part is commonly neglected. The last 2 degrees of freedom are related with the 2 modes of  gravitational waves and they are the only ones that propagate in space.
    
    In principle, one can assume a background metric and with a proper gauge it is possible to construct any perturbed universe. However, it is not possible to recover a unique background space-time from the perturbed one, which has referred as the \textit{fitting problem}. This problem comes from the existence of a \textit{Gauge freedom} in the theory, since in general relativity one has to deal with two spacetime manifolds (a background and a perturbed one) and the map between both geometries is not unique (\citealp{lesgourgues2013tasilecturescosmologicalperturbations}). This introduces a freedom related with the averaging of quantities along different time-slicing. One approach to deal with this freedom is to work with quantities that are gauge invariants, while the other is to fix a gauge and obtain gauge-independent quantities from there.

The simplest gauge fix comes from the requirement of a correspondence with Newtonian dynamics at low scales. The condition for the Newtonian correspondence is traduced as the \textit{Newtonian Gauge} or  \textit{Longitudinal Gauge} which erases 2 scalar degrees of freedom. Ignoring vectorial and tensorial perturbations, the metric of a perturbed flat FLRW universe can be wrote in conformal time $\eta$ as:

\begin{eqnarray*}
    ds^2=-a(\eta)^2 \left [(1+2\Psi)d\eta^2+(1-2\Phi)d\vec{x}^2\right ]
\end{eqnarray*}

It can be shown that $\Psi$ and $\Phi$ are gauge-invariant and are called \textit{Bardeen potentials} \citealp{Bardeen_1980}. More over, it is possible to use this perturbative approach to compute the back-reaction term which according to \citealp{greenwald2011} is negligible. However, there are some possible problems that could makes perturbative approaches in the back-reaction physics misleading (\citealp{verweg2024averagingcosmicstructurecosmological}):

\begin{enumerate}
    \item \textbf{The background cosmology election is unavoidable: } To compute averaged parameters of the perturbed metric $g_{\mu\nu}$ based solely on the perturbation, it is necessary to express the background metric as a function of the perturbation:

    \begin{eqnarray*}
        \bar{g}_{\mu\nu}=\bar{g}_{\mu\nu}(\delta g_{\mu\nu})
    \end{eqnarray*}

    This impose a choice of a background cosmology $\bar{g}_{\mu\nu}$ which affects directly the values of physical quantities and in particular the back-reaction. Then, the value of the back-reaction is different for different background cosmologies.
    \item \textbf{The effect of the back-reaction cannot be obtained solely from the perturbation term: } As it is noticed in \cite{verweg2024averagingcosmicstructurecosmological}, the LSS effects in the average dynamics of the universe has to be included not just in the perturbative terms, but also in the background terms. If not, the averaging of physical quantities will be just the background and any contribution of the back-reaction will be completely ignored.
    \item \textbf{The effect of the back-reaction is multi-scaled: } Usually, perturbations in the metric are parameterized by a scale parameter $\lambda$ in which \citealp{greenwald2011} assumes the limit $\lambda \rightarrow0$. At this limit, perturbations only affect the background via gravitational waves. However, the effects of the back-reaction is multiscaled along a range of $\lambda$ parameters and the formalism developed in \citealp{greenwald2011} doesn't take into acount this.
\end{enumerate}

\subsection{Averaging approach}

To avoid perturbative theory problems, some researchers take a different approach in which they study how to average observable quantities directly from general relativity equations (\citealp{Noonan1984TheGC}, \citealp{Zotov_1992},\citealp{Futamase}). The idea is to divide the universe in compact regions and get the back-reaction as the average effect of local inhomogeneous and anisotropic distribution of matter. However, this topic has been produced a heated debate over the last years.
The core of the back-reaction debate is the \textit{averaging problem} in general relativity (\citealp{Zalaletdinov:2008ts}), in which the averaging over physical quantities depend on the space-time slicing (similar to the Gauge problem in perturbative theory) and its election impacts the cosmological model used. It can be shown that the foliation selected in the averaging could produce artificial back-reaction terms that could be used to explain the cosmic acceleration (\citealp{verweg2024averagingcosmicstructurecosmological}). Using the averaging approach, some researches claim that back-reaction could be indeed the missing dark energy (see e.g \citealp{Buchert_2000}, \citealp{Adamek_2019}, \citealp{Rasanen_2004}). However, those results has been criticized via perturbative arguments and also by the absence of the Newtonian correspondence at low scales in the method. It has been shown that at low scales, Newtonian dynamics cannot fully described the physics of the large-scale structure due to non-linear emergent effects (\citealp{Korzyński_2015}). Moreover, the evolution of the peculiar velocity field is stronger when relativistic effects are included (\citealp{Filippou_2021},\citealp{Tzaprasi2020}) challenging the needing of Newtonian correspondence at low scales. We study this deeply in chapter \ref{pec}.

The back-reaction is one of the most important problems in the connection between the general theory of relativity and averaged observables via different sources of data and , possibly, will be fully solvable when the numerical treatment of general relativity at large scales became completely possible.

\section{Perturbed FLRW luminosity distance}

It is possible to develop a perturbed luminosity distance from a perturbed flat FLRW universe to study anisotropies in the cosmological data, which has been shown to be a gauge-invariant (\citealp{Biern_2017}).  Using the Newtonian Gauge and being $\Phi$ and $\Psi$ the Bardeen potentials, the direction dependent luminosity distance can be decomposed in a velocity part, a lensing part an a gravitational potential part. Ignoring the \textit{observer} terms, the luminosity distance can be written as (\citealp{Sasaki:1987ad},\citealp{Durrer:2008eom}, \citealp{sorrenti2023dipole}):

\begin{equation}
\begin{split}
    d_L(z,\mathbf{n})=\bar{d}_L(z) \left [1+\frac{\mathbf{v}_0\cdot \mathbf{n}}{\mathcal H(z)r(z)}-\left(\frac{1}{\mathcal H(z)r(z)}-1\right ) \left ( \mathbf{v}\cdot \mathbf{n}-\Psi -\int_0^{r(z)}({\Psi'}+\Phi')dr\right ) \right . \\ 
    -\left .\Phi + \int_0^{r(z)}\frac{dr}{r}\left(1-\frac{r(z)-r}{2r(z)}\Delta_\mathbf{n}\right )(\Phi+\Psi)  \right ]
\end{split}
\end{equation}

Where $\mathbf{v}_0$ and $\mathbf{v}$ are respectively the velocity of the observer and the peculiar velocity field in the Newtonian gauge, which contribute to the dipole. $\mathcal{H}(z)$ is the co-moving Hubble parameter, $r(z)$ is the co-moving distance and $\bar{d}_L$ is the luminosity distance of the background. $\Delta_\mathbf{n}$ denotes the Laplacian in radial direction coming from the lensing effect dominant at high redshifts and the primes $'$ are derivatives w.r.t conformal time. One approach commonly used is to fit a dipole using the luminosity distance, considering that a low redshift $z\ll 1$ limit the velocity term highly dominates:

\begin{equation}
    d_L(z,\mathbf{n})=\bar{d}_L(z) \left [1+\frac{\mathbf{u}_0\cdot \mathbf{n}}{\mathcal H(z)r(z)} \right ]
\end{equation}

Note that the term $\mathbf{u}_0=\mathbf{v}_0-\mathbf{v}$ is the difference between the peculiar velocity of the observer and the peculiar velocity of the source. As the source peculiar velocity is impossible to obtain independently of a cosmology used, this has been modeled using Newtonian linear gravitational physics and therefore irrotational fields. However, recent studies suggest that non-Newtonian effects become important in the dynamics of LSS at late times. In (\citealp{sorrenti2023dipole}), the authors take a step further fitting the dipole not assuming a peculiar velocity source, interpreting the best fit for the term $\mathbf{u}_0=\mathbf{v}_0-\mathbf{v}_{bulk}$ as the difference between the velocity of the observer and a common \textit{bulk flow} velocity. They found a disagreement with CMB at low redshift, presumably due to the high correlation between velocities of low-redshift supernovae and the velocity of the observer. Also, the bulk estimated using the CMB value $\mathbf{v}_0$ for the observer velocity gives a value of $\mathbf{v}_{bulk}=\mathbf{v}_0-\mathbf{u}_0$ was way higher than the used in the Pantheon + analysis.

The perturbed luminosity distance also works to fit high order multi-poles when allowing for a quadrupole dependence on the directional luminosity distance (see e.g. \citealp{Sorrenti:2024ztg}), which could be related with the shear of the peculiar velocity field. While the full theoretical expression for the luminosity distance is a gauge invariant, ignoring terms such as the observer contributions in the luminosity distance analysis could break this invariance and induce small errors (\citealp{Biern_2017}). Also, it is worth to mention that this particular perturbed luminosity distance assumes the FLRW metric as the background geometry of the universe and therefore is not enough agnostic to analyse cosmological data.

\section{Cosmographic analysis}

It is desirable in cosmology to remain the most agnostic possible when analyzing the data. The problem here is that researchers have to assume a cosmological model (based on general relativity, the cosmological principle and so on), ending with the cosmological parameters extracted completely related to the model used. To avoid this, we can work with a \textbf{cosmographic approach} which uses very few assumptions (or none) of the universe to extract cosmological parameters, fitting directly kinematical parameters such as $H_0$ and $q_0$ which describe how the universe evolves rather than the dynamics that produces this evolution. The basic cosmographic approach is to assume and isotropic and homogeneous model and compare with the data using a Taylor expansion of the luminosity distance (\citealp{Visser_2004},\citealp{Visser_2005}):

\begin{equation}
    D_L(z)=\frac{z}{H_0}+\frac{(1-q_0)}{2H_0}z^2-\frac{(1-q_0-3q_0^2+j
    _0+\frac{kc^2}{H_0^2})}{6H_0}z^3 + \mathcal{O}(z^4),
\end{equation}

Just like $H_0$ is related with the velocity of the expansion and $q_0$ is related with the acceleration of the expansion, $j_0$ is called the \textit{jerk parameter} and it is related with the third temporal derivative of the expansion. If we expand high order terms, we end having more free parameters such as the \textbf{snap, crackle, pop} and so on, which correspond to higher order temporal derivatives. The main disadvantage of this approach over a full cosmological model is that this is not valid for high redshift. Also, note that if we assume a flat universe ($k=0$) or if $\frac{H_0}{c} \gg 1$, the number of free parameters up to a third order agrees with the number of coefficients and it would be possible to constraint the jerk alone, as in principle it is degenerated with the curvature term $\frac{kc^2}{H_0^2}$.

In the context of this thesis, it is interesting to ask how we can use the agnostic benefits of the cosmographic approach to study the cosmological principle. To do this, extensions of this Taylor expansion has been derived to allow for anisotropies in the cosmological parameters. One simple way is to fit cosmological parameters splitting the data in different directions, areas or portions of the space. This method has been used to extract a possible dipole in SNIA and has been called \textit{hemisphere comparison method} (\citealp{Deng18b},\citealp{Deng18a}). However, another method is to allow for a directional dependence cosmological parameters which has been called \textit{dipole fitting} (see e.g \citealp{Mohayaee20}). The deceleration parameter is is assumed composed by a monopole term and a dipole as:

\begin{eqnarray}
    q_0=q_m+\mathbf{q}_d\cdot \mathbf{\hat{n}} \mathcal{F}(z,s)
\end{eqnarray}

Where $q_m$ es the monopole, $\mathbf{q}_d$ is the dipole and $\mathbf{\hat{n}}$. $ \mathcal{F}(z,S)$ is a function that describes the scale dependence of the dipole based on the $s$ parameter, usually set to an exponential form $\mathcal{F}(z,s)\sim e^{-\frac{z}{S}}$. Due to the large anisotropy of the data and the low quantity when used just low redshifts, it is difficult to fit for the $\mathbf{\hat{n}}$ direction so it is assumed tipically in the direction of the CMB.

The generalization of the Taylor expansion of the luminosity distance up to a third order, allowing for anisotropies of the kinematical parameters has been recently formalized by \citealp{Heinesen:2020bej} in the so-called \textit{Multipole decomposition of the luminosity distance}. Is easy to note that this method could be compared with the multi-polar expansion of the CMB anisotropies. In the future, when more cosmological data will be available, this multipole expansion could be a promising tool to study the cosmological principle. 

Also, it is possible to use the cosmographic approach not only to study anisotropies, but also radial inhomogeneities using a \textit{binned universe}. This method consist in to fit cosmological parameters in different portions of the redshift space (see chapter \ref{chap:reds}). Is interesting to note that mutiple analysis using a binned universe has concluded that cosmological parameters such as $\Omega_m$ and $q_0$ evolve with redshift (\citealp{Perivolaropoulos:2023iqj},\citealp{PASTEN2023101224},\citealp{Colgain2022}).

\section{Tilted cosmologies}\label{sTCM}

\begin{figure}[ht]
    \centering
    \includegraphics[width=13cm]{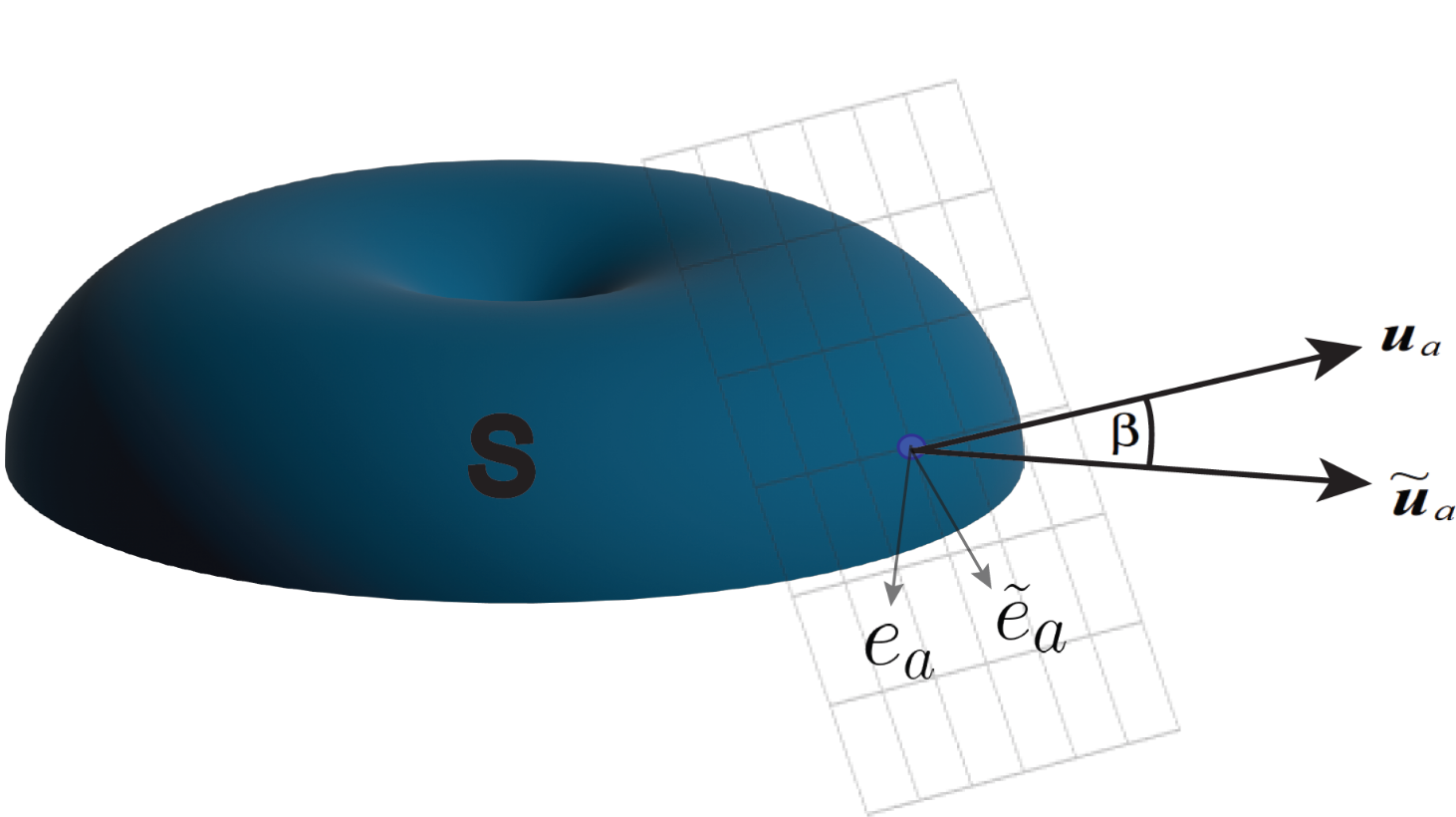}
    \caption{The 4-velocities $u_a$ and $\tilde{u}_a$ define, respectively, the CMB frame of the background expansion and the frame of the peculiar motion, having an hiperbolic tilt angle $\beta$ (with $cosh\beta=-u_a\tilde{u}^a$) between them. The surfaces $S$ is orthogonal to the 4-velocity $u_a$ and define the rest-space of the CMB family of observers.}
    \label{PVSNIA}
\end{figure}

Recently, some studies (\citealp{Tsagas2011}, \citealp{Tsagas2015}) has developed a novel way to understand cosmology from the point of view of an observer moving with a peculiar velocity with respect to the background cosmology. In this section we will provide the reader a theoretical background to understand the cosmological implications of this approach.
Consider a perturbed Friedmann-Robertson-Walker (FRW) universe with two groups of relatively moving observers. Assuming that $u_a$ and $\tilde{u}_a$ are the 4-velocities of these observers and $v_a$ is the (non-relativistic) peculiar velocity of the latter group with respect to the former, we have
\begin{equation}
\tilde{u}_a= u_a+ v_a\,,  \label{4vels}
\end{equation}
to first approximation (with $u_av^a=0$ always). Introducing two sets of observers means that (strictly speaking) there are two temporal directions (along $u_a$ and $\tilde{u}_a$) and two associated 3-spaces (orthogonal $u_a$ to and $\tilde{u}_a$, see e.g figure \ref{PVSNIA}). Then, the corresponding (covariant) differential operators are $\dot{}= u^a\nabla_a$ and ${}^{\prime}=\tilde{u}^a\nabla_a$ for the time derivatives, with ${\rm D}_a=h_a{}^b\nabla_b$ and $\tilde{\rm D}_a=\tilde{h}_a{}^b\nabla_b$ for the spatial gradients. Also, the tensors $h_{ab}=g_{ab}+u_au_b$ and $\tilde{h}_{ab}=g_{ab}+ \tilde{u}_a\tilde{u}_b$ project orthogonal to $u_a$ and $\tilde{u}_a$ respectively. 

The kinematic information of the observers' motion is decoded by decomposing the gradient of their 4-velocity field as follows (see \ref{PVSNIA}):
\begin{equation}
\nabla_bu_a= \frac{1}{3}\,\Theta h_{ab}+ \sigma_{ab}+ \omega_{ab}- A_au_b\,.  \label{Nbua}
\end{equation}
In the above, $\Theta$ is the volume expansion/contraction scalar (when positive/negative respectively), $\sigma_{ab}$ is the shear, $\omega_{ab}$ is the vorticity and $A_a$ is the 4-acceleration (e.g.~see~\citealp{Tsagas2008,Ellis2012}). In an exactly analogous way, the $\tilde{u}_a$-field splits as $\nabla_b\tilde{u}_a=(\tilde{\Theta}/3)\tilde{h}_{ab}+ \tilde{\sigma}_{ab}+\tilde{\omega}_{ab}- \tilde{A}_a\tilde{u}_b$, with the tildas denoting variables evaluated in the tilted frame of the bulk flow. Relative to the same coordinate system, the peculiar-velocity field splits as
\begin{equation}
\tilde{\rm D}_b\tilde{v}_a=  \frac{1}{3}\,\tilde{\theta}\tilde{h}_{ab}+ \tilde{\varsigma}_{ab}+ \tilde{\varpi}_{ab}\,,  \label{tDntva}
\end{equation}
where $\tilde{\theta}$, $\tilde{\varsigma}_{ab}$ and $\tilde{\varpi}_{ab}$ are the volume scalar, the shear and the vorticity of the bulk peculiar motion~\citealp{Tsagas2015}. Of the last three variables, the most important for our purposes is the peculiar volume scalar ($\tilde{\theta}$), which takes positive/negative values in locally expanding/contracting bulk flows respectively.

\begin{figure}[ht]
    \centering
    \includegraphics[width=14cm]{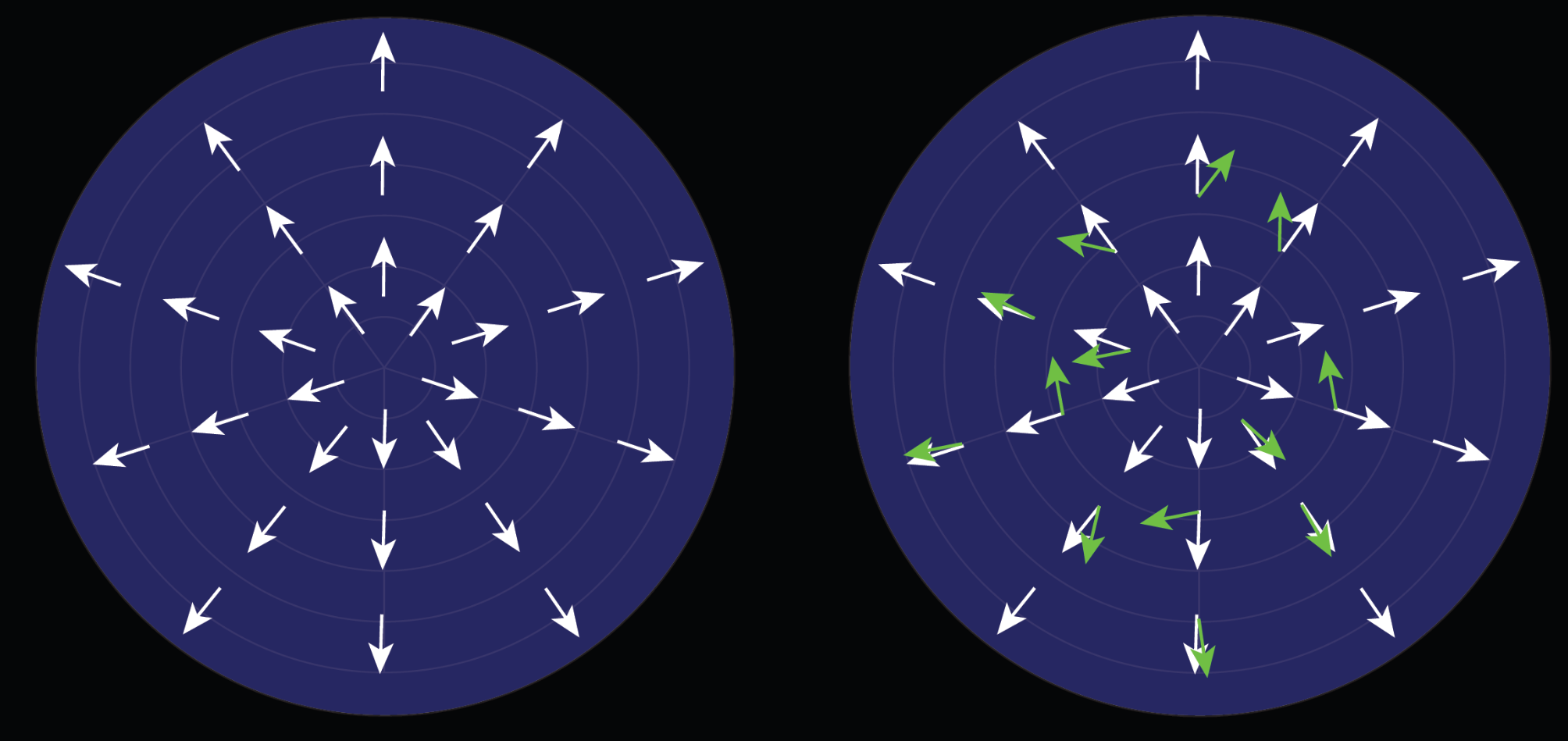}
    \caption{White arrows represent the family of idealized observers that moves with the expansion of the Hubble Flow. Green arrows are the field describing the real observers that doesn't follow the Hubble expansion due to the peculiar movement. Note that, as those families of observers are related to a fluid-like behavior, fluid parameters such as the expansion, vorticity, shear and acceleration doesn't have to me the same for both fields. }
    \label{PVSNIAsq}
\end{figure}

The three kinematic sets defined above are related by lengthy nonlinear expressions (e.g.~see~\citealp{Maartens1998} for the full list). Assuming non-relativistic peculiar motions on an FRW background, we obtain the linear relations
\begin{equation}
\Tilde{\Theta}= \Theta+ \tilde{\theta} \hspace{10mm} {\rm and} \hspace{10mm} \Tilde{\Theta}^{\prime}= \dot{\Theta}+ \tilde{\theta}^{\prime}\,,  \label{Thetas}
\end{equation}
between the volume scalars and between their time derivatives evaluated in the two frames (see figure \ref{PVSNIAsq} for a schematic representation). At this point, we note that $\Theta$ and $\tilde{\Theta}$ monitor the expansion rate of the universe, namely the Hubble parameters, as measured in their corresponding frames (that is $\Theta=3H$ and $\tilde{\Theta}=3\tilde{H}$). Then, equations (\ref{Thetas}) imply that the expansion and the acceleration/deceleration rates measured in the tilted coordinate system differ from those measured in its CMB counterpart solely due to relative-motion effects. In particular, recalling that
\begin{equation}
q= -1- {3\dot{\Theta}\over\Theta^2} \hspace{10mm} {\rm and} \hspace{10mm} \tilde{q}= -1-{3\tilde{\Theta}^{\prime}\over\tilde{\Theta}^2}\,,  \label{qs} 
\end{equation}
define the deceleration parameters in the CMB and the bulk-flow frames respectively, the following useful relation between $\tilde{q}$ and $q$ can be obtained \citealp{Tsagas2015,Tsagas2021}:
\begin{equation}
\tilde{q}= q+ {\tilde{\theta}^{\prime}\over2\dot{H}}\,,  \label{tq1}
\end{equation}
to first approximation. Recall that $\tilde{\Theta}=\Theta=3H$ in the Friedmann background. Also note that, whereas $\tilde{\theta}/H\ll1$ at the linear level, the ratio $\tilde{\theta}^{\prime}/\dot{H}$ of their time derivatives is not always small. Finally, using relativistic linear cosmological perturbation theory, we arrive at:
\begin{equation}
\tilde{q}= q+ {1\over9}\left({\lambda_H\over\lambda}\right)^2 {\tilde{\theta}\over H}\,.  \label{tq2}
\end{equation}
with $\lambda_H=1/H$ and $\lambda$ representing the Hubble horizon and the scale of the bulk flow in question. Note that we have focused on bulk peculiar flows with sizes considerably smaller than the Hubble length (i.e.~$\lambda\ll\lambda_H$ -- see~\citealp{Tsagas2015,Tsagas2021} for the full details of the derivation).\footnote{Expression (\ref{tq2}) has been obtained on an Einstein-de Sitter background, primarily for reasons of mathematical simplicity. It is fairly straightforward to show that the linear result (\ref{tq2}) holds on essentially all FRW and Bianchi backgrounds with conventional equation of state satisfying the strong energy condition (it doen't work, for example, in cases with $\omega=-1$ state equation). ~\citealp{Tsagas2022,tzartinoglou_2024}.}

Following (\ref{tq2}), the deceleration parameter measured locally by the bulk flow observers ($\tilde{q}$) differs from that of the global universe, which by definition coincides with the deceleration parameter measured in the idealised CMB frame ($q$). The difference is entirely due to the peculiar motion of the tilted observer, since $\tilde{q}=q$ when $\tilde{\theta}=0$. Also, the ``correction'' term in (\ref{tq2}) is scale-dependent and it gets stronger on progressively smaller scales (i.e.~for $\lambda\ll\lambda_H$), despite the fact that $\tilde{\theta}/H\ll1$ throughout the linear regime. Moreover, in accord with (\ref{tq2}), the overall impact of relative motion on $\tilde{q}$ is also determined by the sign of the peculiar volume scalar ($\tilde{\theta}$). The latter is positive in locally expanding bulk flows, which means that the deceleration parameter measured by observers residing in them will be larger than that of the actual universe (i.e.~$\tilde{q}>q$ when $\tilde{\theta}>0$). In the opposite case, that is inside locally contracting bulk flows, the local deceleration parameter becomes smaller (i.e.~$\tilde{q}<q$ for $\theta<0$). The latter case is clearly the most intriguing, since it allows for the sign of the deceleration parameter to change, from positive to negative, when measured by observers inside locally contracting bulk flows. Although the sign-change of $\tilde{q}$ is simply an illusion and a local artefact of the observer's relative motion, the affected scales can be large enough (between few to several hundred Mpc, see related Tables in \citealp{Tsagas2021,Tsagas2022,tzartinoglou_2024}) to make it look as a recent global event. If so, an unsuspecting observer may be misled to believe that their universe has recently entered a phase of accelerated expansion. According to (\ref{tq2}), the ``transition scale'', where the local deceleration parameter crosses the $\tilde{q}=0$ threshold is~\citealp{Tsagas2021} 
\begin{equation}
\lambda_T= {1\over3}\,\sqrt{\tilde{|\theta}|\over qH}\,\lambda_H\,,  \label{lambdaT}
\end{equation}
where $q>0$ always (with $q=1/2$ in the case of the Einstein-de Sitter background). In chapter \ref{chap:tilt} we will study how to relate tilted cosmology with observations and explore if the tilted scenario could be considered as a potential explanation for the recent acceleration of the universe.
\begin{figure}[ht!]
    \centering
    \includegraphics[width=14cm]{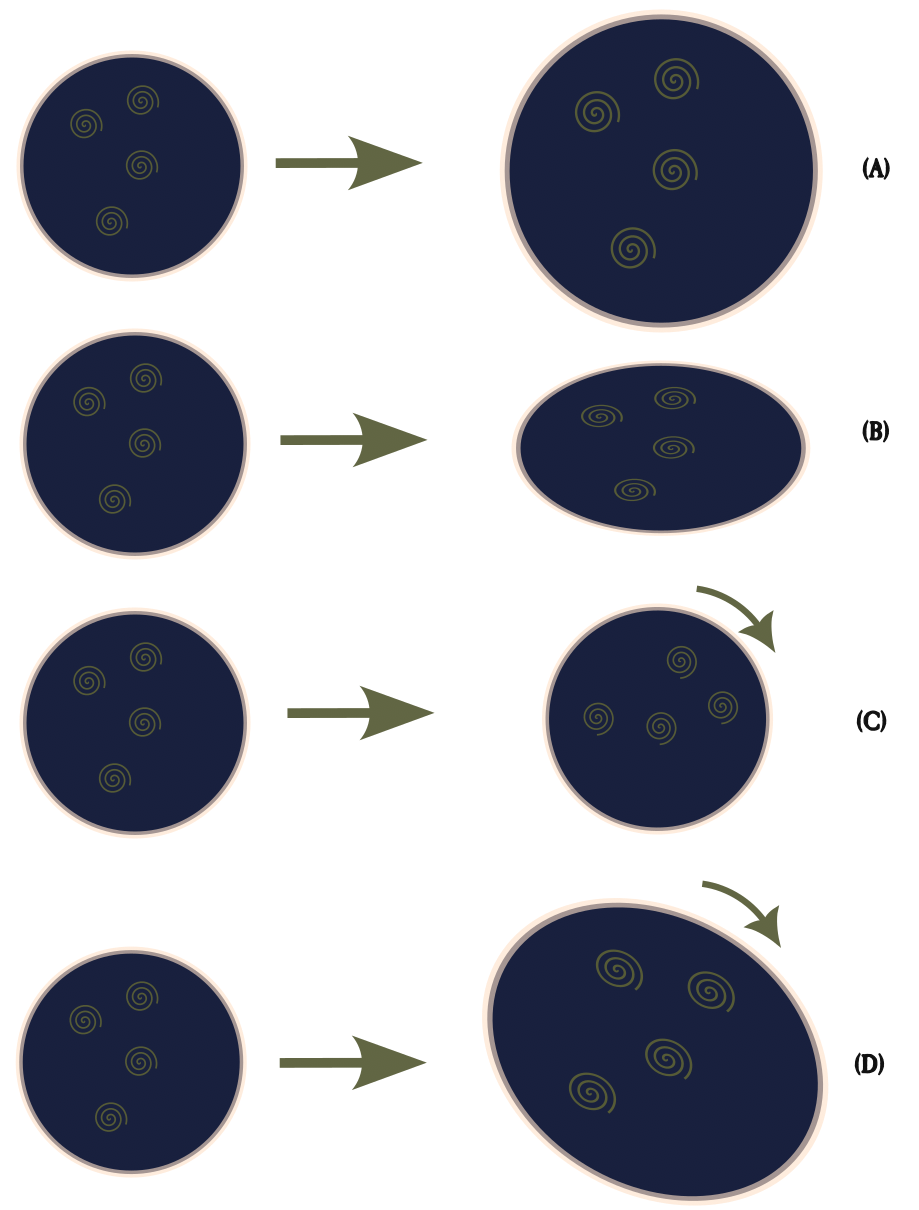}
    \caption{The gradient of the velocity field can be decomposed in (A) a expansion scalar (divergence) which quantifies the change in the volume, (B) a shear tensor that characterizes the change of shape without a change in the volume and (C) a vorticity tensor (curl) measuring the rotational part of the field. A mixture of these 3 features compose all the instantaneous variations that any field could experience (D).}
    \label{PVSNIAgrad}
\end{figure}

\clearchapterpage
\chapter{Modeling the fractal matter structure in cosmology}
\label{chap:fractal}

Although the universe appears to be highly homogeneous and isotropic on large scales, a more detailed examination reveals notable deviations from this idealized view at smaller scales. The existence of intricate structures such as underdense regions \citealp{Keenan13}, fractal-like patterns \citealp{Labini11, Labini98}, and bulk flows that diverge from the Hubble flow rest frame \citealp{Feindt13, Hudson99, Magoulas16} challenges the assumption of perfect global uniformity. These local structures, which occur on smaller cosmological scales, may have a significant impact on observed phenomena such as the apparent acceleration of the universe's expansion. In fact, several studies have proposed that the collective effects of these local inhomogeneities could mimic the influence typically attributed to dark energy \citealp{Celerier06, Enqvist07, Cosmai19, Tsagas2011, Asvesta22}. It is therefore plausible that the contribution of these local structures—whether individually or together—could be far more substantial than previously thought.

In this chapter, we aim to take a closer look at how local structures influence cosmological measurements by specifically investigating the fractal-like nature of the universe. We take a simplified approach, using luminosity distance relations derived from Type Ia Supernova (SNIA) data as a tool to explore this connection. By analyzing these relations, we seek to uncover potential links between the universe's fractal structure and the observed acceleration, offering new insights into the possible role of local inhomogeneities in the broader cosmological context. This chapter is based on the published work \citealp{Pasten_Fractal} made by the author.

\section{Fractal cosmology}

Fractal cosmology is defined as the branch of astrophysics that study the cosmos using fractals as tools, from minimum scales such as Planck to the enormous size of the universe (\citealp{Dickau2009}). Fractals appear in observational and theoretical cosmology, and its definition comes from Benoit Mandelbrot in the book \textit{The fractal geometry of nature}. In this work we adopt the definition based on the \textbf{fractal dimension} $D$ on a distribution of a physical property. Usually the integer dimension of a curve, surface or volume can be related with an homogeneous distribution of matter. For example, if a plane surface has a surface mass density of $\sigma$, then the mass $M$ enclosed in a circle of radius $r$ will be:

\begin{equation}
    M(r) \propto r^2
\end{equation}

In the same sense, a homogeneous density $\rho$ in the space will fix the mass enclosed in a sphere of radius $R$ as:

\begin{equation}
    M(r) \propto r^3
\end{equation}

Here 2 and 3 are the dimensions, respectively, of the surface and the space. We introduce then the fractal dimension $D$, as the exponent in the power function of the mass enclosed in a sphere of radius $r$ as:

\begin{equation}
    M(r) \propto r^D
\end{equation}

In observational astrophysics, a fractal distribution of galaxies was demonstrated to fit the astronomical data at the late 80's (\citealp{PIETRONERO1987257}). Since then, the scale in which the galaxy distribution smooths (in which the fractional dimension $D$ of the matter distribution goes to  an integer value $3$) has been deeply researched and has been interpreted as an homogeneity scale of the universe.

In the previous chapter, we provided the reader with an overview of the theoretical framework of the Lemaître-Tolman-Bondi (LTB) model and in the next section, we will explain its incorporation of fractal structures within cosmology, based in the previous definition of a fractal dimension. Next, we introduce our novel model, which aims to address specific problems and limitations associated with previous approaches. By incorporating both a fractal dimension and a length-scale parameter into the LTB metric, our model provides a more robust and comprehensive framework for understanding the interplay between cosmic acceleration and fractal structure formation. Finally, we present the results of our analysis, where we compare the luminosity distance relations derived from SNIA data with the predictions of our proposed cosmology using MCMC (\textit{Markov chain Monte Carlo}). 

\section{Fractals in LTB Model}

In a recent paper by \citealp{Cosmai19, Cosmai2023}, an intriguing proposal was put forward. The authors demonstrated that a fractal-like function, denoted as $M(r)$, could successfully fit the SNIA data within a non-dark energy LTB model. The expression for $M(r)$ was given as:

\begin{equation}\label{eq1}
    M(r) \propto r^D,
\end{equation}

where $D$ represents the fractal dimension. However, it is important to note that the analysis presented in that paper relied on highly specific assumptions such as an inhomogenous bang-time function, rendering the model an exceptional case with limited generalizability.

Building upon the previous work, we propose an alternative analysis based on the LTB metric, incorporating both a fractal dimension and a length-scale parameter. This approach offers several advantages, addressing the theoretical concerns associated with previous models while ensuring a homogeneous big bang scenario. We follow the treatment of Ref. \citealp{Cosmai19,Cosmai2023}. In that work, the authors argue that from a statistical analysis of galaxy distribution in 3D, the average conditional density can be described locally as
\begin{equation}\label{eq22}
    \langle n(r) \rangle  \sim r^{-\gamma}
\end{equation}
where $\gamma$ is a phenomenological parameter that can take values from $0.2$ to $0.9$ in a range scale from $1$ Mpc/$h$ until $100$ Mpc/$h$. Beyond that scale, a transition to homogeneity ($\gamma=0$) is expected, whose behavior is not completely clear. Then, for the structure of a real galaxy we can approximate the matter density as
\begin{equation}\label{eq23}
    \rho_M(r) \sim \langle n(r) \rangle,
\end{equation}
a behaviour that can be described using a LTB model. In this case, the function $M(r)$ is related to this fractal density function due to equation (\ref{eq6})
\begin{equation}\label{eq24}
    M(r)=\int_0^r \langle n(r) \rangle(4 \pi R'R^2 dr),
\end{equation}
where $M(r)$ can be interpreted as the matter contained inside a sphere of radius $r$. In \citealp{Cosmai19}, it was stated that a fractal-like $M(r)$ function can fit the SNIA data in a pressure-less LTB model given by
\begin{equation}\label{eq25}
    M(r)=\Phi r^D,
\end{equation}
where $D=3-\gamma$ is the fractal dimension and $\Phi$ is the mass-scale. Let us study this model at some detail.

First of all, we have to emphasize a subtlety in the notation. The authors of \citealp{Cosmai19} perform the integration of the equation
 \begin{equation}\label{eqadd1}
    \dot{R}^2+2R\Ddot{R}+k=0,
 \end{equation}
assuming that at $t=0$, the $R$ function takes the value $R_0$ from which leads the definition:
\begin{equation}\label{eq26}
    R(r,0)=R_0(r).
\end{equation}
In the LTB literature the symbol $R_0(r)$ is used to denote the {\it present} scale factor (which is usually a gauge in LTB cosmology when $\Omega_m(r)$ is used instead of $M(r)$), while $R(r,0)=0$ is the initial scale factor in the homogeneous big bang itself (usually settled to $0$). This leads to a confusion with the common definitions in LTB cosmology literature and should be pointed out.

Second, following the previous redefinition for $R_0(r)$, we get an inhomogeneous negative bang time function $t_{BT}(r)<0$ which notably depends on the matter fractal dimension $D$ \citealp{Cosmai2023}
\begin{equation}
    t_{BT}=-\frac{2}{3}\sqrt{\frac{R_0^3(r)}{2GM(r)}}.
\end{equation}
However, an inhomogeneous bang time is physically undesirable if we are trying to understand the influence of the fractal matter distribution in cosmology. A similar problem was pointed out by \citealp{Kenworthy19} discussing the work made in \citealp{Hoscheit18} about the influence of a large void in a LTB cosmology, because an inhomogenous big bang leads to obtain different ages of the universe in difference places of it. This introduces a new fundamental inhomogeneity that requires new physics to explain, in the same sense than the addition of dark energy. If we perform the integration for a purely flat universe with an homogeneous bang time function and use a $M(r)$ with the usual limits ($t\in[0,t],\>R\in[0,R(r,t)]$ we just get the EdS results
\begin{equation}\label{eq27}
    R(r,t)= \left (\frac{9M(r)}{2}\right )^{1/3}t^{2/3}.
\end{equation}
From here, we see that the election made of $M(r)$ with the assumption of global flatness and homogeneous big-bang, it is just a gauge related to the scale of the radial coordinate. Also, the Hubble function can be written as
\begin{eqnarray}\label{eq28}
    H(t)=\frac{\dot{R}}{R}= \frac{2}{3t}, 
\end{eqnarray}
independently of the $M(r)$ function used. We have no divergences for $H_0$ at $r=0$, however the fractal structure described with $M(r)$ cannot be used to explain any $H_0$ tension.

The question now emerges, can we use LTB models with $E(r)=0$ (see \ref{eq10}) to describe a fractal distribution of matter assuming an homogeneous big bang? In the following we describe our proposal to do that.

\subsection{Sharp transition model}

First, we propose to define a function $M(r)$ that can describe both the internal fractal statistical behaviour of the density and also a transition to a FLRW universe (a transition between fractality to non-fractality). The function $M(r)$ can be defined as
\begin{align}\label{eq29-30}
    M_{in}(r)&=\Phi_{in}r^D,  &r<L \\
    M_{out}(r)&=\Phi_{out}r^3, &r>L,
\end{align}
where $M\sim r^3$ is the mass for a FLRW universe. At this point, we have four free parameters in the model: the mass scale $\Phi_{in}$ and $\Phi_{out}$, the fractal dimension $D$ and the scale of the fractal structure $L$. If we demand for continuity at $r=L$ we can reduce to three free parameters
\begin{eqnarray}\label{eq31}
   \Phi_{in}L^D=\Phi_{out}L^3,
\end{eqnarray}
then the model becomes
\begin{align}\label{eq32-33}
    M_{in}(r)&=\Phi_{out}L^3\left (\frac{r}{L}\right)^D, &r<L \\
    M_{out}(r)&=\Phi_{out}r^3, &r>L.
\end{align}
Defining the function in this way, the effects between the internal fractal structure and the external EdS universe should be noticeable. A similar approach was developed in \citealp{Ruffini17}, but the difference is that they fix the scale of transition about $r=2300$ Mpc meanwhile for us this scale is a free parameter to be fixed by the observations.

\begin{figure}[!ht]
\centering
\includegraphics[scale=0.5]{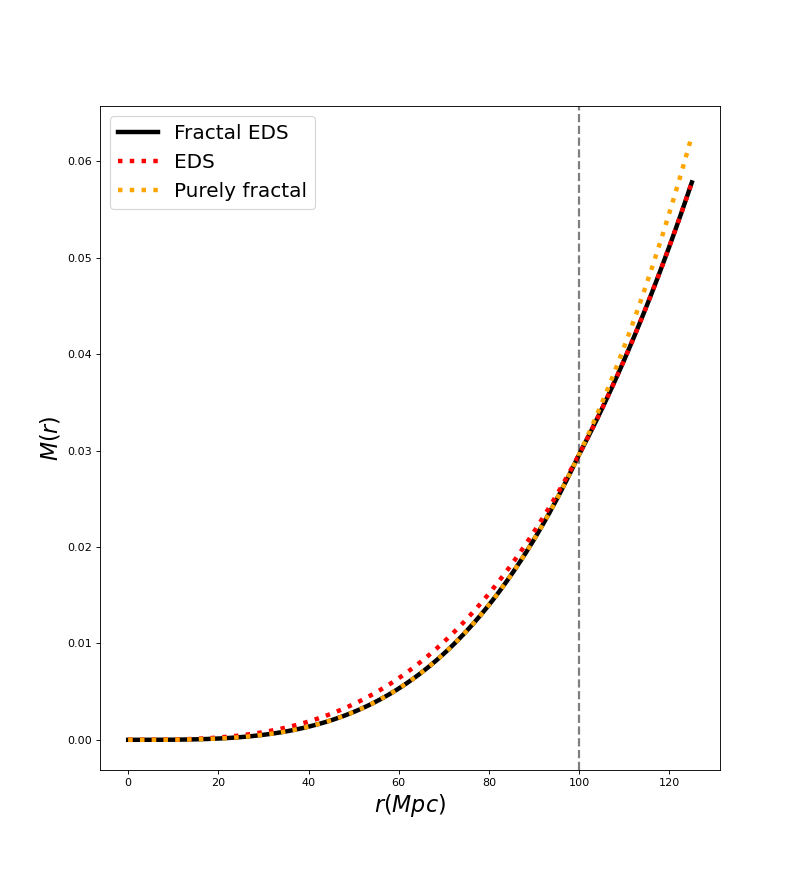}
\caption{$M(r)$ profiles. In black appears the fractal model with a sharp transition to EdS; in red pointed line EdS and in yellow pointed line a purely fractal universe. We use $D=3.36$, $L=100 Mpc$ and $h_{out} \sim 0.73$. The vertical gray line is the lenght $L$ of the transition. The functions $M_1(r)$ and $M_2(r)$ that describes a smooth transition in equations (\ref{add3}) and (\ref{add4}) are not plotted as those are very similar to the black line (see, e.g., fig. 2 of  \citealp{Calcagni_2017}). }
\label{M(R)profiles}
\end{figure}

\begin{figure}[!ht]
\centering
\includegraphics[scale=0.5]{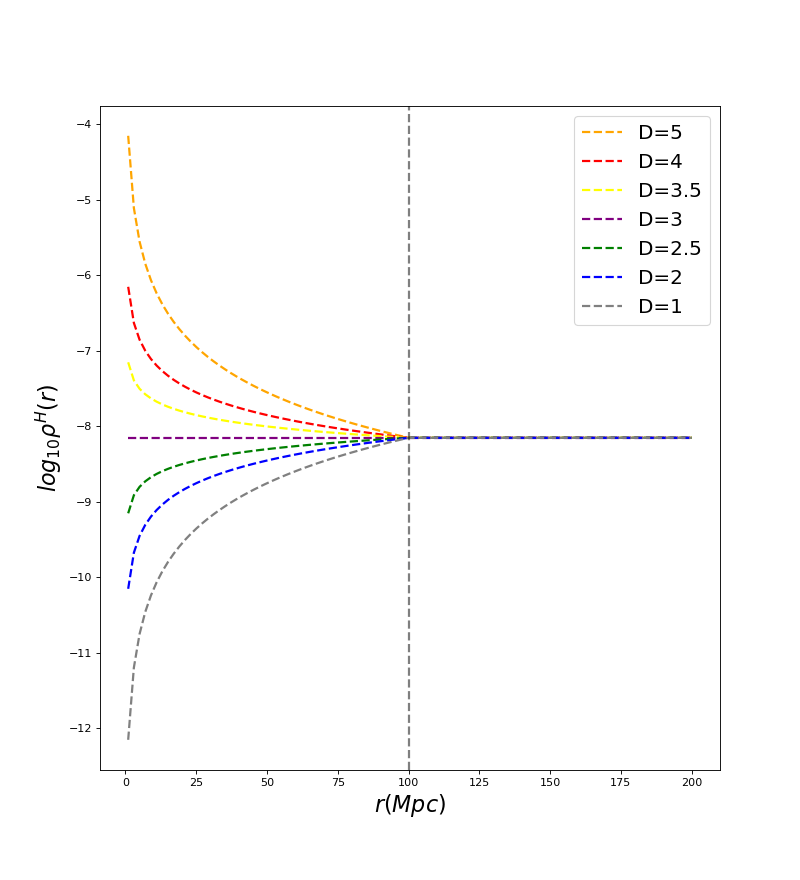}
\caption{Integrated density profiles for different values of $D$. We use $L=100 Mpc$ and $h_{out} \sim 0.73$. The vertical gray line is the length $L$ of the transition. Note that profiles with $D>3$ are similar to an over-dense region and $D<3$ to an under-dense region. $D=3$ correspond to the common constant integrated density profile in EdS universe. }
\label{fig:2}
\end{figure}

\begin{figure}[!ht]
\centering
\includegraphics[scale=0.45]{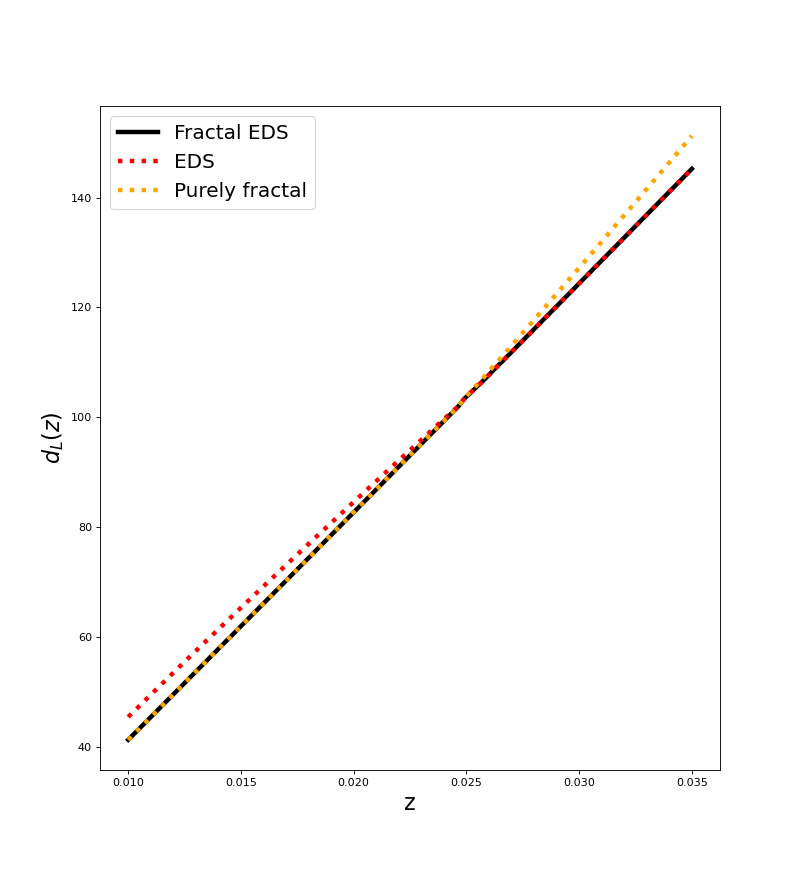}
\caption{Luminosity Distance for the different universes. Same description as in FIG. \ref{M(R)profiles}. Note that we use the purely fractal universe with the scale $\Phi_{in}$ to illustrate.}
\label{fig:3}
\end{figure}

As a first approach, we assume $\Omega_{\Lambda}=0$ and the universe being just EdS beyond the transition. 
We use low redshift type Ia supernova data to estimate the transition scale to fractality. We also use the full sample \footnote{We understand this approach as being a tool to estimate the fractal transition in low redshift data, but also as a possible explanation for dark energy with fractal structure using the full sample. However, this last approach is questionable as it does not exist a clear evidence for fractal structure for scales $r>100$ Mpc}, in which case we have to add another parameter to be fixed. Following \citealp{Alexander09} and considering an EdS universe for $r>L$, we choose for $\Phi_{out}$ to be in agreement with the age of the universe in an EdS universe:
\begin{eqnarray}\label{eq34}
    \Phi_{out}=\frac{4 \pi}{3} \rho_0 = \frac{H_0^2}{2},
\end{eqnarray}
where $H_0 \sim h_{out}/3000$ Mpc is the Hubble parameter outside the fractal region in the FLRW limit. Using this scale we get the radial coordinate in Mpc and just two free parameters: $D$ and $L$. The $R(r,t)$ can then be written as:
\begin{eqnarray}\label{eq35-36}
    R_{in}(r,t) &=& \left(\frac{9\Phi_{in}}{2}\right)^{\frac{1}{3}}L\left (\frac{r}{L}\right)^{D/3}t^{2/3},\>\>r<L \\
    R_{out}(r,t) &=& \left(\frac{9\Phi_{out}}{2}\right)^{\frac{1}{3}} rt^{2/3},\>\>r>L, 
\end{eqnarray}
It is important to note that in this model, the density is surprisingly homogeneous. Using equation (\ref{eq11}) we can obtain the same density profile than an EdS universe:
\begin{eqnarray}\label{eq37}
     \rho_0(r)=\frac{3\Phi_{out}}{4\pi},
\end{eqnarray}
but the integrated density at $t_0$ is
\begin{eqnarray}\label{eq38}
    \rho_0^H(r) \propto r^{D-3}=r^{-\gamma},\>\>r<L 
\end{eqnarray}
which is constant for $r>L$. Using the continuity condition at $L$
we obtain:
\begin{equation}\label{eq39-40}
    \rho_0^H(r)= \left\{
  \begin{array}{lr} 
       \frac{3\Phi_{out}}{4\pi}\left(\frac{r}{L}\right)^{-\gamma} &, r<L \\
      \frac{3\Phi_{out}}{4\pi} &, r>L.
      \end{array}
\right.
\end{equation}
The behaviour of $M(r)$ and the densities are displayed in Fig. \ref{M(R)profiles} and  Fig. \ref{fig:2}. It is interesting to note that the fractal behaviour can be also be defined as
\begin{eqnarray}\label{eq41-42}
 \rho_0^H(r)&=&\rho_0 \left(\frac{r}{L}\right)^{-\gamma} ,\>\>r<L \\ 
 \rho_0^H(r)&=&\rho_0  ,\>\>r>L .
\end{eqnarray}
As simple it is, this model assumes a sharp transition which is not very realistic. The Purely Fractal and Fractal-EdS universe are plotted in Fig. \ref{fig:3}.

\subsection{Smooth transition model}

A smooth slope transition around $r=L$ could be introduced considering a multi-power mass function. Motivated by the work developed in \citealp{Calcagni_2017}, a binomial ``average" mass profile could be useful for this case. If $D<3$ we can write
\begin{align}\label{add3}
    M_1(r)&=\Phi_{out}\left[ L^3\left (\frac{r}{L}\right)^D + r^3 \right].
\end{align}
Note that as $D<3$, for $r \gg L$ the term $r^3$ dominates the mass function, meanwhile for $r \ll L$ the $r^D$ term is stronger. In the opposite case when $D>3$ we can propose:

\begin{align}\label{add4}
    M_2(r)&=\frac{\Phi_{out}}{\frac{1}{L^3}\left (\frac{L}{r}\right)^D+\frac{1}{r^3}},
\end{align}
which follows a similar asymptotic behaviour. We choose the same $\Phi_{out}$ value described in the last section to allow for consistent convergences at $r\gg L$. This approach has the advantage of being numerically friendlier as we have a unique mass profile for all $r$ space.

\subsection{Luminosity Distance in LTB models}

The luminosity distance in LTB models can be obtained from
\begin{equation}\label{eq19}
    d_L = (1+z)^2 R(r,t),
\end{equation}
where $r=r(z)$ and $t=t(z)$ are defined through the geodesic equations
\begin{eqnarray}\label{eq20-21}
\frac{dr}{dz} &=& \frac{\sqrt{1-k}}{(1+z)\dot{R'}}, \\
\frac{dt}{dz} &=& -\frac{R'}{(1+z)\dot{R'}}.
\end{eqnarray}
These equations are necessary to make contact with the observations. We use the Pantheon sample (\citealp{Scolnic18}) that release the redshift $z$, distance modulus $\mu(z)$ and the error $\sigma_{\mu}(z)$ for $1048$ type Ia supernova that can be compared to the theoretical expectation using an LTB model in a fractal-like context:

\begin{equation}\label{eq44}
    \mu(z) = 5\log_{10}\frac{d_L(z)}{10 \text{pc}}.
\end{equation}

\section{Statistical analysis with SNIa data}

To test our models, we use the Pantheon sample comprised by 1048 SNIa in the redshift range $z \in (0.01, 2.3)$. As we are interested in studying a possible transition between fractal behaviour and a FLRW universe, we consider two approaches. First, using only low redshift SNIa, using a cutoff of $z<0.3$ corresponding approximately to $\sim 800 $Mpc. Second, using the full sample. We select $h_{out}\sim 0.73$ in the first case to match local Hubble parameter scale measures (\citealp{Dainotti_2021}) and $h_{out}\sim 0.45$ in the second case, in order to converge for low scales of the Hubble parameter that fits CMB with an EdS universe (\citealp{Alnes_2006}). The full Tripp Formulae for distance modulus is
\begin{equation}\label{eq43}
    \mu_{obs}=m_b^*-M=m_b+\alpha x-\beta c+\Delta_M-M,
\end{equation}
here $m_b$ corresponds to the peak apparent magnitude in the B-band, $M$ is the absolute B-band magnitude of a fiducial SNIa, $x$ and $c$ are light curve shape and color parameters, $\alpha$ is a coefficient of the relation between luminosity and stretch and $\beta$ is a coefficient of the relation between luminosity and colour, and $\Delta_M$ is a corrections based on the mass of the host galaxy. All of these parameters, usually called \textit{nuisance parameters}, are needed to standardize the SNIa data leading to the corrected magnitude $m_b^*$. Those correction are provided in the Pantheon catalogue, and the parameter $M$ can be marginalized. This relation can be used to perform statistical analysis with the distance theoretical modulus for any cosmological model given by Eq.(\ref{eq44}).

\section{Results}

We use the code \texttt{EMCEE} (\citealp{Foreman2013}) to test the models against the SNIa data. This is a pure python implementation of the affine invariant ensemble sampler for Markov chain Monte Carlo proposed by Goodman and Weare (\citealp{Goodman2010}). From this analysis we obtain the best fit values for $D$ and $L$ for each model. We used directly the corrected magnitudes from the catalogue, but also another approach allowing the nuisance parameters $\alpha$ and $\beta$ to vary along the cosmological parameters. In this last approach, we ignore the step mass correction as this does not impact significantly the cosmological fit.

\subsection{Sharp transition}

\begin{figure}[!ht]
\centering
\includegraphics[scale=0.45]{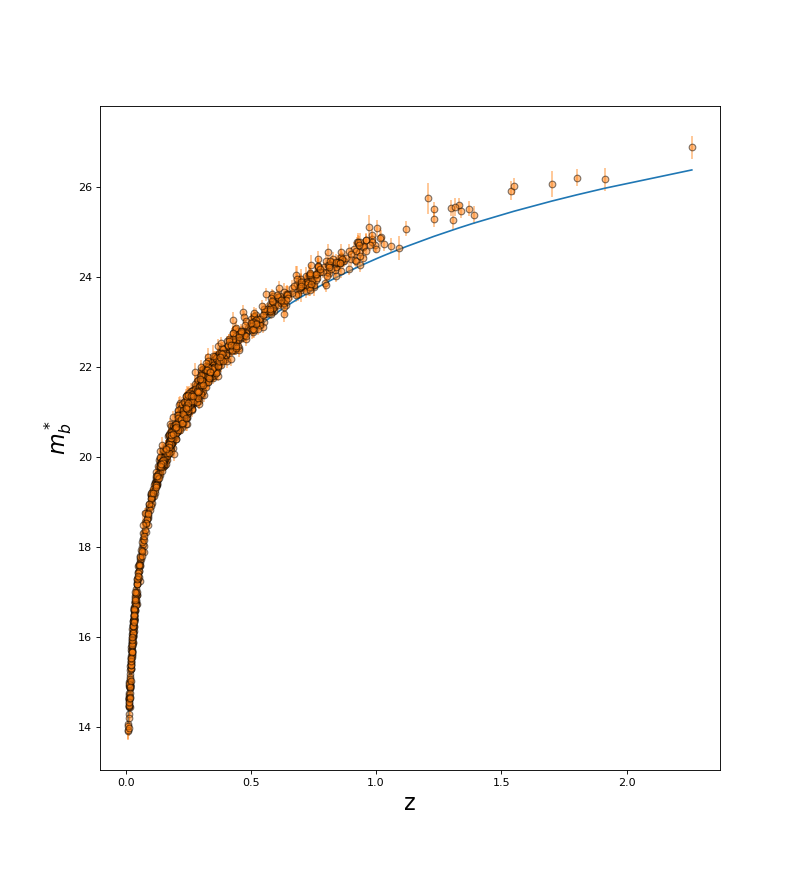}
\caption{Best fit function for corrected magnitudes of the full Pantheon sample using a fractal sharp transition. We get a notably worst fit than previous studies in fractal LTB models that use an inhomogeneous big bang.}
\label{fig:4}
\end{figure}

The results for the sharp transition model case and details are shown in Table \ref{Table:sharp}. We note that using low redshift supernovae data, we can get a relatively good fit, but it does not happens using the full sample. We also note the the errors increase in the parameter estimation if we do not use tight priors, a result possible attributed to the sharp transition at $r=L$. The best fit result using the full sample is displayed in Figure \ref{fig:4}.

To look for insights into the value of the fractal dimension $D$, we also explore the case of considering priors on the parameter $L$, which is our fractal transition scale. Instead of fixing it at a convenient value, as previous works have done, here we introduce Gaussian priors for $L$ based on values coming from large scale observations of the transition scale to homogeneity. For example, in \citealp{Yadav2010} the authors concluded that the scale of homogeneity should be less than $260h^{-1}$ Mpc. Also in \citealp{Yadav2005} and \citealp{Hogg2005} the authors have suggested even a smaller scale of $L=60h^{-1}$Mpc. Let us consider four values for $L$: $L=60h^{-1}$, $L=100h^{-1}$, $L=150h^{-1}$ and $L=200h^{-1}$. Considering Gaussian priors on this values with a 5\% of error, we find the results showed in the last panel in Table \ref{Table:sharp} using low redshift supernovae at $z<0.3$. Again, for the sharp transition model the error on $D$ is very high. This means that a flat LTB fractal model with a sharp transition, an homogeneous big bang and no cosmological constant, does not perform well to fit SNIA data.

\begin{figure}[!ht]
\centering
\includegraphics[scale=0.35]{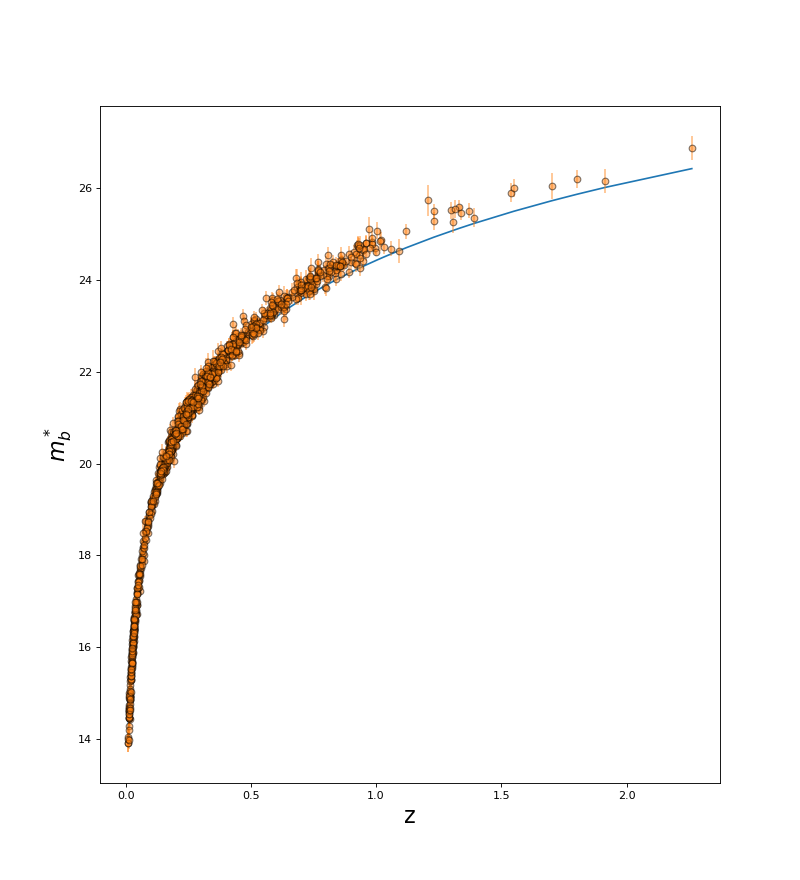}
\includegraphics[scale=0.35]{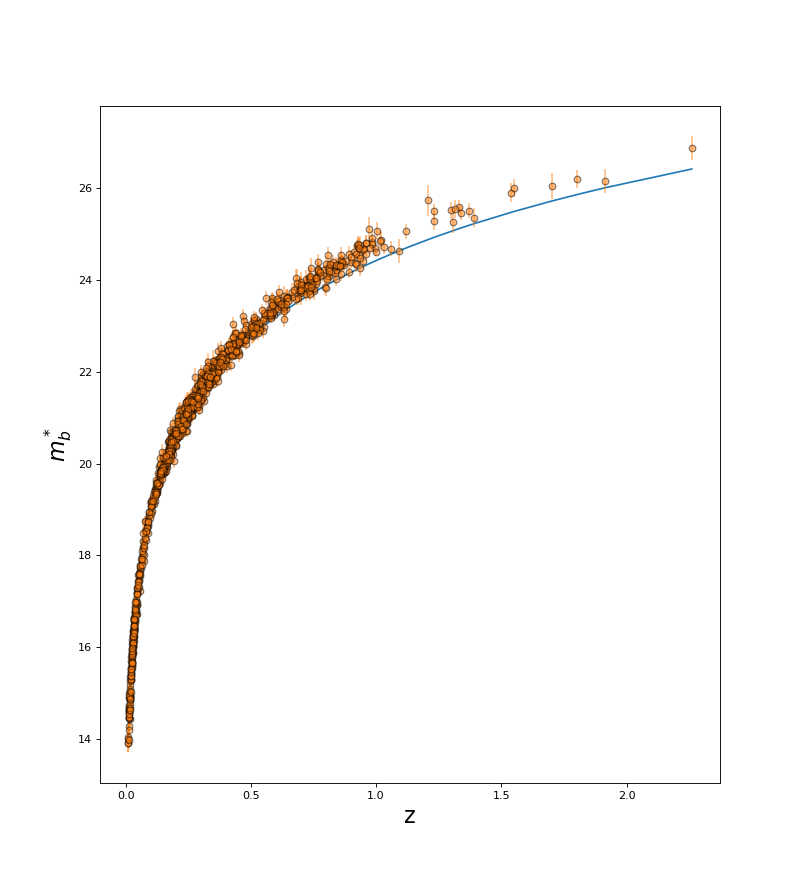}
\caption{Best fit functions for corrected magnitudes of the full Pantheon sample using the two $M(r)$ functions of the fractal smooth transition. The errors in the parameter estimation are much more lower than the sharp case, but still the fit does not perform well. }
\label{fig:6}
\end{figure}


\subsection{Smooth transition}

The results for the best fit for the sharp transition model and details are shown in Table \ref{Table:Smooth}. We note that the errors in the parameter estimation are notably lower once the model is tested against the corrected magnitudes of the Pantheon sample, which support the idea that a smooth function performs much better in the MCMC sampling. However, the $\chi^2_{red}$ values obtained are even worse than in the sharp transition model for all the cases. We conclude that a flat LTB fractal model with a smooth transition, an homogeneous big bang and no cosmological constant neither perform well to fit SNIA data. The best fits are displayed in Figure \ref{fig:6}.

\begin{table}[!ht]
\begin{center}
    \textbf{Corrected magnitudes}\\
\begin{tabular}{ccccc}
 $h_{out}$ &$z_{cut}$ & $D$ & $L (Mpc)$ &$\chi^2_{red}$ \\
\hline
$0.73$ & $z<0.3$ & $3.16^{+0.36}_{-0.63}$ & $696.83^{+113.01}_{-298.50}$ & $1.15$ \\ 
$0.45$ & Full Sample & $3.47^{1.04}_{ -1.38}$ & $1354.17^{232.08}_{-313.73}$
& 1.43 \\
\end{tabular}
\end{center}
\begin{center}
\textbf{Tight priors for $L$ and corrected magnitudes}\\
\begin{tabular}{cccc}
 $L$ Prior & $ D $ & $L(Mpc)$ & $\chi^2_{red}$ \\
\hline
$60h^{-1}$  & $2.97 \substack{+1.26\\-1.07}$ & 85.02 $\substack{+0.53 \\ -0.52}$  & $1.15$ \\ 
$100h^{-1}$ & $2.74 \pm \substack{+1.24\\-1.09}$ & $137.02 \substack{+0.54\\-0.51}$ & $1.15$ \\
$150h^{-1}$ & $2.73 \pm \substack{+1.2\\-1.1}$ & $205.02 \substack{+0.54\\-0.51}$ & $1.16$ \\
$200h^{-1}$ & $2.6 \pm \substack{+1.15\\-1.13}$ & $273.00 \substack{+0.51\\-0.5}$ & $1.16$ \\
\end{tabular}
\end{center}
\caption{\label{Table:sharp}
Best fit values for $D$ and $L$ using a \textbf{sharp} transition, fitting for corrected magnitudes as are given in Pantheon catalogue. Also we perform a fit with a tight prior for $L$ with the corrected magnitudes to look for insights into the value of $D$.}
\end{table}

\begin{table}[!ht]
\begin{center}
\textbf{Corrected magnitudes $M_1(r)$ function}\\
\begin{tabular}{ccccc}
 $h_{out}$ &$z_{cut}$ & $D$ & $L (Mpc)$ &$\chi^2_{red}$ \\
\hline
$0.73$ & $z<0.3$ & $3.16^{+0.018}_{-0.018}$ & $1020.86^{+0.015}_{-0.021}$ & $1.37$ \\ 
$0.45$ & Full Sample & $3.15^{0.017}_{ -0.017}$ & $1020.84^{+0.026}_{-0.014}$
& 1.46 \\
\end{tabular}
\end{center}
\begin{center}
\textbf{Corrected magnitudes $M_2(r)$ function}\\
\begin{tabular}{ccccc}
 $h_{out}$ &$z_{cut}$ & $D$ & $L (Mpc)$ &$\chi^2_{red}$ \\
\hline
$0.73$ & $z<0.3$ & $3.14^{+0.019}_{-0.018}$ & $1020.86^{+0.022}_{-0.020}$ & $1.33$ \\ 
$0.45$ & Full Sample & $3.15^{+0.014}_{ -0.017}$ & $1020.85^{+0.021}_{-0.018}$
& $1.46$\\
\end{tabular}
\end{center}
\caption{\label{Table:Smooth}
Best fit values for $D$ and $L$ using a \textbf{smooth} transition, fitting for corrected magnitudes. Due to the low errors in the parameter estimation we did not extent further this approach.}
\end{table}

\section{Discussion}

We have show that a purely fractal model following $M(r)\sim r^D$ with an homogeneous big bang is not different from an usual EdS universe with another scaling relation. As an alternative to the fractal model proposed in \citealp{Cosmai19}, a more consistent model was developed, allowing for a transition between a fractal scale with $M \sim r^D$ to an homogeneous universe $M \sim r^3$ without requiring the idea of using an inhomogeneous big bang.

We performed two analyses putting in stress our model with the Pantheon data. In the case of a sharp transition model, we got a good fit using low redshift supernovae, but the model fails to describe the full behaviour of the Pantheon data sample. We conclude that a fractal LTB model cannot explain the effects of dark energy. Overall, we looked for insights into the fractal dimension value performing a different analysis using low redshift supernovae with tight priors for the transition scale. It was observed that the errors are still high at $z<0.3$. The results are not very different from the smooth transition case model, giving a slightly worst fit than the sharp case, although the errors were much lower.

The physical validity of our model could be studied more deeply if we summarize the problems that those have. First, we have not used a cosmological constant. This constant could be added to the models in order to fit the full SNIA data looking for effects of the local fractal structure in the cosmological constant value. Also, we assumed that the $M$ scale should converge to a fixed $H_0$ experimental value. Two possibilities accounting for convergence to CMB $h \sim 0.45$ and another to SNIA estimation to $h \sim 0.73$ were used. Those values may be modified allowing for other definitions like the one presented in \citealp{Camarena20} \citealp{Valkenburg13}. In future works, we propose to study how a fractal structure in LTB models can impact $\Lambda$CDM cosmology. We reinforce the importance to take into account the full information about our local large scale structure to the development of a better cosmological description that lead us to stronger constraints on the $\Lambda$CDM model. We highlight that, in principle, the fractality degree of our universe cannot be studied independently from cosmological parameters using cosmological data such as SNIA. However, it could be useful to extract fractal dimension from other sources of data and incorporate it as a correction to the cosmological fit. The procedure to perform this inclusion of fractal dimension in cosmological analysis is still an open question, since the nature of fractal geometry is contrary to the differential nature of general relativity. 
\\
\\
\\

\clearchapterpage
\chapter{Tilted Cosmologies}
\label{chap:tilt}

In Chapter \ref{chap:models}, we introduced cosmologies that challenge the assumption of global homogeneity and isotropy of the universe. However, there is also a class of cosmological models that maintain the hypothesis of homogeneity and isotropy but account for the effects of peculiar motions of observers. These models are referred to as tilted cosmologies.

One such model is the tilted cosmological framework proposed by \citealp{Tsagas2011}. This model offers a natural environment to study the large-scale peculiar motions observed in the universe, allowing for two groups of relatively moving observers. The first group is aligned with the reference frame of the cosmos, identified with the coordinate system of the Cosmic Microwave Background (CMB), where the associated dipole vanishes by construction. The second group consists of real observers, such as those in galaxies like our Milky Way, which are moving relative to the CMB frame (e.g., see \citealp{Tsagas2008, Ellis2012}). In this tilted almost-Friedmann universe, linear relativistic cosmological perturbation theory was used to show that the effects of relative motion can lead to an apparent change in the sign of the deceleration parameter inside locally contracting bulk flows. Although this effect is a local artifact of the observers' peculiar motion, the scales involved can be large enough to have cosmological relevance. In this context, observers within slightly contracting bulk flows could be misled into believing that their universe has recently entered a phase of accelerated expansion. In other words, these observers might incorrectly interpret the local contraction of their bulk flow as the global acceleration of the surrounding universe (see \citealp{Tsagas2015, Tsagas2021, Tsagas2022} for further discussion and details). The goal of this study is to investigate this possibility by comparing the theoretical predictions to observational data.

Other studies have utilized the velocity field reconstruction from the 2M++ galaxy survey \citealp{Carrick_2015}, which provides a pair of data cubes containing the density contrast and velocity vectors in galactic coordinates. This data can be used to apply basic calculus and to make corrections to cosmological data for peculiar velocities, as was done in the latest Pantheon+ SN Ia compilation \citealp{Scolnic_2022, Carr_2022}. In this paper, we estimate the average volume scalar of this local velocity-field reconstruction using various methods and scales. In all cases, the local bulk flow is found to contract on average, leading to negative values for the local deceleration parameter across a range of scales. These results appear to support the tilted cosmological scenario as a plausible alternative explanation for the dark energy problem. However, caution is required in interpreting these results, as the data may contain significant biases due to the reconstruction procedure.

In Section~\ref{sTCM} of Chapter \ref{chap:models}, we provided a brief yet concise description of the tilted cosmological scenario and referred readers to the related literature for further details. In Section~\ref{sobs} of this chapter, we will discuss the observational parameters that can be related to this theoretical framework. In Section~\ref{cfa}, we explain how to relate the parameters obtained from the velocity-field reconstruction to the theory. Section~\ref{2MVF} presents the data used and addresses the potential biases in the data. Finally, in Sections~\ref{Meth} and \ref{Res}, we summarize the method and results obtained from this reconstruction and discuss their implications for cosmology. In addition to cosmology, our analysis has potential applications to astrophysics and the dynamics of local structures.

\section{Observables in Tilted Cosmology}\label{sobs}

Tilted cosmology model has been described briefly in the section \ref{sTCM} of chapter \ref{chap:models}. Although theoretically the model outlined above is well developed, it is not obvious yet how one should relate the tilted cosmological scenario to the observations. Parametrizing the deceleration function as $\tilde{q}=\tilde{q}(z)$ and then using it in (\ref{tq2}), has led to a good fit with the Pantheon SNIA sample~\citealp{Asvesta22}. Also, an apparent (Doppler-like) dipole anisotropy is expected to appear in the observed distribution of the local deceleration parameter ($\tilde{q}$), due to the bulk-flow motion relative to the CMB frame~\citealp{Tsagas2011}. However, which observational frame (heliocentric, geocentric, galactic, or cosmological) should be employed and what peculiar-velocity corrections should be applied to the data, in order to observe the aforementioned dipolar anisotropy, are the subjects of ongoing debate~\citealp{Colin19,Rubin_2020,ColinResponse}. 

The possible astronomical observables in the tilted cosmology context can be divided mainly in two groups. The first one involve parameters that depend on the luminosity distance to be computed using standard candles such as SNIA. The second group of observables can be studied directly from the local large scale structure and represent physical parameters of the bulk flow and the peculiar velocity field.

\subsection{Luminosity distance dependent observables}

The negative deceleration parameter observed when comparing SNIA with FLRW luminosity distance is attributed to a real acceleration of the expanding universe. However, in the tilted cosmology scenario, the deceleration parameter structure should have features that are not present in the standard cosmology. The 2 important anomalies that should be observed are:

\begin{enumerate}
    \item \textbf{Anisotropy in $q_0$ angular distribution:} According to the standard model of cosmology, the anisotropy level of the $q_0$ parameter should not be significant, since LCDM obeys the cosmological principle. There are many studies that claim to detect a significant anisotropy in $q_0$ distribution (\citealp{Colin19,Deng18a,sorrenti2023dipole}). However, there are others that report consistency with LCDM (see e.g.\citealp{Chang19}). Whether the anomaly is real and if it can be explained in standard cosmology scenario is an interesting and contingent topic in observational cosmology.
    \item \textbf{Evolution in $q_0$ with redshift:} In standard cosmology, the value of $q_0$ should be the same no matter the redshift space of the universe observed. However, since in tilted cosmology the negative deceleration parameter is a local artifact of the peculiar motion measured by observers inside a bulk flow, a evolution in $q_0$ with the redshift is expected. This feature has been addressed in multiple studies such as \citealp{PASTEN2023101224,Perivolaropoulos:2023iqj,OCOLGAIN2023101216} with different results. 
\end{enumerate}

The problem with those observational insights, is the needing of the luminosity distance, as SNIA provide the best constraint to the late epoch parameters. However, to use the luminosity distance, the data has to be corrected to the CMB frame of reference as the geodesic equation from which it is derived deals with an observer in rest w.r.t the Hubble flow. Whether it is a valid approach is debatable (\citealp{Colin19,Rubin_2020,ColinResponse}). It is possible that correcting the data to the CMB frame could erase tilted effects, since those are general relativistic in nature. Therefore, the proper way to analyze those features are still unknown.

\subsection{Local large-scale structure observables}

One interesting feature of tilted cosmology is that the values of the cosmological parameters are related to LLSS features (see eq \ref{tq2}) and therefore are extremely weakly dependent of the luminosity distance. Therefore, it is possible to study tilted cosmology physics directly from the surveys of the local universe. In particular, relation \ref{tq2} requires a negative (average) divergence of the bulk flow to explain a negative deceleration parameter. Our aim is to study the dynamical structure of the peculiar velocity field directly from data reconstruction. We choose the velocity field reconstruction of the 2M++ survey \citealp{Carrick_2015}, which has been previously used in cosmology to correct the peculiar velocities of the SNIA data in the last Pantheon+ compilation~\citealp{Scolnic_2022,Carr_2022}. The methods used to characterize this peculiar velocity field and the procedures employed to relate our results with those of the tilted cosmologies are discussed in the next sections.

\section{Classical Fluid approximation}\label{cfa}
The kinematic analysis outlined in the previous section, is straightforwardly adapted to the Newtonian framework as well (e.g.~see~\citealp{Ellis1971,Ellis1990} for further discussion and details). In so doing, one replaces the projector ($h_{ab}$), which also acts as the metric tensor of the 3-space, with the Kronecker delta ($\delta_{ij}$). Also, time derivatives and 3-dimensional covariant gradients are replaced by convective derivatives and by ordinary partial derivatives respectively. Note that, given the near spatial flatness of the observed universe, any curvature corrections due to a nonzero connection ($\Gamma^a{}_{bc}$) will be of the second perturbative order. Then, focusing on the peculiar-velocity field ($\mathbf{v}=v_i$), we have 
${\tilde{\theta}}_{ij}=\nabla\mathbf{v}=\partial_jv_i$ and
\begin{eqnarray}
    \tilde{\theta}_{ij}= \frac{1}{3}\, \tilde{\theta}\delta_{ij}+\tilde{\varsigma}_{ij}+ \tilde{\varpi}_{ij}\,.
\end{eqnarray}
Here, the (local) volume expansion/contraction scalar, the shear tensor and the vorticity tensor of the bulk peculiar flow are respectively defined as
\begin{eqnarray}
    \tilde{\theta}&=&\partial_iv_i= \delta^{ij}\partial_jv_i\,, \\
    \tilde{\varsigma}_{ij}&=&\frac{1}{2} \left(\partial_j v_i+\partial_iv_j\right)- \frac{1}{3}\,\tilde{\theta}\,\delta_{ij}, \\
    \tilde{\varpi}_{ij}&=&\frac{1}{2} \left(\partial_j v_i- \partial_iv_j\right)\,.
\end{eqnarray}

It ts possible to evaluate the gradient tensor of the local peculiar-velocity field using this approach. In particular, the gradient tensor can be reduced to the $3\times3$-matrix of the partial derivatives of the $\tilde{v}_i$-field as:

\begin{eqnarray}
    \tilde{\theta}_{ij}=\partial_j v_i=\begin{pmatrix}
        \frac{\partial v_x}{\partial x} & \frac{\partial v_x}{\partial y} & \frac{\partial v_x}{\partial y} \\
        \frac{\partial v_y}{\partial x} & \frac{\partial v_y}{\partial y} & \frac{\partial v_y}{\partial y} \\
        \frac{\partial v_z}{\partial x} & \frac{\partial v_z}{\partial y} & \frac{\partial v_z}{\partial y} \\
    \end{pmatrix}\,,
\end{eqnarray}
directly relating $\tilde{\theta}_{ij}$ to the Jacobian tensor of the field. 



\begin{figure}[ht]
    \centering
    \includegraphics[width=10cm]{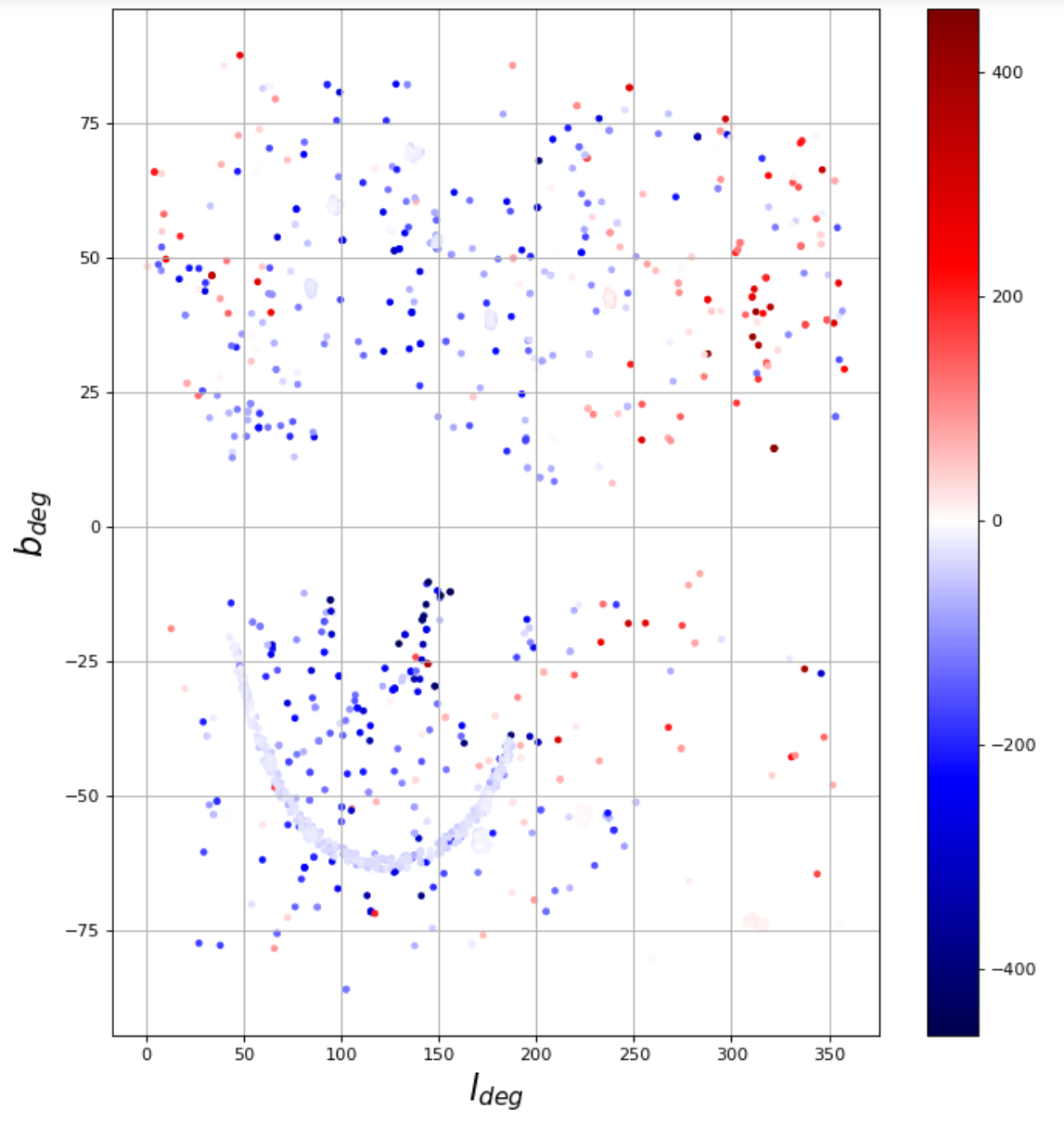}
    \caption{Peculiar velocities of the Pantheon+ SNIA compilation}
    \label{PVpantheon}
\end{figure}

\begin{figure*}[ht]
    \centering
    \includegraphics[width=8cm]{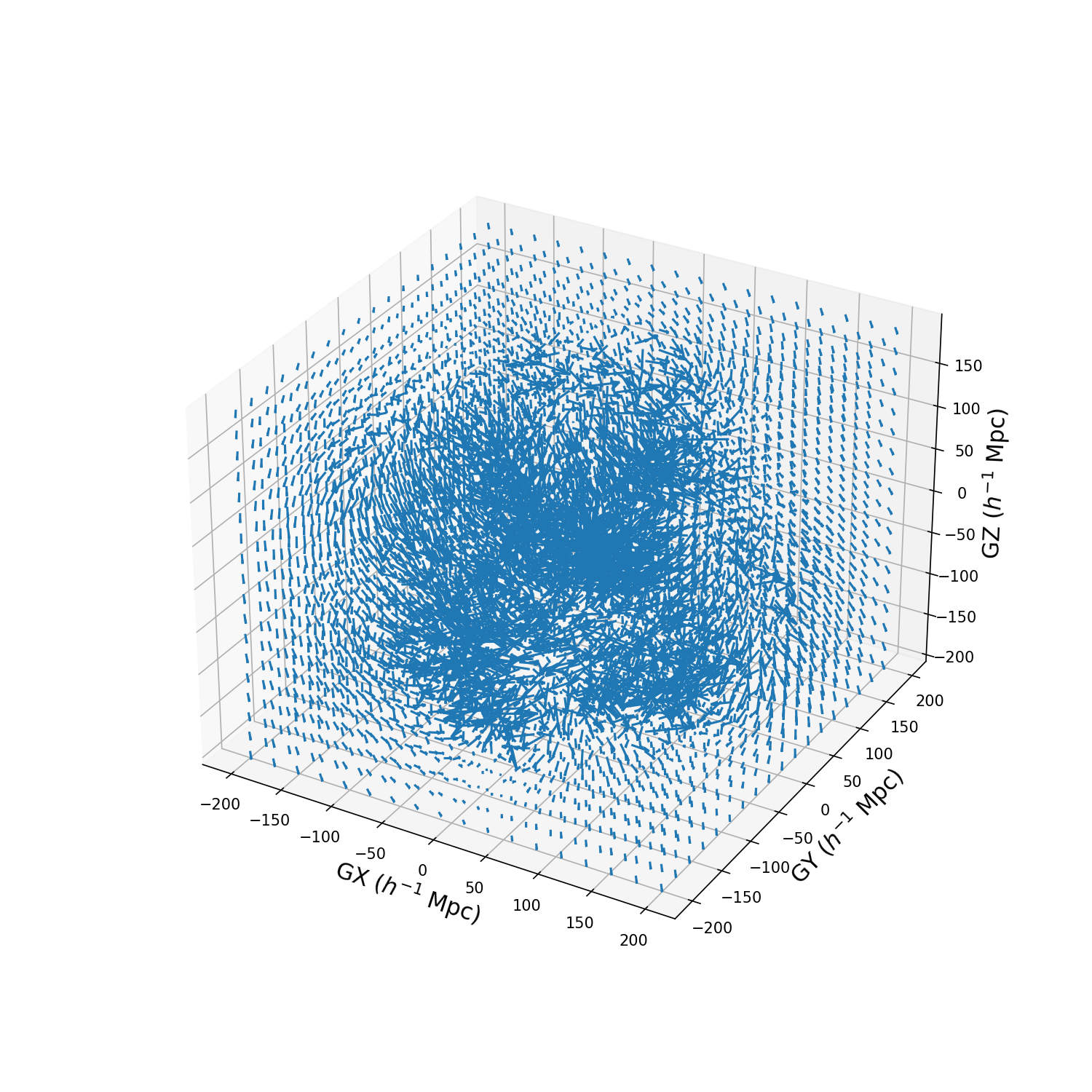}\includegraphics[width=8cm]{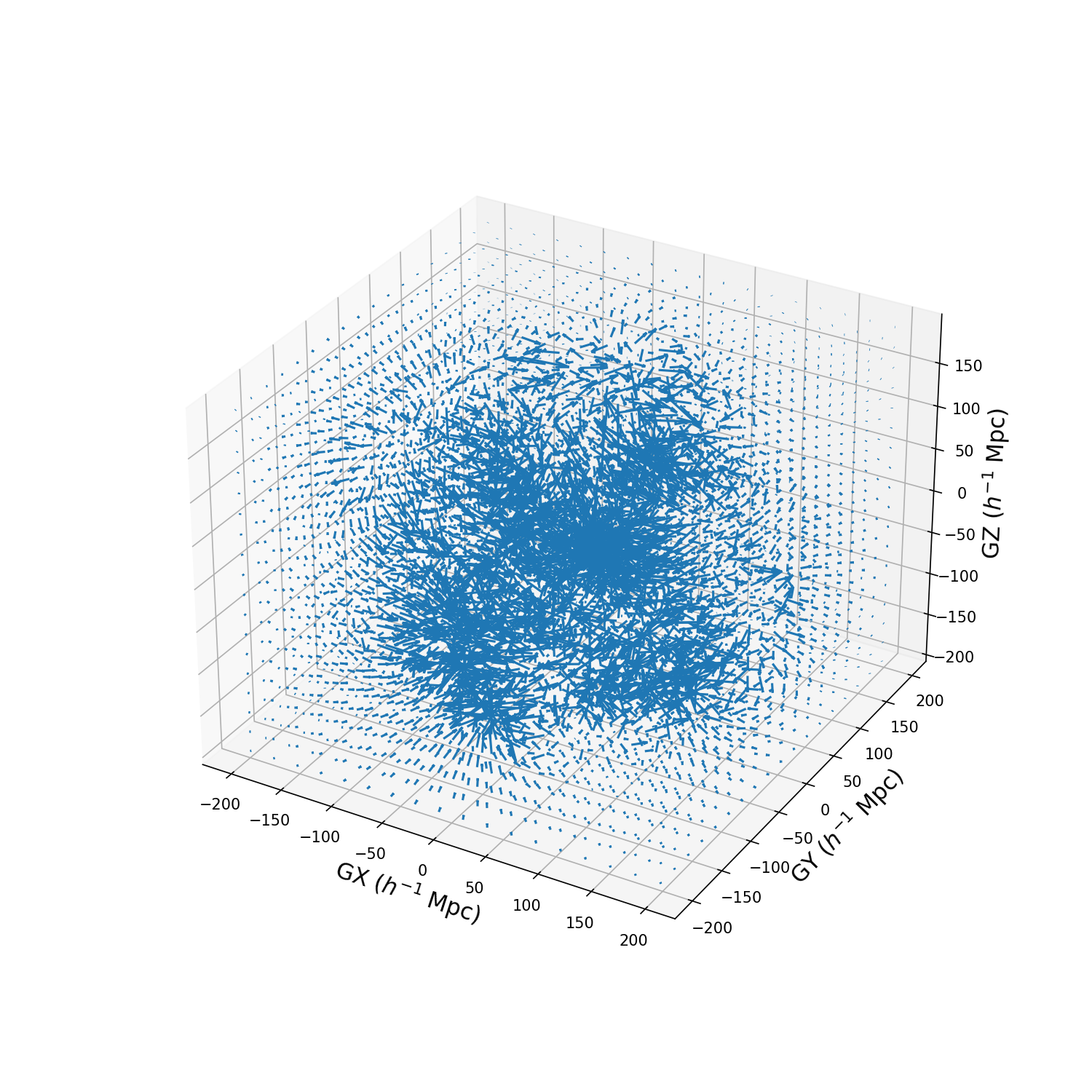}
    \caption{Peculiar Velocity field reconstruction from the 2M++ density field in galactic coordinates. Visualizations in 3D, with (left) and without (right) external dipole component.}
    \label{PV3D}
\end{figure*}

\section{The 2M++ velocity field reconstruction}\label{2MVF}

The last Supernovae IA compilation, namely Pantheon+~\citealp{Scolnic_2022}, was released in 2022 showing a great improvement in the utility of data at low redshifts for cosmological uses. Part of this improvement is due to better corrections of the peculiar velocities of the SNIA data~\citealp{Carr_2022} (see Figure~\ref{PVpantheon}). This was done by using a velocity field reconstruction based on the 2M++ galaxy survey \citealp{Carrick_2015} (see Figure~\ref{PV3D}). The reconstruction procedure can be summarized as follows. If $\delta(\mathbf{r})$ is the density contrast, then the peculiar velocity field can be approximated as proportional to the gravitational acceleration when the fluctuations are small:

\begin{equation}
    \mathbf{v}(\mathbf{r})=\frac{f(\Omega_m)}{4\pi}\int d^3\mathbf{r'}\delta(\mathbf{r'})\frac{\mathbf{r'}-\mathbf{r}}{     \lvert \mathbf{r'}-\mathbf{r}\rvert^3 }\,. \label{Reconstruction}
\end{equation}
Here, $f$ is the growth rate of cosmic structures defined as $f=\Omega_m^\gamma$, where $\gamma=0.5$ for $\Lambda$CDM cosmology (see \ref{VFR} for further details about peculiar velocity treatment on large-scale structure physics). Also, $r=HR$ is measured in $km/s$, with $R$ being the co-moving distance in $Mpc$ and $H$ the Hubble parameter. 

Since the total density perturbation ($\delta$) cannot be directly observed, a bias parameter ($b$) has been introduced to relate the observed density contrast ($\delta_g$) with the real one:
\begin{equation}
    \delta=\frac{\delta_g}{b}\,,
\end{equation}
at the linear level. Therefore, the important parameter in evaluating the velocity field is the ratio $\beta=f/b$, since we can write:
\begin{equation}
    \mathbf{v}(\mathbf{r})=\frac{\beta}{4\pi}\int d^3\mathbf{r'}\delta_g(\mathbf{r'})\frac{\mathbf{r'}-\mathbf{r}}{     \lvert \mathbf{r'}-\mathbf{r}\rvert^3 }\,,
\end{equation}
to relate directly the peculiar velocity with the observed galaxy density. Also, as the observations extend only up to a maximum scale ($R_{max}$), the contribution beyond this length can be added as a constant external velocity parameter ($\mathbf{V}_{ext}$), so that finally:
\begin{equation}
    \mathbf{v}(\mathbf{r})=\frac{\beta}{4\pi}\int^{R_{max}}d^3\mathbf{r'}\delta_g(\mathbf{r'})\frac{\mathbf{r'}-\mathbf{r}}{     \lvert \mathbf{r'}-\mathbf{r}\rvert^3 }+\mathbf{V}_{ext}\,,  \label{velrec}
\end{equation}
where $\beta$ and $\mathbf{V}_{ext}$ are determined empirically from the reconstruction of the density field. {\color{red} }

\begin{figure*}[!ht]
    \centering
    \includegraphics[width=9cm]{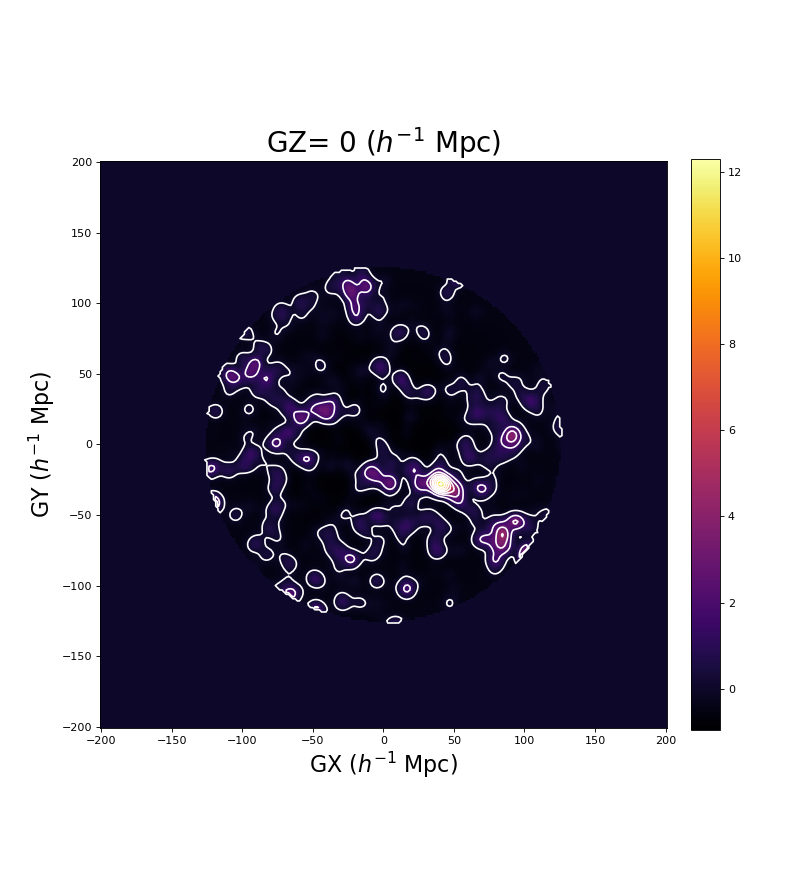}
    \includegraphics[width=9cm]{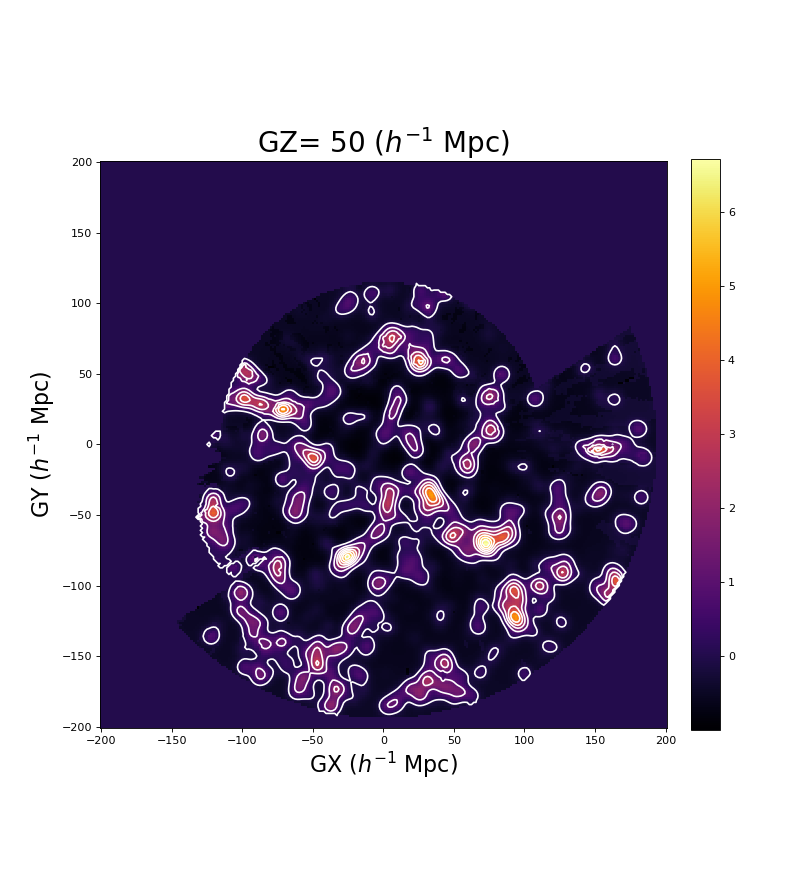}
    \caption{The density contrast projected over the $GZ=0$ and $GZ=50 h^{-1}Mpc$ galactic planes.}
    \label{DensityVel1}
\end{figure*}

\begin{figure*}[!ht]
    \centering
    \includegraphics[width=8cm]{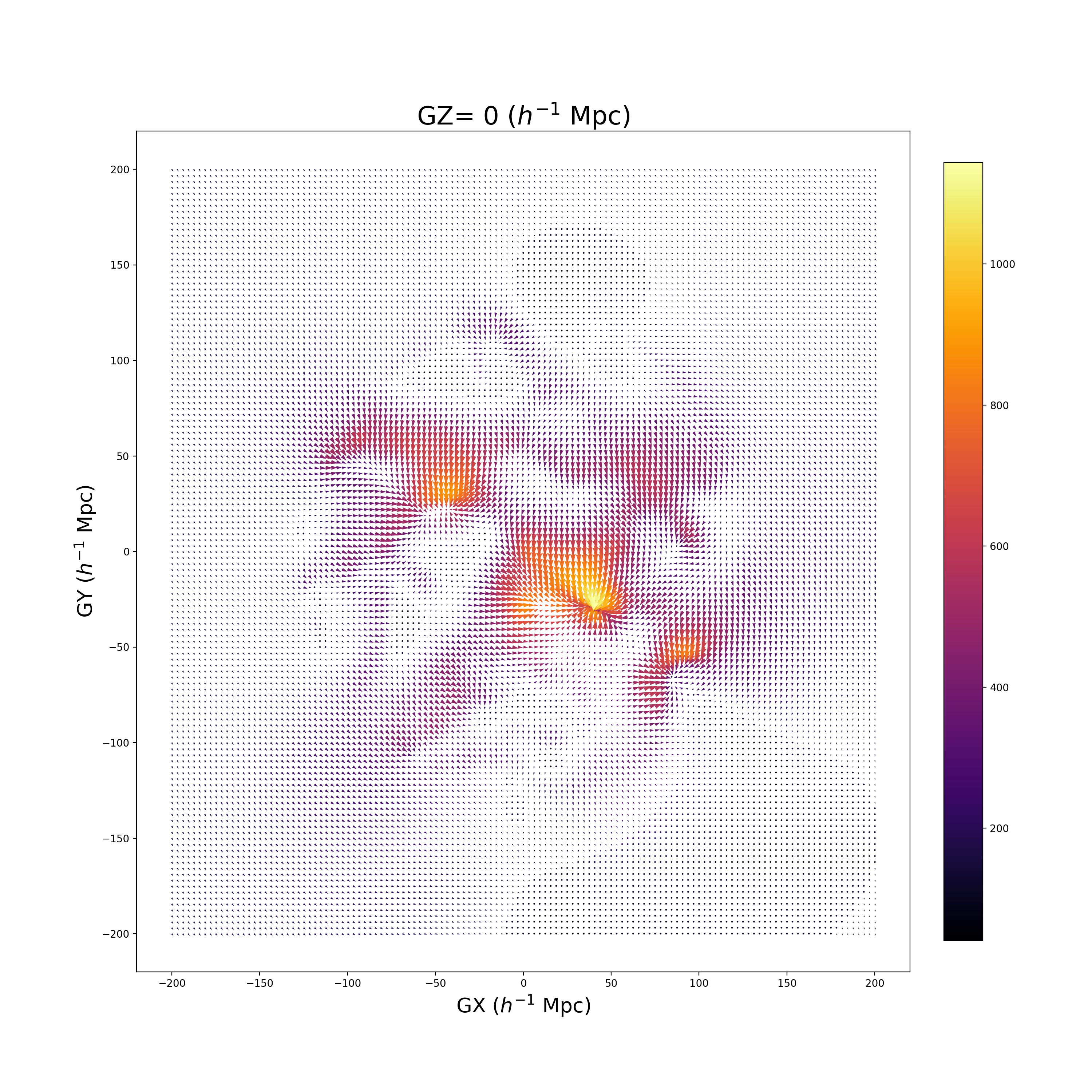}\includegraphics[width=8cm]{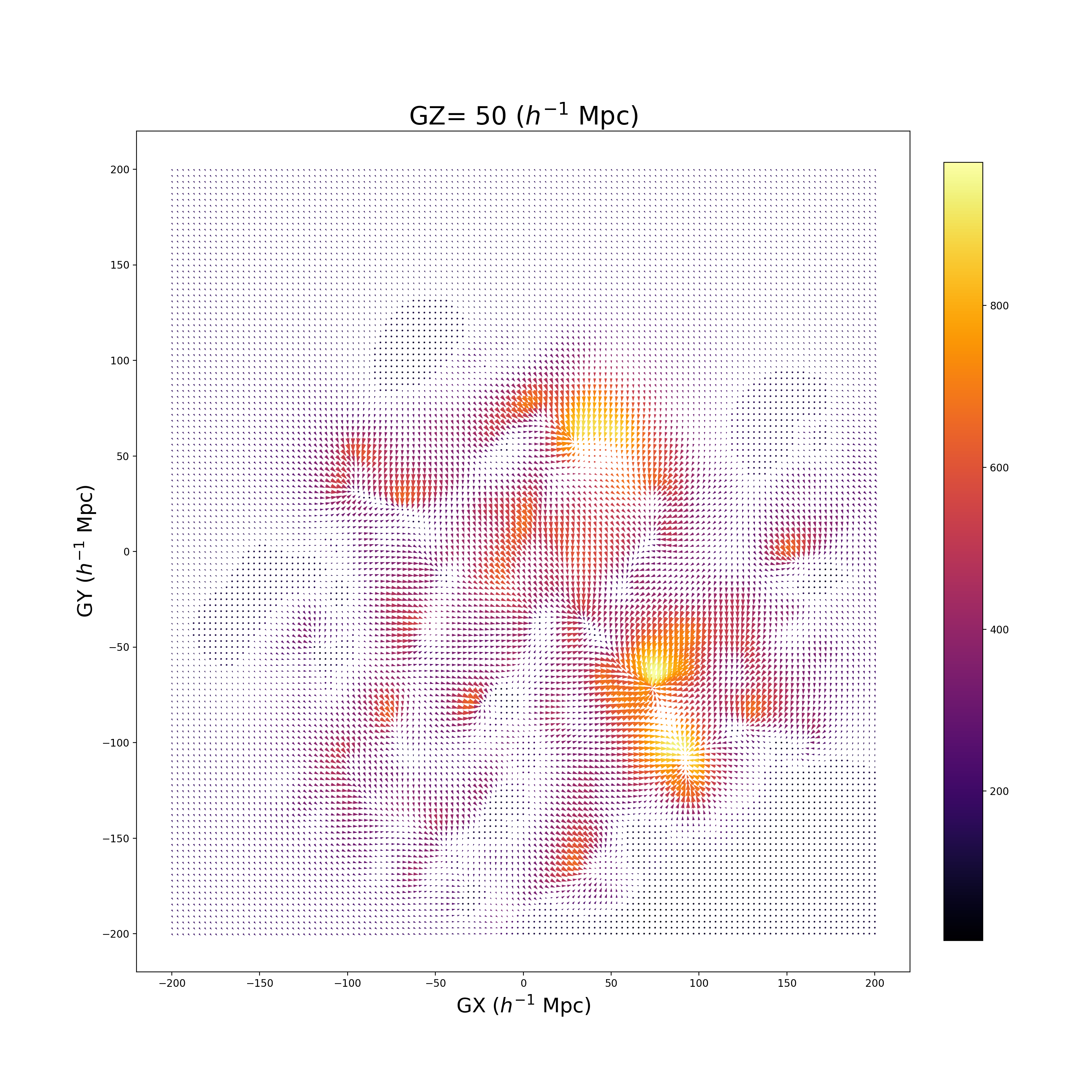}
    \caption{The the vector velocity field projected over the $GZ=0$ and $GZ=50 h^{-1}Mpc$ galactic planes.}
    \label{DensityVel2}
\end{figure*}

We use the density contrast and the velocity field (see Figures ~\ref{DensityVel1} and \ref{DensityVel2})) given by~\citealp{Carrick_2015}, which can be easily downloaded from \href{https://cosmicflows.iap.fr/}{https://cosmicflows.iap.fr/}. There, the authors provide two useful data-cubes containing the density contrast $\delta$ and the velocity field $\mathbf{v}$ using the best-fit parameters $\beta=0.431 \pm 0.021$ and $\mathbf{V}_{ext}=(89\pm21,-131\pm23,17\pm26)\>km/s$ (with $\rvert \mathbf{V}_{ext}\lvert=159\pm23\>km/s$) in galactic Cartesian coordinates. It is also important to note that $\beta$ parameter depends on scale and redshift. To consider this, we use also a different value of $\beta$, namely $\beta=0.341 \substack{+0.031 \\ -0.047}$, which was used to correct the Pantheon+ data~\citealp{Said_20,Carr_2022} combining 6dFGS and SDSS. To cover a wide range of possible values, we also use a lower value for $\beta=0.314 \substack{+0.031 \\ -0.047}$ corresponding to \citealp{Said_20} fit using just SDSS. We expect that this 3 value could be representative of $\beta$ variations. Overall, we can write $\mathbf{v}=\beta\mathbf{v}_{rec}$, where $\mathbf{v}_{rec}$ gives the directions and relative magnitudes of the velocity field. Then, it is easy to use the three values and compare the results.

In order to apply the same corrections to Pantheon+, the whole velocity field was approximated by a radially decaying function along the direction of the bulk flow. The latter is a 200~$Mpc$ sphere, composed by the sum of an external $\mathbf{V}_{ext}$ and a small average internal velocity $\mathbf{v}_{200}$. Interestingly, the external dipole component does not contribute to the gradient as it is a constant.\footnote{Even a radially decaying $\mathbf{V}_{ext}$ function, with a fixed direction, does not affect the average divergence of the velocity field.} Therefore:

\begin{equation}
    \nabla  \mathbf{v}= \nabla  (\beta \mathbf{v}_{rec} + \mathbf{V}_{ext})= \beta \nabla \mathbf{v}_{rec}\,.
\end{equation} 

\subsection{Bias in the velocity field reconstruction} 

The divergence of the reconstructed velocity field could be computed directly as:

\begin{equation*}
    \nabla \cdot \mathbf{v}(\mathbf{r})=-\beta \delta_g(\mathbf{r})\,.
\end{equation*}

This is expected as in linear theory; the divergence of the velocity field is proportional to the density contrast. However, is important to note that according to \citealp{Carrick_2015} this density contrast is normalized over the mean density of a selected scale of the survey denoted by $\bar{\rho}$. Therefore, this could introduce an important bias in the velocity field. To analyse this, we define $\delta$ as the real density contrast:

\begin{equation*}
    \delta=\frac{\rho-\rho_0}{\rho_0},
\end{equation*}

Here, $\rho_0$ refers to the mean density of the universe. It is possible that this density contrast, denoted by $\delta$, is different from the density contrast $\delta_s$ of the reconstruction:

\begin{equation*}
    \delta_s=\frac{\rho-\bar{\rho}}{\bar{\rho}},
\end{equation*}

where $\bar{\rho}$ refers to the mean density of the survey area considered (in this case, over a scale of $200 \frac{Mpc}{h}$ or $z \sim 0.07$). Note that, according to this definition, the averaged density contrast over the entire scale of the survey, and therefore the averaged divergence of the velocity field, should be zero. If we eliminate the real density $\rho$ from the two equations, we can obtain the relation:

\begin{eqnarray}\label{bias}
    (\bar{\delta}+1)\delta_s+\bar{\delta}=\delta
\end{eqnarray}

Where $\bar{\delta}$ is the mean density contrast of the $200 \frac{Mpc}{h}$ scale w.r.t. to the density of the universe:

\begin{eqnarray}
    \bar{\delta}=\frac{\bar{\rho}-\rho_0}{\rho_0}
\end{eqnarray}

Therefore, $\bar{\delta}$ quantify the deviation that $\delta_s$ has from the real value $\delta$. We will consider this possible effect in our analysis.

Overall, is important to note that according to \citealp{Carrick_2015} in section 5.3.4, the authors state that the average density contrast over different regions of the survey is consistent with some observational results in \citealp{Whitbourn2014,Bohringer2014}. Additionally, it is worth remembering that this velocity field has already been used to correct peculiar velocities in the last Pantheon+ compilation, which is crucial for observational cosmology. Therefore, we conclude that while there could be a bias in the data that we need to consider carefully, this reconstruction field is a good starting point for the extraction of interesting parameters for LLSS research.

\begin{figure*}[!ht]
    \centering
    \includegraphics[width=15cm]{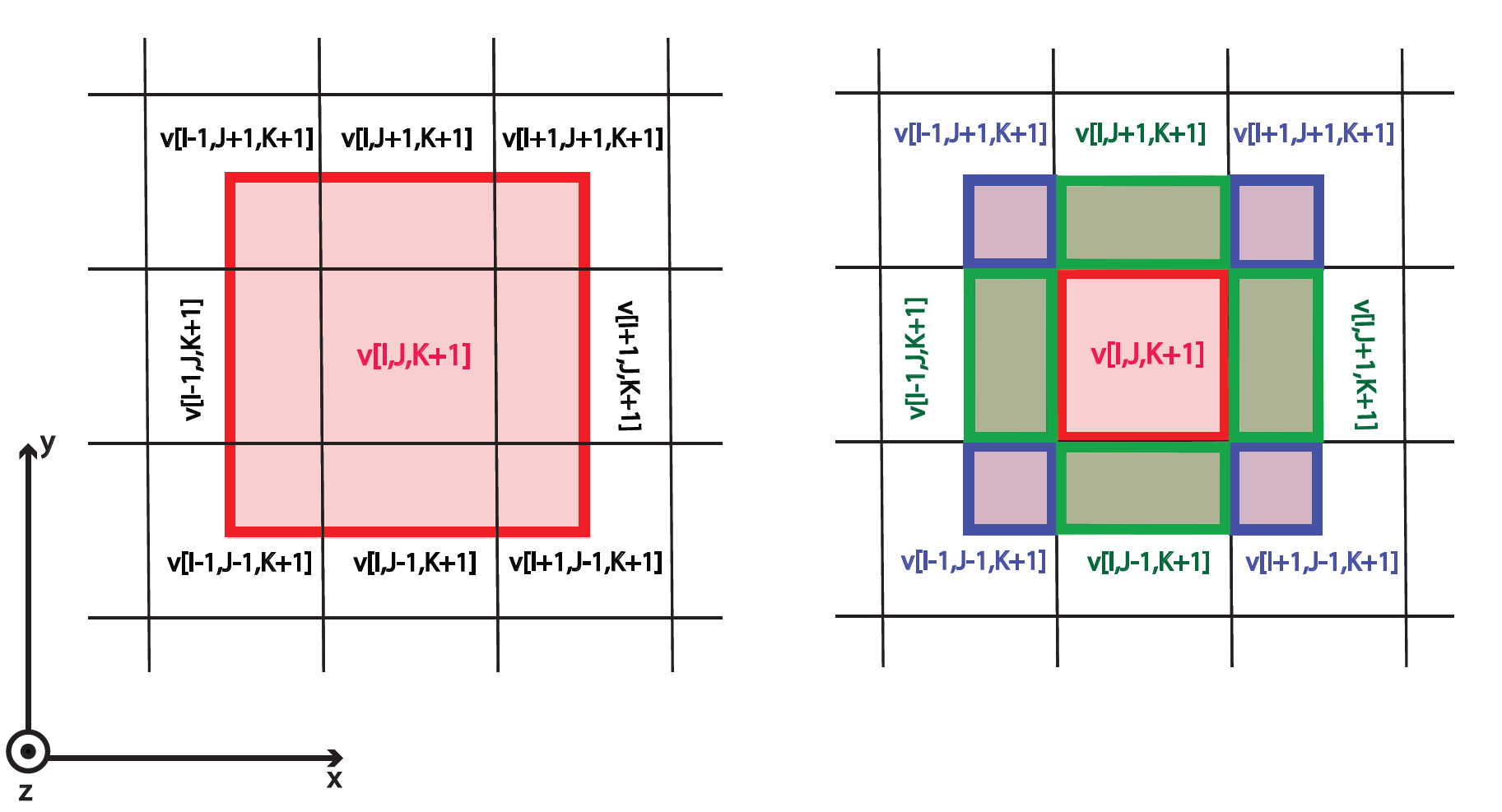}
    \caption{Graphic representation of the upper face of a box with side $2s$ enclosing the pixel $[I,J,K]$. The finite difference method approximate the flux across this surface as simply the contribution of $\mathbf{v}[I,J,K+1]$ over it (left). Meanwhile, the integral approximation technique considers the contributions of the side and of the diagonal pixels as well (right).}
    \label{Grid}
\end{figure*}


%

\subsection{Dependence of the reconstruction in background cosmology}

As the report states, the reconstruction procedure was performed using a complex algorithm using an iterative procedure. This procedure was found to be unbiased in determine the best fitting value of $\beta$ when performed over simulations on $\Lambda$CDM. However, it is important to ask if the data is useful to compute fluid dynamics parameters in backgrounds with values of $\Lambda$ different than the used in the reconstruction. The cosmological dependence of the algorithm comes in the step number (vi) (see section 2.5 of \citealp{Carrick_2015}) which uses the approximation: 

\begin{equation}
    R \sim \frac{c}{H_0}\left (z - \frac{1+q_0}{2}z^2 \right )
\end{equation}

to predict co-moving distance values of $R$. Here $z$ is the cosmological redshift. We ignore which values of $q_0$ and $H_0$ where used to perform the algorithm, but we can study an scale impact of the change between a EdS cosmology with $q_0=0.5$ an a $\Lambda$CDM cosmology with $q_0=-0.3$. Using the the maximum redshift of the survey area as $z\sim 0.07$, we can compute an approximated variation on the distance scale using this 2 cosmologies:

\begin{eqnarray*}
    R_{(0.5)} \propto 0.066 \\
    R_{(-0.3)} \propto 0.068
\end{eqnarray*}

As the divergence scales with magnitude of $\tilde\theta\sim v/R^3$ where $v$ is the peculiar velocity, an approximated variation in the computed divergence value is:

\begin{equation}
    \frac{\tilde{\theta}_{(0.5)}}{\tilde{\theta}_{(-0.3)}}=\frac{0.068^3}{0.066^3}\sim 1.09
\end{equation}

This means that the magnitude of the divergence could be a $9\%$ higher in a Eds universe than a $q_0=-0.3$ one (this is expected as dark energy tend to expand fast the distance scales, lowering the estimations on divergence). As is important to consider this possible bias in the data, note that it affects the magnitude and not the sign of the divergence, which is the most important feature to measure. Even if a model with $q_0=-0.3$ was used in the reconstruction, the values of the divergence should be greater in magnitude in a $EdS$ universe compared with the computed in this work, highlighting our results. We conclude that this reconstruction is useful to extract the approximated values of the divergence of the peculiar velocity field as it is to correct peculiar redshifts on data used to constrain cosmological parameters.

\section{Divergence reconstruction}\label{Meth}

\subsection{Finite differences}

We use the central finite difference method to compute derivatives in each pixel of the data-cube as:
\begin{eqnarray}
    \nabla \mathbf{v}=\partial_j v_i\approx\frac{v_i(x_j+s )-v_i(x_j-s)}{2s}\,, \label{finitedifference}
\end{eqnarray}
where $x_i$ is the central point of a pixel $(I,J,K)$ of the array. Also, $s$ is the physical size of a pixel, so we can use directly the cube data to ensure the right conversion of a pixel to the physical value, which fortunately is the same for each coordinate. For the case of the velocity field used here, with $257^3$ pixels in a datacube of length $400/h$ (where $h$ is the dimensionless normalised Hubble parameter), we have:

\begin{eqnarray}
    s = \frac{400 Mpc}{257h}\,,
\end{eqnarray}

Neglecting the borders of the sample, leads to a $255^3$ array containing each pixel in a $3\times3$ matrix that corresponds to the gradient tensor of the peculiar velocity field. For a central finite difference approximation of a function $f$, one may write:
\begin{eqnarray}
\begin{split}
   f'(x) &= \frac{f(x+s)-f(x-s)}{2s}-\frac{s^2}{6}f'''(\xi)\\ &=f'_1(x)-\epsilon\,,
\end{split}
\end{eqnarray}
for some $\xi \in [x-s,x+s]$. In the above, $f'_1(x)$ is the central approximation for the derivative and $\epsilon \propto s^2f'''(\xi)$ is the truncation error. 



\subsection{Integral Approximations}

A direct approximation of the divergence could be computed recalling the definition of the operator:

\begin{equation}
        \nabla \cdot \mathbf{v}=\lim_{V\rightarrow 0} \frac{1}{V}\oiint_{\partial V}\mathbf{v}\cdot d\mathbf{A}\,.
\end{equation}
In so doing, we choose a box-like volume of size $2s$ around the central point of each pixel and compute the flux of the velocity field through the box. Then, by dividing the flux over the volume, we can extract an approximate value for the divergence. Note the difference with the central finite difference method in equation (\ref{finitedifference}), as this approach ignores the contribution over diagonal pixels (see Figure~\ref{Grid}) .

\subsection{Theoretical estimation}

Following equation (\ref{velrec}) the divergence of the velocity field and the density contrast are also related by the linear expression:
\begin{equation}
    \nabla \cdot \mathbf{v}(\mathbf{r})=\frac{\beta}{4\pi}\int^{R_{max}}d^3\mathbf{r'}\delta_g(\mathbf{r'})\nabla \cdot \frac{\mathbf{r'}-\mathbf{r}}{     \lvert \mathbf{r'}-\mathbf{r}\rvert^3 }=-\beta \delta_g(\mathbf{r})\,.  \label{exactdiv}
\end{equation}
Note that, as the $\mathbf{r}$ coordinate is measured in $km/s$, to express the divergence in $\frac{km/s}{Mpc/h}$ units, we need to multiply this quantity by a factor of $100h^2$. 

\section{Results}\label{Res}

We have computed the gradient matrix using the \texttt{Numpy} package from \texttt{Python} to manipulate matrix and arrays. The following three methods of estimating the volume scalar have been used:
\begin{enumerate}
    \item A decomposition of the full gradient tensor of the velocity field employing finite differences.
     \item A integration approximation for the divergence using a box volume around the pixels.
     \item A theoretical estimation by means of relation (\ref{exactdiv}).
\end{enumerate}

\begin{figure*}[ht]
    \centering
    \includegraphics[width=9cm]{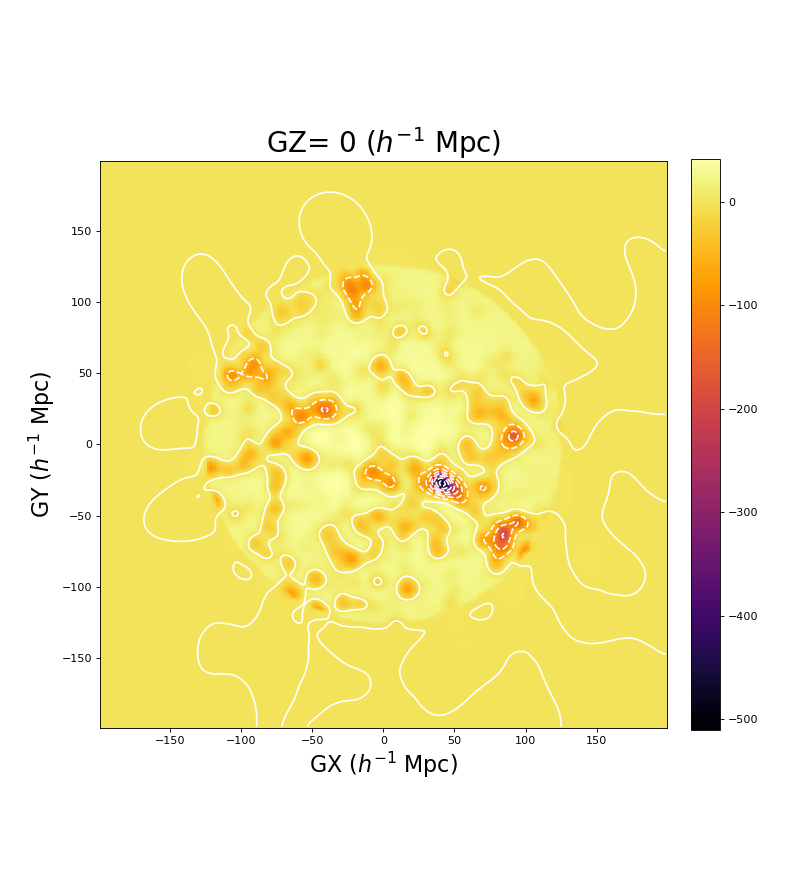}
    \includegraphics[width=9cm]{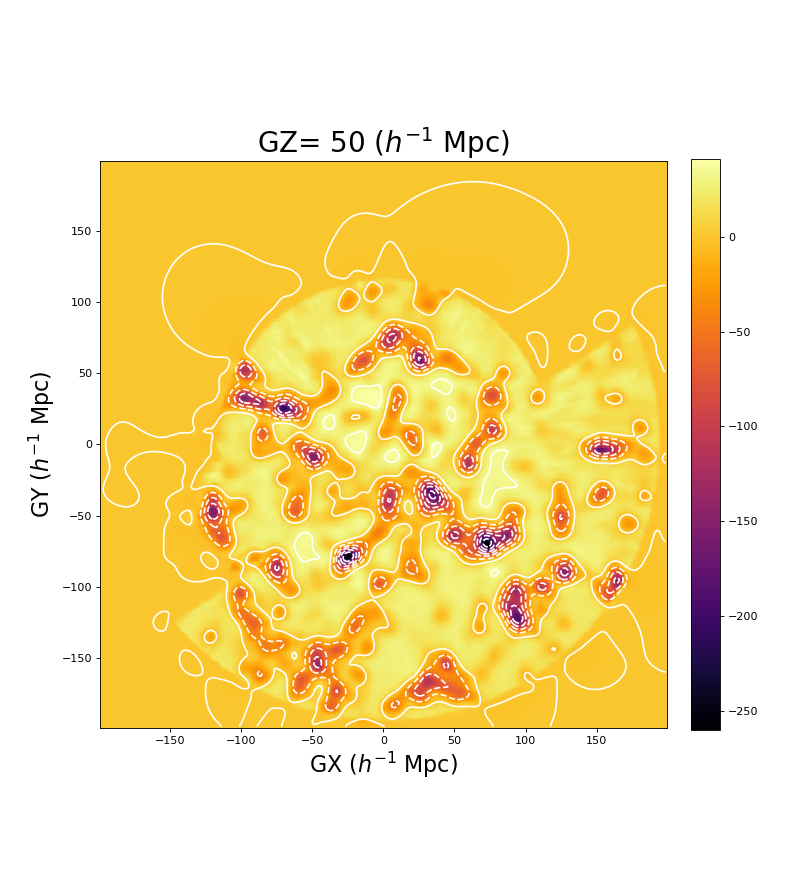}
    \caption{Divergence of the velocity field  in $\frac{km/s}{Mpc/h}$ using the finite difference approximation, projected over $GZ=0$ and $GZ=50 h^{-1}Mpc$ galactic planes.}
    \label{Divergence1}
\end{figure*}

\begin{figure*}[ht]
    \centering
    \includegraphics[width=9cm]{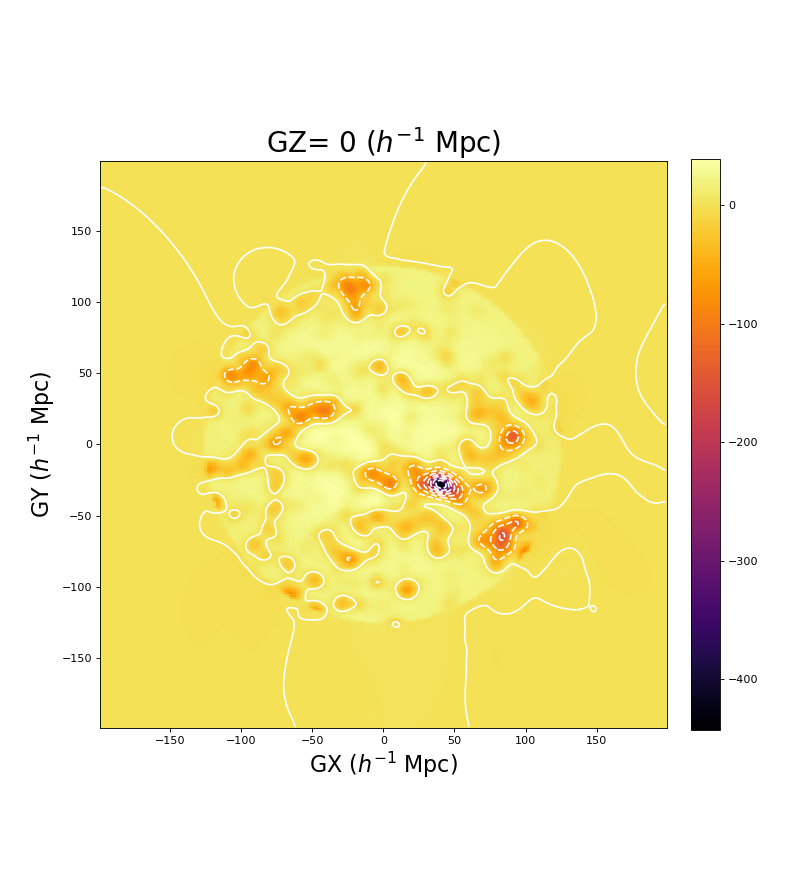}
    \includegraphics[width=9cm]{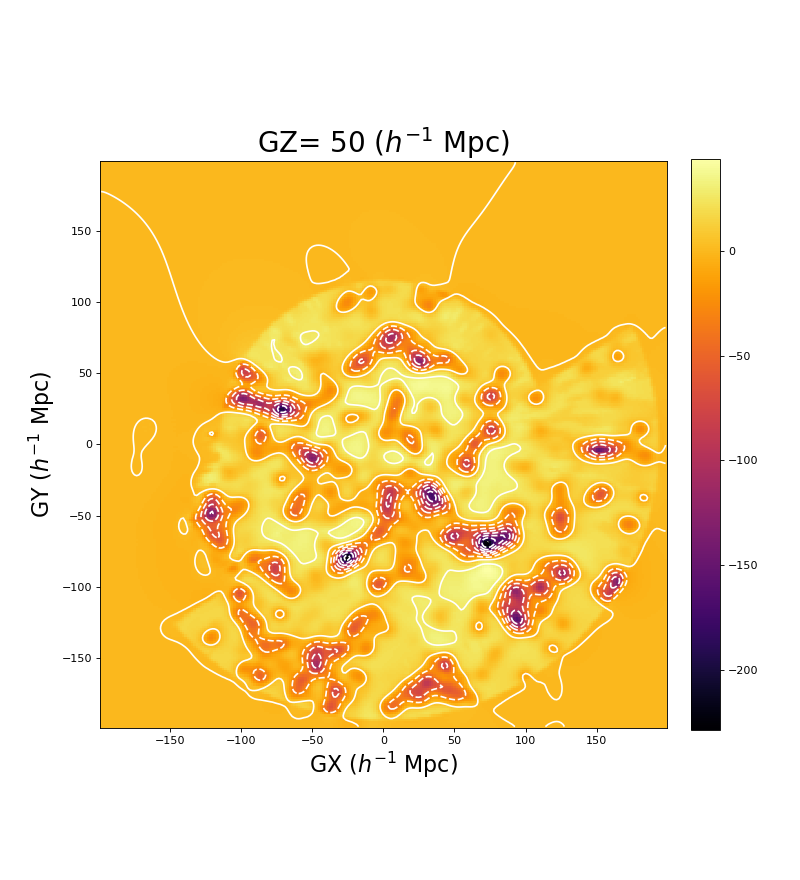}
    \caption{Divergence of the velocity field  in $\frac{km/s}{Mpc/h}$ using the integral approximation projected over $GZ=0$ and $GZ=50 h^{-1}Mpc$ galactic planes.}
    \label{Divergence2}
\end{figure*}

\begin{figure*}[ht]
    \centering
    \includegraphics[width=9cm]{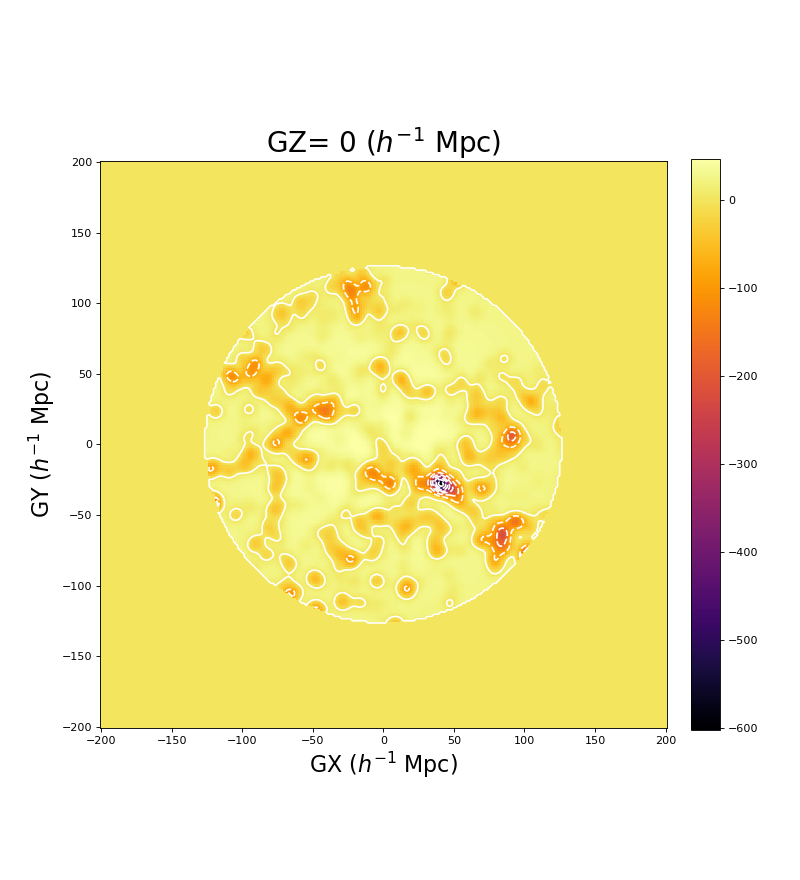}
    \includegraphics[width=9cm]{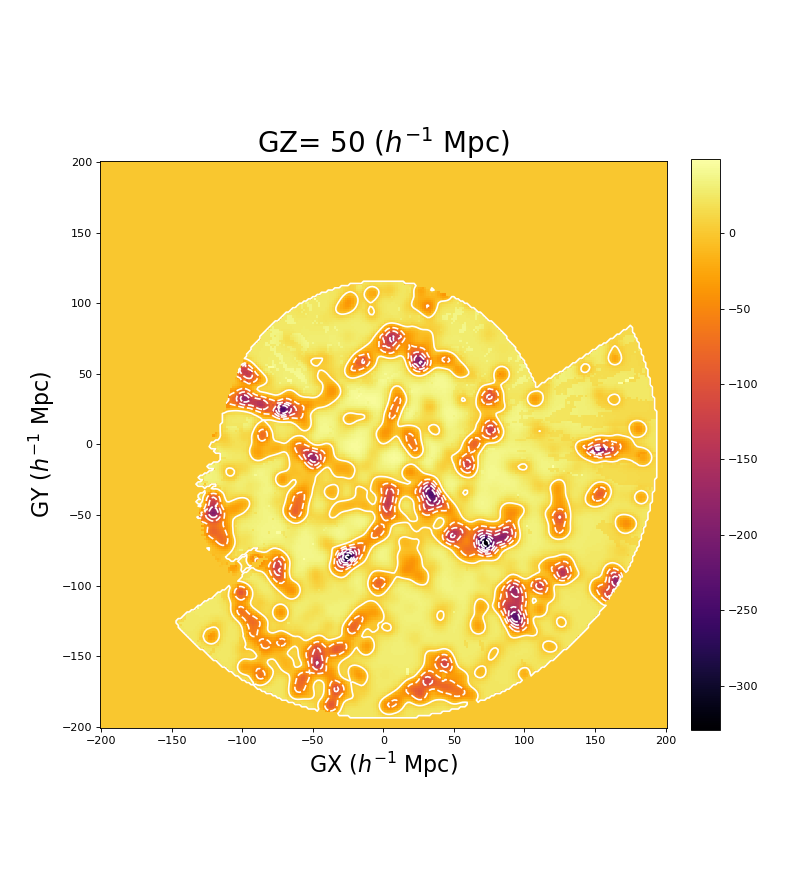}
    \caption{Divergence of the velocity field  in $\frac{km/s}{Mpc/h}$ using theoretical computation, projected over $GZ=0$ and $GZ=50 h^{-1}Mpc$ galactic planes.}
    \label{Divergence3}
\end{figure*}


\begin{figure*}[ht]
    \centering
    \includegraphics[width=9cm]{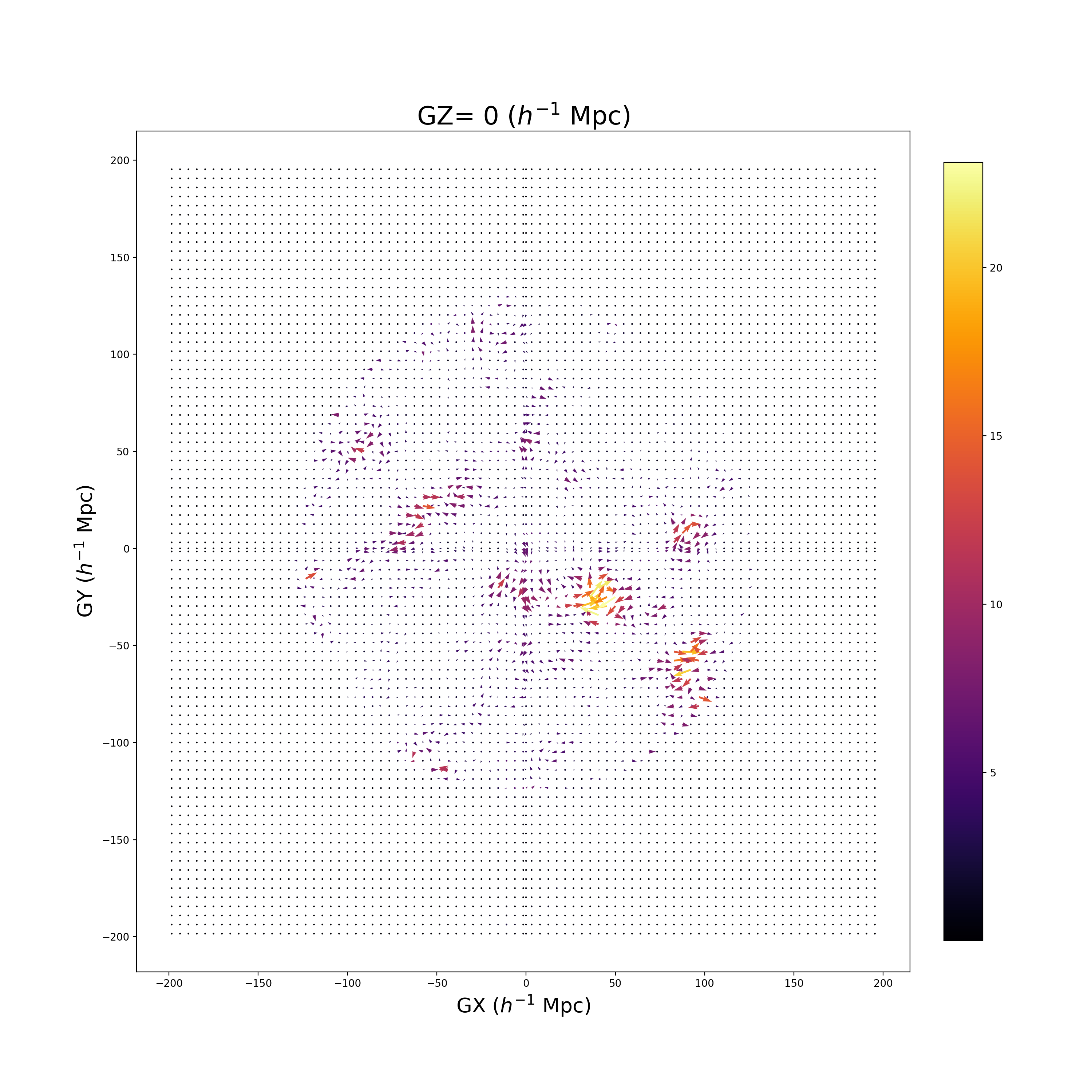}
    \includegraphics[width=9cm]{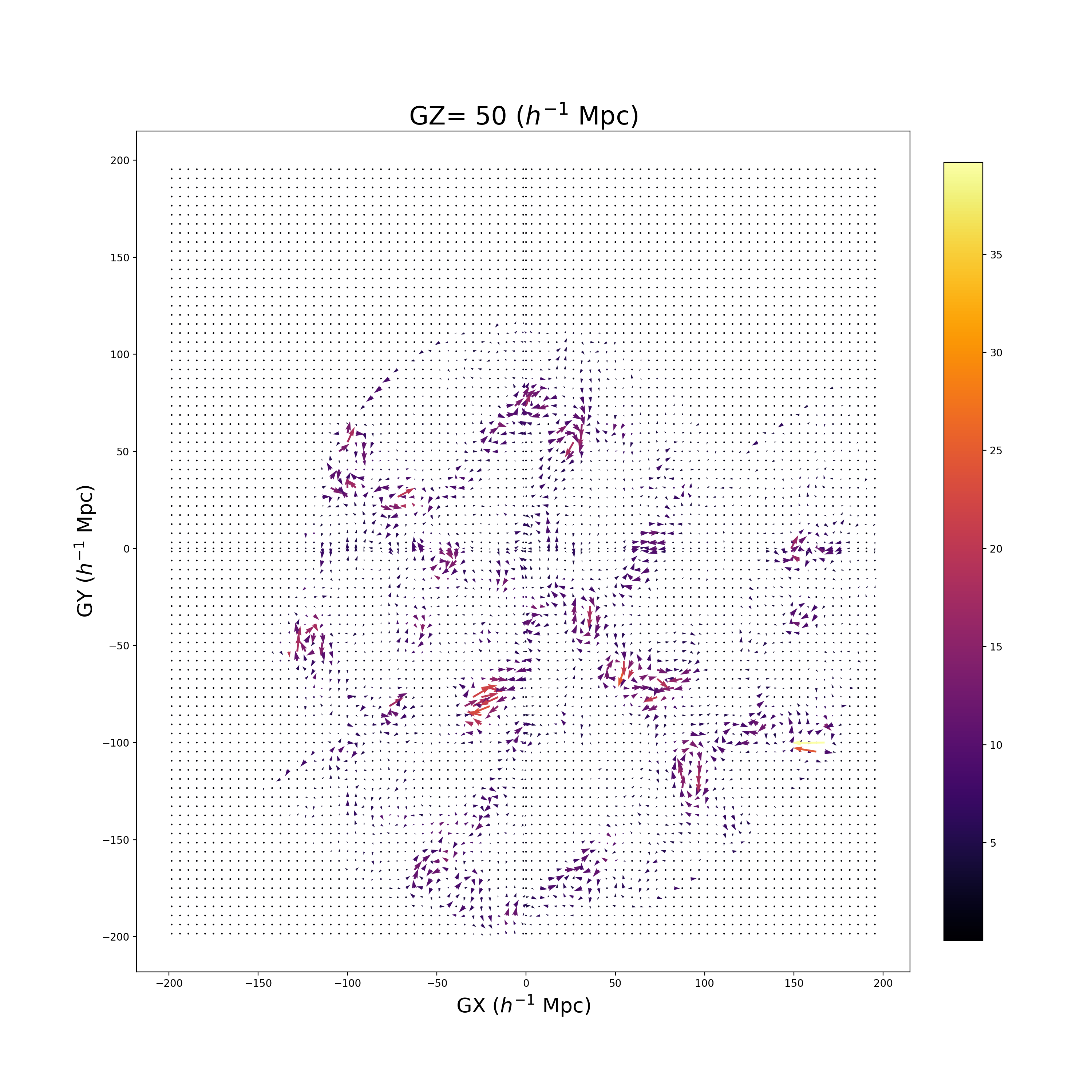}
    \caption{Residual curl vector field in $\frac{km/s}{Mpc/h}$ units from finite difference method, projected over $GZ=0$ and $GZ=50 h^{-1}Mpc$. Note that this is more a numerical residual rather than a physical meaningful result.}
    \label{CurlVec}
\end{figure*}

\begin{figure*}[ht]
    \centering
    \subfloat[X-Plane]{\includegraphics[width=8.5cm]{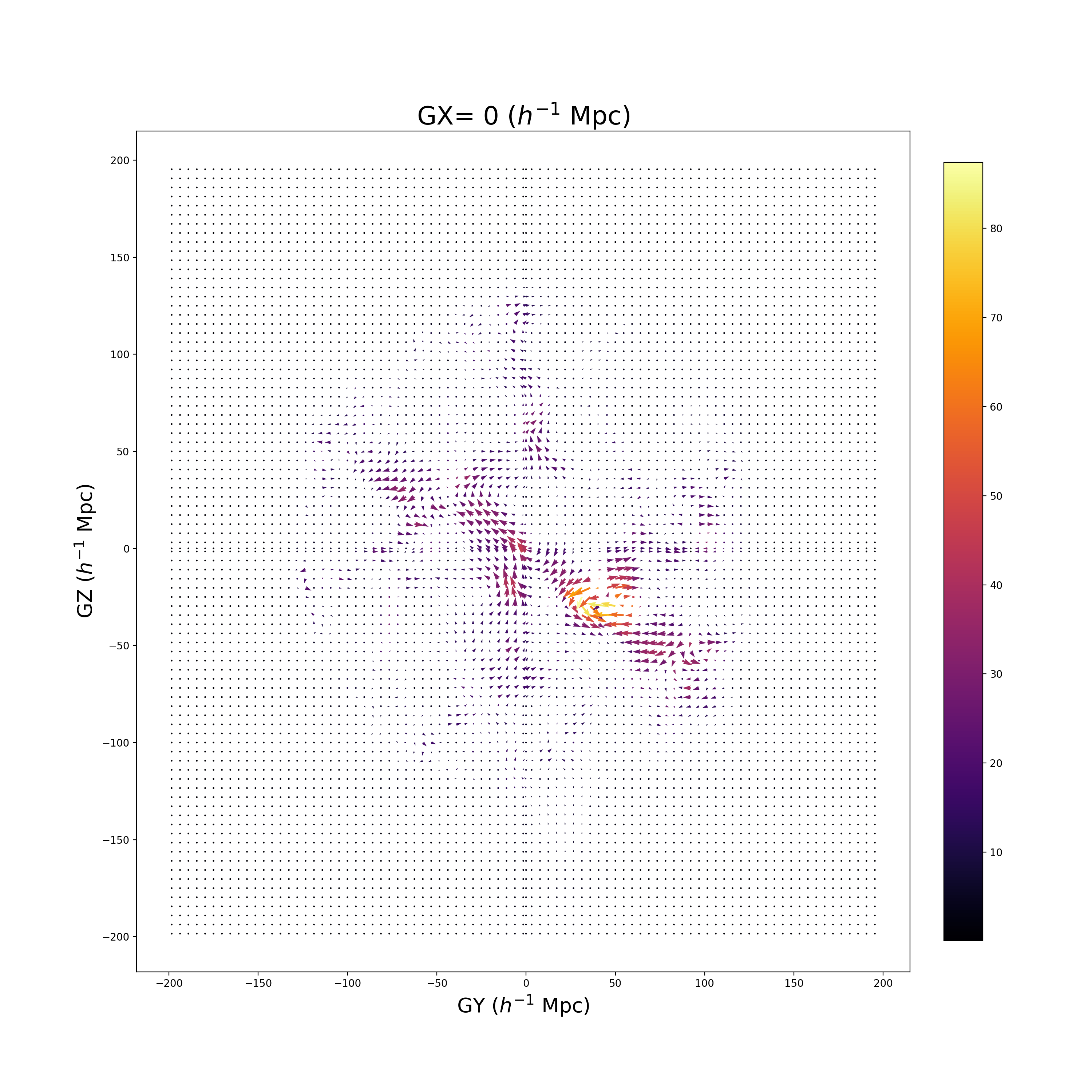}}
    \subfloat[Y-Plane]{\includegraphics[width=8.5cm]{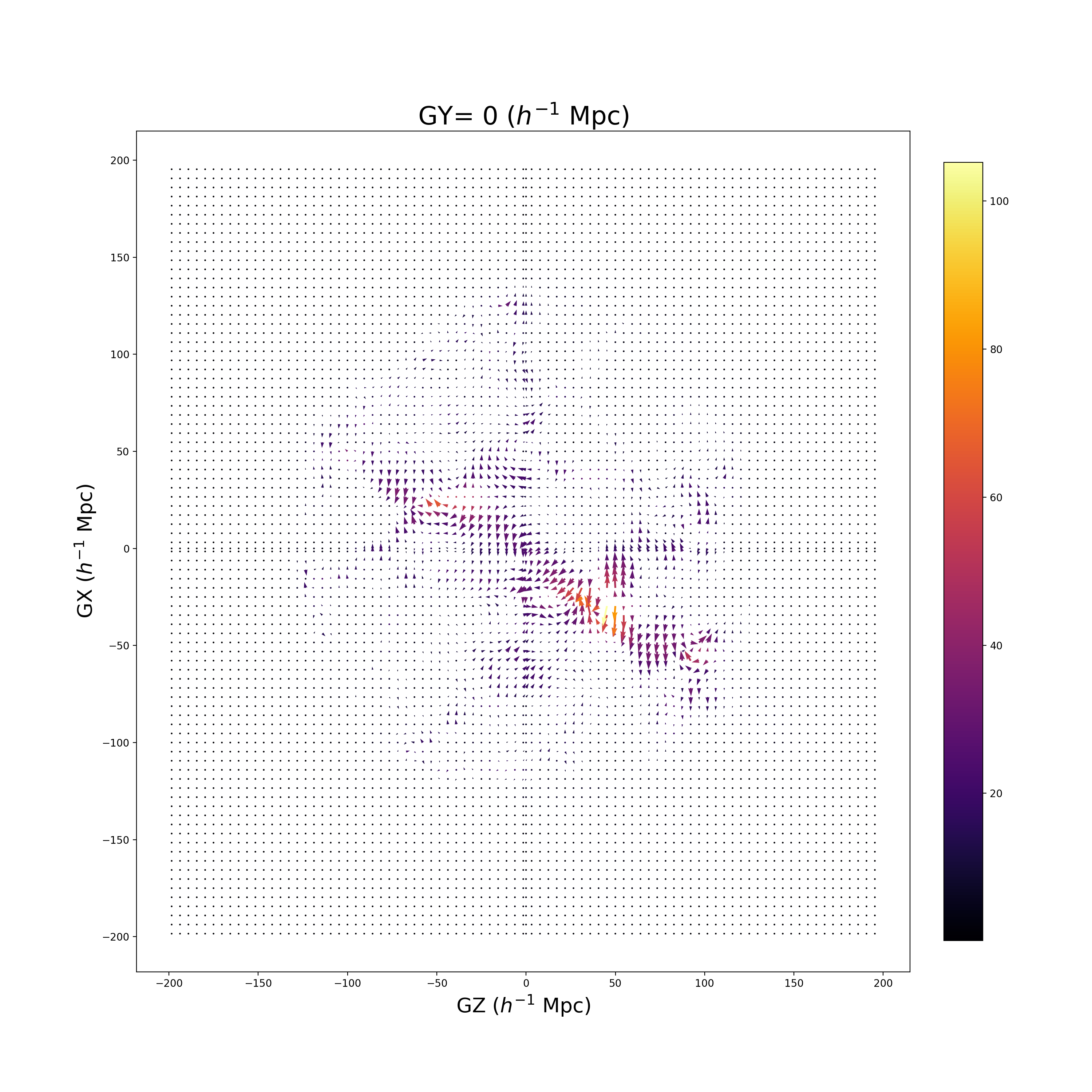}}\hspace{1em}
    \subfloat[Z-Plane]{\includegraphics[width=8.5cm]{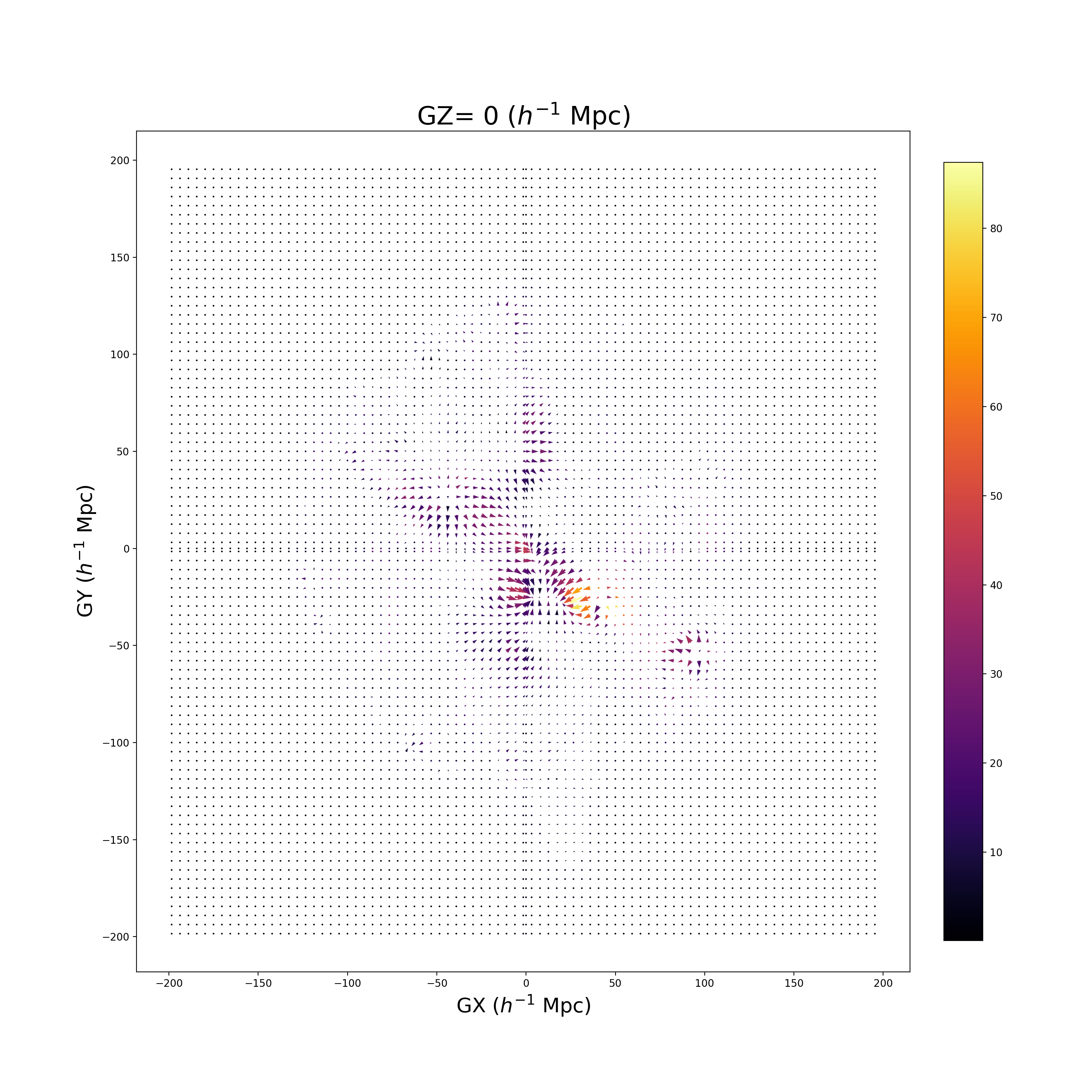}}
    \caption{Projections of the shear tensor in $\frac{km/s}{Mpc/h}$ units estimated with finite difference method over Cartesian planes that pass on the origin.}
    \label{Shear}
\end{figure*}

The results obtained via these different methods are plotted in Figures ~\ref{Divergence1},\ref{Divergence2} and \ref{Divergence3}. To relate the latter with the tilted cosmology scenario, we need to estimate an average value for $\tilde{\theta}$ and thus put the predictions of the tilted model to the test. In our analysis, this corresponds to an average divergence of the entire fluid, which is then averaged over a spherical volume $V=4\pi\lambda^3/3$ as:
\begin{equation}
    \Tilde{\theta} = \frac{1}{V} \int_{V}(\nabla \cdot \mathbf{v}) dV \approx \frac{s^3}{V} \sum_{i}(\nabla \cdot \mathbf{v})_i\,,
\end{equation}
where the sum is over the pixels that reside inside a sphere of radius $\lambda$. 

Substituting $\tilde{\theta}$ into the right-hand side of equation~(\ref{tq2}), we can compute representative estimates of the local deceleration parameter ($\Tilde{q}$) measured by the bulk-flow observers on different scales ($\lambda$). The results, which assign negative values to $\tilde{q}$ on scales up to $200~Mpc$ through all three estimation methods, are summarized in Table~\ref{tabla1}. Note that we have set $h\simeq0.7$ in all cases. Also, although (\ref{tq2}) holds in essentially all tilted FRW models, here  we have assumed an Einstein-de Sitter background (with $q=0.5$) for mathematical simplicity. It seems that the theoretical estimation provides the higher values for $\Tilde{\theta}$, while the integral approximation gives the lowest. What is most important, however, is that all three methods are consistent both in the sign and in the magnitude of $\tilde{\theta}$. 

\subsection{Bias on the averaged density contrast}

According to equation (\ref{bias}), the difference between the mean density used to normalize the survey and the real mean density of the universe could introduce a significant bias in the results. If we express this relation in terms of divergence with a proper unit scaling, we obtain:

\begin{equation}
    \theta=(\bar{\delta}+1)\theta_s-100h^2f\bar{\delta}
\end{equation}

Where $\theta$ is the real divergence, $\theta_s$ is the divergence estimated from the reconstruction, and $f\sim \Omega_m^\gamma$ is the growth factor, which is approximately 1 for an EdS universe. Both divergences are in units of $\frac{km/s}{Mpc/h}$. A positive value of $\bar{\delta}$ indicates that the survey area coverage is inside an over-density region. In this case, the estimated divergence tends to be more negative, which supports the idea of a slightly contracting bulk flow. However, a negative $\bar{\delta}$ would increase the divergence to positive values and therefore should be carefully analyzed.

We can estimate the effects of this possible underdensity by assuming an extreme value for $\bar{\delta}$. For example, according to \citealp{Haslbauer_2020}, for a scale $\sim 300 Mpc$ we can estimate the value of $\bar{\delta} \gtrsim -0.032$ from cosmic variance for a fiducial $\Lambda$CDM cosmology. The deviation $\Delta \theta$ for the estimated divergence from the real value is approximately:

\begin{equation*}
    \Delta \theta \lesssim 1.6 \frac{Mpc}{h}
\end{equation*}

According to Table \ref{tabla1}, cosmic variance could strongly modify the values of the average divergence and even change its sign for scales $\lambda > 125 \frac{Mpc}{h}$. Also, the impact of this possible bias in the values of $q$ is strong. For example for $q$ at $\sim 125 Mpc/h$ computed with Method C:

\begin{equation*}
    \Delta q \lesssim 1.03
\end{equation*}

However, it's important to note that this is an extreme value of cosmic variance, and it's more likely that the deviation with respect to the mean density of the universe $\bar{\delta}$ is more positive than $-0.032$. Furthermore, a value of $\bar{\delta}$ that is much more negative than this cosmic variance could represent a problem for $\Lambda$CDM, as suggested in \citealp{Haslbauer_2020}. Overall, we emphasize the necessity of broader surveys in order to better constrain the values of $\bar{\delta}$ and therefore $\theta$, as the bias is completely dependent of the chosen cosmology and doesn't have a true observational origin yet.

\subsection{Uncertainties}

We have identified both controlled and uncontrolled uncertainties in our estimations. In the former group we have the fit uncertainties for the reconstruction parameters $\beta$ and $\mathbf{V}_{ext}$. Of those two, we are mainly interested in $\beta$, given that $\mathbf{V}_{ext}$ does not enter the gradient calculation. Then, if we define the divergence of the relative velocity field $\mathbf{v}_{rec}$ as $\Tilde{\theta}_{rec}$, we have:
\begin{equation}
\Tilde{\theta}=\beta\Tilde{\theta}_{rec}\,,
\end{equation}
while the uncertainty in $\Tilde{\theta}$ due to the $\beta$ parameter can be written as:
\begin{equation}
\Delta\Tilde{\theta}_\beta=\Tilde{\theta}_{rec} \Delta\beta\,.
\end{equation}
This is the uncertainty recorded in Table~\ref{tabla1}. 

Turning to the uncontrolled uncertainties, we can group different possible systematic effects coming from the reconstruction process, as well as errors between approximations and real values. A detailed summary of the first type can be found in~\citealp{Carrick_2015}.  With regard to the approximation errors, we can estimate the precision of the estimation by comparing to the theoretical result. In this respect, the finite difference method seems more precise than the volume integration method, as it is closer to the theoretically predicted values. Moreover, according to relation (\ref{Reconstruction}), the velocity field should be irrotational as the field is proportional to a Newtonian gravity potential in the linear regime. However, when the anti-symmetric part of the gradient tensor is computed we got a non-zero value, leading to a residual low vorticity term that could be related with a deviation of the finite difference method with respect to theoretical estimation. A symmetric trace-less part of the gradient can also be computed via finite difference method. Residual Curl and projections of the estimated Shear are plotted in Figures \ref{CurlVec} and \ref{Shear}. 

\begin{sidewaystable}[ht]
\centering
\begin{tabular}{@{}lccc@{}}
$\lambda(\frac{Mpc}{h})$ &$\sim\delta(EdS)$ &$\Tilde{\theta}(\frac{km/s}{Mpc/h})$ & $\Tilde{q}$ \\
\hline
\midrule
 \textbf{Method A}\\
 \hline
70  &-&  $-3.36\substack{+0.16\\-0.16}\;(-2.65\substack{+0.37\\-0.24})\;[-2.45\substack{+0.37\\-0.24}]$ & $-6.36\substack{+0.33\\-0.33} (-4.93\substack{+0.75\\-0.49})[-4.50\substack{+0.75\\-0.49}]$\\
100  &-& $-2.77\substack{+0.13\\-0.13}\; (-2.19\substack{+0.30\\-0.20})\;[-2.02\substack{+0.30\\-0.20}]$ & $-2.27\substack{+0.13\\-0.13} (-1.69\substack{+0.30\\-0.20})[-1.52\substack{+0.30\\-0.20}]$\\
125  &-& $-1.48\substack{+0.07\\-0.07}\;(-1.17\substack{+0.16\\-0.11})\;[-1.08\substack{+0.16\\-0.11}]$ & $-0.45 \substack{+0.05\\-0.05}(-0.25\substack{+0.10\\-0.07})[-0.19\substack{+0.10\\-0.07}]$\\
150  &-& $-0.65\substack{+0.03\\-0.03}\;(-0.51\substack{+0.07\\-0.05})\;[-0.47\substack{+0.07\\-0.05}]$ & $+0.21\substack{+0.01\\-0.01} (+0.27\substack{+0.03\\-0.02})[0.29\substack{+0.03\\-0.02}]$\\
200  &-& $-0.24 \substack{+0.01\\-0.01}\;(-0.19\substack{+0.03\\-0.02})\;[-0.17\substack{+0.03\\-0.02}]$ & $+0.44\substack{+0.005\\-0.005} (+0.45\substack{+0.01\\-0.005})[0.46\substack{+0.01\\-0.005}]$\\ 
\hline
\textbf{Method B}\\
\hline
70  &-& $-2.45\substack{+0.12\\-0.12}\;(-1.94\substack{+0.27\\-0.18})\;[-1.78\substack{+0.27\\-0.18}]$ & $-4.5 \substack{+0.24\\-0.24}(-3.46\substack{+0.55\\-0.36})[-3.14\substack{+0.55\\-0.36}]$\\
100  &-& $-1.99\substack{+0.10\\-0.10}\;(-1.57\substack{+0.22\\-0.14})\;[-1.45\substack{+0.22\\-0.14}]$ & $-1.49 \substack{+0.10\\-0.10}(-1.07\substack{+0.22\\-0.14})[-0.95\substack{+0.22\\-0.14}]$\\
125  &-& $-1.13\substack{+0.06\\-0.06}\;(-0.90\substack{+0.12\\-0.08})\;[-0.82\substack{+0.12\\-0.08}]$ & $0.22\substack{+0.04\\-0.04} (-0.07\substack{+0.08\\-0.05})[-0.03\substack{+0.08\\-0.05}]$\\
150  &-& $-0.47\substack{+0.02\\-0.02}\;(-0.37\substack{+0.05\\-0.03})\;[-0.34\substack{+0.05\\-0.03}]$ & $+0.29\substack{+0.01\\-0.01} (+0.33\substack{+0.02\\-0.02})[0.35\substack{+0.02\\-0.02}]$\\
200  &-& $-0.21 \substack{+0.01\\-0.01}\;(-0.17\substack{+0.02\\-0.02})\;[-0.15\substack{+0.02\\-0.02}]$ & $+0.45\substack{+0.005\\-0.005} (+0.46\substack{+0.01\\-0.005})[0.46\substack{+0.01\\-0.005}]$\\
\hline
\textbf{Method C}\\
\hline
70  &$0.080(0.064)[0.059]$ & $-3.94\substack{+0.19\\-0.19}\;(-3.11\substack{+0.43\\-0.28})\;[-2.87\substack{+0.43\\-0.28}]$ & $-7.54 \substack{+0.39\\-0.39}(-5.86\substack{+0.88\\-0.58})[-5.36\substack{+0.88\\-0.058}]$\\
100  &$0.065(0.051)[0.047]$ & $-3.17\substack{+0.15\\-0.15}\;(-2.50\substack{+0.35\\-0.23})\;[-2.31\substack{+0.35\\-0.23}]$ & $-2.67\substack{+0.15\\-0.15} (-2.01\substack{+0.35\\-0.23})[-1.81\substack{+0.35\\-0.23}]$\\
125  &$0.034(0.027)[0.025]$ & $-1.66\substack{+0.08\\-0.08}\;(-1.31\substack{+0.18\\-0.12})\;[-1.21\substack{+0.18\\-0.12}]$ & $-0.56\substack{+0.05\\-0.05} (-0.34\substack{+0.12\\-0.08})[-0.27\substack{+0.12\\-0.08}]$\\
150  & $0.016(0.012)[0.011]$& $-0.76\substack{+0.04\\-0.04}\;(-0.59\substack{+0.08\\-0.05})\;[-0.55\substack{+0.08\\-0.05}]$ & $+0.16\substack{+0.02\\-0.02} (+0.23\substack{+0.04\\-0.02})[0.25\substack{+0.04\\-0.02}]$\\
200  &$0.006(0.005)[0.004]$ & $-0.29 \substack{+0.01\\-0.01}\;(-0.23\substack{+0.03\\-0.02})\;[-0.21\substack{+0.03\\-0.02}]$ & 
$+0.43\substack{+0.005\\-0.005} (+0.44\substack{+0.01\\-0.01})[0.45\substack{+0.01\\-0.01}]$\\
\end{tabular}
\caption{Representative values for $\Tilde{\theta}$ and $\Tilde{q}$ on different scales ($\lambda$) assuming a $EdS$ universe, using $\beta$ as it is in the data-cube with the finite difference approximation, integral approximation and theoretical estimation. In brackets are the values obtained after using 2 different values of $\beta$ from Pantheon+ (round) and the ones in \citealp{Said_20} [square]. Approximations for the density contrast $\delta$ also are shown.  Note that, for numerical simplicity and demonstration purposes, we have set $q=0.5$ in the CMB frame and $h\simeq0.7$.}
\label{tabla1}
\end{sidewaystable}

\subsection{Velocity dispersion}

 In ~\citealp{Carrick_2015}, authors gave a velocity dispersion parameter of $\sigma_v=150 km/s$ for particles and $\sigma_v=250 km/s$ for haloes. Therefore, we have explore how this dispersion impacts the values of the estimated divergence in our 3 different methods. In the 2 first methods we used directly the velocity field reconstructed, so we expect velocity dispersion being important in those cases. In the following we assume that the velocity dispersion estimated is the same in each one of the 3 degrees of freedom.

 If we use a linear approximation for the error in finite difference method, we note that the dispersion contribute 6 times as we subtract 2 velocity values and then we sum the 3 degrees of freedom to compute the divergence. Therefore the uncertainty on the value of the divergence using $\sigma_v\sim 250 km/s$ ended being an humongous value:

 \begin{equation}
     \Delta \tilde\theta =\frac{6\sigma_v}{2s}\sim 360 \frac{km/s}{Mpc}
 \end{equation}

Interestingly, using the same approach with the second method we get the same result: as the dispersion is the same for each cell, we sum this value multiplied over the area $A = (2s)^2$ of the cube 6 times. Then divided by the volume $V=(2s)^3$:

\begin{equation}
     \Delta \tilde\theta =\frac{6\sigma_v A}{V}\sim 360 \frac{km/s}{Mpc}
 \end{equation}

We note that using just a linear approximation of the errors, those values exceed notably the values for the estimated divergence and represent a potential problem for our work. However, for the third method we didn't use the velocity field: the relation \ref{exactdiv} works \textbf{directly with the density contrast and not with the velocity field reconstruction}. As we didn't use directly the velocity field, we expect than \textbf{velocity dispersion doesn't contribute to the error in the estimated divergence using this method} \footnote{As the values estimated by the third method are of the same order of magnitude than the first 2 numerical approximations,
we suspect that velocity dispersion should not be treated directly as a source of error and the inclusion in the numerical methods should be more complex. This is an interesting debate that exceed the scope of this work.}. We can conclude that the main results of this paper are fully trustworthy regardless of the velocity dispersion.

\section{Discussion}
We have estimated the average volume scalar of the reconstructed peculiar velocity of the local universe via different methods. The volume scalar is related to the divergence of the velocity field. This is so because the velocity divergence measures the change in the local volume of the associated bulk flow and therefore its tendency to locally expand or contract. Then, a positive divergence implies that the fluid tends to expand locally, whereas a negative one indicates a contracting region. We have plotted the divergence scalar for different galactic planes in Figure \ref{Divergence1}. There, one can see that the peculiar velocity divergence is highly negative in regions where the density contrast is high, while it is positive in regions where matter content is low. This is to be expected, of course, given the attractive nature of gravity. At this point, it also helps to recall the familiar divergence theorem:
\begin{equation}
        \oint_{V}(\nabla \cdot \mathbf{v}) dV=\oint_{\partial V}\mathbf{v}\cdot d\mathbf{A}\,.
\end{equation}
Integrating the divergence over the region $V$ reveals whether the latter contracts or expands, as the right-hand side of the equation represents the fluid fraction that "enters" or "goes out" of the volume surface $\partial V$ over time. 

Surprisingly, the values of the local volume scalar ($\tilde{\theta}$) associated with the reconstructed peculiar velocity field, were found negative over a range of scales and by means of different estimation methods. This result has direct implications for the tilted cosmological scenario~\citealp{Tsagas2011,Asvesta22}. The latter predicts that observers living in contracting bulk peculiar flows could measure a negative deceleration parameter locally, even when the universe is decelerating globally~\citealp{Tsagas2015,Tsagas2021,Tsagas2022}. Also, as predicted, we found that the impact of the observer's peculiar motion becomes stronger on progressively smaller scales, namely closer to the observer, while it decays away from them (see Table~~\ref{tabla1}). The transition length ($\lambda_T$), that is the maximum scale where the local deceleration parameter appears to cross the $\tilde{q}=0$ mark and turn negative, also depends on the observer's position inside the bulk flow. Following (\ref{lambdaT}), for observes residing within $70/h$~Mpc from the centre of the bulk flow, we find $\lambda_T\gtrsim360$~Mpc, $\lambda_T\gtrsim310$~Mpc and $\lambda_T\gtrsim390$~Mpc, when adopting the Finite Difference method, the Integral Approximation method and the Discrete Density Integration method respectively. Overall, the closer the observer is to the bulk-flow centre, the more negative the local value of $\tilde{\theta}$ and the larger the associated transition length. 
According to \ref{tabla1}, the derived mean deceleration parameter appears to become positive at scales larger than 150 Mpc, corresponding to a redshift of about 0.04. At first, our result seems inconsistent with the findings of \citealp{Asvesta22}, where q becomes positive at higher redshifts ($z > 0.5$) based on a fit to the Pantheon data. However, is important no note that \citealp{Asvesta22} and our results are not directly comparable but they
complement each other as they also seem to point in the same direction. In \citealp{Asvesta22}, the authors fitted the recent evolution of the deceleration parameter, as deduced from the Pantheon data, by adopting a particular profile for $\tilde{\theta}$ (see Eq. (13) in that paper). That profile introduces a parameterised scale/redshift dependence to $\tilde{\theta}$, with its physical motivation discussed just above Eq. (13). In that paper they also assume that $\tilde{\theta}$ is negative. In our paper we are using the density profiles reported on certain scales, to calculate the average values of $\tilde{\theta}$ on those scales and find them negative (a rather important qualitative result). So, by default, this $\tilde{\theta}$ has no scale/redshift dependence. Then, based on these negative values of $\tilde{\theta}$, we have estimated the corresponding average values of the local deceleration parameter on the same scales. Future work should introduce some scale/redshift dependence in our
results, in which case one could compare the theoretically motivated
$\tilde{\theta}$ profile of \citealp{Asvesta22} with the observationally estimated one.

As appealing these results may be, it is important to remain vigilant. It is possible, for example, that the values of the average divergence could change, as more refined surveys and models are developed. The value of the average divergence could be modified due to cosmic variance. which we estimate as $\Delta \theta\sim 1.6 \frac{Mpc}{h}$ for a fiducial $\Lambda$CDM cosmology. Recall that in the reconstruction used here this contribution was approximated by a constant velocity term. In addition, there have been recent claims that we live in a large void extending up to $\sim300~Mpc$. However, a negative expansion scalar is not compatible with the idea of a large void, where one expects to find an expanding bulk flow rather than a contracting one. In this respect, this velocity field reconstruction not seem to support the presence of a large underdensity. 

Finally, peculiar velocities seem unlikely to change the local value of the Hubble parameter appreciably and therefore to solve the $H_0$ tension. One can immediately realise this by looking at the linear relation (\ref{Thetas}a). Indeed, keeping in mind that $|\tilde{\theta}|/\Theta=|\tilde{\theta}|/3H\ll1$ on sufficiently large scales, the impact of the observer's relative motion on the Hubble parameter should be minimal.\footnote{Recall that, although $|\tilde{\theta}|/H\ll1$ always during the linear regime, this is not necessarily the case for the ratio $|\tilde{\theta}^{\prime}|/\dot{H}$.} Instead, there might be other explanations, such as systematics, the evolution of cosmological parameters with redshift, etc (e.g.~see~\citealp{Krishnan_2020, Colgain2022}).

\section{Other evidences for a contracting bulk flow}

As we have seen, characterizing the local peculiar velocity field remains a significant observational challenge in modern cosmology. In principle, it seems nearly impossible to break the degeneracy between the background kinematical cosmological parameters and the properties of the kinematical bulk flow without making global cosmological assumptions. Velocity reconstructions, such as the one used in this work (\citealp{Carr_2022}), rely on Newtonian physics to relate the density contrast to the peculiar velocities, often neglecting contributions from vorticity or possible biases due to local under- or over-densities.

Recently, a blind study of the redshift structure in the Pantheon+ supernova compilation found observational evidence for a locally contracting bulk flow on average (\citealp{Sorrenti:2024ugq}). This result aligns with the findings in this chapter and with the study presented in \citealp{Giani_2024}, which uses a reconstructed velocity field from the Cosmicflows-4 catalog (\citealp{Curtois_2023}). In that study, an ellipsoid-like structure was fitted to the Laniakea velocity structure, revealing a non-zero shear in the local kinematics, but with an average negative scalar expansion. Interestingly, when an isotropic spherical model was applied instead of the ellipsoidal one, the sign of the expansion turned positive. Thus, it is important to analyze the shear degree in the redshift structure of the Pantheon+ compilation to see if it aligns with these recent findings.

To better understand the kinematical properties of the peculiar velocity field, we propose extending the work done in \citealp{Sorrenti:2024ztg}, in which a monopole, quadrupole, and dipole were fitted to the perturbed luminosity distance. In our approach, we interpret the involved parameters as kinematical quantities related to the peculiar velocity field—namely the expansion scalar, bulk flow velocity, and shear. The goal of this study is to assess whether the values extracted from the luminosity distance are consistent with the results found in \citealp{Giani_2024}.

\subsection{Modeling the relation of the peculiar velocity field kinematics between the 2 frames of reference}

We assume 2 frame of reference. One is the centered in Laniakea and is Cartesian in the axis of the ellipsoid $\mathbf{x}_L$ and the other is centered in the solar system $\mathbf{x}_S$ and is Cartesian in the equatorial plane. The first was used in \citealp{Giani_2024} to fit the ellipsoid model while the second is the frame in which SNIA are observed. Consider the solar system is at a position $\mathbf{d}$ from the Laniakea and the equatorial plane is rotated w.r.t to the axis of the ellipsoid. If $\mathbf{R}$ is the rotation matrix associated, then the position of a source in the 2 frames are related by:

\begin{eqnarray}
    \mathbf{x}_L=\mathbf{R}\cdot\mathbf{x}_S+\mathbf{d} \label{positions}
\end{eqnarray}

If we perform the temporal derivative then:

\begin{eqnarray}
    \mathbf{v}_L=\mathbf{R}\cdot \mathbf{v}_S+\mathbf{V}
\end{eqnarray}

Where $\mathbf{v}_L$ and $\mathbf{v}_S$
are the velocities of the source w.r.t Laniakea and the solar system, respectively. $\mathbf{v}_S$ is the velocity of the solar system w.r.t Laniakea. $\mathbf{v}_L$ can be modeled as a ellipsoid-like velocity field with 3 parameters (\citealp{Giani_2024}) which we group in a diagonal tensor $\Delta \mathbf{H}$:

\begin{eqnarray}
    \mathbf{v}_L=\Delta\mathbf{ H} \cdot \mathbf{x}_L 
\end{eqnarray}

Here $\Delta\mathbf{ H}=diag(\Delta H_a, \Delta H_b, \Delta H_b)$. Then, the peculiar velocity field seen from the solar system should be:

\begin{eqnarray}
    \mathbf{v}_S=\mathbf{R}^{-1}\cdot (\Delta\mathbf{ H} \cdot \mathbf{x}_L - \mathbf{V})
\end{eqnarray}

Substituting equation $\ref{positions}$, we have:

\begin{eqnarray}
    \mathbf{v}_S&=&\mathbf{R}^{-1}\cdot [\Delta\mathbf{ H} \cdot (\mathbf{R}\cdot \mathbf{x}_S+\mathbf{d}) - \mathbf{V}] \\
    &=& (\mathbf{R}^{-1}\cdot \Delta\mathbf{ H} \cdot \mathbf{R})\cdot \mathbf{x}_S+\mathbf{R}^{-1}\cdot (\Delta\mathbf{ H} \cdot \mathbf{d}- \mathbf{V})
\end{eqnarray}

The latter relation provides a connection between the peculiar velocity field on both frames. Note that, according to the solar system, the observed peculiar velocity field will be composed by an ellipsoid velocity field plus an extra dipole term.

\subsection{The monopole, dipole and quadrupole of the luminosity distance as kinematical quantities}

The next step is to understand the relation between the monopole, dipole and quadrupole term of the perturbed luminosity distance with the kinematical quantities of the peculiar velocity field measured from the solar system. Let's call by simplicity:

\begin{eqnarray*}
    \mathbf{E}=\mathbf{R}^{-1}\cdot \Delta\mathbf{ H} \cdot \mathbf{R}\\
    \mathbf{P}=\mathbf{R}^{-1}\cdot (\Delta\mathbf{ H} \cdot \mathbf{d}-\mathbf{V})
\end{eqnarray*}

Then the peculiar velocity field in the solar system is:

\begin{eqnarray}
    \mathbf{v}_S=\mathbf{E}\cdot\mathbf{x}_S+\mathbf{P}
\end{eqnarray}

Note that $\mathbf{P}$ is a dipole term in the velocity field while the tensor $\mathbf{E}$ encodes a monopole and a quadrupole term as we will see in the next section.. This means that, even if the bulk flow is subtracted in Laniakea reference frame, the measured dipole in the solar system would not be zero. This is because, as we see in the definition for $\mathbf{P}$, the displacement and the velocity of the solar frame w.r.t to Laniakea modify the measured dipolar structure of the peculiar velocity field, even if the solar system is assumed to be in rest w.r.t Laniakea. If we assume that the velocity of the solar system is well described by the ellipsoid model, then the vector $\mathbf{V}$ can be written as:

\begin{eqnarray}
    \mathbf{V}= \Delta\mathbf{ H} \cdot \mathbf{d}
\end{eqnarray}

Which implies that the vector $\mathbf{P}$ goes to zero:

\begin{eqnarray*}
    \mathbf{P}=\mathbf{0}
\end{eqnarray*}

Note here that if the solar is system well described by Laniakea field, we will not see a relative dipole contribution, since the velocity and the displacement contribution cancel each other. Then, the value of $\mathbf{P}$ indicates how well our velocity w.r.t Laniakea is modeled by the ellipsoid field. Is important to note that to get these results, all supernovae analyzed have to be inside Laniakea. If this is not the case, the relative dipole contribution doesn't have to be zero as we are losing the correlations between the peculiar velocities inside Laniakea. If we want to express the peculiar velocity field $\mathbf{v}_{pec}$ as it should be seen from an observer in rest w.r.t the Hubble flow, we just need to subtract the velocity $\mathbf{v}_0$ of the solar system w.r.t CMB: 

\begin{eqnarray}
    \mathbf{v}_{pec}=\mathbf{E}\cdot\mathbf{x}_S+\mathbf{P}-\mathbf{v_0} \label{peculiarvelocityfield}
\end{eqnarray}

\subsection{The perturbed luminosity distance}

In the analysis of supernovae \citealp{sorrenti2023dipole}, the corrections to the luminosity distance at low redshift are assumed to be solely by the peculiar movements. When a dipole structure is considered, the perturbed luminosity distance can be written as:

\begin{equation}
    d_L(z,\mathbf{n})=\bar{d}_L(z) \left [1+\frac{\mathbf{u}_0\cdot \mathbf{n}}{\mathcal H(z)r(z)} \right ]
\end{equation}

Where $\mathcal H(z)$ is the co-moving Hubble parameter, $\mathbf{n}$ is the source direction vector and $r(z)$ is the proper distance. $\mathbf{u}_0=\mathbf{v}_0-\mathbf{v}_{bulk}$ is the difference between the velocity of the solar system and the peculiar velocity field of source assumed to be fully dipolar $\mathbf{v}_{pec}=\mathbf{v}_{bulk}$, both w.r.t the CMB. A step forward in this analysis made in \citealp{Sorrenti:2024ztg} was to consider a monopole and a quadrupole as well as a dipole in the peculiar velocity field $\mathbf{v}_{pec}$ as:

\begin{equation}
\begin{split}
    d_L(z,\mathbf{n})=\bar{d}_L(z)\left [1+\frac{\mathbf{v}_0\cdot \mathbf{n}}{\mathcal H(z)r(z)} \right .
    \\ \left .- \frac{1}{\mathcal H(z)r(z)}\left ( \gamma+ \mathbf{v}_{bulk}\cdot \mathbf{n}+n^jn^i\alpha_{ij}\right ) \right ] 
\end{split}
\end{equation}

Where $\gamma$ is a monopole contribution and $\alpha_{ij}$ is a traceless simetric tensor corresponding to the quadrupole. We can show that this decomposition is equivalent to a ellipsoid like-structure for the velocity field. Note that we can write the luminosity distance as:

\begin{equation}
\begin{split}
    d_L(z,\mathbf{n})=\bar{d}_L(z)\left [1+\frac{\mathbf{v}_0\cdot \mathbf{n}}{\mathcal H(z)r(z)} \right .
    \\ \left .- \frac{n^{i}n^{j}}{\mathcal H(z)r(z)}\left ( \gamma\delta_{ij}+(\mathbf{v}_{bulk}\cdot \mathbf{n})\delta_{ij} + \alpha_{ij}\right ) \right ] 
\end{split}
\end{equation}

Using the expression \ref{peculiarvelocityfield} for the peculiar ellipsoid-like velocity field as seen from an observer in rest w.r.t the Hubble flow, we can do the correspondence noting that the position in the solar system is related with the proper distance and the direction of the source as:

\begin{eqnarray}
    \mathbf{x}_S=r(z)\mathbf{n}
\end{eqnarray}

Then we can write:

\begin{eqnarray}
    \mathbf{v}_{pec}=r(z)E_{ij} n^j+\mathbf{P}-\mathbf{v_0} \\
    \mathbf{v}_{pec}=r(z)(\frac{1}{3}\theta\delta_{ij}+\sigma_{ij}) n^j+\mathbf{P}-\mathbf{v_0}
\end{eqnarray}

Where we decomposed the $E_{ij}$ tensor in its diagonal and its trace-free part:

\begin{eqnarray}
    \theta&=&\delta^{ij}E_{ij}\\
    \sigma_{ij}&=&E_{ij}-\frac{1}{3}\delta^{ij}\theta
\end{eqnarray}

We can finally write the relations between both, the multi-polar parameters of the luminosity distance and the kinematical parameters of the peculiar velocity field:

\begin{eqnarray}
    \mathbf{v}_{bulk}&=&\mathbf{P}-\mathbf{v}_0 \\
    \gamma&=&\frac{r(z)}{3}\theta  \\
    \alpha_{ij}&=&r(z)\sigma_{ij}
    \label{kinematicalrelations}
\end{eqnarray}

\subsection{Numerical estimations}

We have shown that an ellipsoid structure fitted in \citealp{Giani_2024} to the Laniakea rest frame is composed solely by a monopole an a quadrupole term according to the tensor $\Delta\mathbf{ H}$. However, in the solar system frame, this kinematic is encoded in $\mathbf{E}$ and a dipolar term emerges. The relations \ref{kinematicalrelations} between the kinematical quantities and the multi-polar terms are dependent on the redshift. To explore simple numerical connections between the two sources of data, we adopt an average scale of about $z\sim 0.0175$ which corresponds to a size of approximately $\lambda \sim 75 Mpc$ to ensure that all supernovaes are inside Laniakea. Then, roughly we can approximate:

\begin{eqnarray}
    \mathbf{v}_{bulk}&=&\mathbf{P}-\mathbf{v}_0\\
    \gamma&\sim&\frac{\lambda}{3}\theta  \\
    \alpha_{ij}&=&\lambda\sigma_{ij}
\end{eqnarray}

In section \ref{proposedmethod} we develop a modification to the luminosity distance to include the redshift dependence.

\subsubsection{The monopole term or the expansion scalar}

The monopole term in the luminosity distance is related with the expansion scalar. This is computed from the trace of the gradient of the velocity field. We note that a rotation transformation leaves invariant the trace, so the expansion scalar should be the same in both experiments:

\begin{eqnarray*}
    Tr(\mathbf{R}^{-1}\cdot \Delta\mathbf{ H} \cdot \mathbf{R})=Tr(\Delta\mathbf{ H})
\end{eqnarray*}

Results from \citealp{Giani_2024} follows:

\begin{eqnarray}
    \theta_L \simeq -3.33 \pm 0.27 
\end{eqnarray}

While the estimated from the multi-polar analysis in \citealp{Sorrenti:2024ztg} with a redshift cut of $z_{cut} \sim 0.0175$:

\begin{eqnarray}
    \gamma \simeq -97 \pm 68 [km/s]
\end{eqnarray}

This correspond, as a rough estimation to an expansion scalar of:

\begin{eqnarray}
    \theta_S \simeq -3.88 \pm 2.72
\end{eqnarray}

Both seems to be in agreement with each other. 

\subsubsection{The dipole term or the bulk flow velocity}

The dipole was set to zero in the Laniakea frame to compute the values of the expansion scalar and the shear. However, if we want to compare the values extracted from supernovae with Laniakea, then the $\mathbf{v}_{bulk}$ contribution should be strongly related with the displacement and the velocity of the solar system w.r.t Laniakea. We have shown previously that the dipole component should be:

\begin{eqnarray}
    \mathbf{v}_{bulk}&=&\mathbf{P}-\mathbf{v}_0\\
\end{eqnarray}

If the solar system is well described by Laniakea ellipsoid field, then:

\begin{eqnarray}
    \vert\mathbf{v}_{bulk}\vert&=&\vert \mathbf{v}_0\vert
\end{eqnarray}

According to \citealp{Sorrenti:2024ztg}, the estimated bulk flow for a redshift cut of $z_{cut}\sim 0.0175$ is:

\begin{eqnarray*}
    \vert\mathbf{v}_{bulk}\vert&=&335 \pm 70 km/s\\
    (ra,dec)&=&(216^{17}_{-15},-52.1^{7.2}_{-9.9})
\end{eqnarray*}

 While last Planck results \citealp{PLANCK20}:

 \begin{eqnarray*}
    \vert\mathbf{v}_{0}\vert&=&369.82 \pm 0.9 km/s\\
    (ra,dec)&=&(167.942 \pm 0.007, -6.944 \pm 0.007)
\end{eqnarray*}

The dipoles are in agreement in magnitude, but not in direction. This could indicate that the direction of the motion of the solar system is not well modeled by the ellipsoid field which is expected as this is field is an \textit{average} fit that doesn't identify each object with its exact peculiar velocity. We can also add that, while the velocity of the milky way could be modeled by this peculiar velocity field, the velocity of the solar system is not necessarily pointing in the same direction since it moves around the milky way center. Moreover, we assumed just a redshift cut $z_{cut}$ to do the estimation, but a proper analysis should consider all supernovaes that are inside Laniakea. Finally, we consider the possibility stated in \citealp{sorrenti2023dipole} in which the data is not in agreement with Planck.
 
\subsubsection{The quadrupole term or the shear tensor}

We have shown that the shear of the velocity field is related with the quadrupolar term of the luminosity distance. In the supernova analysis, the quadrupole $\alpha_{ij}$ was diagonalized to extract the eigenvalues and the eigenvector. The shear of the Laniakea peculiar velocity field, in its rest frame could be computed as:

\begin{eqnarray*}
    \sigma^L_{aa}&=&\Delta H_a-\frac{\theta}{3} \simeq -2.29\pm 0.7, \\
    \sigma^L_{bb}&=&\Delta H_b-\frac{\theta}{3} \simeq 0.46 \pm 0.55, \\
    \sigma^L_{cc}&=&\Delta H_c-\frac{\theta}{3} \simeq 1.8 \pm 0.45, \\
\end{eqnarray*}

While in the supernova analysis the eigenvalues for the quadrupole at a $z_{cut} \sim 0.0175$ are:

\begin{eqnarray}
    \lambda_1 &=&-187^{+69}_{-53} [km/s] \\
    \lambda_2 &=&16 \pm 48 [km/s] \\ 
    \lambda_3 &=& 171^{+50}_{-60} [km/s]
\end{eqnarray}

Is easy to note that, if the traceless part of the metric was diagonalized in the frame of the solar system, then this correspond exactly to the diagonal shear of Laniakea, since the diagonalization is actually the application of the rotation matrix. This also could be interpreted as a come back to the Laniakea frame:

\begin{eqnarray*}
\mathbf{R}\cdot\mathbf{E}\cdot\mathbf{R^{-1}}=\Delta\mathbf{ H}
\end{eqnarray*}

When a monopole, a dipole and a quadrupole is fitted, the shear values according to \citealp{Sorrenti:2024ugq} analysis give:

\begin{eqnarray*}
    \sigma^S_{aa}&\simeq& -2.49^{0.92}_{-0.7}, \\
    \sigma^S_{bb}&\simeq& 0.21 \pm 0.64, \\
    \sigma^S_{cc}&\simeq& 2.28^{0.66}_{-0.8}, \\
\end{eqnarray*}

The magnitudes are, surprisingly, in good agreement with the computed from \citealp{Giani_2024}. The eigenvectors of the diagonalization should also be in agreement with the axis of the ellipsoid. However, the eigenvectors related with the monopole, dipole and quadrupole fit are not provided in the paper.

\subsection{Proposed Method}\label{proposedmethod}

We have seen that the rotation and translation of a field composed by a monopole, dipole and quadrupole, maintains the structure in the new frame but adds a dipole term. It follows that if the dipole is set to zero in one frame, it doesn't have to be zero in the other since there is a displacement and a velocity of the solar system with respect to the Laniakea rest frame. Expansion scalars seem to agree roughly in values, as well as the shear tensor, while the relative dipole was found to be non-zero. However, the estimations were used by introducing an average scale $\lambda$ to get rid of the redshift-dependent relations of equations \ref{kinematicalrelations}. To improve the analysis, we propose to use a different relation for the luminosity distance using kinematic parameters accounting for the redshift dependence between the kinematical quantities and the multipolar parameters as:

\begin{equation}
\begin{split}
    d_L(z,\mathbf{n}) = \bar{d}_L(z)\left[1 + \frac{\mathbf{v}_0 \cdot \mathbf{n}}{\mathcal{H}(z) r(z)} \right. \\
    \left. - \frac{1}{\mathcal{H}(z) r(z)} \left( \frac{r(z)}{3} \theta + \mathbf{v}_{\text{bulk}} \cdot \mathbf{n} + r(z) n^j n^i \sigma_{ij} \right) \right]
\end{split}
\end{equation}

In this way, we can obtain the kinematical parameters such as $\theta$ and $\sigma_{ij}$ directly, without having to average over a scale after the fit. The proposition made here by the author of this thesis is currently being worked on together with the authors of \citealp{Sorrenti:2024ztg} and \citealp{Giani_2024}, using the Pantheon + compilation.

\clearchapterpage
\chapter{The non-Newtonian effects in peculiar velocity evolution}
\label{pec}

Over the past fifteen years or so, an increasing number of studies have reported large-scale peculiar flows that are much faster and deeper than theoretical expectations suggest, often defying conventional models. These bulk flows, which involve the motion of galaxies or large-scale structures in the universe, have drawn attention due to their apparent deviations from the standard predictions based on linear theories. However, much of the work done on the linear evolution of peculiar velocities has been grounded in Newtonian theory. Only in recent years have fully relativistic treatments of the problem begun to emerge in the literature, bringing new insights into how these flows may evolve. These relativistic models argue that the peculiar-velocity field grows at a much stronger rate than predicted by Newtonian mechanics, potentially explaining the faster and more pronounced bulk flows observed in cosmological surveys.

The key difference between the Newtonian and relativistic treatments lies in the treatment of the peculiar flux, specifically the energy flux generated by the drift motion of the matter. In Newtonian theory, matter fluxes do not contribute directly to the gravitational field. However, in general relativity, the situation is quite different. Here, the energy fluxes resulting from the motion of matter contribute directly to the energy-momentum tensor, which influences the spacetime curvature and the evolution of the gravitational field. As a result, the gravitational impact of peculiar flux survives at the linear level and feeds into the conservation laws of energy and momentum, leading to modified equations for the evolution of the peculiar-velocity field.

Relativistic treatments of peculiar flows have shown that the gravitational effect of these energy fluxes modifies the relativistic formulae that govern cosmic evolution, leading to a significantly different prediction for the growth rate of the peculiar velocity field compared to the Newtonian case. These differences may explain some of the observed discrepancies in bulk flow measurements. Moreover, the additional contribution of the peculiar flux in relativistic cosmology could have significant implications for understanding large-scale structure formation, galaxy motion, and even the dynamics of dark matter and dark energy.

The goal of this study is to directly compare the Newtonian and relativistic approaches, providing a clear understanding of the reasons behind the markedly different results. By analyzing the discrepancies, we aim to highlight the importance of accounting for relativistic effects when studying large-scale flows in the universe. In the process, we also demonstrate how one could recover the relativistic peculiar growth rate from a Newtonian framework. This can be achieved by selectively incorporating certain terms that are typically ignored in standard Newtonian treatments—namely, source-free terms—into the Poisson equation. Although selective and not uniquely determined, this approach allows us to bridge the gap between Newtonian and relativistic models and refine our understanding of peculiar flows in cosmology.

\section{General relativity and peculiar velocities}\label{section:1}

Large-scale peculiar motions, often referred to as bulk flows, are an established observational fact, confirmed by a variety of surveys over many decades. Although many of the observed bulk peculiar flows have velocities that fall within the theoretical limits of the current cosmological model \citealp{Ma2014, Scrimgeour2016}, a growing number of recent surveys have reported bulk flows that are much faster and deeper than expected \citealp{Watkins2009, Colin2011, Salehi2021, Watkins_2023}. Additionally, there is the controversial issue of "dark flows," which exhibit both size and velocity far exceeding the predictions of the $\Lambda$CDM model \citealp{Kashlinsky2008, Kashlinsky_2011}.

In the case of peculiar motions, the $\Lambda$CDM predictions have traditionally been based on Newtonian theory, which suggests a relatively mild linear growth-rate for the peculiar-velocity field $v \propto t^{1/3}$ \citealp{Peebles1976, Peebles1980}. This result is also consistent with so-called “quasi-Newtonian” treatments \citealp{Maartens1998, Ellis2001}, which, despite their relativistic appearance, ultimately reduce to Newtonian. One reason for this reliance on Newtonian theory was the perceived difficulty of treating the bulk-flow question relativistically. Another factor was the widespread belief that using general relativity for non-relativistic peculiar motions would not substantially alter the established Newtonian picture. As a result, fully relativistic linear studies of peculiar velocities have only recently emerged in the literature \citealp{Tzaprasi2020, Maglara2022, Miliou2024}. These recent studies suggest a much stronger growth-rate for peculiar velocities, with $v \propto t^{4/3}$, which could help explain both the higher velocities and greater depths of the bulk flows observed in \citealp{Watkins2009, Colin2011, Salehi2021, Watkins_2023}. The key difference between the Newtonian/quasi-Newtonian and relativistic studies stems from the significantly different role of the "peculiar flux" in these two theoretical frameworks.

Peculiar flows, by definition, involve matter in motion, and moving matter corresponds to non-zero energy flux. In general relativity, unlike Newtonian gravity, energy fluxes contribute to the stress-energy tensor and, as a result, influence the gravitational field. The gravitational effect of the peculiar flux is not accounted for in either Newtonian or quasi-Newtonian treatments, but for different reasons. In the Newtonian case, this omission is automatic, while in the quasi-Newtonian framework, it occurs due to the constraints imposed by the approximation, which restrict the host spacetime to a form that closely resembles Newtonian gravity \citealp{Ellis2012, Tsagas2024}. The limitations of the quasi-Newtonian approach were known (see, for example, the cautionary comments in \S 6.8.2 of \citealp{Ellis2012}), though the lack of direct comparison with fully relativistic studies meant that this issue was not widely recognized until recently \citealp{Tsagas2024}. Incorporating the gravitational influence of the peculiar flux in a fully relativistic (linear) study feeds into the conservation laws and modifies the linear evolution equations governing the peculiar-velocity field. In practice, the changes to the gravitational field induced by the flux alter the form of the differential equations that describe the growth-rate of peculiar velocities, leading to significant differences between the Newtonian and relativistic solutions. Specifically, general relativity predicts a much stronger linear growth of peculiar velocities during the Einstein-de Sitter phase of the universe. This relativistic field could readily explain the speed and depth of the bulk flows reported in \citealp{Watkins2009, Colin2011, Salehi2021}. Moreover, if the Einstein-de Sitter era transitions into a period of accelerated expansion, the peculiar-velocity field's growth should slow down at later times, as seen in the study of \citealp{Salehi2021} and discussed in \citealp{Maglara2022}.

This work offers a direct comparison between Newtonian and relativistic analyses of linear peculiar velocities in a Friedmann-Robertson-Walker (FRW) background. We begin by setting up the theoretical framework for both approaches, identifying the point at which they diverge, and explaining the underlying reasons for these differences. We then derive alternative linear evolution laws for the peculiar-velocity field, emphasizing the differences between the solutions and discussing their implications for understanding bulk flows. Finally, we present a method for selectively recovering relativistic results from the Newtonian framework by including certain typically ignored source-free terms in Poisson's equation. While this recovery process is not uniquely determined, it provides useful analogies between the two approaches. For instance, neglecting these source-free terms in Newtonian studies can be seen as equivalent to overlooking the gravitational input of the peculiar flux in relativistic models.

We begin our presentation with the traditional Newtonian treatment of cosmological linear peculiar velocities, following the work of \citealp{Peebles1976, Peebles1980} and \citealp{Nusser1991} in Section \S~\ref{sNTPVs}. We then introduce the covariant analogue of the Newtonian analysis in \S~\ref{sNCTPVs} to familiarize the reader with the covariant formalism, connecting the Newtonian framework to the fully relativistic approach discussed in \S~\ref{sRCTPVs}. In the latter section, we explore the physical reasons why Newtonian and quasi-Newtonian studies of linear peculiar velocities yield significantly different results. We also discuss a selective method for bridging these differences by incorporating extra contributions to the Newtonian gravitational potential in \S~\ref{sEGP}, and conclude with a discussion of our findings and their implications in \S~\ref{sD}.

\section{Newtonian treatment of peculiar velocities}\label{sNTPVs}
Let us consider Newtonian theory and use physical coordinates $\left\{\mathbf{r}\right\}$, where $\mathbf{r}=a\mathbf{x}$, with $a = a(t)$ being the cosmological scale factor and $\left\{\mathbf{x}\right\}$ representing the comoving coordinates. In this setup, the velocity ($\mathbf{u}$) of a typical observer in a perturbed (Newtonian) Einstein-de Sitter universe can be expressed as
\begin{equation}
\mathbf{u}= H\mathbf{r}+ \mathbf{v}\,,  \label{vu}
\end{equation}
with $H=\dot{a}/a$ representing the Hubble parameter. Also, $\mathbf{v}=a\dot{\mathbf{x}}$ is the observer's peculiar velocity relative to the universal rest-frame, which has been typically identified with the rest-frame of the Cosmic Microwave Background (CMB). Within the adopted cosmological framework, the linear evolution of the peculiar-velocity field is monitored by the associated Euler equation, namely by~\citealp{Peebles1976}
\begin{equation}
\dot{\mathbf{v}}+ H\mathbf{v}= -{1\over a}\,\mathbf{\nabla}\phi\,,  \label{ldotv}
\end{equation}
where $\phi$ is the peculiar gravitational potential. The latter acts as a linear source of peculiar-velocity perturbations and satisfies Poisson's formula, that is
\begin{equation}
\mathbf{\nabla}^2\phi= 4\pi Ga^2\rho_0\delta\,,  \label{Poisson}
\end{equation}
with $\delta=(\rho-\rho_0)/\rho_0$ being the familiar density contrast, $\rho$ the matter density and $\rho_0$ its background value.

Differentiating Eq.~(\ref{ldotv}) in time, recalling that $H=2/3t$ and $\dot{H}=2/3t^2$ to zero order, while keeping up to linear order terms, leads to
\begin{equation}
\ddot{\mathbf{v}}+ 2H\dot{\mathbf{v}}- {1\over2}H^2\mathbf{v}= \ddot{\mathbf{v}}+ {4\over3t}\,\dot{\mathbf{v}}- {2\over9t^2}\,\mathbf{v}= -{1\over a}\left(\mathbf{\nabla}\phi\right)^{\cdot}\,.  \label{lddotv1}
\end{equation}
One may ignore the right-hand side of the above, by assuming that the gradient of the peculiar gravitational potential varies slowly in time, for example. Then, (\ref{lddotv1}) reduces to
\begin{equation}
9t^2{{\rm d}^2\mathbf{v}\over{\rm d}t^2}+ 12t\,{{\rm d}\mathbf{v}\over{\rm d}t}- 2\,\mathbf{v}= 0\,,  \label{lddotv2}
\end{equation}
which accepts the power-law solution
\begin{equation}
\mathbf{v}= \mathcal{C}_1t^{1/3}+ \mathcal{C}_2t^{-2/3}\,,  \label{bfv1}
\end{equation}
on all scales. Therefore, according to Newtonian physics, linear peculiar velocities ar expected to grow as $\mathbf{v}\propto t^{1/3}$ after recombination.

One can refine solution (\ref{bfv1}) by incorporating to it the source term on the right-hand side of (\ref{lddotv1}). Following~\citealp{Peebles1980}, we start by recalling that the peculiar gravitational acceleration is
\begin{equation}
\mathbf{g}= -{1\over a}\,\mathbf{\nabla}\phi= Ga\rho_0\int\delta{\mathbf{x}^{\prime}-\mathbf{x}\over||\mathbf{x}^{\prime}-\mathbf{x}||}\,{\rm d}^3\mathbf{x}^{\prime}\,,
\end{equation}
Given that $\phi$ and $\delta$ also depend on time, after a relatively straightforward calculation, the temporal derivative of the above leads to the linear auxiliary relation
\begin{equation}
\left(\mathbf{\nabla}\phi\right)^{\cdot}= -H\mathbf{\nabla}\phi- {3\over2}\,aH^2\mathbf{v}\,.  \label{laux1}
\end{equation}
In deriving this relation, we have employed the background expressions $\rho_0 = 3H^2/8\pi G$ and $\dot{\rho}_0 = -3H\rho_0$, along with the linear result $\dot{\delta} = -\mathbf{\nabla}\cdot\mathbf{v}$. This follows from the continuity equation, which at the linear level is written as $\dot{\rho} = -\mathcal{H}\rho$, where $\rho$ and $\mathcal{H}$ represent the perturbed matter density and Hubble parameter, respectively. By setting $\rho = \rho_0(1 + \delta)$ and $\mathcal{H} = H + \mathbf{\nabla}\mathbf{v}$, and substituting these into the linear continuity equation while retaining first-order terms, we obtain the desired relation: $\dot{\delta} = -\mathbf{\nabla}\cdot\mathbf{v}$. Finally, combining equations (\ref{ldotv}), (\ref{lddotv1}), and (\ref{laux1}), we arrive at the differential equation,\begin{equation}
\ddot{\mathbf{v}}+ 3H\dot{\mathbf{v}}- H^2\mathbf{v}= \ddot{\mathbf{v}}+ {2\over t}\,\dot{\mathbf{v}}- {4\over9t^2}\,\mathbf{v}= 0  \label{lddotv3}
\end{equation}
and the associated power-law solution(~\citealp{Peebles1976})
\begin{equation}
\mathbf{v}= \mathcal{C}_1t^{1/3}+ \mathcal{C}_2t^{-4/3}\,,  \label{bfv2}
\end{equation}
on all scales. Comparing solutions (\ref{bfv1}) and (\ref{bfv2}) makes it clear that the inclusion of the nonzero right-hand side of (\ref{lddotv1}) has no physical/practical significance. The rate of the growing mode in the former of the two solutions has remained the same and only the decaying mode changed.

\section{Newtonian covariant treatment of peculiar velocities}\label{sNCTPVs}
Let us now provide the covariant Newtonian treatment of linear peculiar velocities in cosmology. In so doing, we will facilitate the comparison with the standard Newtonian treatment of the previous section, as well as with the subsequent fully relativistic analysis.

\subsection{Newtonian peculiar kinematics}\label{ssNPKs}
Consider a pair of relatively moving observers with velocities $u_{\alpha}$ and $\tilde{u}_{\alpha}$. In Newtonian theory, these two velocity fields are related by the Galilean transformation\footnote{Hereafter, Greek indices will run from 1 to 3, whereas their Latin counterparts (see \S~\ref{sRCTPVs} below) will run from 0 to 3.}
\begin{equation}
\tilde{u}_{\alpha}= u_{\alpha}+ \tilde{v}_{\alpha}\,,  \label{Galilean}
\end{equation}
where $\tilde{v}_{\alpha}$ is the peculiar velocity of the $\tilde{u}_{\alpha}$-field relative to the (reference) $u_{\alpha}$-frame.\footnote{Typical Newtonian studies of peculiar motions introduce physical ($r^{\alpha}$) and comoving ($x^{\alpha}$) coordinates, with $r^{\alpha}=ax^{\alpha}$. The time derivative of the latter leads to $v_t=v_H+v_p$, where $v_t=\dot{r}^{\alpha}$, $v_H=Hr^{\alpha}$ and $v_p=a\dot{x}^{\alpha}$ are the total, the Hubble and the peculiar velocities respectively. On an FRW background, the above velocity relation is equivalent to Eq.~(\ref{Galilean}). Also note that the mean peculiar velocity ($\tilde{V}$) is given by the integral $\tilde{V}=(3/4\pi r^3)\int_{x<r}\tilde{v}{\rm d}x^3$, with $\tilde{v}^2= \tilde{v}_{\alpha}\tilde{v}^{\alpha}$ and with $r$ representing the radius of the moving region.} The irreducible kinematics of the $u_{\alpha}$-field are given by the decomposition(~\citealp{Ellis1990})
\begin{equation}
\partial_{\beta}u_{\alpha}= {1\over3}\,\Theta h_{\alpha\beta}+ \sigma_{\alpha\beta}+ \omega_{\alpha\beta}\,.  \label{Npbua}
\end{equation}
Here, the scalar $\Theta=\partial^{\alpha}u_{\alpha}$ indicating an expanting/contracting $u_{\alpha}$-field, when positive/negative. Also, the symmetric, traceless tensor $\sigma_{\alpha\beta}=\partial_{\langle\beta}u_{\alpha\rangle}$ and its antisymmetric counterpart, $\omega_{\alpha\beta}=\partial_{[\beta}u_{\alpha]}$, are respectively the shear and the vorticity of the $u_{\alpha}$-field. Finally, $h_{\alpha\beta}$, with $h_{\alpha\beta}=h_{(\alpha\beta)}$, $h_{\alpha}{}^{\alpha}=3$ and $h_{\alpha\mu}h^{\mu}{}_{\beta}=h_{\alpha\beta}$, is the metric of the space(~\citealp{Ellis1990}). It goes without saying that an exactly analogous split applies to the tilded $\tilde{u}_{\alpha}$-field (i.e.~$\partial_{\beta}\tilde{u}_{\alpha}= (\tilde{\Theta}/3)h_{\alpha\beta}+\tilde{\sigma}_{\alpha\beta}+ \tilde{\omega}_{\alpha\beta}$). Similarly, the spatial gradient of the peculiar velocity field decomposes as
\begin{equation}
\partial_{\beta}\tilde{v}_{\alpha}= {1\over3}\,\tilde{\vartheta}h_{\alpha\beta}+ \tilde{\varsigma}_{\alpha\beta}+ \tilde{\varpi}_{\alpha\beta}\,,  \label{pbtva}
\end{equation}
with $\tilde{\vartheta}=\partial^{\alpha}\tilde{v}_{\alpha}$, $\tilde{\varsigma}_{\alpha\beta}= \partial_{\langle\beta}\tilde{v}_{\alpha\rangle}$ and $\tilde{\varpi}_{\alpha\beta}= \partial_{[\beta}\tilde{v}_{\alpha]}$. As before, positive/negative values for $\tilde{\vartheta}$ indicate (locally) expanding/contracting peculiar flows. Also, $\tilde{\varsigma}_{\alpha\beta}$ and $\tilde{\varpi}_{\alpha\beta}$ are respectively the shear and vorticity of the peculiar-velocity field. Starting from (\ref{Galilean}), one arrives at the following (nonlinear) relations(~\citealp{Filippou_2021})
\begin{equation}
\tilde{\Theta}= \Theta+ \tilde{\vartheta}\,, \hspace{10mm} \tilde{\sigma}_{\alpha\beta}= \sigma_{\alpha\beta}+\tilde{\varsigma}_{\alpha\beta} \hspace{10mm} {\rm and} \hspace{10mm} \tilde{\omega}_{\alpha\beta}= \omega_{\alpha\beta}+\tilde{\varpi}_{\alpha\beta}\,,  \label{Nrels}
\end{equation}
between the three kinematic sets. In addition, taking the convective derivative of (\ref{Galilean}), with respect to the tilded frame, and then employing Eq.~(\ref{Galilean}) again, we arrive at the expression(~\citealp{Filippou_2021})
\begin{equation}
\tilde{u}_{\alpha}^{\prime}= \dot{u}_{\alpha}+ \tilde{v}_{\alpha}^{\prime}+ {1\over3}\,\Theta\,\tilde{v}_{\alpha}+ \left(\sigma_{\alpha\beta}+\omega_{\alpha\beta}\right) \tilde{v}^{\beta}\,,  \label{tEuler1}
\end{equation}
relating the (inertial) acceleration vectors in the two coordinate systems. Alternatively, one can replace $\tilde{v}_{\alpha}^{\prime}$ on the right-hand side of the above by means of the (nonlinear) expression $\tilde{v}_{\alpha}^{\prime}= \dot{\tilde{v}}_{\alpha}+(\tilde{\vartheta}/3)\tilde{v}_{\alpha} +(\tilde{\varsigma}_{\alpha\beta}+\tilde{\varpi}_{\alpha\beta}) \tilde{v}^{\beta}$, which relates the convective derivatives of the peculiar velocity ($\tilde{v}_{\alpha}$) in the two frames. Then, Eq.~(\ref{tEuler1}) recasts as(~\citealp{Filippou_2021})
\begin{equation}
\tilde{u}_{\alpha}^{\prime}= \dot{u}_{\alpha}+ \dot{\tilde{v}}_{\alpha}+ {1\over3}\,\Theta\,\tilde{v}_{\alpha}+ (\sigma_{\alpha\beta}+\omega_{\alpha\beta})\tilde{v}^{\beta}+ {1\over3}\,\tilde{\vartheta}\tilde{v}_{\alpha}+ (\tilde{\varsigma}_{\alpha\beta}+\tilde{\varpi}_{\alpha\beta}) \tilde{v}^{\beta}\,.  \label{tEuler2}
\end{equation}
Note that primes denote convective derivatives in the tilded frame (i.e.~$\tilde{u}_{\alpha}^{\prime}= \partial_t\tilde{u}_{\alpha}+ \tilde{u}^{\beta}\partial_{\beta}\tilde{u}_{\alpha}$), while overdots indicate convective differentiation in the reference frame (i.e.~$\dot{u}_{\alpha}=\partial_tu_{\alpha}+ u^{\beta}\partial_{\beta}u_{\alpha}$). Following (\ref{tEuler1}) and (\ref{tEuler2}), the presence of relative motion means that (generally) we cannot set $\dot{u}_{\alpha}$ and $\tilde{u}^{\prime}_{\alpha}$ to zero simultaneously. The same is also true for the rest of the kinematic variables (see Eqs.~(\ref{Nrels}a)-(\ref{Nrels}c) above). Finally, we should point out that, in contrast to the relativistic treatment (see \S~\ref{sRCTPVs} below), the matter variables remain unchanged when transforming from one coordinate system to the other.

\subsection{Linear sources of peculiar
velocities}\label{ssLSPVs}
So far all our Newtonian formulae have been nonlinear. Let us now consider a perturbed almost-FRW Newtonian universe, filled with a pressureless perfect fluid (baryonic or/and CDM -- with $p=0= \pi_{\alpha\beta}$) and treat the peculiar velocity field as a linear perturbation on the aforementioned background. Then, by identifying the $u_{\alpha}$-field with the CMB frame, we may set $\dot{u}_{\alpha}=0= \sigma_{\alpha\beta}=\omega_{\alpha\beta}$. On these grounds, expressions (\ref{tEuler1}) and (\ref{tEuler2})  linearise to
\begin{equation}
\tilde{v}_{\alpha}^{\prime}= \dot{\tilde{v}}_{\alpha}= -H\tilde{v}_{\alpha}+ \tilde{u}_{\alpha}^{\prime}\,,  \label{ltEuler1}
\end{equation}
where $H=\dot{a}/a$ is the background Hubble parameter. According to the above, the sole source of linear peculiar velocities is the acceleration, which in the absence of pressure is given by $\tilde{u}_{\alpha}^{\prime}= -\partial_{\alpha}\Phi$. Then, Eq.~(\ref{ltEuler1}) recasts into
\begin{equation}
\dot{\tilde{v}}_{\alpha}= -H\tilde{v}_{\alpha}- \partial_{\alpha}\phi\,.  \label{ltv'}
\end{equation}
Thus, in Newtonian theory, linear peculiar velocities are generated and driven by gradients in the gravitational potential ($\Phi$). Note that relation (\ref{ltv'}) is identical to the one obtained in typical Newtonian studies, provided the equations of the latter are written in physical (rather than comoving) coordinates (e.g.~see~\citealp{Peebles1976,Nusser1991}).

\subsection{Linear evolution of peculiar
velocities}\label{ssLNEPVs}
Taking the convective derivative of (\ref{ltv'}), employing the background relations $\dot{H}=-3H^2/2$, $\partial_{\beta}u_{\alpha}= Hh_{\alpha\beta}$ and keeping up to linear-order terms, leads to
\begin{equation}
\ddot{\tilde{v}}_{\alpha}+ 2H\dot{\tilde{v}}_{\alpha}- {1\over2}\,H^2\tilde{v}_{\alpha}= -\partial_{\alpha}\dot{\phi}\,,  \label{ltv''1}
\end{equation}
since $(\partial_{\alpha}\phi)^{\cdot}=\partial_{\alpha}\dot{\phi}- H\partial_{\alpha}\phi$ to first approximation. To proceed further, we may appeal to \textit{Jeans' swindle} (see e.g. \citealp{Binney_2008}) and write the Poisson equation as $\partial^2\phi=\rho\delta/2$, where $\delta=\delta\rho/\rho$ is the familiar density contrast. Then, a relatively lengthy but fairly straightforward calculation leads to
\begin{equation}
\partial^2\dot{\phi}= -H\partial^2\phi+ {1\over2}\,\rho\dot{\delta}\,.  \label{LapldotPhi1}
\end{equation}
The next step involves the continuity equation, $\dot{\rho}=-\tilde{\Theta}\rho$, starting from which one finds that $\dot{\delta}=-\tilde{\vartheta}$. Substituting this linear result back into the above gives
\begin{equation}
\partial^2\dot{\phi}= -H\partial^2\phi- {3\over2}\,H^2\tilde{\vartheta}\,.  \label{LapldotPhi2}
\end{equation}
given that $\rho=3H^2$ in the background. Taking the divergence of (\ref{ltv'}) and combining the resulting expression of $\tilde{\vartheta}$ with (\ref{LapldotPhi2}), yields
\begin{equation}
\partial_{\alpha}\dot{\phi}= {1\over2}\,H\partial_{\alpha}\phi+ {3\over2}\,H\dot{\tilde{v}}_{\alpha}\,.  \label{graddotPhi}
\end{equation}
Finally, after substituting the latter expression into the right-hand side of (\ref{ltv''1}), one arrives at the homogeneous differential equation
\begin{equation}
\ddot{\tilde{v}}_{\alpha}+ 3H\dot{\tilde{v}}_{\alpha}- H^2\tilde{v}_{\alpha}= 0\,,  \label{ltddotv2}
\end{equation}
with $H=2/3t$ in the background. The above solves to give
\begin{equation}
\tilde{v}= \mathcal{C}_1t^{1/3}+ \mathcal{C}_2t^{-4/3}= \mathcal{C}_3a^{1/2}+ \mathcal{C}_4a^{-2}\,,  \label{lNv2}
\end{equation}
on all scales and in complete agreement with solution (\ref{bfv2}) in \S~\ref{sNTPVs} and also with~\citealp{Peebles1980}.

\section{Relativistic covariant treatment of peculiar velocities}\label{sRCTPVs}
In relativity, the Galilean transformation between two relatively moving frames/observers is replaced by the Lorentz boost, which in the case of non-relativistic peculiar motions reads just as equation \ref{Galilean}:
\begin{equation}
\tilde{u}_a= u_a+ \tilde{v}_a\,,.  \label{nrLboost}
\end{equation}
Here, $u_a$ is the 4-velocity of the reference frame, $\tilde{u}_a$ is the 4-velocity of the tilted frame and $\tilde{v}_a$ is the peculiar velocity of the latter coordinate system relative to the former. Note that $u_au^a=-1=\tilde{u}_a\tilde{u^a}$ and $u_a\tilde{v}^a=0$ by construction. Also, $\tilde{v}^2=\tilde{v}_a\tilde{v}^a\ll1$ when dealing with non-relativistic peculiar flows.

\subsection{Relativistic peculiar kinematics}\label{sRPKs}
In the relativistic analysis, the irreducible kinematics of the reference $u_a$-field is no longer given by decomposition (\ref{Npbua}). Instead, the (covariant) gradient of the 4-velocity splits as(~\citealp{Tsagas2008, Ellis2012})
\begin{equation}
\nabla_bu_a= {1\over3}\,\Theta h_{ab}+ \sigma_{ab}+ \omega_{ab}- A_au_b\,,  \label{rNbua}
\end{equation}
where $\Theta$, $\sigma_{ab}$ and $\omega_{ab}$ are the relativistic analogues of the volume scalar, the shear tensor and the vortiicty tensor respectively (see (\ref{Npbua}) for the Newtonian version of the above). Relative to the latter expression, the new quantity is the 4-acceleration, which is defined by $A_a=u^b\nabla_bu_a$ and represents non-gravitational forces. An exactly analogous decomposition also applies to the gradient of the $\tilde{u}_a$-field, so that $\nabla_b\tilde{u}_a=(\tilde{\Theta}/3)\tilde{h}_{ab}+\tilde{\sigma}_{ab}+\tilde{\omega}_{ab}-\tilde{A}_a\tilde{u}_b$, with $\tilde{h}_{ab}=h_{ab}+2u_{(a}\tilde{v}_{b)}$ at the linear level. Also to linear order, the peculiar-velocity field splits according to
\begin{equation}
{\rm D}_b\tilde{v}_a= {1\over3}\,\tilde{\vartheta}h_{ab}+ \tilde{\varsigma}_{ab}+\tilde{\varpi}_{ab} 
\end{equation}
with ${\rm D}_a=h_a{}^b\nabla_b$ representing the spatial (covariant) derivative operator, while $\tilde{\vartheta}$, $\tilde{\varsigma}_{ab}$ and $\tilde{\varpi}_{ab}$ are the volume scalar, the shear and the vorticity associated with the $\tilde{v}_a$-field. The three kinematic sets defined above are related by(~\citealp{Maartens1998})
\begin{equation}
\begin{split}
\tilde{\Theta}= \Theta+ \tilde{\vartheta}\,, \\ \tilde{\sigma}_{\alpha\beta}= \sigma_{\alpha\beta}+\tilde{\varsigma}_{\alpha\beta}\,, \\ \tilde{\omega}_{\alpha\beta}= \omega_{\alpha\beta}+\tilde{\varpi}_{\alpha\beta} \\ \tilde{A}_a= A_a+ \dot{\tilde{v}}_a+ H\tilde{v}_a\,
\end{split}
\label{lRrels1}
\end{equation}
to first approximation.

\subsection{Peculiar flux and peculiar 4-acceleration}\label{ssPFP4A}
The crucial distinction, at least for our purposes, between the Newtonian and relativistic treatments of peculiar motions arises when the matter variables are involved. Unlike in Newtonian physics, in relativity the matter variables observed by two relatively moving observers generally differ. For example, in a perturbed Friedmann universe filled with a pressureless perfect fluid, we have (\citealp{Maartens1998}):
\begin{equation}
\tilde{\rho}= \rho\,, \hspace{10mm} \tilde{p}= p= 0\,, \hspace{10mm} \tilde{q}_a= q_a- \rho\tilde{v}_a \hspace{10mm} {\rm and} \hspace{10mm} \tilde{\pi}_{ab}= \pi_{ab}= 0\,,  \label{Rrels2}
\end{equation}
to linear order. In the above, $\rho$. $p$, $q_a$ and $\pi_{ab}$ are respectively the density, the isotropic pressure, the energy flux and the viscosity of the matter measured in the reference $u_a$-frame, with their tilded counterparts measured in the coordinate system of the peculiar flow.

At this point, it is important to draw the readers' attention to relation (\ref{Rrels2}c). According to the latter, in the presence of peculiar motions, one cannot set both $q_a$ and $\tilde{q}_a$ to zero simultaneously. This means that the cosmic fluid can no longer be treated as perfect. The imperfection appears in the form of a nonzero \textit{peculiar flux} triggered by the matter flow. Indeed, setting $q_a=0$ ensures that $\tilde{q}_a=-\rho\tilde{v}_a\neq0$, whereas $\tilde{q}_a=0$ guarantees that $q_a=\rho\tilde{v}_a\neq0$. One is free to choose the coordinate system where the energy flux vanishes. In what follows, we will take the traditional view and assume that there is no flux in the frame of the matter (i.e.~we will set $q_a=0$). Reversing this assumption and setting $\tilde{q}_a=0$ does not change the results, since it is only the frames' relative motion that matters. 

In general relativity matter fluxes gravitate, since they contribute to the energy momentum tensor. The gravitational input of the peculiar flux survives at the linear level and modifies the relativistic conservation laws of energy and momentum, which for a pressureless medium assume the linear forms (\citealp{Sah2024}):
\begin{equation}
\dot{\rho}= -\Theta\rho- {\rm D}^aq_a \hspace{10mm} {\rm and} \hspace{10mm} \rho A_a= -\dot{q}_a- 4Hq_a\,,  \label{lRcls}
\end{equation}
respectively. According to (\ref{lRcls}b), the nonzero peculiar flux ensures the presence of a nonzero \textit{peculiar 4-acceleration} even in the absence of pressure. It is also important to note that expression (\ref{lRcls}b) also follows from the linear relation (\ref{lRrels1}d) after setting $q_a=\rho\tilde{v}_a$ and $\tilde{A}_a=0$. Indeed, when dealing with pressure-free matter, one is free to choose the coordinate system where the 4-acceleration vanishes. Here, we will also take the traditional route and set $\tilde{A}_a=0$, in which case the linear relations (\ref{lRrels1}d) and (\ref{lRcls}b) guarantee that the Newtonian evolution law of linear peculiar velocities (see Eq.~(\ref{ltEuler1} in \S~\ref{sNCTPVs} previously) is replaced by\footnote{As with the energy flux before, it makes no practical difference if we were to assume that $A_a=0$}.
\begin{equation}
\dot{\tilde{v}}_{\alpha}= -H\tilde{v}_{\alpha}- A_a\,,  \label{lREuler}
\end{equation}
Therefore, in general relativity, linear peculiar velocities are not sourced by the gravitational potential, which does not exist anyway, but by the non-gravitational forces that the 4-acceleration represents. The exact linear form of the 4-acceleration follows for the linear energy density conservation law, $\dot{\rho}=-\Theta\rho-{\rm D}^aq_a$, the spatial gradient of which leads to (\citealp{Tzaprasi2020,Maglara2022,Miliou2024}):
\begin{equation}
A_a= -\frac{1}{3H}\,{\rm D}_a \tilde{\vartheta}+ \frac{1}{3aH} \left(\dot{\Delta}_a+\mathcal{Z}_a\right)\,,  \label{lR4a}
\end{equation}
where $\Delta_a=(a/\rho){\rm D}_a\rho$ and $\mathcal{Z}_a=a{\rm D}_a\Theta$ monitor the non-gravitational forces triggered by inhomogeneities in the spatial distribution of the matter and in that of the universal expansion.

\subsection{Linear evolution of peculiar
velocities}\label{ssLREPVs}
Taking the time-derivatives of (\ref{lREuler} and (\ref{lR4a}), combining the resulting expressions and keeping up to linear order terms, leads to the inhomogeneous differential equation~\citealp{Tzaprasi2020,Miliou2024}
\begin{equation}
\ddot{\tilde{v}}_a+ {1\over2}\,H\dot{\tilde{v}}_a- 2H^2\tilde{v}_a= {1\over3H}\,{\rm D}_a\dot{\tilde{\vartheta}}+ {1\over3aH}\left(\ddot{\Delta}_a+\dot{\mathcal{Z}}_a\right)\,.  \label{lRddotv}
\end{equation}
There is no analytic solution for the above. We will therefore confine to its homogeneous component, namely to\footnote{According to a well known theorem, the solution of an inhomogeneous differential equation comprises of the general solution of its homogeneous part plus a partial solution of the inhomogeneous equation. This guarantees that the growth rate ($\tilde{v}\propto t^{4.3}$) seen in solution (\ref{lRv}) is the minimum possible.}
\begin{equation}
\ddot{\tilde{v}}_a+ \frac{1}{3t}\dot{\tilde{v}}_a- {8\over9t^2}\tilde{v}_a= 0\,,  \label{diffv} 
\end{equation}
since $H=2/3t$ between equipartition and the onset of the accelerated-expansion phase. Then,~\citealp{Tzaprasi2020,Miliou2024}
\begin{equation}
\tilde{v}= \mathcal{C}_1t^{4/3}+ \mathcal{C}_2t^{-2/3}= \mathcal{C}_3a^2+ \mathcal{C}_4a^{-1}\,,  \label{lRv} 
\end{equation}
the dominant mode of which grows much faster than its Newtonian counterpart (see solution (\ref{bfv2}) in \S~\ref{sNTPVs} before).

\section{An effective gravitational potential}\label{sEGP}
We can compare the two equations which are the origin of the second order velocity equations. In relativistic physics we have:

\begin{eqnarray}
    \ddot{\tilde{v}}_a+ {1\over2}\,H\dot{\tilde{v}}_a- 2H^2\tilde{v}_a= =0
\end{eqnarray}

And in the Newtonian treatment:

\begin{eqnarray}
    \ddot{\tilde{v}}_{\alpha}+ 2H\dot{\tilde{v}}_{\alpha}- {1\over2}\,H^2\tilde{v}_{\alpha}= -\partial_{\alpha}\dot{\phi} \\
\end{eqnarray}

Where we have shown that two evolutions for the gravitational potential are used, namely:

\begin{eqnarray}
    \partial_{\alpha}\dot{\phi}&=&0 \\
    \partial_{\alpha} \dot{\phi}&=& H\dot{\tilde{v}}_a-\frac{1}{2}H^2\tilde{v}_a 
\end{eqnarray}

These both ansatz gives a growing mode for the velocity evolution as $\tilde{v}_a\propto t^{\frac{1}{3}}$. The first represent a slowly evolving gravitational field, meanwhile the second is a consequence of ignore source-free terms on the gravitational potential when the Poisson equation is integrated. This last assumption is described in (\citealp{Peebles1980}) as a way to get rid of source-free gravitational evolution. However, we showed that density perturbations are not the only source of gravitational effects in general relativity. The question now is if we can introduce another ansatz in the Newtonian treatment to obtain general relativity evolution law for peculiar velocities. We propose a general function of the form:

\begin{eqnarray}
    \partial_{\alpha}\dot{\phi}&=& f(H,\dot{H},\tilde{v}_a ,\dot{\tilde{v}}_a) 
\end{eqnarray}

This is well motivated because general relativistic effects are introduced by the flux $q_a=\rho v_a$ as a source of gravity. By simple inspection, we note that this function has to be:

\begin{eqnarray}
    \partial_{\alpha}\dot{\phi}&=& -\frac{3}{2}H\dot{\tilde{v}}_a-\frac{3}{2}H^2\tilde{v}_a
\end{eqnarray}

to reproduce relativistic velocity perturbation equation. The relation can be written also as:

\begin{eqnarray}
    \partial_{\alpha}\dot{\phi}&=& -\frac{3}{2}H\dot{\tilde{v}}_a+\dot{H}\tilde{v}_a =-\frac{3}{2}H(\dot{\tilde{v}}_a+H\tilde{v}_a )
\end{eqnarray}

or:

\begin{eqnarray}
    \partial_{\alpha}\dot{\phi}&=& \frac{3}{2}H\partial_{\alpha}\phi
\end{eqnarray}

It is possible then to approximate linear effects of general relativity in the evolution of peculiar velocities, modifying the \textbf{time evolution of the peculiar gravitational potential gradient} of Newtonian Gravity. The implications of this will be discussed in the next section.

\section{Discussion}\label{sD}
Large-scale peculiar motions, often referred to as bulk flows, are a well-established phenomenon in cosmology, consistently confirmed by numerous surveys. These flows, which currently span domains up to several hundred megaparsecs (Mpc), are believed to have originated as small peculiar-velocity perturbations around the time of recombination. Over time, these perturbations grew larger and faster through the process of structure formation and the increasing inhomogeneity of the universe after recombination. Despite this, the detailed evolution and implications of these bulk peculiar motions remain largely unexplored and poorly understood. The bulk-flow problem has become even more puzzling with the advent of more surveys, which have reported velocities and sizes that exceed the theoretical expectations set by the $\Lambda$CDM cosmological model. This discrepancy is particularly striking when considering the controversial "dark flows," which, with sizes on the order of 1000 Mpc and velocities reaching around 1000 km/s, seem to severely contradict the predictions of the standard model.

Interestingly, the growth rate of peculiar velocities appears to be more pronounced when a covariant general relativistic approach is employed, as opposed to the traditional Newtonian framework. To gain deeper insight into these results, we have attempted to develop a Newtonian model where the growth rates emerge more naturally. A critical aspect of this framework is the term involving the gradient of the time derivative of the peculiar gravitational potential, $\partial_{\alpha} \dot{\phi}$. We demonstrated that in the work of \citealp{Filippou_2021}, this term is set to zero, while in \citealp{Peebles1980}, it is effectively handled by solving the Poisson equation and neglecting the so-called "source-free" terms in its integral solution. Building on this, we suggest that if these "source-free" terms are interpreted as resulting from the peculiar velocity flux, the growth rate of peculiar velocities is likely higher than previously anticipated. The key conclusions derived from this approach are currently under development in collaboration with Professor Christos Tsagas from the Aristotle University of Thessaloniki, with further updates expected by the time of this thesis submission.
\clearchapterpage
\chapter{Studying the cosmological principle using SNIA data}
\label{chap:reds}

Recent studies using the latest \textit{Pantheon}+ SNIA compilation \citealp{Scolnic_2022}, \citealp{Brout_2022} continue to favor a non-zero cosmological constant $\Lambda$, which supports the idea of an accelerating universe with a negative deceleration parameter $q_0$ at present. However, to fully understand the true nature of dark energy (DE), further in-depth investigations are needed to extract meaningful insights from these data.

Theoretical perspectives on inhomogeneous cosmologies (\citealp{Celerier06}, \citealp{Alexander09}, \citealp{Cosmai19}, \citealp{Pasten_Fractal}, \citealp{Clarkson_2011}) and tilted cosmologies (\citealp{Tsagas2011}, \citealp{Tsagas2015}) suggest that the large-scale structure of the universe could significantly affect the interpretation of cosmological data. This challenges the traditional assumptions of the Copernican Principle, which posits that we are located in a typical, homogeneous region of the universe. In reality, we might reside in a region of underdensity, overdensity, or even within a bulk flow that is not at rest relative to the Hubble flow, all of which can influence measurements.

Observational evidence has shown that the distribution of matter in the local universe affects measurements in such a way that the scale of homogeneity appears to be larger than previously thought, potentially extending beyond 300 Mpc. Within the framework of the $\Lambda$CDM model, this implies that these effects should be observable at redshifts below 0.1. In fact, the existence of a so-called Hubble Bubble—a local void—was proposed by the author of \citealp{Zehavi:1998gz}, who used type Ia supernovae data to highlight these localized phenomena. This suggests that data at very low redshifts might not be entirely reliable, as shown in \citealp{Kenworthy19}, where anomalous behavior was observed. Since the Pantheon+ dataset includes observations at redshifts as low as $z=0.0001$, we may expect unusual or anomalous behavior in this region as well.

To fully explore these possibilities, it is crucial to study cosmological data across different redshift ranges and in various directions in order to critically assess the assumptions of the cosmological principle. A significant contribution to this field is the work by \citealp{Asvesta22}, where the authors demonstrated that a tilted Einstein-de Sitter universe could account for the observed acceleration of the universe without invoking dark energy, simply by considering the linear effects of peculiar motions. In the present study, we investigate the local universe through a cosmological differential approach, extracting the cosmological deceleration parameter $q_0$ for different redshift bins of SNIA data. Our results show both qualitative and quantitative deviations from the predictions of the $\Lambda$CDM model. These findings prompt further investigation into the potential causes behind these anomalies. This chapter is based on the published work \citealp{PASTEN2023101224} of the author.

\section{Cosmographic analysis}

The well known cosmographic Taylor expansion (\citealp{Visser_2004,Visser_2005})up to the third term or the luminosity distance is given by:
\begin{equation}
    D_L(z)=\frac{z}{H_0}+\frac{(1-q_0)}{2H_0}z^2-\frac{(1-q_0-3q_0^2+p
    _0)}{6H_0}z^3,
\end{equation}
with $a(t)$ being the scale factor of the FLRW universe, $q_0$ is the \textit{deceleration} parameter $q=-\ddot{a}{a}/\dot{a}^2$ evaluated today and $p_0=j_0-\Omega_k$, where $j=\dot{\ddot{a}}/aH^3$ is the \textit{jerk} which is degenerated with the curvature parameter and cannot be determined separately. Usually it is argued that $q_0 \sim -0.5$ from supernovae data (\citealp{Scolnic18}) and this is interpreted as an acceleration of the universe due to the presence of the mysterious DE. However, this interpretation changes if we consider inhomogeneous or tilted cosmologies, in which case $q_0$ is determined by the inhomogeneties rather than a cosmological constant.

Motivated by ideas on inhomogeneous and tilted cosmologies, we perform a differential study binning the data for different redshift ranges, to determine if the best fit of $q_0$ changes with redshift. In this case we use
\begin{equation}
    D_L^{i}(z)=\frac{z}{H_{0i}}+\frac{(1-q_{0i})}{2H_{0i}}z^2-\frac{(1-q_{0i}-3q_{0i}^2+p
    _{0i})}{6H_{0i}}z^3,
\end{equation}
for the $i$-th bin in the sample. The procedure for binning SNIA data is explained in the following section. To perform a comparison, we use both the cosmographic luminosity distance as well as the one for the flat $\Lambda CDM$ model. In this case the $q_{0i}$ parameter is the same for every redshift given by:
\begin{equation}\label{qcero}
    q_{0i}=\frac{1}{2}\Omega_{mi}-\Omega_{\Lambda i}=\frac{3}{2}\Omega_{mi}-1.
\end{equation}
The main difference is that the fit of $q_0$ from the cosmographic approach needs to be performed for redshifts below $z \simeq 0.3$ as the Taylor approximation fails at higher values, while for the fit of $q_0$ using $\Lambda$CDM we can use all the data for which the quantity of SNIA in those redshift is significant.

We have used $z \sim 0.8$ as an upper redshift limit for $\Lambda$CDM cosmology, as the number of data points per redshift bin for values larger than $z =0.8$ falls rapidly to less than $40$ data points per bin, very low to perform significant statistical analysis. We hope that this redshift threshold increase in the future.

It is important to notice that the cosmographic approach is commonly described as a \textit{fully independent model relation}, but that is not exactly true. This relation comes from the assumption that we are co-moving observers with the so-called CMB rest-frame (which is also assumed to be in rest with respect to the Hubble Flow). This means that possible effects of our relative motion are reduced to redshift corrections on observer and the data due to peculiar velocities.

If the data is corrected due to peculiar velocities, corrections has to been made also over the luminosity distance (\citealp{Davis11}). However, as pointed out in some studies (\citealp{Tsagas2011,Tsagas2015}), strong cosmological implications could be hidden in this relative motion more than just data corrections. If $z_{\odot}$ is the peculiar movement of our solar system moving w.r.t. the CMB and $z_{sn}$ is the peculiar movement of each supernovae, then the luminosity distance should be corrected to: 
\begin{equation}
    \bar{D}_L(z,z_{\odot},z_{sn})=(1+z_{sn})^2 (1+z_{\odot})D_L(z),
\end{equation}
where $\bar{D}_L$ is the corrected luminosity distance and $z$ is the cosmological redshift. This redshift corrections should not modify significantly the cosmological parameters estimation (\citealp{Davis_2019}). However, as we are doing a redshift binned analysis, those corrections could be more important at low redshift bins.

\subsection{SNIA analysis: The Pantheon + Sample} 

The Pantheon + sample (\citealp{Scolnic_2022}, \citealp{Brout_2022}) is the latest compilation of high and low redshifts SNIA and has a lot of improvements on photometric uncertainties and systematic as well as peculiar velocity corrections based on a velocity field reconstruction, improving the quantity of data at low redshift. To extract the information on $q_0$ from the data, we use the full Tripp Formulae for distance modulus:
\begin{equation}
    \mu_{obs}=m_b^*-M=m_b+\alpha x-\beta c+\Delta_M-M
\end{equation}
here, $m_b$ corresponds to the peak magnitude at time of B-band maximum, $M$ is the absolute B-band magnitude of a fiducial SNIA, $x$ and $c$ are light curve shape and color parameters corrections, $\alpha$ is a coefficient of the relation between luminosity and stretch, $\beta$ is a coefficient of the relation between luminosity and colour, and $\Delta_M$ is a corrections based on the mass of the host galaxy. 
This relation is used to compare with the theoretical expectation of a particular model using standard statistical analysis where:
\begin{equation}
    \mu=5\log_{10}{\frac{D_L}{10Mpc}}+25.
\end{equation}
In a $\Lambda$CDM universe we expect to obtain a uniform distribution values for $q_0$ as any part of a homogeneous-isotropic universe should lead to the same `` today" deceleration parameter. This expectation increases as the corrected magnitudes in Pantheon + has been performed even allowing for peculiar velocities corrections and redshift dependent nuisance parameters. As we are interested just in the $q_0$ parameter, we marginalize over the degenerated parameter involving $H_0$ and $M$ according to \citealp{Conley_2010}.

\subsection{Binning}

We follow two different approaches to study the $q_0$ distribution as a function of redshift, searching for cumulative and local effects in the Pantheon+ sample. 
\begin{figure}[!ht]
    \centering
    \includegraphics[width=14cm]{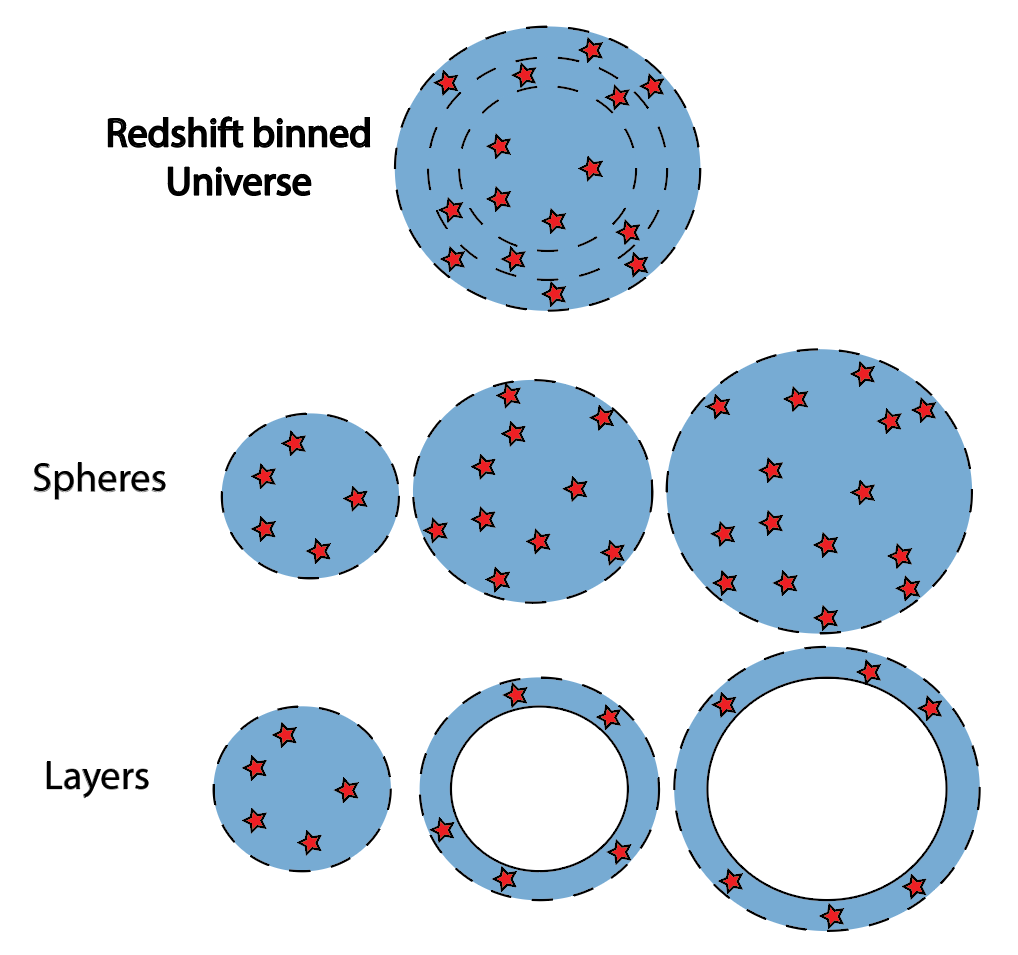}
    \caption{The 2 approaches used to describe $q0$ distribution using redshift bins.}
    \label{scheme}
\end{figure}

\subsubsection{Redshift spheres}

We begin with a sphere centered at $z=0$ of radius $z_{out}=0.01$ and we fit $q_0$ for all supernovae inside this sphere. Then, we increase the radius of the sphere in steps $\Delta z$ repeating the analysis for supernovae inside the new sphere until we reach the radius $z_{out}=0.3$. This method allows us to describe the behaviour of the local deceleration parameter and to search where DE existence becomes noticeable. Also, this procedure has the advantage that the number of data always increases from a initial value which is the number of supernovae with $z \leq 0.01$. \textbf{A similar approach was used in \citealp{Colgain_2019}}. However, we are not able to describe how $q_0$ behaves locally for each redshift as this study search for cumulative effects.

It is well established the presence of a Hubble Bubble in low $z$ (see e.g. \citealp{Kenworthy19}) leading to the usual removal of low redshift supernovae with $z\leq0.023$ in cosmological analysis as we can expect any unusual cosmological inference at those redshift.
 
Because SNIA with low redshift induce a bias in the Hubble diagram we also perform our study avoiding the use of data with $z<0.023$.  Overall, it is difficult to look for effects of the local structure in cosmology if we do not consider the redshifts in which those effects are more powerful. Therefore, we first consider the analysis using all supernovae and then remove the low redshift ones to look for variations.

\subsubsection{Redshift layers}

The second approach is to perform the analysis in layers of redshift.
For example, the studies performed in \citealp{Kazantzidis2020,Dainotti_2021,Dainotti_22}, use this type of layer selection. 
As the previous method, we begin with a sphere this time of radius $z_{out}=0.1$. Then we increase the radius in step of $\Delta z$ but this time we also increase the inner radius in the same step. It means that the second fit is performed in a layer of redshift $z \in [\Delta z,0.1+ \Delta z]$, the second for $z \in [2\Delta z,0.1+2\Delta z]$ and so on. This method is very useful to search for local difference in $q_0$ at different redshift. 

\section{Results}

\begin{figure}[!ht]
    \centering
    \includegraphics[width=14cm]{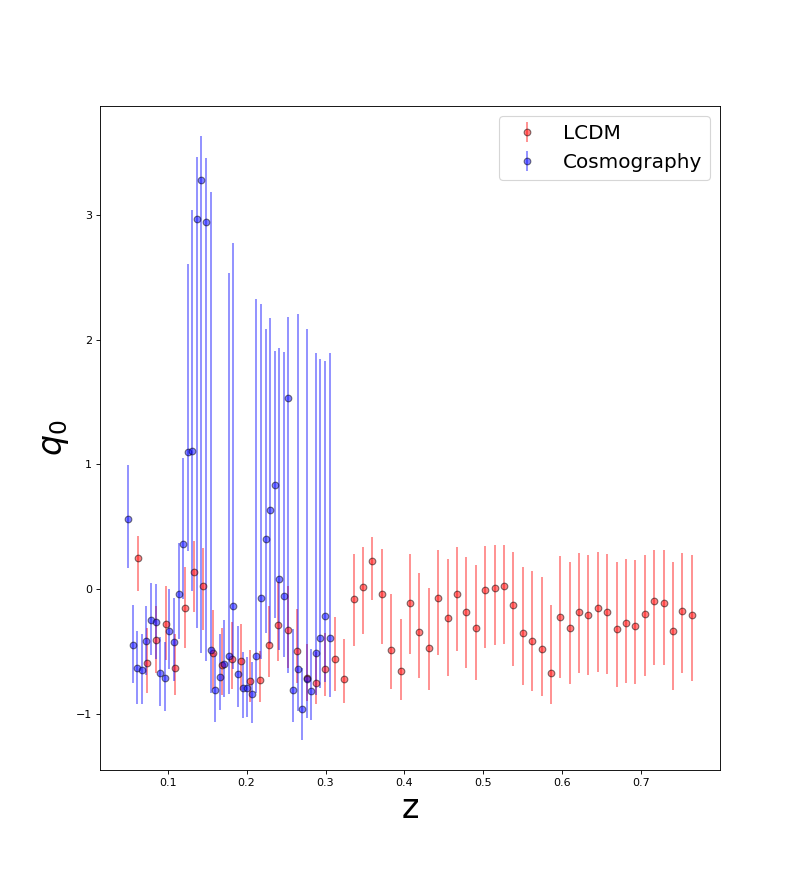}
    \caption{$q_0$ parameter using the layers approach. $z$ here is the mean redshift of the bin.}
    \label{q0varyng}
\end{figure}

\subsection{Initial Qualitative test}

\begin{figure}[!ht]
    \centering
    \includegraphics[width=14cm]{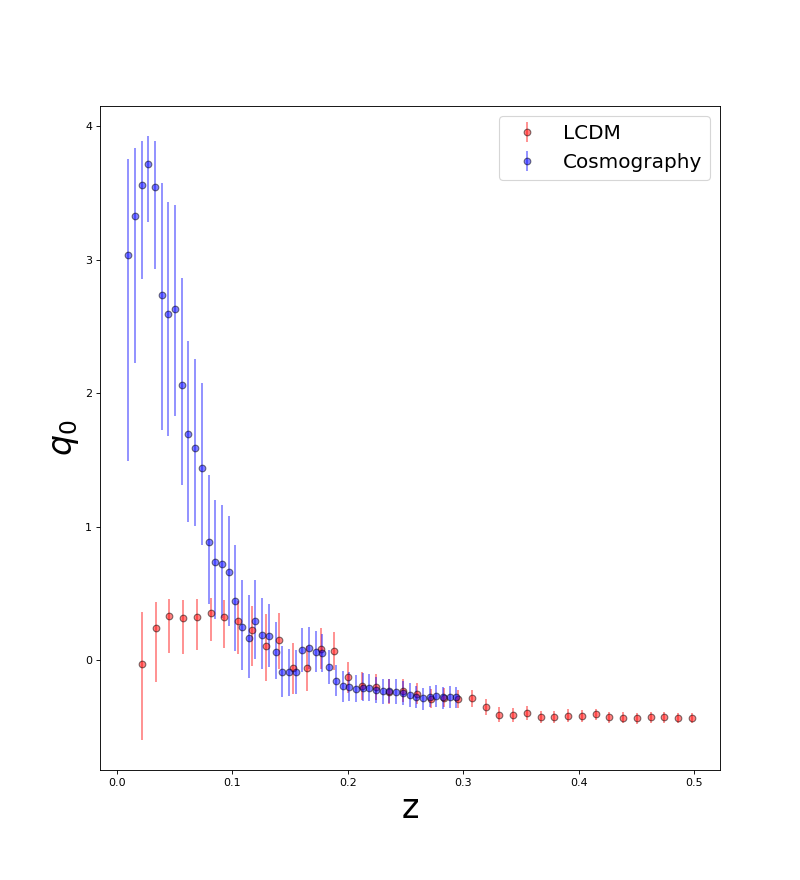}
    \caption{$q_0$ parameter using the spheres approach. $z$ here is the upper redshift of the sphere.}
    \label{q0fixed}
\end{figure}

We use the code \texttt{Emcee} (\citealp{Foreman2013}) to test the model against the data: the Pantheon+ SNIA (\citealp{Brout_2022}). This is a pure python implementation of the affine invariant ensemble sampler for Markov chain Monte Carlo proposed by Goodman and Weare (\citealp{Goodman2010}). The results of fitting $q_0$ using both the cosmographic luminosity and $\Lambda$CDM are shown in Fig. (\ref{q0varyng}) and (\ref{q0fixed}) for the sphere and layer approaches respectively. To look for insights in the cosmographic approach, we use a weakly uninformative uniform prior for $q_0\sim U(-4,4)$ not restricting our analysis to any cosmology. Qualitatively, we can observe a deviation from the uniform distribution for $q_0$ expected from a $\Lambda$CDM universe. Also, it is interesting to note that without a tight prior, MCMC samples tends to prefer higher values for $q_0$ in the cosmographic Taylor approximation than the ones allowed in $\Lambda$CDM at low redshifts.

\subsection{Quantifying possible anomalies in the layer approach}

In the context of the $\Lambda$CDM model, we expect $q_0$ should take the same value independent of $z$. To quantify the possible deviation of this result, we explore by using bins in which qualitatively we see a deviation from a uniform distribution for $q_0$. We have chosen the width of each bin to show the most appreciable differences in the fit of $q_0$. The results are summarized in table \ref{tabla1.2}.

\begin{table}[!ht]
\centering
\begin{tabular}{ccc}
 $z$ Bin & Number of SNIA & $q_0 $ \\
\hline
$0-0.15$  & 826 & $-0.087  \substack{+0.18\\-0.17}$ \\ 
$0.008-0.15$  & 746 & $-0.453 \substack{+0.16\\-0.16}$ \\ 
$0.05-0.2$ & 300 & $-0.457  \substack{0.17\\-0.16}$  \\
$0.1-0.25$ & 346  & $-0.501\substack{0.18\\-0.17}$  \\
$0.15-0.3$ & 381 & $-0.61\substack{0.16\\0.15}$ 
\end{tabular}
\caption{Values for $q_0$ fit in wide redshift bins using the cosmographic Taylor expansion of the luminosity distance.}
\label{tabla1.2}
\end{table}

\begin{figure}[!ht]
    \centering
    \includegraphics[width=14cm]{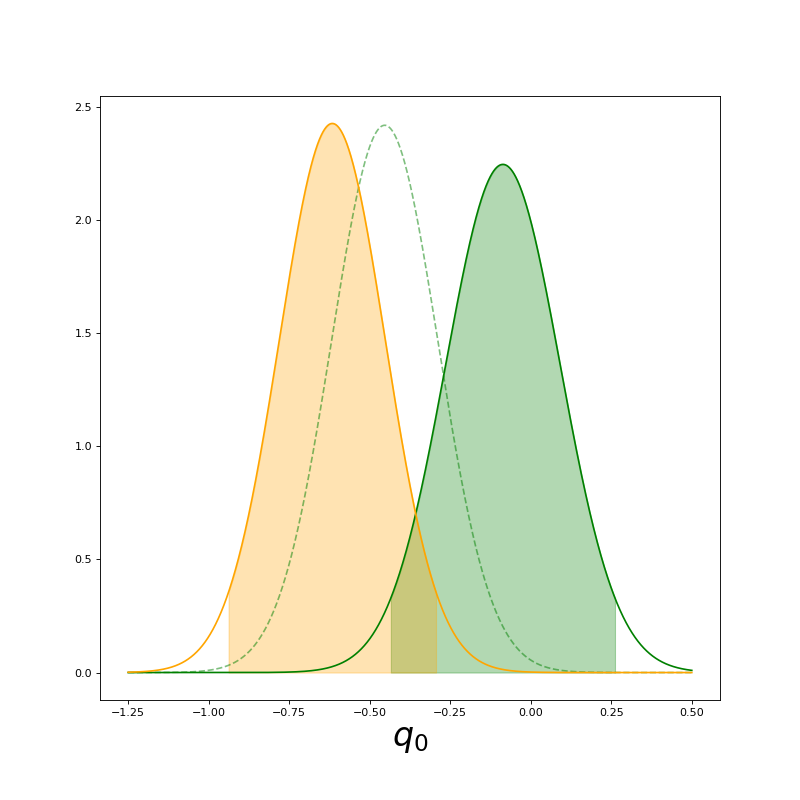}
    \caption{Gaussian distribution of $q_0$ cosmographic parameter for different redshift bins. Between $[0,0.15]$ (green) and $[0.15,0.3]$ (orange) appears a tension of $2.2 \sigma$ considering independent realizations. The tension almost disappear when supernovae below $z<0.008$ are removed, reaching a value $\sim 0.7 \sigma$ (dashed line).}
    \label{sigmas1}
\end{figure}

Assuming independent realizations, we observe that $q_0$ has a tension of $\sim 2.2 \sigma$ between the best fit values from supernovae at $z<0.15$ with those at $z\in[0.15,0.3]$. Overall as data in the two analysis are correlated, this tension could increase. However, if we remove the very low redshift SNIA data at $z<0.08$ the results change, decreasing this tension to zero with a value $\sim 0.7 \sigma$ (Figure \ref{sigmas1}).

\begin{figure}[!ht]
    \centering
    \includegraphics[width=10cm]{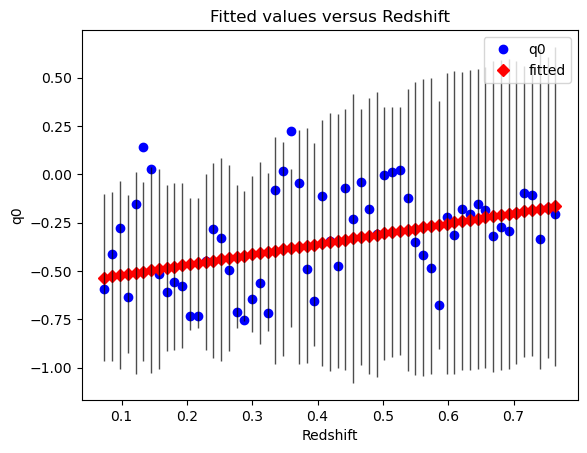}
    \includegraphics[width=10cm]{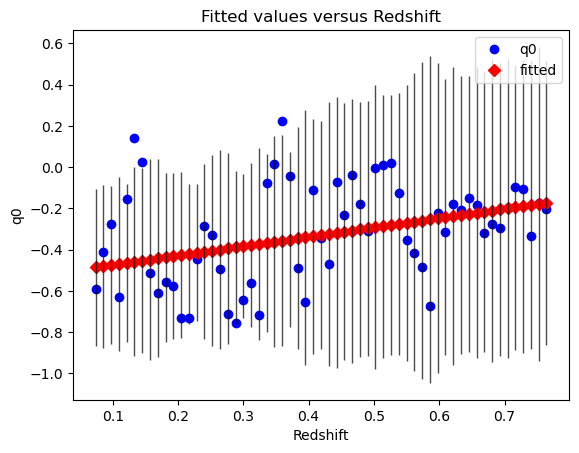}
    \caption{Best general linear model fits for $q_0(z)$ using $\Lambda$CDM luminosity distance with Gaussian error $\sigma$ estimated as the mean of the asymmetric errors from MCMC (Left) and estimated as the maximum value between them (Right).} 
    \label{Fit}
\end{figure}

To check the consistency of the derived $q_0$ values with the expectations in the $\Lambda$CDM model, we also performed a general linear model fit using the package \texttt{STATSMODELS} \citealp{seabold2010statsmodels}. In this case we use:
\begin{equation}
    q_0(z)=Q_m z+Q_0,
\end{equation}
and any deviation from a zero value for $Q_m$ is an indicator of a departure from the $\Lambda$CDM model. For simplicity we approximate each fit for $q_0(z)$ in the $\Lambda$CDM as being gaussian distributed, estimating the error $\sigma$ as the average between the asymmetric errors obtained by \texttt{Emcee} and also estimating $\sigma$ as the maximum between them. We do not consider the first bin for $z \leq 0.008$ data. We also performed a fit with a zero slope $Q_m=0$ model, in order to compare with the general non zero case. Results are summarized in Table \ref{tabla2} and the best fit are plotted in figure \ref{Fit}. According to the statistics, the fit tends to prefer a non-zero slope over a constant value of $q_0$ for all redshifts. Overall, the fit is consistent with a negative mean $q_0$ parameter. 

\begin{table}[ht]
\begin{flushleft}
\begin{tabular}{lcc}
Statistic & Best linear fit of SNIA & Constrained fit with zero slope \\
\hline
$Q_m$ & $0.5341 \pm 0.190$ ($0.4532 \pm     0.173 $) & 0  \\ 
$Q_0$  & $-0.5736 \pm 0.071 $ ($-0.5189 \pm  0.066$)& $-0.3987      \pm 0.037(-0.3699      \pm 0.034)$ \\ 
$\log{\mathcal{L}}$ & $-9.4073(-5.0585)$ & $-13.241(-8.4040)$  \\
Pearson $\chi^2$ & $35.7(26.0)$  & $40.6(29.1)$  \\
Pseudo $R^2$ & $0.1260(0.1101)$ & $0.0001453(0.0001453)$ \\
AIC & $22.81(14.11)$ & $28.48(18.8)$ \\
BIC & $26.96(18.27)$ & $30.55(20.89)$
\end{tabular}
\end{flushleft}
\caption{Best linear fit parameters for $q_0$ using $\Lambda$CDM luminosity distance estimating the Gaussian error $\sigma$ as the average between asymmetric errors from MCMC.  In parenthesis, same values but this time estimating $\sigma$ as the maximum value between them.}
\label{tabla2}
\end{table}

\subsection{Quantifying possible anomalies in the spheres approach}

\begin{figure}[!ht]
    \centering
    \includegraphics[width=12cm]{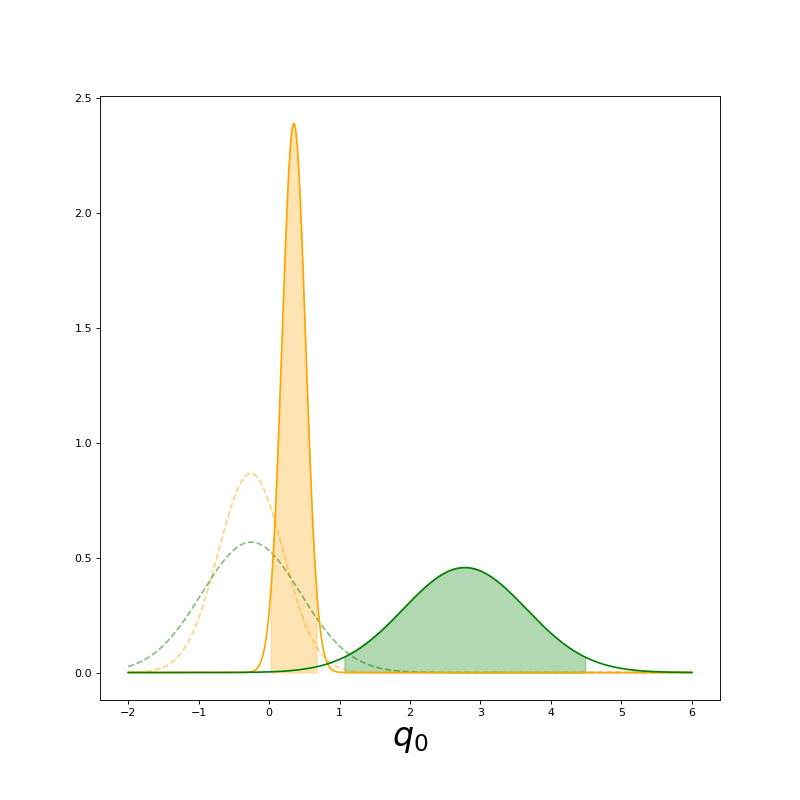}
    \caption{Gaussian distribution of $q_0$ cosmographic luminosity distance (green) and $\Lambda$CDM (orange) up to redshift $0.05$. The cosmographic distance prefer higher values for $q_0$ even when the prior is tight, giving a tension of $\sim 2.8\sigma$ between both distance indicators. This tension disappears when the low redshift $z<0.008$ data is removed (dashed line).}
     \label{sigmas2}
\end{figure}

Using the ``spheres'' method, and from the results displayed in Fig.(\ref{q0fixed}), we observe a noticeable difference between the best fit for $q_0$ from the cosmographic expansion and those from the $\Lambda$CDM model for redshift below $z =0.1$. For example, restricting our analysis to data points below $z = 0.05$, we obtain the best fit values quoted in Table (\ref{tabla3}).
Considering gaussian distributions, we obtain a tension between the cosmographic distance fit and the $\Lambda$CDM one of $\sim 2.8$ (Figure \ref{sigmas2}). Even by decreasing the parameter space for the prior $q_0 \sim U(-1,1)$, the cosmographic distance still prefer a best fit $q_0>1$. Interestingly, in both cases the best fit takes positive values of the deceleration parameter at low redshift $z<0.008$, but the value changes dramatically when they are removed, making the tension disappears.

\begin{table}[ht]
\centering
\begin{tabular}{ccc}
Distance indicator & Number of SNIA & $q_0 $  \\
\hline
Cosmographic & 648(568) & $2.78 \substack{0.98\\-0.87}(-0.25 \substack{0.73\\-0.68})$   \\ 
$\Lambda$CDM & 648(568) & $0.35  \substack{0.11\\-0.22}(-0.26  \substack{0.47\\-0.45})$  \\
\end{tabular}
\caption{Best fit values for $q_0$ in wide redshift bins using the cosmographic Taylor expansion of the luminosity distance at $z<0.05$. In parenthesis we quote the values obtained when we have removed the SNIa data with $z<0.008$}
\label{tabla3}
\end{table}

\subsection{Local Structure Effects}

\begin{figure}[!ht]
    \centering
    \includegraphics[width=14cm]{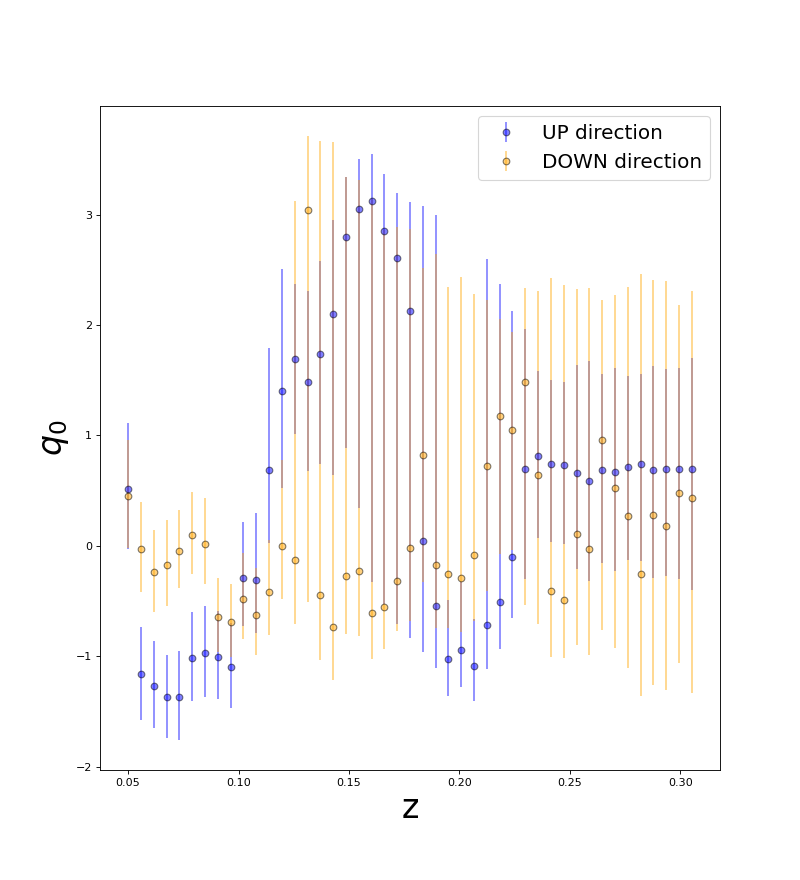}
    \caption{$q0$ dipolar parameter using the layers approach.}
    \label{q0varyngdipole}
\end{figure}

In this section we consider a possible orientation dependence of the best fit value for $q_0$ motivated by anomalies in the CMB dipole direction (see \citealp{Aluri22} for a review). Similar analysis have been performed in the past with different data sets. Usually, a positive result of this orientation dependence is interpreted as a local distribution structure effect on the data. We perform a layer binned analysis with the cosmographic luminosity distance for two hemispheres, oriented to the dipole of the CMB, the one interpreted as our motion respect the CMB rest frame. To look for relations between our movement w.r.t. the CMB and the $q_0$ parameter distribution, we perform a layer binned analysis with the cosmographic luminosity distance for two hemispheres separated by our direction of movement.  We take the direction reported in \citealp{PLANCK20} corresponding to $(263.99,48.26)$ in galactic coordinates to split the data in two groups according to the two hemispheres of the CMB dipole. We call \textit{UP} and \textit{DOWN} respectively to the blue-shifted and red-shifted hemispheres of the dipole. Although there are others possible anomalous directions in cosmology, we selected directly the CMB dipole direction as it is the principal standard cosmological anisotropy (see \citealp{Aluri22} for a review). Also some alternative cosmologies as \textit{tilted cosmologies} predict an anisotropy in the $q_0$ parameter close to this direction (see \citealp{Tsagas2011, Tsagas2015}).

\begin{figure}[!ht]
    \centering
    \includegraphics[width=12cm]{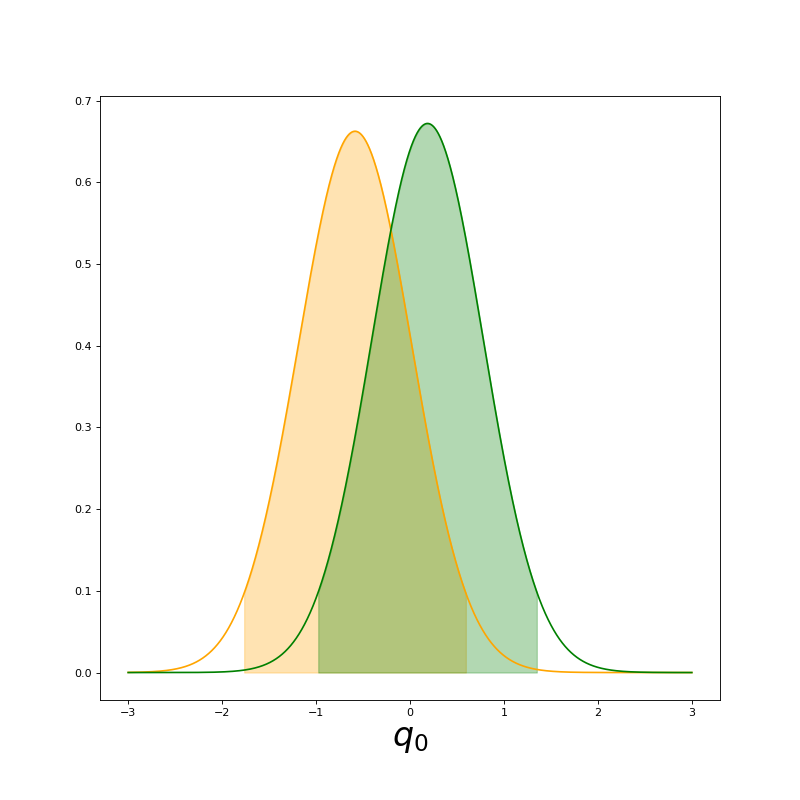}
    \caption{Gaussian distribution of $q_0$ cosmographic luminosity distance at low redshift $[0.008,0.075]$ for the 2 hemispheres. The tension between the fits doesn't reach the unity.}
    \label{sigmas3}
\end{figure}

The results are shown in Figure \ref{q0varyngdipole}. Qualitatively, we can see two possible sources of anomalies. First, a visual discrepancy of $q_0$ values between the hemispheres at low redshift is observed. To analyze this, we fit $q_0$ for the two hemispheres at $z \in [0.008,0.075]$. Results are in Table \ref{tabla4}. For Gaussian distribution of parameters, we can see that the tension between hemispheres is non-existent (Figure \ref{sigmas3}).

Notice that for $z<0.1$ the UP value for $q_0$ is lower than the DOWN one by a factor of $1$, at more than one sigma. In the range $0.11< z < 0.15$ the trend is the opposite, being the UP value higher than the DOWN one. From $z>0.2$ considering the uncertainties, both hemispheres fit  essentially the same value for $q_0$, although with larger uncertainties.

\begin{figure}[!ht]
    \centering
    \includegraphics[width=12cm]{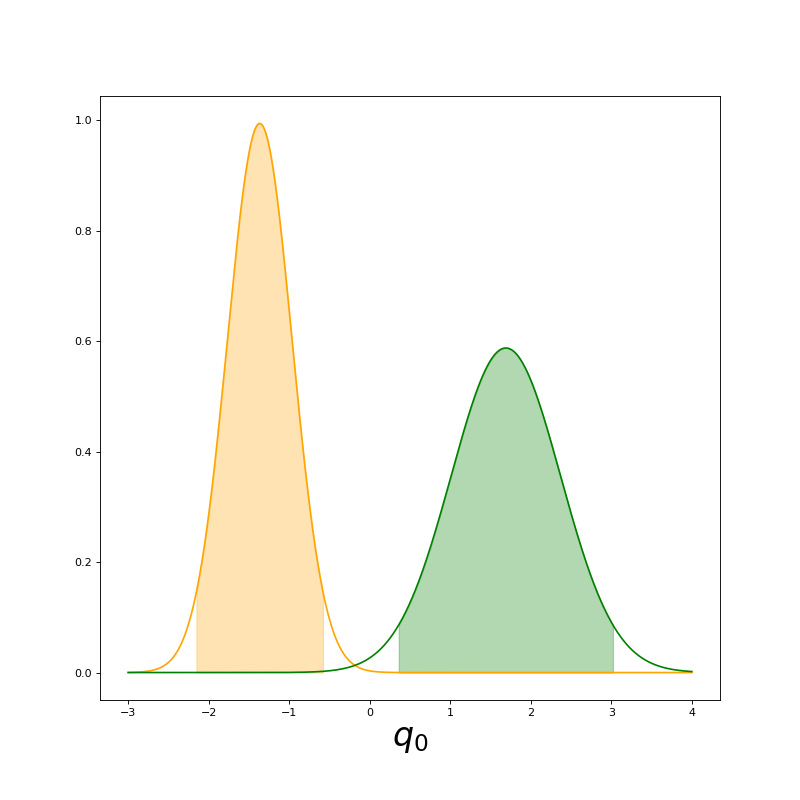}
    \caption{Gaussian distribution of $q_0$ cosmographic luminosity distance at $z \in [0.0232,0.1232]$ and $z \in [0.0754,0.1754]$ for the UP hemispheres. The tension between the posterior paremeter distributions reach 3.8 $\sigma$.}
    \label{sigmas4}
\end{figure}

Also only on the UP hemisphere, we note an increase in the $q_0$ parameter comparing the redshift ranges lower than $z\sim0.075$ and $z\sim 0.11-0.15$. For example, in particular the bins $[0.0232,0.1232]$ and $[0.0754,0.1754]$ show the best fit parameters summarized in Table \ref{tabla5}, giving a tension around $3.8 \sigma$ (Figure \ref{sigmas4}). However, at $z\sim 0.11-0.15$ there are only 23 supernovas, requiring more data to extract significant conclusions. 

\begin{table}[!ht]
\centering
\begin{tabular}{ccc}
$z$ bin & Number of SNIA & $q_0 $  \\
\hline
$0.008-0.075$(UP) & 314 & $-0.59 \substack{0.62\\-0.59}$   \\ 
$0.008-0.075$(DOWN) & 314 & $0.19 \substack{0.62\\-0.57}$  \\
\end{tabular}
\caption{Values for $q_0$ fit at low redshift using the cosmographic Taylor expansion of the luminosity distance for the 2 hemispheres.}
\label{tabla4}
\end{table}

\begin{table}[!ht]
\centering
\begin{tabular}{ccc}
$z$ bin & Number of SNIA & $q_0 $  \\
\hline
$0.0232-0.1232$(UP) & 214 & $-1.37 \substack{ 0.41\\  -0.39}$   \\ 
$0.0754-0.1754$(UP) & 92 & $1.69 \substack{0.68\\-0.67}$  \\
\end{tabular}
\caption{Values for $q_0$ fit in particular redshift bins using the cosmographic Taylor expansion of the luminosity distance on the \textit{UP} hemisphere.}
\label{tabla5}
\end{table}

\section{Discussion}

In this work we have found several unexpected behavior for the distribution of the deceleration parameter $q_0$ values  using the recent type Ia supernova sample Pantheon+.
First, analyzing the low redshift supernovae ($z<0.3$) we obtain evidence for a local decelerating universe using the cosmographic distance and considering the full sample of Pantheon+. However, this behaviour disappears once the SNIA at redshift lower than $z=0.008$ are removed. A tension of $\sim 2.2\sigma$ between the cosmographic expansion fit at different redshift bins are observed when the full sample is considered, and also a discrepancy of $\sim 2.8\sigma$ between the preferred values for $q_0$ with the cosmographic expansion function and the full $\Lambda$CDM one. Again, this behaviour disappears once the lowest redshift supernovae (those with $z<0.008$) are removed.

At this point it is interesting to comment the result of \citealp{Colgain_2019} where the author find, using the Pantheon sample, the fit for $\Omega_{m0}$ varies with the maximum redshift of the data used, finding a very low value between $z \simeq 0.1$ to $z\simeq 0.15$, a result also found in \citealp{Camarena20}, and even wiggles have been found \citealp{Kazantzidis2020}.

Also, assuming the $q_0(z)$ is a varying function 
of redshift, being described by a linear model, the statistical analysis favours a non-zero slope according to AIC and BIC criteria, showing a clear departure from the $\Lambda$CDM universe. Finally, we performed a binned study in the hemispheres defined by the CMB dipole. We did not found evidence for tensions between the hemispheres. However, in the hemisphere with the positive direction of our movement, we found a tension of $3.8 \sigma$ between two particular redshift bins. All results can be summarized as:

\begin{enumerate}
    \item \textbf{When $z<0.008$ supernovaes are considered, a local decelerated universe in all fits is observed, and tensions $\sim 2\sigma-3\sigma$ between $q_0$ parameters at different redshift and between different distance indicator emerge.} According to \citealp{Brout_2022}, the very nearby Hubble Diagram shows a positive bias at $z<0.008$, attributed to measurement errors, unmodeled peculiar velocities and the volumetric bias they induce. Moreover, they suggest to avoid the use of data for $z<0.01$ and also in \citealp{Riess_2022} the author neglect data below $z<0.023$ to measurements of the Hubble Flow. Then, is therefore possible that the tensions found in $q_0$ comes from the same origin. However, it is difficult to perform a deep analysis of local large-structure effects if the lowest SNIA redshift are removed, as it is possible that strong effects for the overall cosmology are ignored with this approach. As is presented qualitatively in the \textit{spheres  binned} analysis, the consideration of the lowest redshift SNIA fit a positive value for $q_0$ until $z\sim 0.09$ for the cosmographic luminosity distance. We conclude that it is important to model the very nearby peculiar velocity field to extract cosmological useful information of the most local universe with Pantheon+ as is impossible to do it with the data as it is.
    \item \textbf{Linear regression favours a non-zero slope for $q_0$ in the layer approach with $\Lambda$CDM luminosity distance when the posterior distributions for $q_0$ are assumed to be gaussians.} This result is important because it is observed even removing the lowest redshift supernovae, concluding that the binned values $q_0(z)$ are not uniform across redshift bins. The $\Lambda$CDM model predicts the same value for $q_0$ independent of redshift bin chosen. We consider that a departure from this behaviour can be considered as an evidence against $\Lambda$CDM. This method would become stronger as the quantity of data increase, reducing the errors in the parameter extraction from each redshift binned universe.
    \item \textbf{Not evidence for tensions between UP and DOWN hemispheres.} As previous studies, we did not find evidence for anisotropies between the hemispheres defined by the CMB dipole. In our analysis we used as redshift $z$ the value $z_{HD}$ provided in the catalogue. However, there is an ongoing debate about which is the correct reference frame to be used to find anisotropies in the data. Overall, we do not suggest to use $z_{hel}$ as the frame of reference, because the current cosmology and distance indicators are constructed for observers that are in rest w.r.t. the Hubble flow, requiring a new theoretical approach to analyse the data directly from observed redshift.
    \item \textbf{$3.8 \sigma$ tension for the UP hemisphere at particular redshift bins.} Intriguingly, in one of the hemispheres appears a high tension for the $q_0$ parameter at different redshift bins. If the data is correct, this could be evidence for a local large structure effects and/or peculiar velocities that are not considered in the current cosmological analysis.
\end{enumerate}

\subsection{Interpretations}

We have summarized the main results of our analysis of the Pantheon+ survey, showing that the $q_0$ parameter exhibits certain anomalies when compared with what we would expect in the standard $\Lambda$CDM model. However, what could these deviations mean from a cosmological perspective?

One possibility is that the universe could still be described as a $\Lambda$CDM universe on a local scale, but the global cosmological parameters ($\Omega_M \sim 0.3$, $\Omega_\Lambda \sim 0.7$, $\Omega_K \sim 0$) may not represent the behavior of the universe everywhere. This idea, however, poses a contradiction. If the $\Lambda$CDM model cannot provide the same parameters across all regions of the universe, where homogeneity is assumed to hold, then the universe is not truly $\Lambda$CDM after all. Variations in the $q_0$ parameter across different regions could suggest that the dynamics of the universe are not uniform everywhere. We must also keep in mind that our analysis is based on a Taylor approximation and assumes a flat $\Lambda$CDM universe, so other parameters—such as curvature, $\Omega_k$—might vary from place to place, potentially introducing biases into our local estimates of $q_0$. Again, this would challenge the cosmological principle and necessitate the development of alternative models.

Evidence supporting such deviations in cosmological parameters with redshift has been observed in previous studies (\citealp{Dainotti_2021}, \citealp{Dainotti_22}, \citealp{Colgain_2019}, \citealp{Wong_2019}, \citealp{Millon_2020}, \citealp{Krishnan_2020}, \citealp{Colgain2022}, \citealp{Camarena20}). These works show a decreasing trend for $H_0$ and an increasing one for $\Omega_{m0}$, which align with the results we present in Figure \ref{Fit}, as $\Omega_{m0}$ is directly related to $q_0$ through Eq. \ref{qcero}, and $H_0$ is anti-correlated with $\Omega_{m0}$ in the $\Lambda$CDM framework.

However, we cannot disregard the possibility that the discrepancies observed in the data are due to systematic errors resulting from the combination of different surveys within the Pantheon sample. According to \citealp{Kazantzidis2020}, there are noticeable “wiggles” in the best-fit values of $\Omega_{m0}$ when derived from Pantheon data. While Pantheon+ has made efforts to reduce these systematic anomalies, it is still plausible that such issues persist in the dataset. These systematic effects could potentially account for the anomalies we observe, particularly since we are comparing different surveys across various redshift bins. It is also crucial to replicate our analysis on mock data to ensure that the observed results are not merely artifacts of the binning procedure. As our analysis involves extensive binning and multiple statistical tests, which require significant computational resources, we are preparing such an analysis for a future study.

Alternatively, if the universe is not $\Lambda$CDM, it raises the question of whether the $q_0$ value extracted from cosmological luminosity distance measurements truly reflects cosmic acceleration. For example, in inhomogeneous cosmological models like Lemaître-Tolman-Bondi (LTB), $q_0$ arises from the radial inhomogeneities in our universe, rather than from a real kinematic acceleration. Moreover, some models of cosmological Taylor approximations predict $q_0$ values outside the range expected by the $\Lambda$CDM cosmology ($q_0 \in [-1, 0.5]$), but further data is needed to confidently constrain these values and fully understand the implications.

It is also important to consider the possibility that the data itself has not been properly processed. There has been significant criticism regarding the treatment of type Ia supernova data, particularly in not accounting for potential magnitude evolution or introducing redshift-dependent nuisance parameters to address correlations between SNIA properties and redshift. These correlations are often attributed solely to intrinsic properties of the supernovae, rather than being related to cosmological factors. As such, we must be cautious in drawing definitive conclusions from the current dataset. Further, additional data is needed to better constrain the potential deviations in $q_0$ from the expectations of the $\Lambda$CDM cosmology.

In conclusion, while our results highlight some intriguing deviations, a conclusive understanding of these anomalies requires much more detailed data and further theoretical investigation. Until we have more comprehensive and refined datasets, any strong assertions regarding the nature of $q_0$ and the validity of $\Lambda$CDM are premature.

\section{Conclusions}

We proposed two approaches to extract information about the local deceleration parameter $q_0$ from different redshift binned universe, using the well known cosmological Taylor expansion of the luminosity distance and also a flat $\Lambda$CDM universe. The first accounting for a cumulative effect (\textit{spheres}) and the second to study differences in cosmological parameter for different redshift bins (\textit{layers}). We found that when $z<0.008$ supernovae are added to cosmological analysis, tensions $\sim 2\sigma-3\sigma$ emerge between different redshift bins and different luminosity distance indicator. We suggest not to use lowest Supernovae IA in cosmological analysis, but we also state the importance to improve the scientific utility of the data at lower redshift as are necessary to understand the influence of the local large-scale structure in cosmology. Also, using the cosmological Taylor expansion of the luminosity distance, we observe a strong tension $\sim 3.9$ between particular redshift bins, when we analyze the data on the hemisphere that the CMB dipole points. However, significant anisotropies between the 2 hemispheres were not observed. Finally, performing a simple general linear model fit against the redshift binned $\Lambda$CDM parameters, we found that the $q_0(z)$ function prefers a non-zero slope over a model with zero slope expected from a $\Lambda$CDM universe. Those results have to be treated with caution as the quantity of data is not that big to state definitive conclusions. We proposed this method as a test for $\Lambda$CDM cosmology, which could be improved as more data is release. 

Finally, we conclude that binned cosmology will be a promising tool to analyze deeply the structure of our universe without assuming the cosmological principle. Our analysis is similar to a recent paper \citealp{Perivolaropoulos:2023iqj} -- that use the Pantheon + sample -- where the authors have shown a $2$ to $3$ $\sigma$ tension between the low and high redshift best fit for the SNIA absolute magnitude $M$. However they did not find variations in $\Omega_{m0}$ and therefore not in $q_0$. As the quantity of data increase, this tool could be performed jointly with anisotropic cosmology in order to get the best insights of the deviations of our universe from $\Lambda$CDM model. 


\section{On the inhomogeneity of redshift bins}

\begin{figure}[!ht]
    \centering
    \includegraphics[width=12cm]{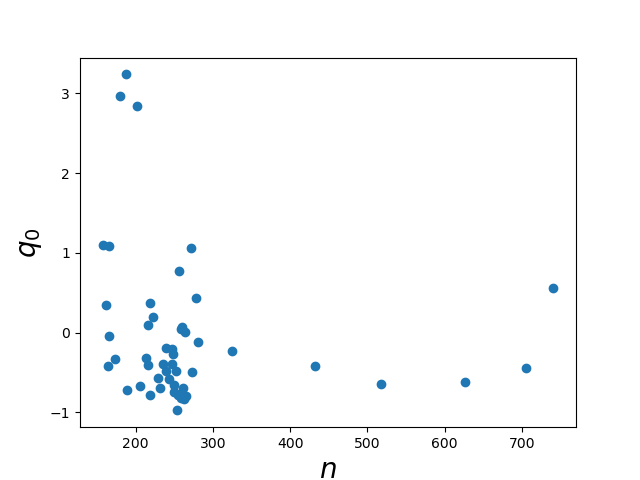}
    \includegraphics[width=12cm]{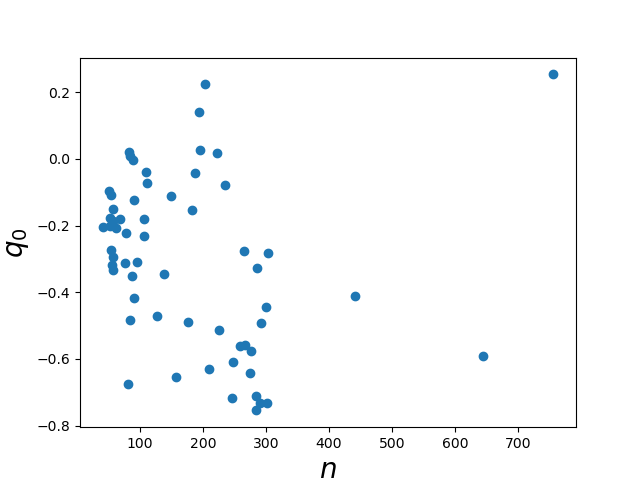}
    \caption{$q_0$ values against the number $n$ of data in each redshift bin for the Cosmographic (left) and the $\Lambda$CDM (right) approach. The correlation coefficients are $R=-0.168$ and $R=-0.201$ respectively.} 
    \label{Correlations}
\end{figure}

\begin{figure}[!ht]
    \centering
    \includegraphics[width=9cm]{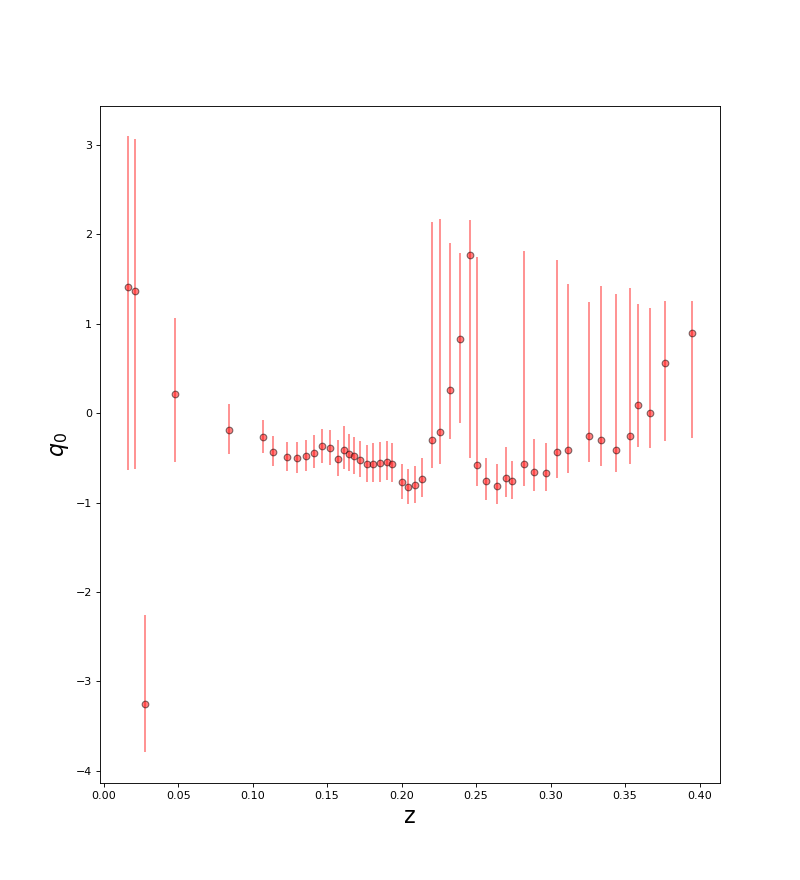}
    \includegraphics[width=9cm]{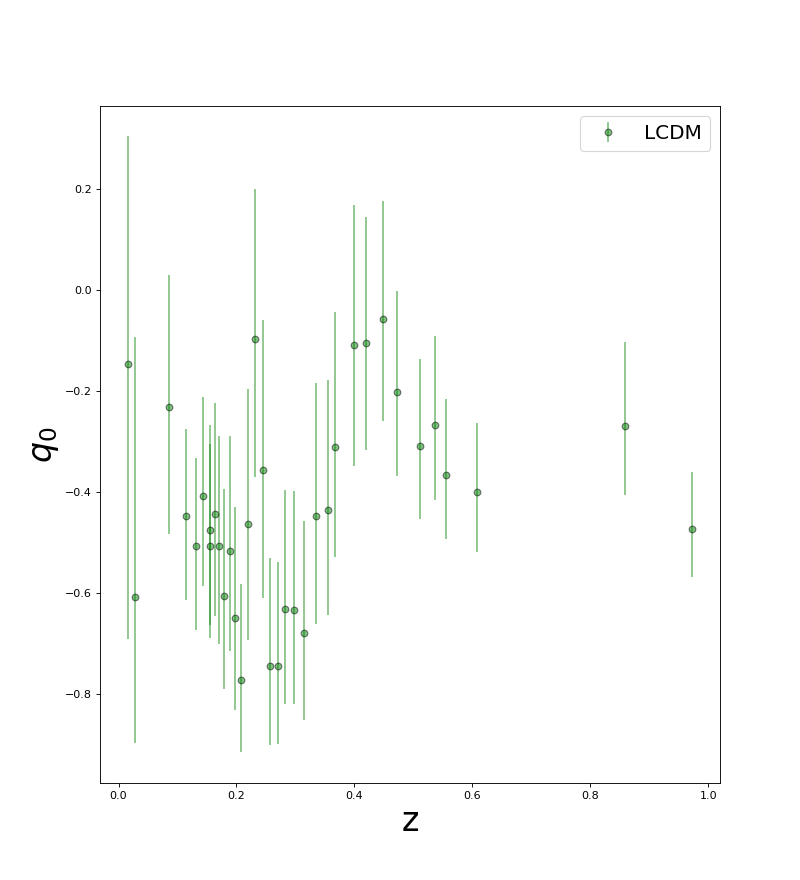}
    \caption{$q_0$ values against $z$ for cosmographic and $\Lambda$CDM luminosity distance in the layer approach using redshift bins with exactly $300$ supernovaes each one. $z$ here is the mean redshift value of the bin.  Note that the bins are not equispaced anymore and the presence of the outlier value around $z\sim 0.03$ possible attributed to the influence of lower redshift supernovaes even when $z<0.008$ are removed. } 
    \label{Homogenous}
\end{figure}

\begin{figure}[!ht]
    \centering
    \includegraphics[width=12cm]{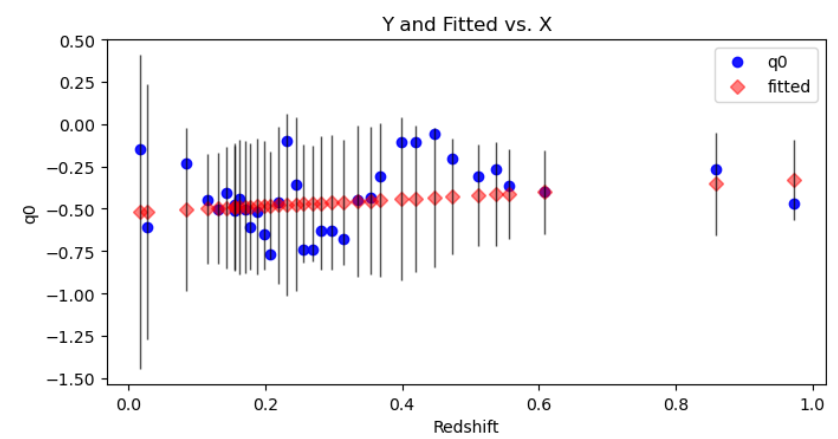}
    \includegraphics[width=12cm]{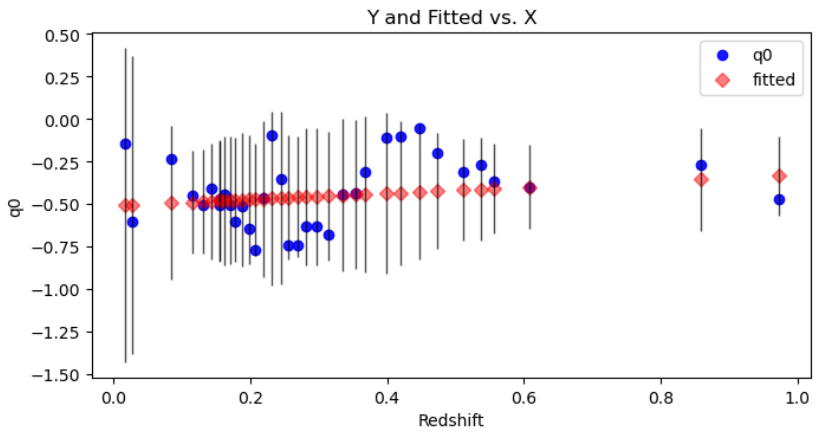}
    \caption{Best general linear model fits for $q_0(z)$ using homogeneous redshift bins, each one with $300$ SNIA. We used $\Lambda$CDM luminosity distance with Gaussian error $\sigma$ estimated as the mean of the asymmetric errors from MCMC (Left) and estimated as the maximum value between them (Right).} 
    \label{Fit2}
\end{figure}

\begin{figure}[!ht]
    \centering
    \includegraphics[width=14cm]{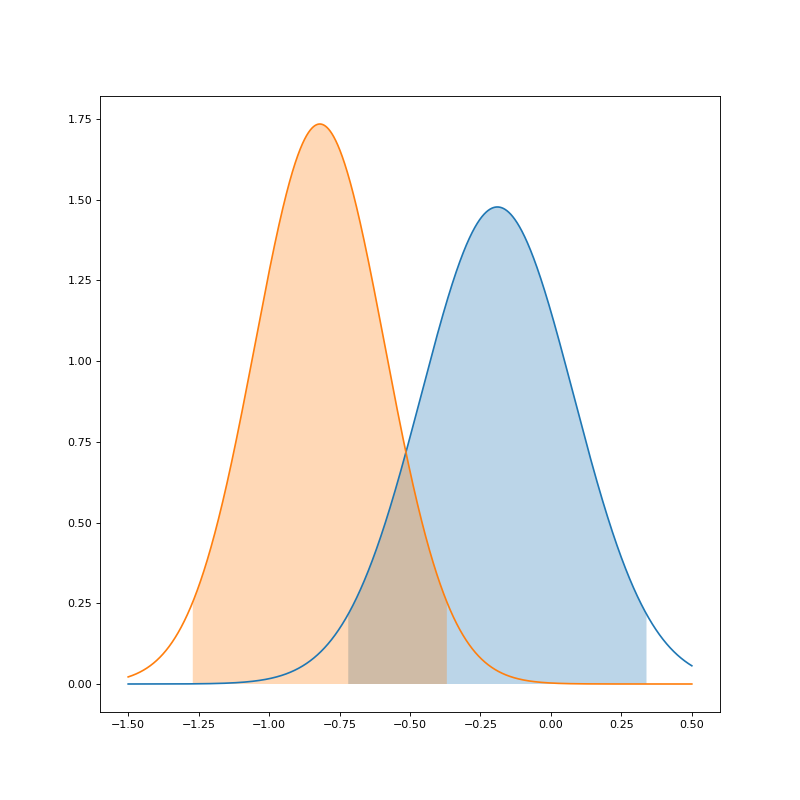}
    \caption{Gaussian distribution of $q_0$ cosmographic parameter for different redshift homogeneous bins, each one of 300 SNIA. Between $[0.0312,0.1364]$ (blue) and $[0.2052,0.3221]$ (orange) appears a tension of $1.78 \sigma$ considering independent realizations.}
    \label{sigmasultima}
\end{figure}

There is the possibility that the inhomogeneity in the quantity of data in each redshift bin affect our results. To study this concern, we seek for correlations between $q_0$ best values and the quantity of data in each redshift bin for Cosmographic and $\Lambda$CDM approach. Plots of $q_0$ against the quantity of data in each bin $n$ appears in Figure \ref{Correlations} for the 2 approaches. The correlation coefficients are $R=-0.168$ and $R=-0.201$ respectively, which pointed a slightly negative but statistically negligible correlation. We can address the inhomogeneity in bins further studying $q_0$ over redshift bins with $z>0.008$ each one having $300$ supernovaes. Note that despite of an homogeneous distribution of data across redshift bins, each redshift space is not homogeneous anymore. To perform a good comparison, we select the lower limits of the bin to be the same obtained in the layer approach for cosmographic and $\Lambda$CDM luminosity distance, varying therefore the upper limit. Results are plotted in Figure \ref{Homogenous}. The principal problem with this approach, is that we are not able to study higher redshift data isolated as those bins are far wider. For example the last bin with 300 SNIA data is $z \in [0.40,1.54]$, covering more than $\Delta z \sim 1$ space. Indeed, if we try to fit the general linear model over $q_0(z)$ function, we cannot discriminate between zero or non-zero slope. Results are plotted in Figure \ref{Fit2} and best statistics in Table \ref{tablafinal}. Finally, we tried to find tensions at low redshift data with the cosmographic Taylor expansion approach analogous as the results shown in Table \ref{tabla1.2} and Figure \ref{sigmas1}, but this time using the homogeneous redshift bin. We found a tension of $\sim 1.7\sigma$ between 2 particular redshifts. Results are plotted in Figure \ref{sigmasultima} and details in Table \ref{tablaultima}. Note that SNIA with $z<0.008$ are not influencing this result.

\begin{table}[!ht]
\centering
\begin{tabular}{lcc}
Statistic & Best linear fit of SNIA & Constrained fit with zero slope \\
\hline
$Q_m$ & $0.1970 \pm 0.110$ ($0.1817 \pm     0.108 $) & 0  \\ 
$Q_0$  & $-0.5220 \pm 0.053  $ ($-0.5109   \pm  0.052$)& $-0.4429     \pm 0.030(-0.4383    \pm 0.030)$ \\ 
Log-Likelihood & $10.061(10.356)$ & $8.4402(8.9112)$  \\
Pearson Chi-Square & $ 27.4(22.9)$  & $30.1(24.8)$  \\
Pseudo R-Square & $0.08897(0.07949)$ & $0.0004160(0.0004160)$ \\
AIC & $-16.12(-16.71)$ & $-14.88(-15.82)$ \\
BIC & $-13.01(1-13.60)$ & $-13.32(-14.27)$
\end{tabular}
\caption{ Best statistics for general linear model fits of $q_0$ using $\Lambda$CDM luminosity distance, estimating the Gaussian error $\sigma$ as the average between asymmetric errors from MCMC.  In parenthesis, same values but this time estimating $\sigma$ as the maximum value between them.}

\label{tablafinal}
\end{table}
\begin{table}[!ht]
\centering
\begin{tabular}{ccc}
 $z$ Bin & Number of SNIA & $q_0 $ \\
\hline
$0.0312-0.1364$  & 300 & $-0.19 \substack{+0.27\\-0.29}$ \\ 
$0.0544-0.20549$ & 300 & $-0.50  \substack{+0.16\\-0.17}$  \\
$0.1066-0.23684 $ & 300  & $-0.52\substack{+0.19\\-0.21}$  \\
$0.2052-0.3221$ & 300 & $-0.815\substack{+0.25\\-0.20}$ 
\end{tabular}
\caption{Values for $q_0$ fit in wide homogeneous redshift bins using the cosmographic Taylor expansion of the luminosity distance.}
\label{tablaultima}
\end{table}

\clearchapterpage
\chapter{Global discussion and conclusions}
\label{conc}

In this work, I provided different observational and theoretical contributions to the science behind the cosmological principle and the large-scale structure evolution. First, I summarized the state of the art of the theoretical framework in which the cosmological principle is currently studied, classifying alternative inhomogeneous and/or anisotropic metric tensors, perturbative approaches, averaging approaches, cosmographic analysis, and tilted cosmological scenarios. As a first approach to inhomogeneous cosmology, I used a simple LTB radially inhomogeneous metric to study the potential connections between the fractal distribution of matter and dark energy, allowing for a fractal dimension and a transition scale. Next, the tilted cosmology scenario has been deeply studied, in which the deceleration parameter could be explained by a contracting bulk flow and the possible related observables that could be extracted from the data. I estimated the divergence of the local peculiar velocity field using a Newtonian linear reconstruction and concluded that there is a high chance that this is, in average, contracting. Following the same covariant formulation of general relativity used in tilted cosmology, I explored some results in which the peculiar velocity field grows faster in time when relativistic effects are included in the dynamics of large-scale structure, showing that this effect could also be included by modifying the temporal derivative of the peculiar gravitational potential. Finally, a statistical analysis over the last SNIA Pantheon+ compilation was performed, showing that the deceleration parameter fitted across redshift bins is not fully in agreement with a $\Lambda$CDM universe.

\section{Main conclusions}
    
In the following, I provide what I consider to be the most significant implications of this thesis to the science of the cosmological principle.

\subsection{The magnitude of the effects of the LLSS in cosmology could be greater than what is commonly assumed}

There exists a significant tension between different claims regarding the effects of the large-scale structure (LSS) on cosmological parameters. On one hand, some studies argue that voids, under-densities, bulk flows, and even non-linear structure dynamics have minimal to no impact on the estimation of cosmological parameters. On the other hand, a different faction within the astrophysics community suggests that these very features—or possibly the sum of them all—could explain the observed negative deceleration parameter in cosmological data, thereby challenging the need for dark energy. According to this view, cosmic acceleration may be an artifact of our observational perspective, with us being non-typical observers within the universe. From the perspective of the author, these two claims seem to be two opposing sides of a see-saw, where each one negates the other. 

Based on the results presented in this work, I conclude that \textbf{the effects of the LSS on cosmology could be currently misunderstood, irrespective of whether they are the origin of dark energy or not}. This conclusion is supported by the following arguments:

\begin{enumerate}
    \item \textbf{The modeling of the large-scale structure is degenerated w.r.t to cosmology:} Corrections in cosmological data assume a distribution of peculiar velocities. Currently, the peculiar velocity field could not be fully separated from the selection of a cosmological model or the ignoring of GR effects. Commonly, the peculiar velocity structure properties are obtained by measuring densities of matter and reconstructing a Newtonian velocity field, such as in \citealp{Carrick_2015, Giani_2024}, while the other approach is to study the redshift structure of the data assuming a background homogeneous and isotropic cosmology as in \citealp{Sorrenti:2024ugq}. The first approach ignores general relativistic effects such as vorticity, while the second assumes a background cosmology that fully agrees with the cosmological principle. Even when the consensus is that the effects of those approximations are small, in light of the recent theoretical and observational discoveries, those effects should not be ignored.
    \item \textbf{The back-reaction problem still holds:} There is no consensus on how to perform the average of a clumpy distribution of matter to account for back-reaction effects without using perturbative approaches. However, arguments such as those presented in \citealp{verweg2024averagingcosmicstructurecosmological} seem to challenge previous constructions based on scale-dependent perturbations that reduce the back-reaction effect to gravitational waves \citealp{greenwald2011}. Without complete solutions to the field equations of general relativity with a complex non-linear structure, it is not possible to strongly claim that the back-reaction is negligible.

    \item \textbf{Metric approaches are not the best to study the LSS in cosmology:} In \ref{chap:fractal}, we used a toy-like LTB model developed in \citealp{Cosmai19}, in which a fractal parameter $D$ was included in the distribution of matter. Following the observations that contradict a fractal universe at large scales, we imposed a transition scale to homogeneity. The goal of this study was to explore possible connections between the fractality and dark energy by contrasting it with the SNIA data, but the results were negative, making it impossible to explain cosmic acceleration without introducing unphysical assumptions such as an inhomogeneous Big Bang. However, the analysis presented here relies on a simple metric, which is far from accurately describing the true fractal distribution of matter. The inclusion of the true fractal distribution in the Einstein field equations remains an open research topic that involves two types of mathematics that are not compatible. In such approaches, the differentiability of GR has to be abandoned to introduce fractal manifolds and fractal calculus, which presents a theoretical challenge to this day (see e.g. \citealp{KHALILIGOLMANKHANEH2024105081}).
    \item \textbf{General relativity is still not fully understood:} Although a significant portion of the scientific community is actively working on extensions of general relativity (GR), the theory itself—without modifications—remains incompletely understood in several key areas. General relativity, as formulated by Einstein, is a well-established theory that has been confirmed through numerous experiments and observations, from the bending of light by gravity to the precise orbits of planets and the recent observations of gravitational waves. However, despite its success, GR has limitations and open questions that continue to drive research. One of the central issues is the integration of GR with quantum mechanics. While GR excellently describes gravity on large scales (such as planets, stars, and galaxies), it breaks down at the quantum level, where the effects of quantum mechanics become significant. Current formulations of GR do not account for quantum effects, and there is no fully consistent theory of quantum gravity that has been universally accepted. This discrepancy is particularly evident in black hole physics, where the event horizon and singularities predicted by GR seem to conflict with the principles of quantum mechanics. The famous "information paradox" arising from the behavior of matter at the singularity is one example of unresolved issues. Furthermore, GR also faces challenges when applied to extreme conditions, such as within the early universe or at very high densities. The theory breaks down at the Planck scale, and the singularities present in black holes and the Big Bang suggest that new physics, beyond GR, is required to fully understand these regimes. The theory’s inability to describe such phenomena suggests that GR, though highly successful, is incomplete in its description of the universe.

Moreover, the effect of the peculiar velocities, and how they interact with gravitational fields at large scales, has yet to be fully incorporated into the general relativistic treatment of LSS. Standard models use Newtonian approximations, but these neglect relativistic corrections, such as the influence of vorticity or the time evolution of the peculiar velocity field, which can influence the rate of expansion and the growth of structures. The role of dark energy, dark matter, and their interactions with LSS is another complex issue where GR's implications remain unclear. For example, models such as tilted cosmologies propose that peculiar motions could lead to misinterpretations of cosmological acceleration, challenging the need for dark energy as an explanation for cosmic acceleration.

To address these questions, we need a more robust general relativistic treatment of the evolution of LSS, one that properly incorporates inhomogeneities, relativistic effects, and non-linear dynamics. This is crucial for refining our understanding of the universe's expansion, structure formation, and the potential need for new physics beyond GR. Given the difficulty in fully solving the Einstein field equations for an inhomogeneous universe, the field is still evolving, and a comprehensive understanding of GR's role in large-scale structure evolution is far from complete. 
In the next section, this last issue will be explored further.
\end{enumerate}

It is important to highlight that, regardless of the preconceived notions scientists may have about the universe—biases that are often naturally influenced by human nature—the true effects of large-scale structure (LSS) on cosmology remain an open field of study. Whether these effects are significant enough to explain dark energy, or whether other explanations are needed, is a question that will be addressed through further research. Over time, and independent of personal biases, the collective efforts of diverse scientific teams will clarify this issue. The evolution of scientific understanding relies on continued investigation, cross-validation of results, and the application of new methodologies to ensure that our conclusions are as objective and accurate as possible.

\subsection{Local effects of General Relativity in cosmology and in the evolution of LSS could being underestimated}

In Chapter \ref{chap:models}, we explored tilted cosmology as a potential explanation for the observed negative deceleration parameter, considering the role of peculiar observers in a contracting bulk flow. Under such conditions, it was shown that the deceleration parameter could be measured as negative locally, even when the global background cosmology suggests a positive deceleration. In Chapter \ref{pec}, we extended this analysis to the evolution of the peculiar velocity field in the large-scale structure (LSS), using a relativistic framework that led to a stronger growth rate than those predicted by Newtonian and semi-Newtonian approaches. This difference in growth rates was attributed to the omission of general relativistic effects, such as energy flux in the dynamics of the peculiar velocity field, encapsulated in the concept of the "effective peculiar gravitational potential evolution." These findings may have significant implications, especially in addressing the high-velocity bulk flows observed, which are currently incompatible with standard cosmological models, such as those discussed by Tully et al. (2023).

From these results, it is evident that the effects of General Relativity on large-scale structure evolution are still not fully understood. Importantly, this is not a call for modifications to General Relativity itself, which has been the focus of much research over the years, but rather a recognition that the standard, Einsteinian formulation of General Relativity may still hold key insights yet to be fully explored. The possibility that these relativistic effects could explain cosmic acceleration warrants serious consideration. Even if this is not the case, phenomena such as back-reaction where inhomogeneities affect the cosmological evolution remain crucial areas for further study. As such, these issues represent unresolved aspects of General Relativity that should continue to be examined in greater depth.

\subsection{The geometry and the dynamics of the local-large scale structure needs to be elucidated, independently of the background cosmology}

We have demonstrated that the study of the local large-scale structure (LSS) dynamics cannot be fully separated from the background cosmology, as the redshift structure of astronomical objects combines local effects with cosmological effects. The most widely used method to describe the peculiar velocity field to date is through the reconstruction of the velocity field from mass density surveys, relying on Newtonian linear dynamics. However, this approach completely neglects relativistic effects and introduces biases, as the observed luminous mass does not necessarily represent the total amount of matter. In Chapter \ref{pec}, we explored the workings of these reconstructions and used one of them to conclude that the local peculiar velocity field is indeed contracting. This result aligns with findings from other studies (\citealp{Giani_2024}, \citealp{Sorrenti:2024ztg}), which employed similar methods. Nevertheless, these studies are all dependent on either a Newtonian linear reconstruction or a background cosmology framework. This highlights the need for the development of new methods to study the local peculiar velocity field, a topic that will be further discussed in the \textit{future prospects} section.

\subsection{Blind and cosmographic statistical analysis on cosmological data should be preferred over preselected cosmological models}

Since the establishment of standard cosmology and the cosmological principle, the $\Lambda$CDM model has been widely used as the benchmark to test against observational data. However, the use of this model is not necessarily driven by an accurate understanding of the true structure of the universe, but rather by its simplicity. The preference for a particular cosmological model, such as $\lambda$CDM, can introduce community bias, where the focus shifts from uncovering the true dynamics of the universe to fitting observations within the confines of the established model. As we discussed in Chapter \ref{chap:models}, it is possible to approach cosmology without predefined assumptions about the universe's dynamics. Some of these methods, such as cosmography, were applied in Chapter \ref{chap:reds} to study the latest supernova compilation, and the results appear to contradict the predictions of the $\Lambda$CDM model. A recent study by \citealp{Seifert2024} demonstrated how a model other than $\Lambda$CDM, based on \textit{timescape cosmology}, can fit the data even more effectively. This highlights the importance of adopting a blind approach to data analysis—such as cosmography—rather than relying on specific models. Without knowing the true dynamics of the universe, we risk drawing incorrect conclusions or following misguided paths where researchers attempt to force the universe to fit the model, rather than allowing the data to guide our understanding of the universe.

\section{Future prospects}

Here, I propose several future projects to build upon the work presented in this thesis regarding the cosmological principle and the large-scale structure (LSS) from both observational and theoretical perspectives:

\subsection{The covariant formulation of gravity and its effects in Cosmology and LSS evolution}

We have shown in chapters \ref{chap:tilt} and \ref{pec} that relativistic covariant cosmology can reveal gravitational effects that were previously ignored using the metric approach. In the first of these chapters, we observed that a negative deceleration parameter can appear within contracting peculiar velocity fields, even when the deceleration parameter outside is positive or zero. In the second chapter, we demonstrated that the growth of peculiar velocities increases when gravitational effects, such as the flow of energy due to the movement of matter in a peculiar velocity field, are incorporated. If these effects are indeed real, the central question is how these theories relate to cosmological and astrophysical observables. The key is to deeply study the physical meaning behind these theoretical results to interpret them correctly. Specifically, some study objectives may be:

\begin{enumerate}
    \item Revisit the theoretical considerations behind the derivation of luminosity distance from the point of view of an observer moving with a certain peculiar velocity. It is likely that, if the effects of tilted cosmology are real and some assumptions in this derivation are modified, the negative deceleration parameter would disappear.
    \item Develop methods for reconstructing peculiar velocities that include general relativity effects, using approximations such as those proposed in chapter \ref{pec}. This is the most direct way to study relativistic effects on large-scale structure velocities, as simulations using general relativity are still extremely demanding. Furthermore, by using this method, it is possible to estimate parameters related to general relativity by directly contrasting it with the data. 
    \item Derive other parameters of interest in the theory of tilted cosmology, such as the divergence of the peculiar velocity field studied in chapter \ref{chap:tilt}. It is quite possible that other measurable quantities can be extracted from the theoretical formulation of tilted cosmology if this is studied deeply.
    \item Elucidate the correct frame of reference in which the degree of anisotropy in the deceleration parameter should be analyzed. This anisotropy is predicted by tilted cosmology. However, a heated debate has arisen recently, as it remains unclear which frame of reference is appropriate for analyzing the data without losing important information. Once again, the key to resolving this discussion may lie in a deeper study of the tilted-luminous distance derivation.
\end{enumerate}

\subsection{The elucidation of the LLSS dynamics via different sources of data and different methods}

One of the main challenges in analyzing the peculiar velocities of the local large-scale structure is that they cannot be separated from the dynamics of the background universe. This requires assuming the cosmological principle to study the various multi-polar contributions of the velocity field, attributing each deviation to a local rather than cosmological origin. However, since it is impossible to distinguish which contributions are truly due to local structure and which may arise from deviations from the cosmological principle, analyzing the anisotropy of the universe from this data becomes unfeasible. A key contribution is the analysis in \citealp{Sorrenti:2024ztg}, where a monopole, dipole, and quadrupole are fitted to the supernova redshift structure without assuming a cosmological model. However, interpreting all of these contributions as arising from the local large-scale peculiar velocity field, rather than a cosmological origin, is heavily dependent on the a priori assumption of the cosmological principle.

One approach to mitigate cosmological dependence is to reconstruct the velocities using mass density surveys and Newtonian physics. However, this approach ignores possible relativistic effects that may significantly influence the dynamics. As discussed in the previous section, there is a need to develop methods that incorporate relativistic effects into these reconstructions. Specifically, the work presented in Chapter \ref{pec} moves in this direction.

\subsection{The development of alternative cosmological sources of information to study the cosmological principle}

Currently, the main sources of cosmological information are the Cosmic Microwave Background (CMB) (\citealp{PLANCK20}), Baryon Acoustic Oscillations (BAO) (\citealp{desicollaboration2024desi2024vicosmological}), and Type Ia Supernovae (SNIA) data (\citealp{Scolnic_2022}), with the latter being the most important for measuring the acceleration of the universe. Other recently developed observational fields include TRGB stars (\citealp{Freedman_2020}), JAGB stars (\citealp{Madore_2020}), cosmic chronometers (\citealp{Borghi_2022}), and gravitational lensing (\citealp{Shajib_2024}). Gravitational waves have also gained attention, particularly due to the results from the NANOGRAV collaboration in pulsar array experiments (\citealp{Agazie_2023}). On the other hand, the cosmic neutrino background (\citealp{Scott_2024}) remains a theoretical predicted observable, still without an experimental approach. The expansion of data sources provides a greater number of independent methods for studying cosmology and, by extension, the cosmological principle. Thus, it is crucial to stimulate the study of methods such as TRGB or gravitational lensing, which can serve as arbiters in the debates and tensions surrounding more classical observables like the CMB and supernovae. Moreover, it is possible that additional cosmological methods and observables are still under development. Notably, the work on quasars (\citealp{Marziani_2014, Watson_2011}) as potential indicators of cosmological parameters seems particularly promising.

The author of this thesis is also a part of the DESC collaboration, an international scientific group focused on making high-accuracy measurements of fundamental cosmological parameters using data from the LSST (\citealp{DESC}). This collaboration offers increased opportunities for the author to engage in work related to other cosmological observables.

\subsection{The importance of practicing pedagogy for a scientist}

Finally, I would like to highlight the significant impact that actively engaging in pedagogy during my PhD has had on my development as a researcher. Throughout my doctoral studies, I had the privilege of teaching a range of undergraduate courses in physics at both the University of Valparaiso and the Federico Santa María University. Among the courses I taught were Optics, Stellar Structure and Evolution, Astroparticles, Electrodynamics, and Cosmology. Teaching Cosmology, in particular, was incredibly important, as it required me to study the fundamentals of physical cosmology—from general relativity to thermal history—thus deepening my understanding and enhancing the quality of this thesis.

I believe that research work during a PhD is often prioritized over pedagogical work, with teaching relegated to a secondary or optional activity. However, in the future, it is highly likely that a PhD in physics will also involve substantial time dedicated to teaching and interacting with students. This shift in responsibilities should be acknowledged by both professionals and academic institutions, as educational work is as crucial as scientific research for doctoral physicists. It is essential to foster and support development in both areas, ensuring that future researchers also excel as educators.

Personally, I aspire to dedicate a significant portion of my career to teaching. I have been fortunate to find a wonderful team of colleagues at the Federico Santa María University’s physics department, who are equally enthusiastic about this field. Together, we are advancing the concept of active pedagogy, which focuses on engaging students through active participation in the learning process. This pedagogical approach has been fundamental to shaping my teaching methods, and I have drawn inspiration from it to enhance the effectiveness of my lectures and create a more interactive learning experience.

\clearchapterpage
\appendix


\chapter{Appendix}
\label{app}


\section{Physics of Newtonian velocity field reconstruction and insights of structure evolution}
\label{VFR}

We study the physics behind the velocity field reconstruction used in \autoref{chap:tilt}. The matter is treated as a pessure-less ideal fluid. $\mathbf{u}$ is the velocity referred to an inertial system with proper position $\mathbf{r}$ in physical units. $t$ is invariant between observers. We assume 3 fundamental equations to model the velocity field: mass conservation (matter), Euler equation (fluid) and Poisson equation (gravity).

\begin{eqnarray}
\left ( \frac{\partial \rho}{\partial t}\right )_{\mathbf{r}}+\nabla_{\mathbf{r}}(\rho\mathbf{u})=0 \\
\left ( \frac{\partial \mathbf{u}}{\partial t}\right )_{\mathbf{r}}+(\mathbf{u}\cdot\nabla_{\mathbf{r}})\mathbf{u}= - \nabla_\mathbf{r} \Phi\\
\nabla_\mathbf{r}^2 \Phi = 4 \pi G \rho
\end{eqnarray}

\subsubsection{Co-moving coordinates}

We introduce the co-moving coordinate  $\mathbf{x}$ which expand with the Hubble flow. This coordinate is related with the physical $\mathbf{r}$ coordinate as:
\begin{eqnarray}
\mathbf{x}=\frac{\mathbf{r}}{a(t)}
\end{eqnarray}
The velocity $\mathbf{u}=\dot{\mathbf{r}}$ can be writed as:
\begin{eqnarray}
\mathbf{u}=\dot a \mathbf{x} + a \dot{\mathbf{x}}=H\mathbf{r}+\mathbf{v}
\end{eqnarray}
Where $H$ is the Hubble parameter and $\mathbf{v}$ the peculiar velocity. Consider a function in co-moving coordinates $f(t,\mathbf{x})=f(t,\mathbf{r}/a(t))$. We can compute the partial time derivative at a fixed physical location $\mathbf{r}$ in terms of fixed co-moving location $\mathbf{x}$:
\begin{eqnarray}
\left (\frac{\partial f}{\partial t} \right )_\mathbf{r}= \left (\frac{\partial f}{\partial t} \right )_\mathbf{x}-\frac{\dot{a}}{a} \mathbf{x}\cdot (\nabla_{\mathbf{x}} f)
\end{eqnarray}
Note that $\nabla_\mathbf{x}=a\nabla_\mathbf{r}$. In this section, we use the notation:
\begin{eqnarray}
\nabla_\mathbf{x}=\nabla\\ \frac{\partial}{\partial t}_\mathbf{x}=\frac{\partial}{\partial t}
\end{eqnarray}

\subsubsection{Mass conservation}

In  co-moving coordinates, the mass conservation can be written as:

\begin{eqnarray}
\left (\frac{\partial \rho}{\partial t} \right )-\frac{\dot{a}}{a} \mathbf{x}\cdot (\nabla \rho)+\frac{1}{a}\nabla\cdot(\rho\mathbf{u})=0 
\end{eqnarray}

Replacing the density contrast:

\begin{eqnarray}
\left ( \frac{\partial }{\partial t}-\frac{\dot{a}}{a} \mathbf{x}\cdot \nabla\right)\bar{\rho}(1+\delta)+\frac{\bar{\rho}}{a}\nabla\cdot[(1+\delta)(\dot{a}\mathbf{x}+\mathbf{v})]=0 
\end{eqnarray}

Expanding and noting that $\nabla \cdot \mathbf{x}=3$, we can finally write:

\begin{eqnarray}
\frac{\partial \delta}{\partial t}+\frac{1}{a}\nabla\cdot[(1+\delta)\mathbf{v}]=0 
\end{eqnarray}

\subsubsection{Poisson equation}

We can write Poisson equation in co-moving coordinates as:

\begin{eqnarray}
\frac{1}{a^2}\nabla^2\Phi=4\pi G \bar{\rho}(1+\delta)-\Lambda
\end{eqnarray}

Where we have introduced the cosmological constant $\Lambda$. Defining the gravitational potential as:

\begin{eqnarray}
\Phi=\phi+\frac{2}{3}\pi G \bar{\rho}a^2\mathbf{x}^2-\frac{1}{6}\Lambda a^2 \mathbf{x}^2
\end{eqnarray}

Removing perturbations, we can write:

\begin{eqnarray}
\frac{1}{a^2}\nabla^2\phi=4\pi G \bar{\rho}\delta
\end{eqnarray}

\subsubsection{Euler equation}

Finally, the euler equation can be written as:

\begin{align}
    \left ( \frac{\partial }{\partial t}-\frac{\dot{a}}{a} \mathbf{x}\cdot \nabla\right ) (\dot{a}\mathbf{x}+\mathbf{v})+\frac{1}{a}[(\dot{a}\mathbf{x}+\mathbf{v})\cdot \nabla](\dot{a}\mathbf{x}+\mathbf{v}) \notag \\ =-\frac{1}{a}\phi-\left (\frac{4}{3}\pi G \bar{\rho}-\frac{1}{3}\Lambda  \right ) a\mathbf{x}
\end{align}

Recalling the acceleration equation for a pressure-less universe:

\begin{eqnarray}
   \frac{\ddot a}{a}=-\frac{4}{3}\pi G \bar{\rho}+\frac{\Lambda}{3}
\end{eqnarray}

We can write for a zero cosmological constant:

\begin{eqnarray}
    \left ( \frac{\partial }{\partial t}-\frac{\dot{a}}{a} \mathbf{x}\cdot \nabla\right ) (\dot{a}\mathbf{x}+\mathbf{v})+\frac{1}{a}[(\dot{a}\mathbf{x}+\mathbf{v})\cdot \nabla](\dot{a}\mathbf{x}+\mathbf{v})=-\frac{1}{a}\phi+\ddot{a}\mathbf{x}
\end{eqnarray}

Which results in:

\begin{eqnarray}
    \frac{\partial \mathbf{v}}{\partial t}+\frac{\dot{a}}{a}\mathbf{v}+\frac{1}{a}(\mathbf{v} \cdot \nabla)\mathbf{v}=-\frac{1}{a}\nabla\phi
\end{eqnarray}

\subsection{Linear Regime}

Removing $v\delta$ or $v^2$ terms, the linear mass conservation and Euler equation are:

\begin{eqnarray}
    \frac{\partial \delta}{\partial t}+\frac{1}{a}\nabla\cdot\mathbf{v}=0 \\
    \frac{\partial \mathbf{v}}{\partial t}+\frac{\dot{a}}{a}\mathbf{v}+\frac{1}{a}\nabla\phi=0
\end{eqnarray}

We can derivate the first equation w.r.t time and compute the divergence of the second to get:

\begin{eqnarray}
    \frac{\partial^2 \delta}{\partial t^2}+\frac{1}{a}\nabla\cdot\frac{\partial \mathbf{v}}{\partial t}-\frac{\dot a}{a^2}\nabla\cdot \mathbf{v}=0 \\
    \nabla\cdot\frac{\partial \mathbf{v}}{\partial t}+\frac{\dot{a}}{a}\nabla\cdot\mathbf{v}+\frac{1}{a}\nabla^2\phi=0
\end{eqnarray}

Subtracting both:

\begin{eqnarray}
    \frac{\partial^2 \delta}{\partial t^2}-2\frac{\dot{a}}{a^2}\nabla\cdot\mathbf{v}-\frac{1}{a^2}\nabla^2\phi=0
\end{eqnarray}

Finally we get:

\begin{eqnarray}
    \frac{\partial^2 \delta}{\partial t^2}+2\frac{\dot{a}}{a}\frac{\partial \delta}{\partial t}=4\pi G \bar{\rho}\delta
\end{eqnarray}

For EdS universe, $a\propto t^\frac{2}{3}$ and $t=\frac{1}{\sqrt{6\pi G\bar{\rho}}}$. Then the equation becomes:

\begin{eqnarray}
    \frac{\partial^2 \delta}{\partial t^2}+\frac{4}{3t}\frac{\partial \delta}{\partial t}=\frac{2}{3t^2}\delta
\end{eqnarray}

With the solution:

\begin{eqnarray}
    \delta =A(\mathbf{x})t^\frac{2}{3}+B(\mathbf{x})t^{-1}
\end{eqnarray}

Where the growing mode is identified by the term $\delta=A(\mathbf{x})t^\frac{2}{3}$ and governs the evolution of the density contrast in time. Suppose that $\delta $ evolves with a growth mode $\delta=A(\mathbf{x})D(t)$. Then the linear mass conservation equation goes as:

\begin{eqnarray}
    \nabla \mathbf{v}=-a\frac{\partial \delta}{\partial t}=-a A \dot{D}=-a\delta\frac{\dot{D}}{D} \label{divergence1}
\end{eqnarray}

Considering the velocity field as irrotational, we can integrate this Poisson equation to get finally:

\begin{eqnarray}
    \mathbf{v}(\mathbf{x})=\frac{a\dot{D}}{4\pi D}\int d^3\mathbf{x'}\delta(\mathbf{x'})\frac{\mathbf{x'}-\mathbf{x}}{     \lvert \mathbf{x'}-\mathbf{x}\rvert^3 }
\end{eqnarray}

This equation can be rewritten in the form:

\begin{eqnarray}
    \mathbf{v}(\mathbf{x})=\frac{afH}{4\pi}\int d^3\mathbf{x'}\delta(\mathbf{x'})\frac{\mathbf{x'}-\mathbf{x}}{     \lvert \mathbf{x'}-\mathbf{x}\rvert^3 }
\end{eqnarray}

Where $H=\frac{\dot a}{a}$ and $f$ is the dimensionless \textit{Growth factor}:

\begin{eqnarray}
    f=\frac{1}{H}\frac{\dot D}{D}
\end{eqnarray}

Note that we can write also the growth factor as:

\begin{eqnarray}
    f=\frac{d \ln\delta}{d \ln a}
\end{eqnarray}

Therefore, we can write the velocity field in Hubble units as:

\begin{eqnarray}
    \mathbf{v}(\mathbf{r})=\frac{f}{4\pi}\int d^3\mathbf{r'}\delta(\mathbf{r'})\frac{\mathbf{r'}-\mathbf{r}}{     \lvert \mathbf{r'}-\mathbf{r}\rvert^3 }
\end{eqnarray}

According the equation \ref{divergence1}, the divergence of the velocity field is proportional to the density contrast $\delta$ as:

\begin{eqnarray}
    \nabla \mathbf{v}=-afH \delta
\end{eqnarray}

These relations were used to compute the divergence of the local peculiar velocity field in chapter \autoref{chap:tilt}.

\subsection{The Growth Factor and $\delta$ evolution}

For and EdS universe:

\begin{eqnarray}
    H=\frac{2}{3t}
\end{eqnarray}

And therefore, if $\delta \sim  t^{2/3}$ we have:

\begin{eqnarray}
    \frac{\dot{D}}{D}=\frac{2}{3t}
\end{eqnarray}

And $f=1$. However, if $\delta \sim t^{\alpha}$ we get:

\begin{eqnarray}
    \frac{\dot{D}}{D}=\frac{\alpha}{t}
\end{eqnarray}

The growth factor $f$ for an EdS universe now becomes:

\begin{eqnarray}
    f=\frac{3}{2}\alpha
\end{eqnarray}

This is an important result. If $\alpha \geq \frac{2}{3}$, the density perturbation growth faster over time and the peculiar velocities are bigger than was previously expected.

\subsection{The Growth Factor as a parametrization of $\Omega_m$ }

We follow the treatment in \citealp{Wang_1998}. Using the acceleration equation with a quintessential form for a pressure-less matter content:

\begin{eqnarray}
   \frac{\ddot{a}}{a}=-\frac{4\pi G}{3}\left ( \bar{\rho}+\rho_Q + 3p  \right )=-\frac{8\pi G}{3}(\frac{\bar{\rho}+\rho_Q}{2} + \frac{3}{2}p_Q )
\end{eqnarray}

The state equation $p_Q=w\rho_Q$:

\begin{eqnarray}
   \frac{\ddot{a}}{a}=-\frac{8\pi G}{3}(\frac{\bar{\rho}+\rho_Q}{2} + \frac{3}{2}w\rho_Q )
\end{eqnarray}

Recalling that the density parameter $\Omega_{m}(a)$ can be defined as:

\begin{eqnarray}
    \Omega_m(a)=\frac{8\pi G \bar{\rho}}{3H^2} \\
    \Omega_Q(a)=\frac{8\pi G \rho_Q}{3H^2}
\end{eqnarray}

We get:

\begin{eqnarray}
   \frac{\ddot{a}}{a}=-H^2[\frac{\Omega_m(a)}{2}+(\frac{3}{2}w +\frac{1}{2})\Omega_\Lambda(a)]
\end{eqnarray}

Finally:

\begin{eqnarray}
   \frac{\ddot{a}a}{\dot{a}^2}=-\frac{1}{2}[1+3w(1-\Omega_m(a))]
\end{eqnarray}

\subsubsection{Energy conservation}

Using the energy conservation:

\begin{eqnarray}
   d(\rho a^3)=-p d(a^3)
\end{eqnarray}

And the relation:

\begin{eqnarray}
   \frac{\bar{\rho}}{\Omega_m}=\bar{\rho}+\rho_\Lambda=\rho
\end{eqnarray}

We can write:

\begin{eqnarray}
   a^3d\rho+3\rho a^2 da =-w\rho_Q 3a^2 da \\
   a\left (\frac{d\bar{\rho}}{\Omega_m}-\frac{\bar{\rho} d\Omega_m}{\Omega_m^2}\right )=3[-(w+1)\rho_Q-\bar{\rho}]da\\
\end{eqnarray}

We can expand the $d\bar{\rho}$ term as:

\begin{eqnarray}
   d\bar{\rho}=-\frac{3\bar{\rho}}{a}da=-\frac{9H^2\Omega_m}{8\pi G a}da
\end{eqnarray}

Then:

\begin{eqnarray}
   -a\left (\frac{9H^2}{8\pi G a}da+\frac{3H^2 d\Omega_m}{8\pi G\Omega_m}\right )=3[-(w+1)\frac{3H^2(1-\Omega_m)}{8\pi G}-\frac{3H^2\Omega_m}{8\pi G\Omega_m}]da\\
\end{eqnarray}

Rearranging terms and simplifying:

\begin{eqnarray}
  3w\Omega_m(1-\Omega_m)=\frac{ d\Omega_m}{d \ln a}\\
\end{eqnarray}

\subsubsection{Perturbation equation}

The linear perturbation equation for density contrast is:

\begin{eqnarray}
    \frac{\partial^2 \delta}{\partial t^2}+2\frac{\dot{a}}{a}\frac{\partial \delta}{\partial t}=4\pi G \bar{\rho}\delta
\end{eqnarray}

We can write the following logarithmic derivatives  as:

\begin{eqnarray}
    \frac{d }{d \ln a}=a\frac{d }{d a}=\frac{a}{\dot a}\frac{d}{dt}
\end{eqnarray}

Then:

\begin{eqnarray}
    \frac{d \ln \delta}{d \ln a}&=&\frac{a}{\dot{a}}\frac{\dot{\delta}}{\delta}\\
    \frac{d^2 \ln \delta}{d \ln a^2}&=&\frac{a}{\dot a}\left [\frac{\dot{\delta}}{\delta}\frac{d}{dt}\left ( \frac{a}{\dot{a}} \right )+ \frac{a}{\dot{a}}\frac{d}{dt}\left ( \frac{\dot{\delta}}{\delta} \right )\right ]
\end{eqnarray}

We have for the derivatives:

\begin{eqnarray}
    \frac{d}{dt}\left ( \frac{a}{\dot{a}} \right )=\frac{\dot{a}^2-a\ddot{a}}{\dot{a}^2}\\
    \frac{d}{dt}\left ( \frac{\dot{\delta}}{\delta} \right )=\frac{\delta\ddot{\delta}-\dot{\delta}^2}{\delta^2}\\
\end{eqnarray}

And for the second order logarithmic derivative:

\begin{eqnarray}
    \frac{d^2 \ln \delta}{d \ln a^2}=\frac{a}{\dot a}\left [\frac{\dot{\delta}}{\delta}\left ( \frac{\dot{a}^2-a\ddot{a}}{\dot{a}^2} \right )+ \frac{a}{\dot{a}}\left ( \frac{\delta\ddot{\delta}-\dot{\delta}^2}{\delta^2} \right )\right ]
\end{eqnarray}

This can be written as:

\begin{eqnarray}
    \frac{d^2 \ln \delta}{d \ln a^2}=\frac{d \ln \delta}{d \ln a}\left ( 1+\frac{1}{2}[1+3w(1-\Omega_m)] \right )+ \frac{1}{H^2}\frac{\ddot{\delta}}{\delta}-\left (\frac{d \ln \delta}{d \ln a} \right )^2
\end{eqnarray}

Then:

\begin{eqnarray}
    \ddot{\delta}&=& H^2\delta \left [\frac{d^2 \ln \delta}{d \ln a^2}-\frac{d \ln \delta}{d \ln a}\left ( \frac{3}{2}+\frac{3}{2}w(1-\Omega_m) \right )+\left (\frac{d \ln \delta}{d \ln a} \right )^2 \right ] \\
    2\frac{\dot{a}}{a}\dot{\delta}&=&2H^2\delta \frac{d \ln \delta}{d \ln a}
\end{eqnarray}

The perturbation equation can be written in a intersting form:

\begin{eqnarray}
      \frac{d^2 \ln \delta}{d \ln a^2}+\frac{d \ln \delta}{d \ln a}\left ( \frac{1}{2}-\frac{3}{2}w(1-\Omega_m) \right )+\left (\frac{d \ln \delta}{d \ln a} \right )^2  = \frac{3  \Omega_m}{2}
\end{eqnarray}

\subsubsection{Growth factor definition}

With the definition:

\begin{eqnarray}
    f=\frac{d \ln\delta}{d \ln a}
\end{eqnarray}

And the result

\begin{eqnarray}
    3w\Omega_m(1-\Omega_m)=\frac{ d\Omega_m}{d \ln a}
\end{eqnarray}

We can use the derivative of the growth factor as:

\begin{eqnarray}
    \frac{df}{d\Omega}=\frac{d^2 \ln\delta}{d \ln a^2}\frac{d \ln a}{d \Omega_m}
\end{eqnarray}

And the perturbation equation could be transformed in a relation between $f$ and $\Omega_m$:

\begin{eqnarray}
    3w\Omega_m(1-\Omega_m)\frac{df}{d\Omega}+f\left ( \frac{1}{2}-\frac{3}{2}w(1-\Omega_m) \right )+f^2 = \frac{3  \Omega_m}{2}
\end{eqnarray}

\subsubsection{Parametrization}

Usually, $f$ can be parametrized with a power law function:

\begin{eqnarray}
    f=\Omega_m^{\gamma(\Omega_m)}
\end{eqnarray}

And the equation becomes:

\begin{eqnarray}
    3w(1-\Omega_m)\gamma +3w\Omega_m(1-\Omega_m)\frac{d\gamma}{d\Omega_m}\ln &\Omega_m&
    \\
    +\frac{1}{2}-\frac{3}{2}w(1-\Omega_m) +\Omega_m^{\gamma} -\frac{3  \Omega_m^{1-\gamma}}{2}&=&0 
\end{eqnarray}

This equation can be resolved numerically for a slowly varying state equation ($\frac{dw}{d\Omega_m}\sim 0$) to get:

\begin{eqnarray}
    \gamma = \frac{3}{5-w/(1-w)}+\frac{3}{125}\frac{(1-w)(1-3w/2)}{(1-6w/5)^3}+\mathcal{O}[(1-w)^2]
\end{eqnarray}

For $\Lambda CDM$ with $w=-1$:

\begin{eqnarray}
    f\sim \Omega_m^{6/11}
\end{eqnarray}

Which is a well known result.

\subsection{Some time-evolution insights of velocity components}

The Divergence of the velocity field can be wrote in the linear regime as:

\begin{eqnarray}
    \nabla \mathbf{v}=-a\frac{\partial \delta}{\partial t}
\end{eqnarray}

If $\delta \propto t^{2/3}$ in a Eds Universe, then the expansion $\theta$:

\begin{eqnarray}
    \theta=\frac{1}{a}\nabla \mathbf{v} \propto  t^{-1/3}
\end{eqnarray}

We can derive a similar result with the linearized Euler equation written in tensorial notation as:

\begin{eqnarray}
    \partial_tv_i+Hv_i+\frac{1}{a}(\nabla \phi)_i=0
\end{eqnarray}

Taking the gradient of this expresion:

\begin{eqnarray}\partial_t(\partial_kv_i)+H(\partial_kv_i)+\frac{1}{a}\partial_k(\nabla \phi)_i=0,
\end{eqnarray}

we can get the expansion evolution equation $a\theta=\nabla \cdot\mathbf{v}=\partial_kv_k$:

\begin{eqnarray}
\dot{\theta}=-2H\theta-\frac{1}{a^2}\nabla^2 \phi
\end{eqnarray}

For the shear $a\sigma_{ki}=\partial_{(k}v_{i)}$=

\begin{eqnarray}
\dot{\sigma_{ki}}=-2H\sigma_{ki}-\frac{1}{a^2}\partial_{(k}\partial_{i)}\phi
\end{eqnarray}

And for the vorticity $a\omega_{ki}=\partial_{[k}v_{i]}$:

\begin{eqnarray}
\dot{\omega_{ki}}=-2H\omega_{ki}-\frac{1}{a^2}\partial_{[k}\partial_{i]}\phi
\end{eqnarray}

where the last term on the right side is zero due to the conservative nature of the newtonian potential. The solution of this equation in an EdS universe is then:

\begin{eqnarray}
\dot{\omega_{ki}}=-2\frac{\dot{a}}{a}\omega_{ki}=-\frac{4}{3t}\omega_{ki}\\
\omega_{ki} \propto t^{-4/3}
\end{eqnarray}

This result comes from the Newtonian nature of the gravity potential. Using the Poisson equation the divergence equation results in:

\begin{eqnarray}
\dot{\theta}=-2H\theta-4\pi G\bar{\rho}\delta \\
\dot{\theta}=-2H\theta-\delta\frac{3H^2\Omega_m}{2}
\end{eqnarray}

If we set  $\delta = A(\mathbf{x})D(t)$:

\begin{eqnarray}
\dot{\theta}=-2H\theta-A(\mathbf{x})D(t)\frac{3H^2\Omega_m}{2}
\end{eqnarray}

And for a Eds Universe $\delta\propto t^{2/3}$:

\begin{eqnarray}
\dot{\theta}=-\frac{4}{3t}\theta-\frac{8}{3t^{4/3}}A(\mathbf{x})
\end{eqnarray}

Which solves as:

\begin{eqnarray}
\frac{d(\theta t^{4/3})}{dt}&=&-\frac{8}{3}A(x)\\
\theta(t)t^{4/3}&=&-\frac{8}{3}A(x)t+C(\mathbf{x})\\
\theta(t)&=&-\frac{8}{3}A(x)t^{-1/3}+C(\mathbf{x})t^{-4/3}
\end{eqnarray}

As this equation is linear of first order, we can get the expansion of the velocity field evolution formulae in any FRWL cosmology. The general evolution of the divergence equation can be written as:

\begin{eqnarray}
\dot{\theta}+2H\theta=-\delta\frac{3H^2\Omega_m}{2}
\end{eqnarray}

Note that the integrating factor of this equation is:
\begin{eqnarray}
    I=e^{\int 2H dt }=e^{2da/a }=a^2
\end{eqnarray}

Then:

\begin{eqnarray}
\frac{d(\theta a^2)}{dt}=-\delta a^2\frac{3H^2\Omega_m}{2}\\
\theta(t)=-\frac{3}{2a^2}\int\dot{a}^2\Omega_m\delta dt+\frac{C}{a^2} \\
\end{eqnarray}

\subsubsection{Second order differential equations}

The temporal derivative of the euler equation is:

\begin{eqnarray}
    \frac{\partial^2\mathbf{v}}{\partial t^2}+\dot{H}\mathbf{v}+H\dot{\mathbf{v}}-\frac{H}{a}\nabla\phi+\frac{1}{a}\nabla\dot{\phi}=0 \\
    \frac{\partial^2\mathbf{v}}{\partial t^2}+\dot{H}\mathbf{v}+H\dot{\mathbf{v}}+H\left (\dot{\mathbf{v}}+H\mathbf{v} \right )+\frac{1}{a}\nabla\dot{\phi}=0 
\end{eqnarray}

Ignoring the term $\nabla \dot{\phi}$ we can write:

\begin{eqnarray}
    \frac{\partial^2\mathbf{v}}{\partial t^2}+2H\dot{\mathbf{v}}+(\dot{H}+H^2)\mathbf{v}=0 
\end{eqnarray}

For EdS universe $\dot{H}=-\frac{2}{3t^2}=-\frac{3H^2}{2}$. Then:

\begin{eqnarray}
    \frac{\partial^2\mathbf{v}}{\partial t^2}+2H\dot{\mathbf{v}}-\frac{H^2}{2}\mathbf{v}=0 
\end{eqnarray}

And we can construct a differential equation for EdS universe:

\begin{eqnarray}
    \frac{\partial^2\mathbf{v}}{\partial t^2}+\frac{4}{3t}\dot{\mathbf{v}}-\frac{2}{9t^2}\mathbf{v}=0\\
    9t^2\frac{\partial^2\mathbf{v}}{\partial t^2}+12t\dot{\mathbf{v}}-2\mathbf{v}=0
\end{eqnarray}

With the solutions:

\begin{eqnarray}
    \mathbf{v}(t)=\mathbf{C}_1t^{1/3}+\mathbf{C}_2t^{-2/3} 
\end{eqnarray}

Note that the growing mode for the evolution of the peculiar velocity field is $\propto t^{1/3}$. The use of other ansatz for $\nabla{\dot{\phi}}$ is deeply studied in chapter \ref{pec}. We can construct also a second order differential equation for the divergence equation taking the temporal derivative :

\begin{eqnarray}
    \ddot{\theta}+2\dot{H}\theta+2\dot{\theta}H-\frac{2\dot{a}}{a^3}\nabla^2\phi+\frac{1}{a^2}\nabla^2\dot{\phi}=0\\
\end{eqnarray}

Ignoring the term $\nabla^2\dot{\phi}$ and using the euler equation we can write:

\begin{eqnarray}
    \ddot{\theta}+2\dot{H}\theta+2\dot{\theta}H+\frac{2\dot{a}}{a^2}(\nabla\dot{\mathbf{v}}+H\nabla\mathbf{v})=0\\
    \ddot{\theta}+2\dot{H}\theta+2\dot{\theta}H+2H(\frac{\nabla\dot{\mathbf{v}}}{a}+H\theta)=0
\end{eqnarray}

Now we can remember the definition for the physical expansion $\theta$, and therefore its temporal derivative:

\begin{eqnarray}
    \dot{\theta}=\frac{\partial}{\partial t}\left (\frac{\nabla\mathbf{v}}{a} \right )=\frac{a\nabla \dot{\mathbf{v}}-\nabla{\mathbf{v}}\dot{a}}{a^2}\\
    \frac{\nabla\mathbf{\dot{v}}}{a}=\dot{\theta}+H\theta 
\end{eqnarray}

Then:

\begin{eqnarray}
\ddot{\theta}+4H\dot{\theta}+H^2\theta=0\\
\end{eqnarray}

Which results in the differential equation:

\begin{eqnarray}
    \ddot{\theta}+\frac{8}{3t}\dot{\theta}+\frac{4}{9t^2}\theta=0\\
    9t^2\ddot{\theta}+24t\dot{\theta}+4\theta=0
\end{eqnarray}

With the solutions:

\begin{eqnarray}
    \theta(t)=C_1t^{1/3}+C_2t^{-4/3} 
\end{eqnarray}

\clearchapterpage

%
%
\section{1+3 covariant formalism for cosmology}\label{s1+3CF}
The covariant approach uses a local splitting of the spacetime
into \textit{time} and \textit{space}, by introducing a family of time-like observers.
As the metric doesn't provide a covariant description, the covariant formalism uses instead the observers kinematic
variables, the energy-momentum tensor and the
gravito-electromagnetic components of the Weyl tensor, leading to a set of equations that describe the kinematic and dynamic evolution
of the space-time. In this appendix I will review the basic features
of the covariant formalism in order to understand better the basis of tilted cosmologies. 

\subsection{The 1+3 covariant description}\label{ss1+3CD}
In this appendix, we follow the amazing review of \citealp{TSAGAS200861}, keeping the enough information needed to understand the work done in this thesis. The key equations in the covariant description of cosmology are the Ricci and Bianchi identities, applied to the fluid 4-velocity vector, while Einstein's equations are incorporated via algebraic relations between the Ricci and the
energy-momentum tensor. Following \citealp{TSAGAS200861}, latin indices vary between $0$ and $3$ and refer to arbitrary coordinate, meanwhile greek indices run from $1$ to $3$. We will use the unit convention $c=1=8\pi G$.

\subsubsection{Local spacetime splitting}\label{sssLStS}
Consider a general spacetime with a Lorentzian metric $g_{ab}$ of
signature $(-,\,+,\,+,\,+)$ and a family of observers with
world-lines tangent to the time-like 4-velocity vector:
\begin{equation}
u^a=\frac{{\rm d}x^a}{{\rm d}\tau}\,,  \label{ua}
\end{equation}
where $\tau$ is the proper time. The vector $u_a$ determines
the \textbf{time direction}, while the \textit{projection tensor} defined by:

\begin{eqnarray}
    h_{ab}=g_{ab}+u_au_b
\end{eqnarray} 

projects orthogonal to $u_a$ into the observers instantaneous rest space at each event. In the absence of vorticity, $h_{ab}$ is the metric of the dimensional spatial sections orthogonal to $u_a$.

The spacetime can be decomposed in time-like and space-like parts using the 4-velocity $u_a$ and the projection tensor $h_{ab}$.
These objects can also be used to define the covariant time of any tensor $T_{ab\cdots}{}^{cd\cdots}$ as:

\begin{equation}
\dot{T}_{ab\cdots}{}^{cd\cdots}=
u^e\nabla_eT_{ab\cdots}{}^{cd\cdots} \hspace{5mm} 
\end{equation}

while the covariant spatial derivative is:

\begin{equation}
\hspace{5mm} {\rm D}_eT_{ab\cdots}{}^{cd\cdots}=
h_e{}^sh_a{}^fh_b{}^ph_q{}^ch_r{}^d\cdots
\nabla_sT_{fp\cdots}{}^{qr\cdots}\,,  
\end{equation}

The contraction of the spacetime volume element ($\eta_{abcd}$) along the time direction gives the volume element in the observer instantaneous rest frame:

\begin{equation}
\varepsilon_{abc}=\eta_{abcd}u^d\,.  \label{veps1}
\end{equation}

\subsubsection{The gravitational field}\label{sssGrF}
In general relativity, energy modify the curvature of the spacetime as the latter dictates the motion of the matter. The Einstein field equations represent this interaction:

\begin{equation}
G_{ab}\equiv R_{ab}- {1\over2}\,Rg_{ab}= T_{ab}- \Lambda g_{ab}\,,
\label{EFE1}
\end{equation}
where $G_{ab}$ is the Einstein tensor, $R_{ab}=R_{acb}{}^c$ is the Ricci tensor, $R$ is the trace of the Ricci tensor, $T_{ab}$ is the energy-momentum tensor and $\Lambda$ is the
cosmological constant. The twice contracted Bianchi identities
guarantee that $\nabla^bT_{ab}=0$ and total energy-momentum
conservation.

The Ricci tensor describes the local gravitational field, while the long-range gravitational field (mediated via gravitational waves and tidal forces) is encoded in
the \textit{Weyl conformal curvature tensor $C_{abcd}$}. The splitting of the gravitational field is given by the decomposition of the Riemann tensor as:

\begin{equation}
R_{abcd}= C_{abcd}+ {1\over2}\left(g_{ac}R_{bd}+g_{bd}R_{ac}
-g_{bc}R_{ad}-g_{ad}R_{bc}\right)
-{1\over6}\,R\left(g_{ac}g_{bd}-g_{ad}g_{bc}\right)\,,
\label{Riemann}
\end{equation}
The Weyl tensor shares all the symmetries of the Riemann
tensor and is also trace-free. The conformal curvature tensor decomposes into its irreducible parts:

\begin{equation}
E_{ab}= C_{acbd}u^cu^d \hspace{15mm} {\rm and} \hspace{15mm}
H_{ab}= {1\over2}\,\varepsilon_a{}^{cd}C_{cdbe}u^e\,.
\label{EabHab}
\end{equation}
Then,
\begin{equation}
C_{abcd}=
\left(g_{abqp}g_{cdsr}-\eta_{abqp}\eta_{cdsr}\right)u^qu^sE^{pr}-
\left(\eta_{abqp}g_{cdsr}+g_{abqp}\eta_{cdsr}\right)u^qu^sH^{pr}\,,
\label{Weyl}
\end{equation}
where 

\begin{equation}
    g_{abcd}=g_{ac}g_{bd}-g_{ad}g_{bc}
\end{equation}

The spatial, symmetric and trace-free tensors $E_{ab}$ and $H_{ab}$
are known as the \textbf{electric} and \textbf{magnetic} Weyl components. The electric
part represent the well known tidal tensor of the Newtonian gravitational potential, but $H_{ab}$ has no Newtonian counterpart. 

The dynamics of the Weyl tensor are given by the Bianchi identities which gives:

\begin{equation}
\nabla^dC_{abcd}= \nabla_{[b}R_{a]c}+
{1\over6}\,g_{c[b}\nabla_{a]}R\,,  \label{Bianchi}
\end{equation}
This equation can be splitted into a set of two propagation and two constraint equations, which
govern the dynamics of the electric and magnetic Weyl components.

\subsubsection{Matter fields}\label{sssMFs}
The energy-momentum tensor of an imperfect fluid decomposes into its irreducible parts as:

\begin{equation}
T_{ab}= \rho u_au_b+ ph_{ab}+ 2q_{(a}u_{b)}+ \pi_{ab}\,.
\label{Tab1}
\end{equation}
where

\begin{equation}
    \rho=T_{ab}u^au^b
\end{equation}

is the matter energy density, while

\begin{equation}
    p=T_{ab}h^{ab}/3
\end{equation}
is the effective isotropic pressure of the fluid and

\begin{eqnarray}
    q_a=-h_a{}^bT_{bc}u^c
\end{eqnarray}
is the energy-flux vector. The tensor

\begin{equation}
    \pi_{ab}=h_{\langle a}{}^{c} h_{b\rangle}{}^dT_{cd}
\end{equation}
is the symmetric and trace-free anisotropic stress tensor. We used angled brackets denoting the symmetric and trace-free part of spatially projected second-rank tensors and the projected part of vectors according to
\begin{equation}
S_{\langle ab \rangle}= h_{\langle a}{}^{c}h_{b\rangle}{}^dS_{cd}=
h_{(a}{}^{c}h_{b)}{}^dS_{cd}- {1\over3}\,h^{cd}S_{cd}h_{ab}
\hspace{10mm} {\rm
and} \hspace{10mm} {V}_{\langle a\rangle}= h_a{}^bV_b\,,
\label{angled}
\end{equation}
respectively (with $S_{\langle ab \rangle}h^{ab}=0$). 

When the fluid
is perfect, there is a unique 4-velocity, relative
to which $q_a$, $\pi_{ab}$ are identically zero and the effective pressure reduces to the equilibrium one:

\begin{equation}
T_{ab}= \rho u_au_b+ ph_{ab}\,.  \label{pfTab}
\end{equation}

If we additionally assume dust-like fluid with $p=0$, we have the simplest case, which includes baryonic matter
(after decoupling) and cold dark matter. Otherwise, we need to
determine $p$ as a function of $\rho$ and potentially of other
thermodynamic variables in a equation of state. In general, the equation of state takes the
form $p=p\,(\rho,s)$, where $s$ is the specific entropy. For a
barotropic fluid we have the simple and well known relation $p=p\,(\rho)$.

Expression~(\ref{Tab1}) describes any type of matter, including
electromagnetic radiation and other scalars and vectorial fields. Since $R=4\Lambda-T$, with
$T=T_a{}^a$, Einstein's equations can be written as:

\begin{equation}
R_{ab}= T_{ab}- {1\over2}\,Tg_{ab}+ \Lambda g_{ab}\,.
\label{EFE2}
\end{equation}
The successive contraction of the above leads to a set of useful relations:

\begin{eqnarray}
R_{ab}u^au^b&=& {1\over2}\,(\rho+3p)- \Lambda\,, \label{EFE3}\\
h_a{}^bR_{bc}u^c&=& -q_a\,, \label{EFE4}\\ h_a{}^ch_b{}^dR_{cd}&=&
{1\over2}\,(\rho-p)h_{ab}+ \Lambda h_{ab}+ \pi_{ab}\,.  \label{EFE5}
\end{eqnarray}

\subsection{Covariant relativistic cosmology}\label{ssCRC}
There are various physical choices in cosmology for the fundamental 4-velocity field that defines the $1+3$ splitting of spacetime. The two common used frames are the one in which the dipole of the CMB
anisotropy vanishes and the other one is the local rest-frame of the matter. In specific situations, it may be appropriate to
choose the frame that simplifies the physics. Once $u_a$ is specified, its integral curves define the
worldlines of the fundamental observers.

\subsubsection{Kinematics}\label{sssKs}
The motion of an observer is characterized by the irreducible kinematic quantities associated with the $u_a$ velocity field, which arise from the covariant decomposition of the 4-velocity gradient:

\begin{equation}
\nabla_bu_a=\sigma_{ab}+ \omega_{ab}+ {1\over3}\,\Theta h_{ab}-
A_au_b\,,  \label{Nbua}
\end{equation}

where $\sigma_{ab}$ is the shear tensor, defined as $\sigma_{ab}={\rm D}_{\langle b}u_{a\rangle}$, $\omega_{ab}$ is the vorticity tensor, defined as $\omega_{ab}={\rm D}_{[b}u_{a]}$, $\Theta$ is the volume expansion scalar, with $\Theta=\nabla^au_a={\rm D}^au_a$, and $A_a$ is the 4-acceleration vector, $A_a=\dot{u}a=u^b\nabla_bu_a$. The shear tensor $\sigma{ab}$ describes the deformation of a fluid element’s shape without changing its volume, while $\omega_{ab}$ represents the rotation (vorticity), $\Theta$ represents the expansion or contraction of volume, and $A_a$ accounts for non-gravitational forces acting on the observer. It’s important to note that $A_a$ vanishes when motion is solely governed by gravity.

By construction, we have:

\begin{equation}
     \sigma_{ab}u^a=0= \omega_{ab}u^a=A_au^a
\end{equation}

The tensor $v_{ab}={\rm D}_bu_a=\sigma_{ab}+\omega_{ab}+
(\Theta/3)h_{ab}$ describes the relative motion between two nearby observers. Specifically, $v_a=v_{ab}\chi^b$ represents the relative velocity between the worldlines of the two observers, where $\chi_a$ is the relative position vector between them.

The volume expansion scalar $\Theta$ governs the average separation between neighboring observers and is used to define a representative length scale $a$ by the relation $\dot{a}/a=\Theta/3$, which corresponds to the Hubble parameter. The vorticity causes changes in the orientation of a fluid element without altering its volume or shape, whereas the shear affects the shape but leaves the volume unchanged.

The covariant kinematics are determined by a set of equations, which are purely geometrical
in origin and essentially independent of the Einstein equations.
Both sets emerge after applying the Ricci identities:

\begin{equation}
2\nabla_{[a}\nabla_{b]}u_c= R_{abcd}u^d\,,  \label{Ris}
\end{equation}
to the fundamental 4-velocity vector defined in (\ref{ua}).
Substituting in from (\ref{Nbua}), using decompositions
(\ref{Riemann}), (\ref{Weyl}) and the auxiliary relations
(\ref{EFE3})-(\ref{EFE5}), the timelike and spacelike parts of the
resulting expression lead to a set of three propagation and three
constraint equations. The former contains \textbf{Raychaudhuri's formula}:
\begin{equation}
\dot{\Theta}= -{1\over3}\,\Theta^2- {1\over2}\,(\rho+3p)-
2(\sigma^2-\omega^2)+ {\rm D}^aA_a+ A_aA^a+ \Lambda\,, \label{Ray}
\end{equation}
for the time evolution of $\Theta$; the shear propagation equation:
\begin{equation}
\dot{\sigma}_{\langle ab\rangle}= -{2\over3}\,\Theta\sigma_{ab}-
\sigma_{c\langle a}\sigma^c{}_{b\rangle}- \omega_{\langle
a}\omega_{b\rangle}+ {\rm D}_{\langle a}A_{b\rangle}+ A_{\langle
a}A_{b\rangle}- E_{ab}+ {1\over2}\,\pi_{ab}\,, \label{sigmadot}
\end{equation}
which describes kinematical anisotropies; and the evolution
equation of the vorticity:
\begin{equation}
\dot{\omega}_{\langle a\rangle}= -{2\over3}\,\Theta\omega_a-
{1\over2}\, curl\> A_a+ \sigma_{ab}\omega^b\,. \label{omegadot}
\end{equation}
Here, $\sigma^2=\sigma_{ab}\sigma^{ab}/2$ and
$\omega^2=\omega_{ab}\omega^{ab}/2=\omega_a\omega^a$ are
respectively the scalar square magnitudes of the shear and the
vorticity, while $E_{ab}$ is the electric component of the Weyl tensor. 


The spacelike component of (\ref{Ris}) leads to a set of three
complementary constraints. These are the shear constraint
\begin{equation}
{\rm D}^b\sigma_{ab}= {2\over3}\,{\rm D}_a\Theta+ curl \> \omega_a+
2\varepsilon_{abc}A^b\omega^c- q_a\,, \label{shearcon}
\end{equation}
the vorticity-divergence identity
\begin{equation}
{\rm D}^a\omega_a= A_a\omega^a\,,  \label{omegacon}
\end{equation}
and the magnetic Weyl equation
\begin{equation}
H_{ab}= curl \> \sigma_{ab}+ {\rm D}_{\langle a}\omega_{b\rangle}+
2A_{\langle a}\omega_{b\rangle}\,.  \label{Hcon}
\end{equation}

Raychaudhuri's equation is fundamental to the study of gravitational collapse, as it describes how the average separation between two neighboring observers changes over time. For this reason, Equation~(\ref{Ray}) has been central to all singularity theorems. Negative terms on the right-hand side of (\ref{Ray}) lead to contraction, while positive terms oppose collapse. This implies that conventional (non-phantom) matter is always attractive, unless the pressure satisfies $p < -\frac{\rho}{3} $.

\subsubsection{Conservation laws}\label{sssCLs}
The twice contracted Bianchi identities guarantee the conservation of the total energy-momentum tensor. This constraint splits into a timelike and a spacelike part, which respectively lead to the energy and the momentum conservation laws. When dealing with a general imperfect fluid, the former is
\begin{equation}
\dot{\rho}= -\Theta(\rho+p)- {\rm D}^a q_a- 2A^a q_a-
\sigma^{ab}\pi_{ab}\,,  \label{edc1}
\end{equation}
while the latter satisfies the expression
\begin{equation}
(\rho+p)A_a= -{\rm D}_a p- \dot{q}_{\langle a\rangle}-
{4\over3}\,\Theta q_a- (\sigma_{ab}+\omega_{ab})q^b- {\rm
D}^b \pi_{ab}- \pi_{ab} A^b\,.  \label{mdc1}
\end{equation}
When the fluid is perfect, the energy-momentum tensor is given by (\ref{pfTab}) and the above reduce to
\begin{equation}
\dot{\rho}= -\Theta(\rho+p) \hspace{15mm} {\rm and} \hspace{15mm}
(\rho+p)A_a= -{\rm D}_a p\,,  \label{pfcls}
\end{equation}
respectively. It follows, from (\ref{pfcls}b), that the sum $\rho+p$
describes the relativistic total inertial mass of the medium.

\clearchapterpage

\section{Marginalization of parameters in cosmological analysis}
\label{Marg}

\subsection{Maximum likelihood method}

In the scientific context, frequentist statistics allows us to calculate the probability of observing the result of an experiment given a theoretical model. Thus, if $C$ is the theoretical model (cause) and $E$ is the result of the experiment (effect), the probability of observing $E$ given that $C$ occurs is given by the conditional probability:

\begin{equation*}
    P(E/C)=\frac{P(E\cap C)}{P(C)}
\end{equation*}

\textbf{Bayes' theorem} allows us to invert this relationship, so we can estimate \textbf{the probability that a certain theoretical model $C_\theta$ is correct, given that we observe the results of an experiment}. This theorem is stated as:

\begin{equation*}
    P(C_\theta/E)=\frac{P(E/C_\theta)P(C_\theta)}{P(E)}=\frac{P(E/C_\theta)P(C_\theta)}{\sum_\theta P(E/C_\theta)P(C_\theta)}
\end{equation*}

We will identify the following elements in Bayes' theorem:

\begin{enumerate}
    \item $P(E/C_\theta)$ is the \textbf{conditional probability} of observing the data given a model. If the model $C_\theta$ is defined by the set of parameters $\theta$ and $x$ are the observed values of a certain physical variable, then this conditional probability is a \textbf{probability distribution of the variable $x$ given the parameters $\theta$}:
    
    \begin{equation*}
        P(E=x/C_\theta)=f(x;\theta)
    \end{equation*}

    If we consider that in each realization of the experiment a value $x_i$ was obtained, then the total probability of observing the set of values $\mathbf{x}$ given a parameter $\theta$, if the realizations are independent is:

    \begin{equation*}
        f(\mathbf{x};\theta)=\prod_i f(x_i,\theta)
    \end{equation*}

    Commonly, the probability distribution of each experiment run whose error is $\sigma_i$ is assumed Gaussian. We will then say that $x_i$ is distributed in a Gaussian manner, where the average $\mu(\theta)$ is the expected theoretical value of the variable $x$ given the theoretical model $C_\theta$:

    \begin{equation*}
        f(\mathbf{x};\theta)= \prod_i^N \frac{1}{\sqrt{2\pi\sigma_i^2}}e^{-\frac{1}{\sigma_i^2}[x_i-\mu(\theta)]^2}
    \end{equation*}

    Where $N$ is the amount of data.

    \item $P(C_\theta)$ is called \textbf{a priori probability} and represents the prior information we know about the parameters of the theoretical model. In the language of statistical analysis, this term is a probability distribution of $\theta$ that is called \textit{prior} $p(\theta)$ and if we do not know much a priori information about the possible parameters $\theta$ , we can assign a uniform probability to each possible value. Thus, if $\theta$ moves in an interval $[\theta_1,\theta_2]$:

    \begin{equation*}
        p(\theta)=\frac{1}{\theta_2-\theta_1}
    \end{equation*}
    
    If a large range of parameters is used, then the prior is known as a \textbf{weakly informative prior}.

    \item $P(E)$ is the \textbf{total probability} of observing the result of the experiment given the entire space of possible theoretical models. This term is known as \textit{marginal probability} and in practice it is not known. However, since it is a numerical value, we can ignore it and assign it a normalization constant $Z$. Thus, Bayes' theorem is written:

    \begin{equation*}
        P(C_\theta/E)=\frac{1}{Z}P(E/C_\theta)P(C_\theta)
    \end{equation*}

    \item $P(C_\theta/E)$ is the \textbf{a posteriori probability} of the model $C_\theta$ given that the outcome $E$ of the experiment was observed. This term is known as \textbf{Likelihood} $\mathcal{L}(\theta ; \mathbf{x})$:

    \begin{equation*}
        P(C_\theta/E)=\mathcal{L}(\theta ; \mathbf{x})
    \end{equation*}

    Which is the basis of the most used technique in cosmology to estimate cosmological parameters: \textbf{The maximum likelihood method}. Given a parameter space $\theta$ for a given model, we can find the best values for these parameters if we can \textbf{maximize the posterior probability}.
    
\end{enumerate}

Given the above information, we can write Bayes' theorem as:

\begin{equation*}
    \mathcal{L}(\theta ; x) =\frac{1}{Z} f(\mathbf{x};\theta)p(\theta)
\end{equation*}

We can simplify the problem a little if we take the natural logarithm of the previous expression, since the logarithm is an increasing function. We define the \textbf{LogLikelihood} $\mathcal{L}\mathcal{L}$ as:

\begin{equation*}
    \mathcal{L}\mathcal{L}=\ln\left (\frac{1}{Z}\right )+\ln{f(x;\theta)}+\ln p(\theta)
\end{equation*}

Which in the case of Gaussian realizations is:

\begin{eqnarray}
    \mathcal{L}\mathcal{L}&=&\ln\left (\frac{1}{Z}\right )+\ln p(\theta)+\sum_i^N\ln{\left (\frac{1}{\sqrt{2\pi\sigma_i^2}}\right )}-\sum_i^N\left(\frac
    {x_i-\mu(\theta)}{\sigma_i}\right)^2
\end{eqnarray}

We will then maximize this function with respect to the parameter $\theta$. Note that the derivative of the first 3 terms on the right with respect to $\theta$ becomes 0, since they are all constants for a uniform prior. Thus, maximizing likelihood is equivalent to minimizing the factor $\chi^2$:

\begin{equation*}
    \chi^2=\sum_i^N\left(\frac
    {x_i-\mu(\theta)}{\sigma_i}\right)^2
\end{equation*}


\subsection{Distance modulus}

The distance modulus of an object with apparent magnitude $m$ and absolute magnitude $M$ can be written as:

\begin{equation}
    \mu=m-M
\end{equation}

If we consider supernova data, each data provides an apparent magnitude $m_i$ at a redshift $z_i$. If we assume that they are standard candles, the value of $M$ must be the same for all of them. Thus, we can define the distance modulus $\mu_i$ for a redshift supernova $z_i$ as:

\begin{equation}
    \mu_i=m_i-M
\end{equation}

Furthermore, the luminous distance for a model of a universe dominated by matter and cosmological constant (model $\Lambda$CDM) as a function of the redsfhit $z$ depends on two parameters $H_0$ and $\Omega_{0,M}$. For a redshift $z_i$, this can be written as:

\begin{equation}
    d^{\Lambda CDM}_L(z_i;H_0,\Omega_{0,M})=c\frac{1+z_i}{H_0}\int_0^{z_i}\frac{dz'}{\sqrt{E(z')}}
\end{equation}

Where $E(z)=\sqrt{\Omega_{0,M}(1+z)^3+(1-\Omega_{0,M})}$. In this way, the distance module as a function of the parameters $H_0$ and $\Omega_{0,M}$ can be written as:

\begin{equation}
    \mu(z_i;\Omega_{0,M},H_0)=5\log_{10}{\frac{d^{\Lambda CDM}_L}{10Mpc}}+25.
\end{equation}

Alternatively, the luminous distance can also be defined for low redshift based on the parameters $H_0$ and $q_0$ which is known as \textbf{cosmographic luminous distance}:

\begin{equation}
    d^{C}_L(z_i;H_0,q_0)=\frac{c}{H_0}[z_i+\frac{(1-q_0)}{2}z_i^2]
\end{equation}

Therefore, the distance module would be defined by:

\begin{equation}
    \mu(z_i;H_0,q_0)=5\log_{10}{\frac{d^{C}_L}{10Mpc}}+25
\end{equation}

In any case, the function to be minimized will be:

\begin{equation*}
    \chi^2=\sum_i^N\left(\frac
    {\Delta_i\mu}{\sigma_i}\right)^2
\end{equation*}

Where:

\begin{equation}
    \Delta_i \mu= \mu_i-\mu(z_i;\hat{\theta})=m_i-5\log_{10}{\frac{d_L(z_i,\hat{\theta})}{10Mpc}}-(M+25)
\end{equation}

The vector $\hat{\theta}$ represents the parameters of the luminous distance used, either $\Lambda$CDM ($\hat{\theta}=H_0,\Omega_{0,M})$ or the distance cosmographic ($\hat{\theta}=H_0,q_0)$. 

\subsection{Marginalization}

We can see that in the two luminous distance functions, the parameter $H_0$ appears as a factor of the form:

\begin{equation}
    d_L(z;H_0,\alpha)=\frac{1}{H_0}F(z;\alpha)
\end{equation}

Where $\alpha$ is either the decelaration parameter $q_0$ or the matter density parameter $\Omega_{0,M}$ depending on the luminous distance we want to use. This implies that in the function $\Delta_i\mu$, there will be a \textbf{degeneration} of the parameters $M$ and $H_0$:

\begin{eqnarray}
    \Delta_i \mu&=&m_i-5\log_{10}{\frac{F(z_i;\alpha)}{10Mpc}}-(M-5\log_{10}H_0+25)\\&=&m_i-5\log_{10}{\frac{F(z_i;\alpha)}{10Mpc}}-\mathcal{M}\\&=&\Delta m_i(\alpha)-\mathcal{M}
\end{eqnarray}

Where we have defined the parameter $\mathcal{M}=M-5\log_{10}H_0+25$ and the function $\Delta m_i(\alpha)=m_i-5\log_{10}{\frac{F(z_i, \theta)}{10Mpc}}$. This means that we cannot determine $H_0$ and $M$ independently using only the supernova data, since there are infinitely many combinations of $M$ and $H_0$ that result in the value of the parameter $\mathcal{M}$. To determine $H_0$ with only supernova data, we need other standard candles to act as calibrators (such as Cepheids). These calibrators allow us to determine the value of $M$ independently and thus break the degeneration in $\mathcal{M}$. In any case, only with supernovae can the value of $\Omega_{0,M}$ and $q_0$ be determined, which is very important since they give us information about the value of the cosmological constant. To do this, it is possible to perform the minimization of $\chi^2$ on the parameters $\mathcal{M}$ and $\alpha$. However, there is a way to \textbf{marginalize} the parameter $\mathcal{M}$ to only work with the parameter $\alpha$, integrating over the entire space of possible values of $\mathcal{M}$. If we integrate the Likelihood in the parameter $\mathcal{M}$, given a uniform prior $p$ we can obtain the likelihood of the cosmological parameter $\alpha$ as:

\begin{equation*}
    \mathcal{L}(\alpha)=\int \mathcal{L}(\alpha,\mathcal{M}) d\mathcal{M} =\frac{1}{Z}p({\alpha;\mathcal{M}})\left ( \prod_i^N \frac{1}{\sqrt{2\pi\sigma_i^2}}\right )\int e^{-\sum_i^N\frac{1}{\sigma_i^2}[\Delta m_i(\alpha)-\mathcal{M}]^2}d\mathcal{M}
\end{equation*}

Developing the square of the Gaussian term:

\begin{eqnarray}
    e^{-\sum_i^N\frac{1}{\sigma_i^2}[\Delta m_i(\alpha)-\mathcal{M}]^2}&=& e^{-\sum_i^N\frac{\Delta m_i(\alpha)^2}{\sigma_i^2}}e^{\left [2\mathcal{M}\sum_i^N\frac{\Delta m_i(\alpha)}{\sigma_i^2}-\mathcal{M}^2\sum_i^N\frac{1}{\sigma_i^2}\right ]}\\&=&e^{-\chi_0^2}e^{-[\mathcal{M}^2B-2\mathcal{M}A]}
\end{eqnarray}

Where we have defined:

\begin{eqnarray}
    \chi_0^2&=&\sum_i^N\frac{\Delta m_i(\alpha)^2}{\sigma_i^2}\\
    A&=& \sum_i^N\frac{\Delta m_i(\alpha)}{\sigma_i^2}\\
    B&=& \sum_i^N\frac{1}{\sigma_i^2}
\end{eqnarray}

Let's notice that we can write:

\begin{eqnarray}
    (\sqrt{B}\mathcal{M}-\frac{A}{\sqrt{B}})^2-\frac{A^2}{B}=\mathcal{M}^2B-2\mathcal{M}A
\end{eqnarray}

Therefore it is convenient to change the variable:

\begin{eqnarray}
    u=\sqrt{B}\mathcal{M}-\frac{A}{\sqrt{B}} \\
    du=\sqrt{B}d\mathcal{M}
\end{eqnarray}

The result of the integral is then:

\begin{equation*}
    \int e^{-\sum_i^N\frac{1}{\sigma_i^2}[\Delta m_i(\alpha)-\mathcal{M}]^2}d\mathcal{M} = \frac{1}{\sqrt{B}}e^{-(\chi_0^2-\frac{A^2}{B})}\int e^{-u^2}du=\sqrt{\frac{\pi}{B}}e^{-(\chi_0^2-\frac{A^2}{B})}
\end{equation*}  

And therefore, the marginalized Likelihood can be written as:

\begin{eqnarray}
    \mathcal{L}(\alpha)=\frac{1}{Z}p({\alpha;\mathcal{M}})\left ( \prod_i^N \frac{1}{\sqrt{2B\sigma_i^2}}\right ) e^{-(\chi_0^2-\frac{A^2}{B})}
\end{eqnarray}

Finally, we take the natural logarithm of the previous expression:

\begin{eqnarray}
    \mathcal{LL}(\alpha)=\ln\left [\frac{1}{Z}p({\alpha;\mathcal{M}})\left ( \prod_i^N \frac{1}{\sqrt{2B\sigma_i^2}}\right )\right ]-\left [\chi_0^2-\frac{A^2}{B}\right ]
\end{eqnarray}

And assuming a uniform prior, we see that only the second term on the right will depend on the parameter $\alpha$. In this way, maximizing the marginalized Likelihood will be the same as minimizing the marginalized $\chi^2$ function:

\begin{eqnarray}
    \chi^2=\chi_0^2-\frac{A^2}{B}
\end{eqnarray}

\subsection{Non-diagonal covariance matrix}

The previous analysis can be extended easily for a non-diagonal covariance matrix (see \citealp{Conley_2010}). In this case, if we denote the magnitude vector data as $\mathbf{m}$,  the covariance matrix as $\mathbf{C}$ and $\mathbf{I}$ being the identity matrix, then:

\begin{eqnarray}
    \chi^2=\chi_0^2-\frac{A^2}{B}
\end{eqnarray}

With the definitions:

\begin{eqnarray}
    \chi_0^2&=&\Delta \mathbf{m}^{T} \cdot \mathbf{C}^{-1} \cdot \Delta \mathbf{m}\\
    A&=& \Delta \mathbf{m}^{T} \cdot \mathbf{C}^{-1} \cdot \mathbf{I}\\
    B&=& \mathbf{I}^{T} \cdot \mathbf{C}^{-1} \cdot \mathbf{I}
\end{eqnarray}

\clearchapterpage

\bibliographystyle{mnras}
\bibliography{8-references}

@article{Riess_1998,
doi = {10.1086/300499},
url = {https://dx.doi.org/10.1086/300499},
year = {1998},
month = {sep},
publisher = {},
volume = {116},
number = {3},
pages = {1009},
author = {Adam G. Riess and Alexei V. Filippenko and Peter Challis and Alejandro Clocchiatti and Alan Diercks and Peter M. Garnavich and Ron L. Gilliland and Craig J. Hogan and Saurabh Jha and Robert P. Kirshner and B. Leibundgut and M. M. Phillips and David Reiss and Brian P. Schmidt and Robert A. Schommer and R. Chris Smith and J. Spyromilio and Christopher Stubbs and Nicholas B. Suntzeff and John Tonry},
title = {Observational Evidence from Supernovae for an Accelerating Universe and a Cosmological Constant},
journal = {The Astronomical Journal}
}

@article{ PLANCK20,
	author = {{Planck Collaboration} and {Aghanim, N.} and {Akrami, Y.} and {Arroja, F.} and {Ashdown, M.} and {Aumont, J.} and {Baccigalupi, C.} and {Ballardini, M.} and {Banday, A. J.} and {Barreiro, R. B.} and {Bartolo, N.} and {Basak, S.} and {Battye, R.} and {Benabed, K.} and {Bernard, J.-P.} and {Bersanelli, M.} and {Bielewicz, P.} and {Bock, J. J.} and {Bond, J. R.} and {Borrill, J.} and {Bouchet, F. R.} and {Boulanger, F.} and {Bucher, M.} and {Burigana, C.} and {Butler, R. C.} and {Calabrese, E.} and {Cardoso, J.-F.} and {Carron, J.} and {Casaponsa, B.} and {Challinor, A.} and {Chiang, H. C.} and {Colombo, L. P. L.} and {Combet, C.} and {Contreras, D.} and {Crill, B. P.} and {Cuttaia, F.} and {de Bernardis, P.} and {de Zotti, G.} and {Delabrouille, J.} and {Delouis, J.-M.} and {Désert, F.-X.} and {Di Valentino, E.} and {Dickinson, C.} and {Diego, J. M.} and {Donzelli, S.} and {Doré, O.} and {Douspis, M.} and {Ducout, A.} and {Dupac, X.} and {Efstathiou, G.} and {Elsner, F.} and {Enßlin, T. A.} and {Eriksen, H. K.} and {Falgarone, E.} and {Fantaye, Y.} and {Fergusson, J.} and {Fernandez-Cobos, R.} and {Finelli, F.} and {Forastieri, F.} and {Frailis, M.} and {Franceschi, E.} and {Frolov, A.} and {Galeotta, S.} and {Galli, S.} and {Ganga, K.} and {Génova-Santos, R. T.} and {Gerbino, M.} and {Ghosh, T.} and {González-Nuevo, J.} and {Górski, K. M.} and {Gratton, S.} and {Gruppuso, A.} and {Gudmundsson, J. E.} and {Hamann, J.} and {Handley, W.} and {Hansen, F. K.} and {Helou, G.} and {Herranz, D.} and {Hildebrandt, S. R.} and {Hivon, E.} and {Huang, Z.} and {Jaffe, A. H.} and {Jones, W. C.} and {Karakci, A.} and {Keihänen, E.} and {Keskitalo, R.} and {Kiiveri, K.} and {Kim, J.} and {Kisner, T. S.} and {Knox, L.} and {Krachmalnicoff, N.} and {Kunz, M.} and {Kurki-Suonio, H.} and {Lagache, G.} and {Lamarre, J.-M.} and {Langer, M.} and {Lasenby, A.} and {Lattanzi, M.} and {Lawrence, C. R.} and {Le Jeune, M.} and {Leahy, J. P.} and {Lesgourgues, J.} and {Levrier, F.} and {Lewis, A.} and {Liguori, M.} and {Lilje, P. B.} and {Lilley, M.} and {Lindholm, V.} and {López-Caniego, M.} and {Lubin, P. M.} and {Ma, Y.-Z.} and {Macías-Pérez, J. F.} and {Maggio, G.} and {Maino, D.} and {Mandolesi, N.} and {Mangilli, A.} and {Marcos-Caballero, A.} and {Maris, M.} and {Martin, P. G.} and {Martinelli, M.} and {Martínez-González, E.} and {Matarrese, S.} and {Mauri, N.} and {McEwen, J. D.} and {Meerburg, P. D.} and {Meinhold, P. R.} and {Melchiorri, A.} and {Mennella, A.} and {Migliaccio, M.} and {Millea, M.} and {Mitra, S.} and {Miville-Deschênes, M.-A.} and {Molinari, D.} and {Moneti, A.} and {Montier, L.} and {Morgante, G.} and {Moss, A.} and {Mottet, S.} and {Münchmeyer, M.} and {Natoli, P.} and {Nørgaard-Nielsen, H. U.} and {Oxborrow, C. A.} and {Pagano, L.} and {Paoletti, D.} and {Partridge, B.} and {Patanchon, G.} and {Pearson, T. J.} and {Peel, M.} and {Peiris, H. V.} and {Perrotta, F.} and {Pettorino, V.} and {Piacentini, F.} and {Polastri, L.} and {Polenta, G.} and {Puget, J.-L.} and {Rachen, J. P.} and {Reinecke, M.} and {Remazeilles, M.} and {Renault, C.} and {Renzi, A.} and {Rocha, G.} and {Rosset, C.} and {Roudier, G.} and {Rubiño-Martín, J. A.} and {Ruiz-Granados, B.} and {Salvati, L.} and {Sandri, M.} and {Savelainen, M.} and {Scott, D.} and {Shellard, E. P. S.} and {Shiraishi, M.} and {Sirignano, C.} and {Sirri, G.} and {Spencer, L. D.} and {Sunyaev, R.} and {Suur-Uski, A.-S.} and {Tauber, J. A.} and {Tavagnacco, D.} and {Tenti, M.} and {Terenzi, L.} and {Toffolatti, L.} and {Tomasi, M.} and {Trombetti, T.} and {Valiviita, J.} and {Van Tent, B.} and {Vibert, L.} and {Vielva, P.} and {Villa, F.} and {Vittorio, N.} and {Wandelt, B. D.} and {Wehus, I. K.} and {White, M.} and {White, S. D. M.} and {Zacchei, A.} and {Zonca, A.}},
	title = {Planck 2018 results - I. Overview and the cosmological legacy of Planck},
	DOI= "10.1051/0004-6361/201833880",
	url= "https://doi.org/10.1051/0004-6361/201833880",
	journal = {A\&A},
	year = 2020,
	volume = 641,
	pages = "A1",
}

@article{Brout_2022,
	doi = {10.3847/1538-4357/ac8e04},
  
	url = {https://doi.org/10.3847%2F1538-4357%2Fac8e04},
  
	year = 2022,
	month = {oct},
  
	publisher = {American Astronomical Society},
  
	volume = {938},
  
	number = {2},
  
	pages = {110},
  
	author = {Dillon Brout and Dan Scolnic and Brodie Popovic and Adam G. Riess and Anthony Carr and Joe Zuntz and Rick Kessler and Tamara M. Davis and Samuel Hinton and David Jones and W. D'Arcy Kenworthy and Erik R. Peterson and Khaled Said and Georgie Taylor and Noor Ali and Patrick Armstrong and Pranav Charvu and Arianna Dwomoh and Cole Meldorf and Antonella Palmese and Helen Qu and Benjamin M. Rose and Bruno Sanchez and Christopher W. Stubbs and Maria Vincenzi and Charlotte M. Wood and Peter J. Brown and Rebecca Chen and Ken Chambers and David A. Coulter and Mi Dai and Georgios Dimitriadis and Alexei V. Filippenko and Ryan J. Foley and Saurabh W. Jha and Lisa Kelsey and Robert P. Kirshner and Anais Möller and Jessie Muir and Seshadri Nadathur and Yen-Chen Pan and Armin Rest and Cesar Rojas-Bravo and Masao Sako and Matthew R. Siebert and Mat Smith and Benjamin E. Stahl and Phil Wiseman},
  
	title = {The Pantheon$\mathplus$ Analysis: Cosmological Constraints},
  
	journal = {The Astrophysical Journal}
}

@misc{desicollaboration2024desi2024vicosmological,
      title={DESI 2024 VI: Cosmological Constraints from the Measurements of Baryon Acoustic Oscillations}, 
      author={DESI and A. G. Adame and J. Aguilar and S. Ahlen and S. Alam and D. M. Alexander and M. Alvarez and O. Alves and A. Anand and U. Andrade and E. Armengaud and S. Avila and A. Aviles and H. Awan and B. Bahr-Kalus and S. Bailey and C. Baltay and A. Bault and J. Behera and S. BenZvi and A. Bera and F. Beutler and D. Bianchi and C. Blake and R. Blum and S. Brieden and A. Brodzeller and D. Brooks and E. Buckley-Geer and E. Burtin and R. Calderon and R. Canning and A. Carnero Rosell and R. Cereskaite and J. L. Cervantes-Cota and S. Chabanier and E. Chaussidon and J. Chaves-Montero and S. Chen and X. Chen and T. Claybaugh and S. Cole and A. Cuceu and T. M. Davis and K. Dawson and A. de la Macorra and A. de Mattia and N. Deiosso and A. Dey and B. Dey and Z. Ding and P. Doel and J. Edelstein and S. Eftekharzadeh and D. J. Eisenstein and A. Elliott and P. Fagrelius and K. Fanning and S. Ferraro and J. Ereza and N. Findlay and B. Flaugher and A. Font-Ribera and D. Forero-Sánchez and J. E. Forero-Romero and C. S. Frenk and C. Garcia-Quintero and E. Gaztañaga and H. Gil-Marín and S. Gontcho A Gontcho and A. X. Gonzalez-Morales and V. Gonzalez-Perez and C. Gordon and D. Green and D. Gruen and R. Gsponer and G. Gutierrez and J. Guy and B. Hadzhiyska and C. Hahn and M. M. S Hanif and H. K. Herrera-Alcantar and K. Honscheid and C. Howlett and D. Huterer and V. Iršič and M. Ishak and S. Juneau and N. G. Karaçaylı and R. Kehoe and S. Kent and D. Kirkby and A. Kremin and A. Krolewski and Y. Lai and T. -W. Lan and M. Landriau and D. Lang and J. Lasker and J. M. Le Goff and L. Le Guillou and A. Leauthaud and M. E. Levi and T. S. Li and E. Linder and K. Lodha and C. Magneville and M. Manera and D. Margala and P. Martini and M. Maus and P. McDonald and L. Medina-Varela and A. Meisner and J. Mena-Fernández and R. Miquel and J. Moon and S. Moore and J. Moustakas and N. Mudur and E. Mueller and A. Muñoz-Gutiérrez and A. D. Myers and S. Nadathur and L. Napolitano and R. Neveux and J. A. Newman and N. M. Nguyen and J. Nie and G. Niz and H. E. Noriega and N. Padmanabhan and E. Paillas and N. Palanque-Delabrouille and J. Pan and S. Penmetsa and W. J. Percival and M. M. Pieri and M. Pinon and C. Poppett and A. Porredon and F. Prada and A. Pérez-Fernández and I. Pérez-Ràfols and D. Rabinowitz and A. Raichoor and C. Ramírez-Pérez and S. Ramirez-Solano and C. Ravoux and M. Rashkovetskyi and M. Rezaie and J. Rich and A. Rocher and C. Rockosi and N. A. Roe and A. Rosado-Marin and A. J. Ross and G. Rossi and R. Ruggeri and V. Ruhlmann-Kleider and L. Samushia and E. Sanchez and C. Saulder and E. F. Schlafly and D. Schlegel and M. Schubnell and H. Seo and A. Shafieloo and R. Sharples and J. Silber and A. Slosar and A. Smith and D. Sprayberry and T. Tan and G. Tarlé and P. Taylor and S. Trusov and L. A. Ureña-López and R. Vaisakh and D. Valcin and F. Valdes and M. Vargas-Magaña and L. Verde and M. Walther and B. Wang and M. S. Wang and B. A. Weaver and N. Weaverdyck and R. H. Wechsler and D. H. Weinberg and M. White and J. Yu and Y. Yu and S. Yuan and C. Yèche and E. A. Zaborowski and P. Zarrouk and H. Zhang and C. Zhao and R. Zhao and R. Zhou and T. Zhuang and H. Zou},
      year={2024},
      eprint={2404.03002},
      archivePrefix={arXiv},
      primaryClass={astro-ph.CO},
      url={https://arxiv.org/abs/2404.03002}, 
}

@article{Peebles_2022,
   title={Anomalies in physical cosmology},
   volume={447},
   ISSN={0003-4916},
   url={http://dx.doi.org/10.1016/j.aop.2022.169159},
   DOI={10.1016/j.aop.2022.169159},
   journal={Annals of Physics},
   publisher={Elsevier BV},
   author={Peebles, P.J.E.},
   year={2022},
   month=dec, pages={169159} }

@article{Lakhal_2019,
doi = {10.1088/1742-6596/1269/1/012017},
url = {https://dx.doi.org/10.1088/1742-6596/1269/1/012017},
year = {2019},
month = {jul},
publisher = {IOP Publishing},
volume = {1269},
number = {1},
pages = {012017},
author = {Bahia Si Lakhal and Ameur Guezmir},
title = {The Horizon Problem},
journal = {Journal of Physics: Conference Series}
}

@article{Guth_1981,
  title = {Inflationary universe: A possible solution to the horizon and flatness problems},
  author = {Guth, Alan H.},
  journal = {Phys. Rev. D},
  volume = {23},
  issue = {2},
  pages = {347--356},
  numpages = {0},
  year = {1981},
  month = {Jan},
  publisher = {American Physical Society},
  doi = {10.1103/PhysRevD.23.347},
  url = {https://link.aps.org/doi/10.1103/PhysRevD.23.347}
}

@article{Hollands_2002,
   title={Essay: An Alternative to Inflation},
   volume={34},
   ISSN={1572-9532},
   url={http://dx.doi.org/10.1023/A:1021175216055},
   DOI={10.1023/a:1021175216055},
   number={12},
   journal={General Relativity and Gravitation},
   publisher={Springer Science and Business Media LLC},
   author={Hollands, Stefan and Wald, Robert M.},
   year={2002},
   month=dec, pages={2043–2055} }

@article{BRANDENBERGER_2011,
   title={ALTERNATIVES TO THE INFLATIONARY PARADIGM OF STRUCTURE FORMATION},
   volume={01},
   ISSN={2010-1945},
   url={http://dx.doi.org/10.1142/S2010194511000109},
   DOI={10.1142/s2010194511000109},
   journal={International Journal of Modern Physics: Conference Series},
   publisher={World Scientific Pub Co Pte Lt},
   author={Brandenberger, Robert H.},
   year={2011},
   month=jan, pages={67–79} }

@article{Penrose_1988mg,
    author = "Penrose, R.",
    editor = "Fenyves, E. J.",
    title = "{Difficulties with inflationary cosmology}",
    doi = "10.1111/j.1749-6632.1989.tb50513.x",
    journal = "Annals N. Y. Acad. Sci.",
    volume = "571",
    pages = "249--264",
    year = "1989"
}

@article{Sol__2013,
   title={Cosmological constant and vacuum energy: old and new ideas},
   volume={453},
   ISSN={1742-6596},
   url={http://dx.doi.org/10.1088/1742-6596/453/1/012015},
   DOI={10.1088/1742-6596/453/1/012015},
   journal={Journal of Physics: Conference Series},
   publisher={IOP Publishing},
   author={Solà, Joan},
   year={2013},
   month=aug, pages={012015} }

@article{Freese_2009,
   title={Review of Observational Evidence for Dark Matter in the Universe and in upcoming searches for Dark Stars},
   volume={36},
   ISSN={1638-1963},
   url={http://dx.doi.org/10.1051/eas/0936016},
   DOI={10.1051/eas/0936016},
   journal={EAS Publications Series},
   publisher={EDP Sciences},
   author={Freese, K.},
   year={2009},
   pages={113–126} }

@article{Feng_2010,
   title={Dark Matter Candidates from Particle Physics and Methods of Detection},
   volume={48},
   ISSN={1545-4282},
   url={http://dx.doi.org/10.1146/annurev-astro-082708-101659},
   DOI={10.1146/annurev-astro-082708-101659},
   number={1},
   journal={Annual Review of Astronomy and Astrophysics},
   publisher={Annual Reviews},
   author={Feng, Jonathan L.},
   year={2010},
   month=aug, pages={495–545} }

@misc{milgrom2001mondapedagogicalreview,
      title={MOND--a pedagogical review}, 
      author={Mordehai Milgrom},
      year={2001},
      eprint={astro-ph/0112069},
      archivePrefix={arXiv},
      primaryClass={astro-ph},
      url={https://arxiv.org/abs/astro-ph/0112069}, 
}

@article{MARTENS2020237,
title = {Dark matter = modified gravity? Scrutinising the spacetime–matter distinction through the modified gravity/ dark matter lens},
journal = {Studies in History and Philosophy of Science Part B: Studies in History and Philosophy of Modern Physics},
volume = {72},
pages = {237-250},
year = {2020},
issn = {1355-2198},
doi = {https://doi.org/10.1016/j.shpsb.2020.08.003},
url = {https://www.sciencedirect.com/science/article/pii/S135521982030109X},
author = {Niels C.M. Martens and Dennis Lehmkuhl}
}

@article{Hossenfelder_2020,
   title={The Milky Way’s rotation curve with superfluid dark matter},
   volume={498},
   ISSN={1365-2966},
   url={http://dx.doi.org/10.1093/mnras/staa2594},
   DOI={10.1093/mnras/staa2594},
   number={3},
   journal={Monthly Notices of the Royal Astronomical Society},
   publisher={Oxford University Press (OUP)},
   author={Hossenfelder, S and Mistele, T},
   year={2020},
   month=aug, pages={3484–3491} }

@article{Weinberg_1988,
    author = "Weinberg, Steven",
    editor = "Hsu, Jong-Ping and Fine, D.",
    title = "{The Cosmological Constant Problem}",
    reportNumber = "UTTG-12-88",
    doi = "10.1103/RevModPhys.61.1",
    journal = "Rev. Mod. Phys.",
    volume = "61",
    pages = "1--23",
    year = "1989"
}

@article{Mistele_2023,
   title={Superfluid dark matter in tension with weak gravitational lensing data},
   volume={2023},
   ISSN={1475-7516},
   url={http://dx.doi.org/10.1088/1475-7516/2023/09/004},
   DOI={10.1088/1475-7516/2023/09/004},
   number={09},
   journal={Journal of Cosmology and Astroparticle Physics},
   publisher={IOP Publishing},
   author={Mistele, T. and McGaugh, S. and Hossenfelder, S.},
   year={2023},
   month=sep, pages={004} }

@article{BIRD2023101231,
title = {Snowmass2021 Cosmic Frontier White Paper: Primordial black hole dark matter},
journal = {Physics of the Dark Universe},
volume = {41},
pages = {101231},
year = {2023},
issn = {2212-6864},
doi = {https://doi.org/10.1016/j.dark.2023.101231},
url = {https://www.sciencedirect.com/science/article/pii/S2212686423000651},
author = {Simeon Bird and Andrea Albert and Will Dawson and Yacine Ali-Haïmoud and Adam Coogan and Alex Drlica-Wagner and Qi Feng and Derek Inman and Keisuke Inomata and Ely Kovetz and Alexander Kusenko and Benjamin V. Lehmann and Julian B. Muñoz and Rajeev Singh and Volodymyr Takhistov and Yu-Dai Tsai}
}

@book{dicke1970gravitation,
  title={Gravitation and the Universe},
  author={Dicke, R.H.},
  isbn={9780871690784},
  lccn={78107344},
  series={American Philosophical Society: Memoirs of the American Philosophical Society},
  url={https://books.google.cl/books?id=k4FGAAAAYAAJ},
  year={1970},
  publisher={American Philosophical Society}
}

@ARTICLE{Helbig_2021,
       author = {{Helbig}, Phillip},
        title = "{Arguments against the flatness problem in classical cosmology: a review}",
      journal = {European Physical Journal H},
         year = 2021,
        month = dec,
       volume = {46},
       number = {1},
          eid = {10},
        pages = {10},
          doi = {10.1140/epjh/s13129-021-00006-9},
       adsurl = {https://ui.adsabs.harvard.edu/abs/2021EPJH...46...10H}
}

@article{Velten_2014,
   title={Aspects of the cosmological “coincidence problem”},
   volume={74},
   ISSN={1434-6052},
   url={http://dx.doi.org/10.1140/epjc/s10052-014-3160-4},
   DOI={10.1140/epjc/s10052-014-3160-4},
   number={11},
   journal={The European Physical Journal C},
   publisher={Springer Science and Business Media LLC},
   author={Velten, H. E. S. and vom Marttens, R. F. and Zimdahl, W.},
   year={2014},
   month=nov }

@article{Schander_2021,
   title={Backreaction in Cosmology},
   volume={8},
   ISSN={2296-987X},
   url={http://dx.doi.org/10.3389/fspas.2021.692198},
   DOI={10.3389/fspas.2021.692198},
   journal={Frontiers in Astronomy and Space Sciences},
   publisher={Frontiers Media SA},
   author={Schander, S. and Thiemann, T.},
   year={2021},
   month=jul }

@article{greenwald2011,
  title = {New framework for analyzing the effects of small scale inhomogeneities in cosmology},
  author = {Green, Stephen R. and Wald, Robert M.},
  journal = {Phys. Rev. D},
  volume = {83},
  issue = {8},
  pages = {084020},
  numpages = {27},
  year = {2011},
  month = {Apr},
  publisher = {American Physical Society},
  doi = {10.1103/PhysRevD.83.084020},
  url = {https://link.aps.org/doi/10.1103/PhysRevD.83.084020}
}

@article{Buchert_2000,
   title={On Average Properties of Inhomogeneous Fluids in General Relativity: Dust Cosmologies},
   volume={32},
   ISSN={1572-9532},
   url={http://dx.doi.org/10.1023/A:1001800617177},
   DOI={10.1023/a:1001800617177},
   number={1},
   journal={General Relativity and Gravitation},
   publisher={Springer Science and Business Media LLC},
   author={Buchert, Thomas},
   year={2000},
   month=jan, pages={105–125} }

@article{Schulz:2014gqa,
    author = "Schulz, Benjamin",
    title = "{Review on the quantization of gravity}",
    eprint = "1409.7977",
    archivePrefix = "arXiv",
    primaryClass = "gr-qc",
    month = "9",
    year = "2014",
    journal = {arXiv e-prints}
}

@article{Abdalla:2022yfr,
    author = "Abdalla, Elcio and others",
    title = "{Cosmology intertwined: A review of the particle physics, astrophysics, and cosmology associated with the cosmological tensions and anomalies}",
    eprint = "2203.06142",
    archivePrefix = "arXiv",
    primaryClass = "astro-ph.CO",
    reportNumber = "FERMILAB-CONF-22-192-SCD",
    doi = "10.1016/j.jheap.2022.04.002",
    journal = "JHEAp",
    volume = "34",
    pages = "49--211",
    year = "2022"
}

@article{DES2022,
  title = {Dark Energy Survey year 3 results: Constraints on cosmological parameters and galaxy-bias models from galaxy clustering and galaxy-galaxy lensing using the redMaGiC sample},
  author = {Pandey, S. and Krause, E. and DeRose, J. and MacCrann, N. and Jain, B. and Crocce, M. and Blazek, J. and Choi, A. and Huang, H. and To, C. and Fang, X. and Elvin-Poole, J. and Prat, J. and Porredon, A. and Secco, L. F. and Rodriguez-Monroy, M. and Weaverdyck, N. and Park, Y. and Raveri, M. and Rozo, E. and Rykoff, E. S. and Bernstein, G. M. and S\'anchez, C. and Jarvis, M. and Troxel, M. A. and Zacharegkas, G. and Chang, C. and Alarcon, A. and Alves, O. and Amon, A. and Andrade-Oliveira, F. and Baxter, E. and Bechtol, K. and Becker, M. R. and Camacho, H. and Campos, A. and Carnero Rosell, A. and Carrasco Kind, M. and Cawthon, R. and Chen, R. and Chintalapati, P. and Davis, C. and Di Valentino, E. and Diehl, H. T. and Dodelson, S. and Doux, C. and Drlica-Wagner, A. and Eckert, K. and Eifler, T. F. and Elsner, F. and Everett, S. and Farahi, A. and Fert\'e, A. and Fosalba, P. and Friedrich, O. and Gatti, M. and Giannini, G. and Gruen, D. and Gruendl, R. A. and Harrison, I. and Hartley, W. G. and Huff, E. M. and Huterer, D. and Kovacs, A. and Leget, P. F. and McCullough, J. and Muir, J. and Myles, J. and Navarro-Alsina, A. and Omori, Y. and Rollins, R. P. and Roodman, A. and Rosenfeld, R. and Sevilla-Noarbe, I. and Sheldon, E. and Shin, T. and Troja, A. and Tutusaus, I. and Varga, T. N. and Wechsler, R. H. and Yanny, B. and Yin, B. and Zhang, Y. and Zuntz, J. and Abbott, T. M. C. and Aguena, M. and Allam, S. and Annis, J. and Bacon, D. and Bertin, E. and Brooks, D. and Burke, D. L. and Carretero, J. and Conselice, C. and Costanzi, M. and da Costa, L. N. and Pereira, M. E. S. and De Vicente, J. and Dietrich, J. P. and Doel, P. and Evrard, A. E. and Ferrero, I. and Flaugher, B. and Frieman, J. and Garc\'{\i}a-Bellido, J. and Gaztanaga, E. and Gerdes, D. W. and Giannantonio, T. and Gschwend, J. and Gutierrez, G. and Hinton, S. R. and Hollowood, D. L. and Honscheid, K. and James, D. J. and Jeltema, T. and Kuehn, K. and Kuropatkin, N. and Lahav, O. and Lima, M. and Lin, H. and Maia, M. A. G. and Marshall, J. L. and Melchior, P. and Menanteau, F. and Miller, C. J. and Miquel, R. and Mohr, J. J. and Morgan, R. and Palmese, A. and Paz-Chinch\'on, F. and Petravick, D. and Pieres, A. and Plazas Malag\'on, A. A. and Sanchez, E. and Scarpine, V. and Serrano, S. and Smith, M. and Soares-Santos, M. and Suchyta, E. and Tarle, G. and Thomas, D. and Weller, J.},
  collaboration = {DES Collaboration},
  journal = {Phys. Rev. D},
  volume = {106},
  issue = {4},
  pages = {043520},
  numpages = {34},
  year = {2022},
  month = {Aug},
  publisher = {American Physical Society},
  doi = {10.1103/PhysRevD.106.043520},
  url = {https://link.aps.org/doi/10.1103/PhysRevD.106.043520}
}

@article{DiValentino21,
	doi = {10.3847/2041-8213/abe1c4},
  
	url = {https://doi.org/10.3847%2F2041-8213%2Fabe1c4},
  
	year = 2021,
	month = {feb},
  
	publisher = {American Astronomical Society},
  
	volume = {908},
  
	number = {1},
  
	pages = {L9},
  
	author = {Eleonora Di Valentino and Alessandro Melchiorri and Joseph Silk},
  
	title = {Investigating Cosmic Discordance},
  
	journal = {The Astrophysical Journal Letters}
}

@article{Giani_2024,
doi = {10.1088/1475-7516/2024/01/071},
url = {https://dx.doi.org/10.1088/1475-7516/2024/01/071},
year = {2024},
month = {jan},
publisher = {IOP Publishing},
volume = {2024},
number = {01},
pages = {071},
author = {Leonardo Giani and Cullan Howlett and Khaled Said and Tamara Davis and Sunny Vagnozzi},
title = {An effective description of Laniakea: impact on cosmology and the local determination of the Hubble constant},
journal = {Journal of Cosmology and Astroparticle Physics}
}

@misc{Colgain2022,
  doi = {10.48550/ARXIV.2206.11447},
  
  url = {https://arxiv.org/abs/2206.11447},
  
  author = {Colgain, Eoin O and Sheikh-Jabbari, M. M. and Solomon, Rance and Dainotti, Maria G. and Stojkovic, Dejan},
  
  title = {Putting Flat LCDM In The (Redshift) Bin},
  
  publisher = {arXiv},
  
  year = {2022},
  
  copyright = {arXiv.org perpetual, non-exclusive license}
}

@article{Deng18b,
	doi = {10.1140/epjc/s10052-018-6159-4},
  
	url = {https://doi.org/10.1140%2Fepjc%2Fs10052-018-6159-4},
  
	year = 2018,
	month = {sep},
  
	publisher = {Springer Science and Business Media {LLC}
},
  
	volume = {78},
  
	number = {9},
  
	author = {Hua-Kai Deng and Hao Wei},
  
	title = {Null signal for the cosmic anisotropy in the Pantheon supernovae~data},
  
	journal = {The European Physical Journal C}
}

@article{Chiocchetta_2021,
doi = {10.1088/1475-7516/2021/08/015},
url = {https://dx.doi.org/10.1088/1475-7516/2021/08/015},
year = {2021},
month = {aug},
publisher = {IOP Publishing},
volume = {2021},
number = {08},
pages = {015},
author = {Chiocchetta, C. and Gruppuso, A. and Lattanzi, M. and Natoli, P. and Pagano, L.},
title = {Lack-of-correlation anomaly in CMB large scale polarisation maps},
journal = {Journal of Cosmology and Astroparticle Physics}
}

@article{Oliveira2004,
  title = {Significance of the largest scale CMB fluctuations in WMAP},
  author = {de Oliveira-Costa, Ang\'elica and Tegmark, Max and Zaldarriaga, Matias and Hamilton, Andrew},
  journal = {Phys. Rev. D},
  volume = {69},
  issue = {6},
  pages = {063516},
  numpages = {12},
  year = {2004},
  month = {Mar},
  publisher = {American Physical Society},
  doi = {10.1103/PhysRevD.69.063516},
  url = {https://link.aps.org/doi/10.1103/PhysRevD.69.063516}
}

@article{Land2007,
    author = {Land, Kate and Magueijo, João},
    title = {The Axis of Evil revisited},
    journal = {Monthly Notices of the Royal Astronomical Society},
    volume = {378},
    number = {1},
    pages = {153-158},
    year = {2007},
    month = {05},
    issn = {0035-8711},
    doi = {10.1111/j.1365-2966.2007.11749.x},
    url = {https://doi.org/10.1111/j.1365-2966.2007.11749.x},
    eprint = {https://academic.oup.com/mnras/article-pdf/378/1/153/3964774/mnras0378-0153.pdf},
}

@article{Vielva_2004,
   title={Detection of Non‐Gaussianity in theWilkinson Microwave Anisotropy ProbeFirst‐Year Data Using Spherical Wavelets},
   volume={609},
   ISSN={1538-4357},
   url={http://dx.doi.org/10.1086/421007},
   DOI={10.1086/421007},
   number={1},
   journal={The Astrophysical Journal},
   publisher={American Astronomical Society},
   author={Vielva, P. and Martinez‐Gonzalez, E. and Barreiro, R. B. and Sanz, J. L. and Cayon, L.},
   year={2004},
   month=jul, pages={22–34} }

@article{Cruz_2006,
   title={The non-Gaussian cold spot in Wilkinson Microwave Anisotropy Probe: significance, morphology and foreground contribution},
   volume={369},
   ISSN={1365-2966},
   url={http://dx.doi.org/10.1111/j.1365-2966.2006.10312.x},
   DOI={10.1111/j.1365-2966.2006.10312.x},
   number={1},
   journal={Monthly Notices of the Royal Astronomical Society},
   publisher={Oxford University Press (OUP)},
   author={Cruz, M. and Tucci, M. and Martínez-González, E. and Vielva, P.},
   year={2006},
   month=may, pages={57–67} }

@article{Pelgrims_2015,
   title={Polarization alignments of quasars from the JVAS/CLASS 8.4-GHz surveys},
   volume={450},
   ISSN={0035-8711},
   url={http://dx.doi.org/10.1093/mnras/stv917},
   DOI={10.1093/mnras/stv917},
   number={4},
   journal={Monthly Notices of the Royal Astronomical Society},
   publisher={Oxford University Press (OUP)},
   author={Pelgrims, V. and Hutsemékers, D.},
   year={2015},
   month=may, pages={4161–4173} }

@article{Friday_2022,
   title={Correlated orientations of the axes of large quasar groups on Gpc scales},
   volume={511},
   ISSN={1365-2966},
   url={http://dx.doi.org/10.1093/mnras/stac269},
   DOI={10.1093/mnras/stac269},
   number={3},
   journal={Monthly Notices of the Royal Astronomical Society},
   publisher={Oxford University Press (OUP)},
   author={Friday, Tracey and Clowes, Roger G and Williger, Gerard M},
   year={2022},
   month=jan, pages={4159–4178} }

@article{Secrest_2021,
   title={A Test of the Cosmological Principle with Quasars},
   volume={908},
   ISSN={2041-8213},
   url={http://dx.doi.org/10.3847/2041-8213/abdd40},
   DOI={10.3847/2041-8213/abdd40},
   number={2},
   journal={The Astrophysical Journal Letters},
   publisher={American Astronomical Society},
   author={Secrest, Nathan J. and Hausegger, Sebastian von and Rameez, Mohamed and Mohayaee, Roya and Sarkar, Subir and Colin, Jacques},
   year={2021},
   month=feb, pages={L51} }

@article{Migkas2018,
	author = {{Migkas, Konstantinos} and {Reiprich, Thomas H.}},
	title = {Anisotropy of the galaxy cluster X-ray luminosity–temperature relation},
	DOI= "10.1051/0004-6361/201731222",
	url= "https://doi.org/10.1051/0004-6361/201731222",
	journal = {A\&A},
	year = 2018,
	volume = 611,
	pages = "A50",
}

@misc{Sah2024,
      title={Anisotropy in Pantheon+ supernovae}, 
      author={Animesh Sah and Mohamed Rameez and Subir Sarkar and Christos Tsagas},
      year={2024},
      eprint={2411.10838},
      archivePrefix={arXiv},
      primaryClass={astro-ph.CO},
      url={https://arxiv.org/abs/2411.10838}, 
}

@article{Roldan_2016,
   title={Interpreting the CMB aberration and Doppler measurements: boost or intrinsic dipole?},
   volume={2016},
   ISSN={1475-7516},
   url={http://dx.doi.org/10.1088/1475-7516/2016/06/026},
   DOI={10.1088/1475-7516/2016/06/026},
   number={06},
   journal={Journal of Cosmology and Astroparticle Physics},
   publisher={IOP Publishing},
   author={Roldan, Omar and Notari, Alessio and Quartin, Miguel},
   year={2016},
   month=jun, pages={026–026} }

@article{ Horvath2015,
	author = {{Horváth, István} and {Bagoly, Zsolt} and {Hakkila, Jon} and {Tóth, L. V.}},
	title = {New data support the existence of the Hercules-Corona Borealis Great Wall⋆},
	DOI= "10.1051/0004-6361/201424829",
	url= "https://doi.org/10.1051/0004-6361/201424829",
	journal = {A\&A},
	year = 2015,
	volume = 584,
	pages = "A48",
	month = "",
}

@article{Balazs2015,
    author = {Balázs, L. G. and Bagoly, Z. and Hakkila, J. E. and Horváth, I. and Kóbori, J. and Rácz, I. I. and Tóth, L. V.},
    title = {A giant ring-like structure at 0.78 \&lt; z \&lt; 0.86 displayed by GRBs},
    journal = {Monthly Notices of the Royal Astronomical Society},
    volume = {452},
    number = {3},
    pages = {2236-2246},
    year = {2015},
    month = {07},
    issn = {0035-8711},
    doi = {10.1093/mnras/stv1421},
    url = {https://doi.org/10.1093/mnras/stv1421},
    eprint = {https://academic.oup.com/mnras/article-pdf/452/3/2236/4909856/stv1421.pdf},
}

@article{Mao_2017,
doi = {10.3847/1538-4357/835/2/161},
url = {https://dx.doi.org/10.3847/1538-4357/835/2/161},
year = {2017},
month = {jan},
publisher = {The American Astronomical Society},
volume = {835},
number = {2},
pages = {161},
author = {Mao, Qingqing and Berlind, Andreas A. and Scherrer, Robert J. and Neyrinck, Mark C. and Scoccimarro, Román and Tinker, Jeremy L. and McBride, Cameron K. and Schneider, Donald P. and Pan, Kaike and Bizyaev, Dmitry and Malanushenko, Elena and Malanushenko, Viktor},
title = {A Cosmic Void Catalog of SDSS DR12 BOSS Galaxies},
journal = {The Astrophysical Journal}
}

@article{Keenan13,
	doi = {10.1088/0004-637x/775/1/62},
	url = {https://doi.org/10.1088/0004-637x/775/1/62},
	year = 2013,
	month = {sep},
	publisher = {American Astronomical Society},
	volume = {775},
	number = {1},
	pages = {62},
	author = {R. C. Keenan and A. J. Barger and L. L. Cowie},
	title = {{EVIDENCE} {FOR} A $\sim$300 {MEGAPARSEC} {SCALE} {UNDER}-{DENSITY} {IN} {THE} {LOCAL} {GALAXY} {DISTRIBUTION}},
	journal = {The Astrophysical Journal},
}

@article{Iguchi2002,
    author = {Iguchi, Hideo and Nakamura, Takashi and Nakao, Ken-ichi},
    title = {Is Dark Energy the Only Solution to the Apparent Acceleration of the Present Universe?},
    journal = {Progress of Theoretical Physics},
    volume = {108},
    number = {5},
    pages = {809-818},
    year = {2002},
    month = {11},
    issn = {0033-068X},
    doi = {10.1143/PTP.108.809},
    url = {https://doi.org/10.1143/PTP.108.809},
    eprint = {https://academic.oup.com/ptp/article-pdf/108/5/809/5468231/108-5-809.pdf},
}

@article{Yoo2008,
    author = {Yoo, Chul-Moon and Kai, Tomohiro and Nakao, Ken-ichi},
    title = {Solving the Inverse Problem with Inhomogeneous Universes},
    journal = {Progress of Theoretical Physics},
    volume = {120},
    number = {5},
    pages = {937-960},
    year = {2008},
    month = {11},
    issn = {0033-068X},
    doi = {10.1143/PTP.120.937},
    url = {https://doi.org/10.1143/PTP.120.937},
    eprint = {https://academic.oup.com/ptp/article-pdf/120/5/937/5207322/120-5-937.pdf},
}

@article{Mazurenko2023,
    author = {Mazurenko, Sergij and Banik, Indranil and Kroupa, Pavel and Haslbauer, Moritz},
    title = {A simultaneous solution to the Hubble tension and observed bulk flow within 250 h−1 Mpc},
    journal = {Monthly Notices of the Royal Astronomical Society},
    volume = {527},
    number = {3},
    pages = {4388-4396},
    year = {2023},
    month = {11},
    issn = {0035-8711},
    doi = {10.1093/mnras/stad3357},
    url = {https://doi.org/10.1093/mnras/stad3357},
    eprint = {https://academic.oup.com/mnras/article-pdf/527/3/4388/53894662/stad3357.pdf},
}

@article{Haslbauer_2020,
	doi = {10.1093/mnras/staa2348},
  
	url = {https://doi.org/10.1093%2Fmnras%2Fstaa2348},
  
	year = 2020,
	month = {oct},
  
	publisher = {Oxford University Press ({OUP})},
  
	volume = {499},
  
	number = {2},
  
	pages = {2845--2883},
  
	author = {Moritz Haslbauer and Indranil Banik and Pavel Kroupa},
  
	title = {The {KBC} void and Hubble tension contradict $\{upLambda}${CDM} on a{\hspace{0.167em}
}Gpc scale - Milgromian dynamics as a possible solution}

@article{Kenworthy19,
	doi = {10.3847/1538-4357/ab0ebf},
  
	url = {https://doi.org/10.3847%2F1538-4357%2Fab0ebf},
  
	year = 2019,
	month = {apr},
  
	publisher = {American Astronomical Society},
  
	volume = {875},
  
	number = {2},
  
	pages = {145},
  
	author = {W. D'Arcy Kenworthy and Dan Scolnic and Adam Riess},
  
	title = {The Local Perspective on the Hubble Tension: Local Structure Does Not Impact Measurement of the Hubble Constant},
  
	journal = {The Astrophysical Journal}
}

@misc{Camarena22,
  doi = {10.48550/ARXIV.2205.05422},
  
  url = {https://arxiv.org/abs/2205.05422},
  
  author = {Camarena, David and Marra, Valerio and Sakr, Ziad and Clarkson, Chris},
  
  title = {A void in the Hubble tension? The end of the line for the Hubble bubble},
  
  publisher = {arXiv},
  
  year = {2022},
  
  copyright = {arXiv.org perpetual, non-exclusive license}
}

@ARTICLE{Hoscheit18,
       author = {{Hoscheit}, Benjamin L. and {Barger}, Amy J.},
        title = "{The KBC Void: Consistency with Supernovae Type Ia and the Kinematic SZ Effect in a {\ensuremath{\Lambda}}LTB Model}",
      journal = {\apj},
         year = 2018,
        month = feb,
       volume = {854},
       number = {1},
          eid = {46},
        pages = {46},
          doi = {10.3847/1538-4357/aaa59b},
archivePrefix = {arXiv},
       eprint = {1801.01890},
 primaryClass = {astro-ph.CO},
       adsurl = {https://ui.adsabs.harvard.edu/abs/2018ApJ...854...46H}
}

@misc{huterer2023enoughlocalvoidsolve,
      title={Not empty enough: a local void cannot solve the $H_0$ tension}, 
      author={Dragan Huterer and Hao-Yi Wu},
      year={2023},
      eprint={2309.05749},
      archivePrefix={arXiv},
      primaryClass={astro-ph.CO},
      url={https://arxiv.org/abs/2309.05749}, 
}

@article{Bouwens_2023,
    author = {Bouwens, Rychard and Illingworth, Garth and Oesch, Pascal and Stefanon, Mauro and Naidu, Rohan and van Leeuwen, Ivana and Magee, Dan},
    title = {UV luminosity density results at z \&gt; 8 from the first JWST/NIRCam fields: limitations of early data sets and the need for spectroscopy},
    journal = {Monthly Notices of the Royal Astronomical Society},
    volume = {523},
    number = {1},
    pages = {1009-1035},
    year = {2023},
    month = {04},
    issn = {0035-8711},
    doi = {10.1093/mnras/stad1014},
    url = {https://doi.org/10.1093/mnras/stad1014},
    eprint = {https://academic.oup.com/mnras/article-pdf/523/1/1009/50488630/stad1014.pdf},
}

@article{Donnan_2022,
    author = {Donnan, C T and McLeod, D J and Dunlop, J S and McLure, R J and Carnall, A C and Begley, R and Cullen, F and Hamadouche, M L and Bowler, R A A and Magee, D and McCracken, H J and Milvang-Jensen, B and Moneti, A and Targett, T},
    title = {The evolution of the galaxy UV luminosity function at redshifts z ≃ 8 – 15 from deep JWST and ground-based near-infrared imaging},
    journal = {Monthly Notices of the Royal Astronomical Society},
    volume = {518},
    number = {4},
    pages = {6011-6040},
    year = {2022},
    month = {11},
    issn = {0035-8711},
    doi = {10.1093/mnras/stac3472},
    url = {https://doi.org/10.1093/mnras/stac3472},
    eprint = {https://academic.oup.com/mnras/article-pdf/518/4/6011/47887826/stac3472.pdf},
}

@article{Peterson_2022,
doi = {10.3847/1538-4357/ac4698},
url = {https://dx.doi.org/10.3847/1538-4357/ac4698},
year = {2022},
month = {oct},
publisher = {The American Astronomical Society},
volume = {938},
number = {2},
pages = {112},
author = {Erik R. Peterson and W. D’Arcy Kenworthy and Daniel Scolnic and Adam G. Riess and Dillon Brout and Anthony Carr and Hélène Courtois and Tamara Davis and Arianna Dwomoh and David O. Jones and Brodie Popovic and Benjamin M. Rose and Khaled Said},
title = {The Pantheon+ Analysis: Evaluating Peculiar Velocity Corrections in Cosmological Analyses with Nearby Type Ia Supernovae},
journal = {The Astrophysical Journal}
}

@article{Carr_2022,
	doi = {10.1017/pasa.2022.41},
  
	url = {https://doi.org/10.1017%2Fpasa.2022.41},
  
	year = 2022,
	publisher = {Cambridge University Press ({CUP})},
  
	volume = {39},
  
	author = {Anthony Carr and Tamara M. Davis and Dan Scolnic and Khaled Said and Dillon Brout and Erik R. Peterson and Richard Kessler},
  
	title = {The Pantheon$\mathplus$ analysis: Improving the redshifts and peculiar velocities of Type Ia supernovae used in cosmological analyses},
  
	journal = {Publications of the Astronomical Society of Australia}
}

@article{ Curtois_2023,
	author = {{Courtois, H. M.} and {Dupuy, A.} and {Guinet, D.} and {Baulieu, G.} and {Ruppin, F.} and {Brenas, P.}},
	title = {Gravity in the local Universe: Density and velocity fields using CosmicFlows-4},
	DOI= "10.1051/0004-6361/202245331",
	url= "https://doi.org/10.1051/0004-6361/202245331",
	journal = {A\&A},
	year = 2023,
	volume = 670,
	pages = "L15",
}

@article{TSAGAS200861,
title = {Relativistic cosmology and large-scale structure},
journal = {Physics Reports},
volume = {465},
number = {2},
pages = {61-147},
year = {2008},
issn = {0370-1573},
doi = {https://doi.org/10.1016/j.physrep.2008.03.003},
url = {https://www.sciencedirect.com/science/article/pii/S0370157308000884},
author = {Christos G. Tsagas and Anthony Challinor and Roy Maartens}
}

@article{Raychaudhuri55,
    author = "Raychaudhuri, Amalkumar",
    title = "{Relativistic cosmology. 1.}",
    doi = "10.1103/PhysRev.98.1123",
    journal = "Phys. Rev.",
    volume = "98",
    pages = "1123--1126",
    year = "1955"
}

@misc{pittphilsci9062,
            year = {2012},
           month = {March},
          author = {Helge Kragh},
           title = {?The most philosophically of all the sciences?: Karl Popper and physical cosmology
},
             url = {https://philsci-archive.pitt.edu/9062/}
}

@article{Tully_2023,
   title={Cosmicflows-4},
   volume={944},
   ISSN={1538-4357},
   url={http://dx.doi.org/10.3847/1538-4357/ac94d8},
   DOI={10.3847/1538-4357/ac94d8},
   number={1},
   journal={The Astrophysical Journal},
   publisher={American Astronomical Society},
   author={Tully, R. Brent and Kourkchi, Ehsan and Courtois, Hélène M. and Anand, Gagandeep S. and Blakeslee, John P. and Brout, Dillon and Jaeger, Thomas de and Dupuy, Alexandra and Guinet, Daniel and Howlett, Cullan and Jensen, Joseph B. and Pomarède, Daniel and Rizzi, Luca and Rubin, David and Said, Khaled and Scolnic, Daniel and Stahl, Benjamin E.},
   year={2023},
   month=feb, pages={94} }

@article{Neyman2018,
    author = {Neyman, Jerzy and Scott, Elizabeth L.},
    title = {Statistical Approach to Problems of Cosmology},
    journal = {Journal of the Royal Statistical Society: Series B (Methodological)},
    volume = {20},
    number = {1},
    pages = {1-29},
    year = {2018},
    month = {12},
    issn = {0035-9246},
    doi = {10.1111/j.2517-6161.1958.tb00272.x},
    url = {https://doi.org/10.1111/j.2517-6161.1958.tb00272.x},
    eprint = {https://academic.oup.com/jrsssb/article-pdf/20/1/1/49172844/jrsssb\_20\_1\_1.pdf},
}

@ARTICLE{Yadav2005,
       author = {{Yadav}, Jaswant and {Bharadwaj}, Somnath and {Pandey}, Biswajit and {Seshadri}, T.~R.},
        title = "{Testing homogeneity on large scales in the Sloan Digital Sky Survey Data Release One}",
      journal = {\mnras},
         year = 2005,
        month = dec,
       volume = {364},
       number = {2},
        pages = {601-606},
          doi = {10.1111/j.1365-2966.2005.09578.x},
archivePrefix = {arXiv},
       eprint = {astro-ph/0504315},
 primaryClass = {astro-ph},
       adsurl = {https://ui.adsabs.harvard.edu/abs/2005MNRAS.364..601Y}
}

@article{Melia2023,
    author = {Melia, Fulvio},
    title = "{The scale of homogeneity in the Rh = ct universe}",
    journal = {Monthly Notices of the Royal Astronomical Society},
    volume = {525},
    number = {3},
    pages = {3248-3253},
    year = {2023},
    month = {08},
    issn = {0035-8711},
    doi = {10.1093/mnras/stad2496},
    url = {https://doi.org/10.1093/mnras/stad2496},
    eprint = {https://academic.oup.com/mnras/article-pdf/525/3/3248/51330927/stad2496.pdf},
}

@article{EINASTO2001355,
title = {Large scale structure},
journal = {New Astronomy Reviews},
volume = {45},
number = {4},
pages = {355-372},
year = {2001},
note = {Understanding the Universe at the close of the 20th century},
issn = {1387-6473},
doi = {https://doi.org/10.1016/S1387-6473(00)00158-5},
url = {https://www.sciencedirect.com/science/article/pii/S1387647300001585},
author = {Jaan Einasto}
}

@article{Springel_2006,
   title={The large-scale structure of the Universe},
   volume={440},
   ISSN={1476-4687},
   url={http://dx.doi.org/10.1038/nature04805},
   DOI={10.1038/nature04805},
   number={7088},
   journal={Nature},
   publisher={Springer Science and Business Media LLC},
   author={Springel, Volker and Frenk, Carlos S. and White, Simon D. M.},
   year={2006},
   month=apr, pages={1137–1144} }

@inbook{Coil_2013,
   title={The Large-Scale Structure of the Universe},
   ISBN={9789400756090},
   url={http://dx.doi.org/10.1007/978-94-007-5609-0_8},
   DOI={10.1007/978-94-007-5609-0_8},
   booktitle={Planets, Stars and Stellar Systems},
   publisher={Springer Netherlands},
   author={Coil, Alison L.},
   year={2013},
   pages={387–421} }

@misc{verweg2024averagingcosmicstructurecosmological,
      title={Averaging over cosmic structure: Cosmological backreaction and the gauge problem}, 
      author={Dave B. H. Verweg and Bernard J. T. Jones and Rien van de Weygaert},
      year={2024},
      eprint={2310.19451},
      archivePrefix={arXiv},
      primaryClass={astro-ph.CO},
      url={https://arxiv.org/abs/2310.19451}, 
}

@inbook{Brandenberger2000,
author = { Robert H.   Brandenberger },
title = {BACK REACTION OF COSMOLOGICAL PERTURBATIONS},
booktitle = {COSMO-99},
chapter = {},
pages = {198-206},
doi = {10.1142/9789812792129_0031},
URL = {https://www.worldscientific.com/doi/abs/10.1142/9789812792129_0031},
year={2000},
eprint = {https://www.worldscientific.com/doi/pdf/10.1142/9789812792129_0031}

}

@article{Lemaitre33,
    author = "Lemaitre, G.",
    title = "{The expanding universe}",
    doi = "10.1023/A:1018855621348",
    journal = "Annales Soc. Sci. Bruxelles A",
    volume = "53",
    pages = "51--85",
    year = "1933"
}

@article{Tolman34,
author = {Richard C. Tolman },
title = {Effect of Inhomogeneity on Cosmological Models},
journal = {Proceedings of the National Academy of Sciences},
volume = {20},
number = {3},
pages = {169-176},
year = {1934},
doi = {10.1073/pnas.20.3.169},
URL = {https://www.pnas.org/doi/abs/10.1073/pnas.20.3.169},
eprint = {https://www.pnas.org/doi/pdf/10.1073/pnas.20.3.169}
}

@article{Bondi47,
    author = {Bondi, H.},
    title = "{Spherically Symmetrical Models in General Relativity}",
    journal = {Monthly Notices of the Royal Astronomical Society},
    volume = {107},
    number = {5-6},
    pages = {410-425},
    year = {1947},
    month = {12},
    issn = {0035-8711},
    doi = {10.1093/mnras/107.5-6.410},
    url = {https://doi.org/10.1093/mnras/107.5-6.410},
    eprint = {https://academic.oup.com/mnras/article-pdf/107/5-6/410/8072561/mnras107-0410.pdf},
}

@article{Enqvist07,
	doi = {10.1007/s10714-007-0553-9},
  
	url = {https://doi.org/10.1007%2Fs10714-007-0553-9},
  
	year = 2007,
	month = {dec},
  
	publisher = {Springer Science and Business Media {LLC}
},
  
	volume = {40},
  
	number = {2-3},
  
	pages = {451--466},
  
	author = {Kari Enqvist},
  
	title = {Lemaitre{\textendash}Tolman{\textendash}Bondi model and accelerating expansion},
  
	journal = {General Relativity and Gravitation}
}

@article{GarciaBellido08,
	doi = {10.1088/1475-7516/2008/04/003},
	url = {https://doi.org/10.1088%2F1475-7516%2F2008%2F04%2F003},
	year = 2008,
	month = {apr},
	publisher = {{IOP} Publishing},
	volume = {2008},
	number = {04},
	pages = {003},
	author = {Juan Garcia-Bellido and Troels Haugb{\o}lle},
	title = {Confronting Lemaitre{\textendash}Tolman{\textendash}Bondi models with observational cosmology},
	journal = {Journal of Cosmology and Astroparticle Physics}
}

@article{Isidro16,
	doi = {10.1088/1475-7516/2016/05/003},
  
	url = {https://doi.org/10.1088%2F1475-7516%2F2016%2F05%2F003},
  
	year = 2016,
	month = {may},
  
	publisher = {{IOP} Publishing},
  
	volume = {2016},
  
	number = {05},
  
	pages = {003--003},
  
	author = {Eddy G. Chirinos Isidro and Cristofher Zu{\~{n}
}iga Vargas and Winfried Zimdahl},
  
	title = {Simple inhomogeneous cosmological (toy) models},
  
	journal = {Journal of Cosmology and Astroparticle Physics}
}

@misc{Celerier99,
      title={Do we really see a cosmological constant in the supernovae data ?}, 
      author={Marie-Noëlle Célérier},
      year={2007},
      eprint={astro-ph/9907206},
      archivePrefix={arXiv},
      primaryClass={astro-ph},
      url={https://arxiv.org/abs/astro-ph/9907206}, 
}

@article{Vanderveld06,
	doi = {10.1103/physrevd.74.023506},
	url = {https://doi.org/10.1103%2Fphysrevd.74.023506},
	year = 2006,
	month = {jul},
	publisher = {American Physical Society ({APS})},
	volume = {74},
	number = {2},
	author = {R. Ali Vanderveld and {\'{E}}anna {\'{E}}. Flanagan and Ira Wasserman},
	title = {Mimicking dark energy with Lema{\^{\i}}tre-Tolman-Bondi models: Weak central singularities and critical points},
	journal = {Physical Review D}
}

@ARTICLE{Bianchi89,
       author = {{Bianchi}, Luigi},
        title = "{On the three-dimensional spaces which admit a continuous group of motions}",
      journal = {Memorie di Matematica e di Fisica della Societ{\`a} Italiana delle Scienze},
         year = 1898,
        month = jan,
       volume = {11},
        pages = {267-352},
       adsurl = {https://ui.adsabs.harvard.edu/abs/1898MMFSI..11..267B}
}

@misc{valent2023axialbianchiixlemaitrehubble,
      title={Axial Bianchi IX and its Lema{\^i}tre-Hubble diagram}, 
      author={Galliano Valent and André Tilquin and Thomas Schücker},
      year={2023},
      eprint={2211.04844},
      archivePrefix={arXiv},
      primaryClass={gr-qc},
      url={https://arxiv.org/abs/2211.04844}, 
}

@article{Fleury_2015,
   title={Light propagation in a homogeneous and anisotropic universe},
   volume={91},
   ISSN={1550-2368},
   url={http://dx.doi.org/10.1103/PhysRevD.91.043511},
   DOI={10.1103/physrevd.91.043511},
   number={4},
   journal={Physical Review D},
   publisher={American Physical Society (APS)},
   author={Fleury, Pierre and Pitrou, Cyril and Uzan, Jean-Philippe},
   year={2015},
   month=feb }

@misc{schücker2024axialbianchiimeets,
      title={Axial Bianchi I meets drifting extragalactic sources}, 
      author={Thomas Schücker},
      year={2024},
      eprint={2407.02889},
      archivePrefix={arXiv},
      primaryClass={astro-ph.CO},
      url={https://arxiv.org/abs/2407.02889}, 
}

@article{KantowskiSachs,
    author = {Kantowski, R. and Sachs, R. K.},
    title = {Some Spatially Homogeneous Anisotropic Relativistic Cosmological Models},
    journal = {Journal of Mathematical Physics},
    volume = {7},
    number = {3},
    pages = {443-446},
    year = {1966},
    month = {03},
    issn = {0022-2488},
    doi = {10.1063/1.1704952},
    url = {https://doi.org/10.1063/1.1704952},
    eprint = {https://pubs.aip.org/aip/jmp/article-pdf/7/3/443/19249867/443\_1\_online.pdf},
}

@article{Bolejko_2011,
doi = {10.1088/0264-9381/28/16/164002},
url = {https://dx.doi.org/10.1088/0264-9381/28/16/164002},
year = {2011},
month = {aug},
publisher = {},
volume = {28},
number = {16},
pages = {164002},
author = {Krzysztof Bolejko and Marie-Noëlle Célérier and Andrzej Krasiński},
title = {Inhomogeneous cosmological models: exact solutions and their applications},
journal = {Classical and Quantum Gravity}
}

@article{Szekeres:1974ct,
    author = "Szekeres, P.",
    title = "{A Class of Inhomogeneous Cosmological Models}",
    reportNumber = "PRINT-74-0911 (ADELAIDE)",
    doi = "10.1007/BF01608547",
    journal = "Commun. Math. Phys.",
    volume = "41",
    pages = "55",
    year = "1975"
}

@article{Krasinski12,
  title = {Repeatable light paths in the conformally flat cosmological models},
  author = {Krasi\ifmmode \acute{n}\else \'{n}\fi{}ski, Andrzej},
  journal = {Phys. Rev. D},
  volume = {86},
  issue = {6},
  pages = {064001},
  numpages = {3},
  year = {2012},
  month = {Sep},
  publisher = {American Physical Society},
  doi = {10.1103/PhysRevD.86.064001},
  url = {https://link.aps.org/doi/10.1103/PhysRevD.86.064001}
}

@misc{bolejko2006cosmologicalapplicationsszekeresmodel,
      title={Cosmological applications of the Szekeres model}, 
      author={Krzysztof Bolejko},
      year={2006},
      eprint={astro-ph/0607130},
      archivePrefix={arXiv},
      primaryClass={astro-ph},
      url={https://arxiv.org/abs/astro-ph/0607130}, 
}

@article{Godel,
  title = {An Example of a New Type of Cosmological Solutions of Einstein's Field Equations of Gravitation},
  author = {G\"odel, Kurt},
  journal = {Rev. Mod. Phys.},
  volume = {21},
  issue = {3},
  pages = {447--450},
  numpages = {0},
  year = {1949},
  month = {Jul},
  publisher = {American Physical Society},
  doi = {10.1103/RevModPhys.21.447},
  url = {https://link.aps.org/doi/10.1103/RevModPhys.21.447}
}

@article{Buser_2013,
doi = {10.1088/1367-2630/15/1/013063},
url = {https://dx.doi.org/10.1088/1367-2630/15/1/013063},
year = {2013},
month = {jan},
publisher = {IOP Publishing},
volume = {15},
number = {1},
pages = {013063},
author = {M Buser and E Kajari and W P Schleich},
title = {Visualization of the Gödel universe},
journal = {New Journal of Physics}
}

@misc{lesgourgues2013tasilecturescosmologicalperturbations,
      title={TASI Lectures on Cosmological Perturbations}, 
      author={Julien Lesgourgues},
      year={2013},
      eprint={1302.4640},
      archivePrefix={arXiv},
      primaryClass={astro-ph.CO},
      url={https://arxiv.org/abs/1302.4640}, 
}

@article{Noonan1984TheGC,
  title={The gravitational contribution to the stress-energy tensor of a medium in general relativity},
  author={Thomas W. Noonan},
  journal={General Relativity and Gravitation},
  year={1984},
  volume={16},
  pages={1103-1118},
  url={https://api.semanticscholar.org/CorpusID:121156830}
}

@article{Zotov_1992,
doi = {10.1088/0264-9381/9/4/017},
url = {https://dx.doi.org/10.1088/0264-9381/9/4/017},
year = {1992},
month = {apr},
publisher = {},
volume = {9},
number = {4},
pages = {1023},
author = {N V Zotov and  W R Stoeger},
title = {Averaging Einstein's equations},
journal = {Classical and Quantum Gravity}
}

@article{Futamase,
  title = {Approximation Scheme for Constructing a Clumpy Universe in General Relativity},
  author = {Futamase, Toshifumi},
  journal = {Phys. Rev. Lett.},
  volume = {61},
  issue = {19},
  pages = {2175--2178},
  numpages = {0},
  year = {1988},
  month = {Nov},
  publisher = {American Physical Society},
  doi = {10.1103/PhysRevLett.61.2175},
  url = {https://link.aps.org/doi/10.1103/PhysRevLett.61.2175}
}

@article{Zalaletdinov:2008ts,
    author = "Zalaletdinov, Roustam",
    editor = "Cianfrani, Francesco and Montani, Giovanni and Ruffini, Remo",
    title = "{The Averaging Problem in Cosmology and Macroscopic Gravity}",
    eprint = "0801.3256",
    archivePrefix = "arXiv",
    primaryClass = "gr-qc",
    doi = "10.1142/S0217751X08040032",
    journal = "Int. J. Mod. Phys. A",
    volume = "23",
    pages = "1173--1181",
    year = "2008"
}

@article{Adamek_2019,
doi = {10.1088/1361-6382/aaeca5},
url = {https://dx.doi.org/10.1088/1361-6382/aaeca5},
year = {2018},
month = {dec},
publisher = {IOP Publishing},
volume = {36},
number = {1},
pages = {014001},
author = {Julian Adamek and Chris Clarkson and David Daverio and Ruth Durrer and Martin Kunz},
title = {Safely smoothing spacetime: backreaction in relativistic cosmological simulations},
journal = {Classical and Quantum Gravity}
}

@article{Rasanen_2004,
doi = {10.1088/1475-7516/2004/02/003},
url = {https://dx.doi.org/10.1088/1475-7516/2004/02/003},
year = {2004},
month = {feb},
publisher = {},
volume = {2004},
number = {02},
pages = {003},
author = {Syksy Räsänen},
title = {Dark energy from back-reaction},
journal = {Journal of Cosmology and Astroparticle Physics}
}

@article{Korzyński_2015,
doi = {10.1088/0264-9381/32/21/215013},
url = {https://dx.doi.org/10.1088/0264-9381/32/21/215013},
year = {2015},
month = {oct},
publisher = {IOP Publishing},
volume = {32},
number = {21},
pages = {215013},
author = {Mikołaj Korzyński},
title = {Nonlinear effects of general relativity from multiscale structure},
journal = {Classical and Quantum Gravity}
}

@article{Filippou_2021,
   title={Large-scale peculiar velocity fields: Newtonian vs relativistic treatment},
   volume={366},
   ISSN={1572-946X},
   url={http://dx.doi.org/10.1007/s10509-020-03912-4},
   DOI={10.1007/s10509-020-03912-4},
   number={1},
   journal={Astrophysics and Space Science},
   publisher={Springer Science and Business Media LLC},
   author={Filippou, Konstantinos and Tsagas, Christos G.},
   year={2021},
   month=jan }

@ARTICLE{Tzaprasi2020,
       author = {{Tsaprazi}, Eleni and {Tsagas}, Christos G.},
        title = "{Relativistic approach to the kinematics of large-scale peculiar motions}",
      journal = {European Physical Journal C},
         year = 2020,
        month = aug,
       volume = {80},
       number = {8},
          eid = {757},
        pages = {757},
          doi = {10.1140/epjc/s10052-020-8312-0},
archivePrefix = {arXiv},
       eprint = {1906.05164},
 primaryClass = {astro-ph.CO},
       adsurl = {https://ui.adsabs.harvard.edu/abs/2020EPJC...80..757T}
}

@article{Biern_2017,
   title={Gauge-invariance and infrared divergences in the luminosity distance},
   volume={2017},
   ISSN={1475-7516},
   url={http://dx.doi.org/10.1088/1475-7516/2017/04/045},
   DOI={10.1088/1475-7516/2017/04/045},
   number={04},
   journal={Journal of Cosmology and Astroparticle Physics},
   publisher={IOP Publishing},
   author={Biern, Sang Gyu and Yoo, Jaiyul},
   year={2017},
   month=apr, pages={045–045} }

@article{Sasaki:1987ad,
    author = "Sasaki, Misao",
    title = "{The Magnitude - Redshift relation in a perturbed Friedmann universe}",
    reportNumber = "RRK-87-1",
    journal = "Mon. Not. Roy. Astron. Soc.",
    volume = "228",
    pages = "653--669",
    year = "1987"
}

@book{Durrer:2008eom,
    author = "Durrer, Ruth",
    title = "{The Cosmic Microwave Background}",
    doi = "10.1017/CBO9780511817205",
    isbn = "978-0-511-81720-5",
    publisher = "Cambridge University Press",
    address = "Cambridge",
    year = "2008"
}

@misc{sorrenti2023dipole,
      title={The Dipole of the Pantheon+SH0ES Data}, 
      author={Francesco Sorrenti and Ruth Durrer and Martin Kunz},
      year={2023},
      eprint={2212.10328},
      archivePrefix={arXiv},
      primaryClass={astro-ph.CO}
}

@article{Visser_2005,
	doi = {10.1007/s10714-005-0134-8},
  
	url = {https://doi.org/10.1007%2Fs10714-005-0134-8},
  
	year = 2005,
	month = {sep},
  
	publisher = {Springer Science and Business Media {LLC}
},
  
	volume = {37},
  
	number = {9},
  
	pages = {1541--1548},
  
	author = {Matt Visser},
  
	title = {Cosmography: Cosmology without the Einstein equations},
  
	journal = {General Relativity and Gravitation}
}

@article{Visser_2004,
	doi = {10.1088/0264-9381/21/11/006},
  
	url = {https://doi.org/10.1088%2F0264-9381%2F21%2F11%2F006},
  
	year = 2004,
	month = {apr},
  
	publisher = {{IOP} Publishing},
  
	volume = {21},
  
	number = {11},
  
	pages = {2603--2615},
  
	author = {Matt Visser},
  
	title = {Jerk, snap and the cosmological equation of state},
  
	journal = {Classical and Quantum Gravity}
}

@article{Deng18a,
	doi = {10.1103/physrevd.97.123515},
  
	url = {https://doi.org/10.1103%2Fphysrevd.97.123515},
  
	year = 2018,
	month = {jun},
  
	publisher = {American Physical Society ({APS})},
  
	volume = {97},
  
	number = {12},
  
	author = {Hua-Kai Deng and Hao Wei},
  
	title = {Testing the cosmic anisotropy with supernovae data: Hemisphere comparison and dipole fitting},
  
	journal = {Physical Review D}
}

@misc{Mohayaee20,
  doi = {10.48550/ARXIV.2003.10420},
  
  url = {https://arxiv.org/abs/2003.10420},
  
  author = {Mohayaee, Roya and Rameez, Mohamed and Sarkar, Subir},
  
  title = {The impact of peculiar velocities on supernova cosmology},
  
  publisher = {arXiv},
  
  year = {2020},
  
  copyright = {Creative Commons Attribution 4.0 International}
}

@article{Heinesen:2020bej,
    author = "Heinesen, Asta",
    title = "{Multipole decomposition of the general luminosity distance 'Hubble law' -- a new framework for observational cosmology}",
    eprint = "2010.06534",
    archivePrefix = "arXiv",
    primaryClass = "astro-ph.CO",
    doi = "10.1088/1475-7516/2021/05/008",
    journal = "JCAP",
    volume = "05",
    pages = "008",
    year = "2021"
}

@ARTICLE{Tsagas2011,
       author = {{Tsagas}, Christos G.},
        title = "{Peculiar motions, accelerated expansion, and the cosmological axis}",
      journal = {\prd},
         year = 2011,
        month = sep,
       volume = {84},
       number = {6},
          eid = {063503},
        pages = {063503},
          doi = {10.1103/PhysRevD.84.063503},
archivePrefix = {arXiv},
       eprint = {[arXiv:1107.4045]},
 primaryClass = {astro-ph.CO},
       adsurl = {https://ui.adsabs.harvard.edu/abs/2011PhRvD..84f3503T}
}

@ARTICLE{Tsagas2015,
       author = {{Tsagas}, Christos G. and {Kadiltzoglou}, Miltiadis I.},
        title = "{Deceleration parameter in tilted Friedmann universes}",
      journal = {\prd},
         year = 2015,
        month = aug,
       volume = {92},
       number = {4},
          eid = {043515},
        pages = {043515},
          doi = {10.1103/PhysRevD.92.043515},
archivePrefix = {arXiv},
       eprint = {[arXiv:1507.04266]},
 primaryClass = {gr-qc},
       adsurl = {https://ui.adsabs.harvard.edu/abs/2015PhRvD..92d3515T}
}

@ARTICLE{Tsagas2008,
       author = {{Tsagas}, Christos G. and {Challinor}, Anthony and {Maartens}, Roy},
        title = "{Relativistic cosmology and large-scale structure}",
      journal = {Phys. Rep.},
         year = 2008,
        month = aug,
       volume = {465},
       number = {2-3},
        pages = {61-147},
          doi = {10.1016/j.physrep.2008.03.003},
archivePrefix = {arXiv},
       eprint = {[arXiv:0705.4397]},
 primaryClass = {astro-ph},
       adsurl = {https://ui.adsabs.harvard.edu/abs/2008PhR...465...61T}
}

@BOOK{Ellis2012,
       author = {{Ellis}, George F.~R. and {Maartens}, Roy and {MacCallum}, Malcolm A.~H.},
        title = "{Relativistic Cosmology}",
    publisher = "{Cambridge University Press, Cambridge}",
         year = 2012,
       adsurl = {https://ui.adsabs.harvard.edu/abs/2012reco.book.....E}
}

@ARTICLE{Maartens1998,
       author = {{Maartens}, Roy},
        title = "{Covariant velocity and density perturbations in quasi-Newtonian cosmologies}",
      journal = {\prd},
         year = 1998,
        month = dec,
       volume = {58},
       number = {12},
          eid = {124006},
        pages = {124006},
          doi = {10.1103/PhysRevD.58.124006},
archivePrefix = {arXiv},
       eprint = {[arXiv:astro-ph/9808235]},
 primaryClass = {astro-ph},
       adsurl = {https://ui.adsabs.harvard.edu/abs/1998PhRvD..58l4006M}
}

@ARTICLE{Tsagas2021,
       author = {{Tsagas}, Christos G.},
        title = "{The peculiar Jeans length}",
      journal = {Eur. Phys. J. C},
         year = 2021,
        month = aug,
       volume = {81},
       number = {8},
          eid = {753},
        pages = {753},
          doi = {10.1140/epjc/s10052-021-09515-9},
archivePrefix = {arXiv},
       eprint = {[arXiv:2103.15884]},
 primaryClass = {gr-qc},
       adsurl = {https://ui.adsabs.harvard.edu/abs/2021EPJC...81..753T}
}

@ARTICLE{Tsagas2022,
       author = {{Tsagas}, Christos G.},
        title = "{The deceleration parameter in `tilted' universes: generalising the Friedmann background}",
      journal = {Eur. Phys. J. C},
         year = 2022,
        month = jun,
       volume = {82},
       number = {6},
          eid = {521},
        pages = {521},
          doi = {10.1140/epjc/s10052-022-10452-4},
archivePrefix = {arXiv},
       eprint = {[arXiv:2112.04313]},
 primaryClass = {gr-qc},
       adsurl = {https://ui.adsabs.harvard.edu/abs/2022EPJC...82..521T}
}

@article{Labini98,
	doi = {10.1016/s0370-1573(97)00044-6},
  
	url = {https://doi.org/10.1016%2Fs0370-1573%2897%2900044-6},
  
	year = 1998,
	month = {jan},
  
	publisher = {Elsevier {BV}
},
  
	volume = {293},
  
	number = {2-4},
  
	pages = {61--226},
  
	author = {F.Sylos Labini and M. Montuori and L. Pietronero},
  
	title = {Scale-invariance of galaxy clustering},
  
	journal = {Physics Reports}
}

@article{Labini11,
	doi = {10.1088/0264-9381/28/16/164003},
  
	url = {https://doi.org/10.1088%2F0264-9381%2F28%2F16%2F164003},
  
	year = 2011,
	month = {aug},
  
	publisher = {{IOP} Publishing},
  
	volume = {28},
  
	number = {16},
  
	pages = {164003},
  
	author = {Francesco Sylos Labini},
  
	title = {Inhomogeneities in the universe},
  
	journal = {Classical and Quantum Gravity}
}

@article{Feindt13,
	author = {Feindt, U. et al.},
	title = {Measuring cosmic bulk flows with Type Ia supernovae  from the Nearby Supernova Factory},
	DOI= "10.1051/0004-6361/201321880",
	url= "https://doi.org/10.1051/0004-6361/201321880",
	journal = {A\&A},
	year = 2013,
	volume = 560,
	pages = "A90",
	month = "",
}

@article{Hudson99,
	doi = {10.1086/311883},
  
	url = {https://doi.org/10.1086%2F311883},
  
	year = 1999,
	month = {feb},
  
	publisher = {American Astronomical Society},
  
	volume = {512},
  
	number = {2},
  
	pages = {L79--L82},
  
	author = {Michael J. Hudson and Russell J. Smith and John R. Lucey and David J. Schlegel and Roger L. Davies},
  
	title = {A Large-scale Bulk Flow of Galaxy Clusters},
  
	journal = {The Astrophysical Journal}
}

@INPROCEEDINGS{Magoulas16,
       author = {{Magoulas}, Christina and {Springob}, Christopher and {Colless}, Matthew and {Mould}, Jeremy and {Lucey}, John and {Erdo{\u{g}}du}, Pirin and {Jones}, D. Heath},
        title = "{Measuring the cosmic bulk flow with 6dFGSv}",
    booktitle = {The Zeldovich Universe: Genesis and Growth of the Cosmic Web},
         year = 2016,
       editor = {{van de Weygaert}, R. and {Shandarin}, S. and {Saar}, E. and {Einasto}, J.},
       volume = {308},
        month = oct,
        pages = {336-339},
          doi = {10.1017/S1743921316010115},
       adsurl = {https://ui.adsabs.harvard.edu/abs/2016IAUS..308..336M}
}

@misc{Celerier06,
      title={Accelerated-like expansion: inhomogeneities versus dark energy}, 
      author={Marie-Noelle Celerier},
      year={2006},
      eprint={astro-ph/0609352},
      archivePrefix={arXiv},
      primaryClass={astro-ph},
      url={https://arxiv.org/abs/astro-ph/0609352}, 
}

@article{Cosmai19,
	doi = {10.1088/1361-6382/aae8f7},
  
	url = {https://doi.org/10.1088%2F1361-6382%2Faae8f7},
  
	year = 2019,
	month = {jan},
  
	publisher = {{IOP} Publishing},
  
	volume = {36},
  
	number = {4},
  
	pages = {045007},
  
	author = {L Cosmai and G Fanizza and F Sylos Labini and L Pietronero and L Tedesco},
  
	title = {Fractal universe and cosmic acceleration in a Lema{\^{\i}
}tre{\textendash}Tolman{\textendash}Bondi scenario},
  
	journal = {Classical and Quantum Gravity}
}

@article{Asvesta22,
    author = "Asvesta, Kerkyra and Kazantzidis, Lavrentios and Perivolaropoulos, Leandros and Tsagas, Christos G.",
    title = "{Observational constraints on the deceleration parameter in a tilted universe}",
    eprint = "[arXiv:2202.00962]",
    archivePrefix = "arXiv",
    primaryClass = "astro-ph.CO",
    doi = "10.1093/mnras/stac922",
    journal = "Mon. Not. R. Astron. Soc.",
    volume = "513",
    number = "2",
    pages = "2394--2406",
    year = "2022"
}

@article{Dickau2009,
title = {Fractal cosmology},
journal = {Chaos, Solitons \& Fractals},
volume = {41},
number = {4},
pages = {2103-2105},
year = {2009},
issn = {0960-0779},
doi = {https://doi.org/10.1016/j.chaos.2008.07.056},
url = {https://www.sciencedirect.com/science/article/pii/S0960077908003846},
author = {Jonathan J. Dickau}
}

@article{PIETRONERO1987257,
title = {The fractal structure of the universe: Correlations of galaxies and clusters and the average mass density},
journal = {Physica A: Statistical Mechanics and its Applications},
volume = {144},
number = {2},
pages = {257-284},
year = {1987},
issn = {0378-4371},
doi = {https://doi.org/10.1016/0378-4371(87)90191-9},
url = {https://www.sciencedirect.com/science/article/pii/0378437187901919},
author = {L. Pietronero}
}

@misc{Cosmai2023,
  doi = {10.48550/ARXIV.2302.04679},
  
  url = {https://arxiv.org/abs/2302.04679},
  
  author = {Cosmai, Leonardo and Fanizza, Giuseppe and Labini, Francesco Sylos and Pietronero, Luciano and Tedesco, Luigi},
  
  title = {Comment on "A fractal LTB model cannot explain Dark Energy''},
  
  publisher = {arXiv},
  
  year = {2023},
  
  copyright = {Creative Commons Attribution 4.0 International}
}

@inproceedings{Ruffini17,
    author = "Ruffini, Remo and Stahl, Cl\'ement",
    title = "{Cosmological fractal matter with an upper cutoff}",
    booktitle = "{14th Italian-Korean Symposium on Relativistic Astrophysics}",
    doi = "10.1142/9789813226609_0595",
    year = "2017"
}

@article{Alexander09,
	doi = {10.1088/1475-7516/2009/09/025},
	url = {https://doi.org/10.1088%2F1475-7516%2F2009%2F09%2F025},
	year = 2009,
	month = {sep},
	publisher = {{IOP} Publishing},
	volume = {2009},
	number = {09},
	pages = {025--025},
	author = {Stephon Alexander and Tirthabir Biswas and Alessio Notari and Deepak Vaid},
	title = {Local void vs dark energy: confrontation with {WMAP} and type Ia supernovae},
	journal = {Journal of Cosmology and Astroparticle Physics}
}

@article{Calcagni_2017,
	doi = {10.1007/jhep03(2017)138},
  
	url = {https://doi.org/10.1007%2Fjhep03%282017%29138},
  
	year = 2017,
	month = {mar},
  
	publisher = {Springer Science and Business Media {LLC}
},
  
	volume = {2017},
  
	number = {3},
  
	author = {Gianluca Calcagni},
  
	title = {Multifractional theories: an unconventional review},
  
	journal = {Journal of High Energy Physics}
}

@article{Scolnic18,
doi = {10.3847/1538-4357/aab9bb},
url = {https://dx.doi.org/10.3847/1538-4357/aab9bb},
year = {2018},
month = {may},
publisher = {The American Astronomical Society},
volume = {859},
number = {2},
pages = {101},
author = {Scolnic, D. M. and Jones, D. O. and Rest, A. and Pan, Y. C. and Chornock, R. and Foley, R. J. and Huber, M. E. and Kessler, R. and Narayan, G. and Riess, A. G. and Rodney, S. and Berger, E. and Brout, D. J. and Challis, P. J. and Drout, M. and Finkbeiner, D. and Lunnan, R. and Kirshner, R. P. and Sanders, N. E. and Schlafly, E. and Smartt, S. and Stubbs, C. W. and Tonry, J. and Wood-Vasey, W. M. and Foley, M. and Hand, J. and Johnson, E. and Burgett, W. S. and Chambers, K. C. and Draper, P. W. and Hodapp, K. W. and Kaiser, N. and Kudritzki, R. P. and Magnier, E. A. and Metcalfe, N. and Bresolin, F. and Gall, E. and Kotak, R. and McCrum, M. and Smith, K. W.},
title = {The Complete Light-curve Sample of Spectroscopically Confirmed SNe Ia from Pan-STARRS1 and Cosmological Constraints from the Combined Pantheon Sample},
journal = {The Astrophysical Journal}
}

@article{Dainotti_2021,
	doi = {10.3847/1538-4357/abeb73},
  
	url = {https://doi.org/10.3847%2F1538-4357%2Fabeb73},
  
	year = 2021,
	month = {may},
  
	publisher = {American Astronomical Society},
  
	volume = {912},
  
	number = {2},
  
	pages = {150},
  
	author = {M. G. Dainotti and B. De Simone and T. Schiavone and G. Montani and E. Rinaldi and G. Lambiase},
  
	title = {On the Hubble Constant Tension in the {SNe} Ia Pantheon Sample},
  
	journal = {The Astrophysical Journal}
}

@article{Alnes_2006,
	doi = {10.1103/physrevd.73.083519},
  
	url = {https://doi.org/10.1103%2Fphysrevd.73.083519},
  
	year = 2006,
	month = {apr},
  
	publisher = {American Physical Society ({APS})},
  
	volume = {73},
  
	number = {8},
  
	author = {H{\aa}vard Alnes and Morad Amarzguioui and {\O}yvind Gr{\o}n},
  
	title = {Inhomogeneous alternative to dark energy?},
  
	journal = {Physical Review D}
}

@ARTICLE{Foreman2013,
       author = {{Foreman-Mackey}, Daniel and {Hogg}, David W. and {Lang}, Dustin and {Goodman}, Jonathan},
        title = "{emcee: The MCMC Hammer}",
      journal = {\pasp},
         year = 2013,
        month = mar,
       volume = {125},
       number = {925},
        pages = {306},
          doi = {10.1086/670067},
archivePrefix = {arXiv},
       eprint = {1202.3665},
 primaryClass = {astro-ph.IM},
       adsurl = {https://ui.adsabs.harvard.edu/abs/2013PASP..125..306F}
      }

@ARTICLE{Goodman2010,
       author = {{Goodman}, Jonathan and {Weare}, Jonathan},
        title = "{Ensemble samplers with affine invariance}",
      journal = {Communications in Applied Mathematics and Computational Science},
         year = 2010,
        month = jan,
       volume = {5},
       number = {1},
        pages = {65-80},
          doi = {10.2140/camcos.2010.5.65},
       adsurl = {https://ui.adsabs.harvard.edu/abs/2010CAMCS...5...65G}
}

@ARTICLE{Yadav2010,
       author = {{Yadav}, Jaswant K. and {Bagla}, J.~S. and {Khandai}, Nishikanta},
        title = "{Fractal dimension as a measure of the scale of homogeneity}",
      journal = {\mnras},
         year = 2010,
        month = jul,
       volume = {405},
       number = {3},
        pages = {2009-2015},
          doi = {10.1111/j.1365-2966.2010.16612.x},
archivePrefix = {arXiv},
       eprint = {1001.0617},
 primaryClass = {astro-ph.CO},
       adsurl = {https://ui.adsabs.harvard.edu/abs/2010MNRAS.405.2009Y}
}

@ARTICLE{Hogg2005,
       author = {{Hogg}, David W. and {Eisenstein}, Daniel J. and {Blanton}, Michael R. and {Bahcall}, Neta A. and {Brinkmann}, J. and {Gunn}, James E. and {Schneider}, Donald P.},
        title = "{Cosmic Homogeneity Demonstrated with Luminous Red Galaxies}",
      journal = {\apj},
         year = 2005,
        month = may,
       volume = {624},
       number = {1},
        pages = {54-58},
          doi = {10.1086/429084},
archivePrefix = {arXiv},
       eprint = {astro-ph/0411197},
 primaryClass = {astro-ph},
       adsurl = {https://ui.adsabs.harvard.edu/abs/2005ApJ...624...54H}
}

@article{Camarena20,
	doi = {10.1103/physrevresearch.2.013028},
  
	url = {https://doi.org/10.1103%2Fphysrevresearch.2.013028},
  
	year = 2020,
	month = {jan},
  
	publisher = {American Physical Society ({APS})},
  
	volume = {2},
  
	number = {1},
  
	author = {David Camarena and Valerio Marra},
  
	title = {Local determination of the Hubble constant and the deceleration parameter},
  
	journal = {Physical Review Research}
}

@article{Valkenburg13,
	doi = {10.1093/mnrasl/slt140},
  
	url = {https://doi.org/10.1093%2Fmnrasl%2Fslt140},
  
	year = 2013,
	month = {nov},
  
	publisher = {Oxford University Press ({OUP})},
  
	volume = {438},
  
	number = {1},
  
	pages = {L6--L10},
  
	author = {W. Valkenburg and V. Marra and C. Clarkson},
  
	title = {Testing the Copernican principle by constraining spatial homogeneity},
  
	journal = {Monthly Notices of the Royal Astronomical Society: Letters}
}

@article{Scolnic_2022,
	doi = {10.3847/1538-4357/ac8b7a},
  
	url = {https://doi.org/10.3847%2F1538-4357%2Fac8b7a},
  
	year = 2022,
	month = {oct},
  
	publisher = {American Astronomical Society},
  
	volume = {938},
  
	number = {2},
  
	pages = {113},
  
	author = {Dan Scolnic and Dillon Brout and Anthony Carr and Adam G. Riess and Tamara M. Davis and Arianna Dwomoh and David O. Jones and Noor Ali and Pranav Charvu and Rebecca Chen and Erik R. Peterson and Brodie Popovic and Benjamin M. Rose and Charlotte M. Wood and Peter J. Brown and Ken Chambers and David A. Coulter and Kyle G. Dettman and Georgios Dimitriadis and Alexei V. Filippenko and Ryan J. Foley and Saurabh W. Jha and Charles D. Kilpatrick and Robert P. Kirshner and Yen-Chen Pan and Armin Rest and Cesar Rojas-Bravo and Matthew R. Siebert and Benjamin E. Stahl and WeiKang Zheng},
  
	title = {The Pantheon$\mathplus$ Analysis: The Full Data Set and Light-curve Release},
  
	journal = {The Astrophysical Journal}
}

@article{Colin19,
	doi = {10.1051/0004-6361/201936373},
  
	url = {https://doi.org/10.1051%2F0004-6361%2F201936373},
  
	year = 2019,
	month = {nov},
  
	publisher = {{EDP} Sciences},
  
	volume = {631},
  
	pages = {L13},
  
	author = {Jacques Colin and Roya Mohayaee and Mohamed Rameez and Subir Sarkar},
  
	title = {Evidence for anisotropy of cosmic acceleration},
  
	journal = {Astronomy {\&} Astrophysics}
}

@article{Rubin_2020,
	doi = {10.3847/1538-4357/ab7a16},
  
	url = {https://doi.org/10.3847%2F1538-4357%2Fab7a16},
  
	year = 2020,
	month = {may},
  
	publisher = {American Astronomical Society},
  
	volume = {894},
  
	number = {1},
  
	pages = {68},
  
	author = {D. Rubin and J. Heitlauf},
  
	title = {Is the Expansion of the Universe Accelerating? All Signs Still Point to Yes: A Local Dipole Anisotropy Cannot Explain Dark Energy},
  
	journal = {The Astrophysical Journal}
}

@misc{ColinResponse,
  doi = {10.48550/ARXIV.1912.04257},
  
  url = {https://arxiv.org/abs/1912.04257},
  
  author = {Colin, Jacques and Mohayaee, Roya and Rameez, Mohamed and Sarkar, Subir},
  
  title = {A response to Rubin {\&}; Heitlauf: "Is the expansion of the universe accelerating? All signs still point to yes"},
  
  publisher = {arXiv},
  
  year = {2019},
  
  copyright = {arXiv.org perpetual, non-exclusive license}
}

@article{Chang19,
	doi = {10.1088/1674-1137/43/12/125102},
	url = {https://doi.org/10.1088/1674-1137/43/12/125102},
	year = 2019,
	month = {dec},
	publisher = {{IOP} Publishing},
	volume = {43},
	number = {12},
	pages = {125102},
	author = {Zhe Chang and Dong Zhao and Yong Zhou},
	title = {Constraining the anisotropy of the Universe with the Pantheon supernovae sample {\ast}},
	journal = {Chinese Physics C}
}

@article{PASTEN2023101224,
title = {Testing ΛCDM cosmology in a binned universe: Anomalies in the deceleration parameter},
journal = {Physics of the Dark Universe},
volume = {40},
pages = {101224},
year = {2023},
issn = {2212-6864},
doi = {https://doi.org/10.1016/j.dark.2023.101224},
url = {https://www.sciencedirect.com/science/article/pii/S2212686423000584},
author = {Erick Pastén and Víctor H. Cárdenas}
}

@article{Perivolaropoulos:2023iqj,
    author = "Perivolaropoulos, Leandros and Skara, Foteini",
    title = "{On the homogeneity of SnIa absolute magnitude in the Pantheon+~sample}",
    eprint = "2301.01024",
    archivePrefix = "arXiv",
    primaryClass = "astro-ph.CO",
    doi = "10.1093/mnras/stad451",
    journal = "Mon. Not. Roy. Astron. Soc.",
    volume = "520",
    number = "4",
    pages = "5110--5125",
    year = "2023"
}

@article{OCOLGAIN2023101216,
title = {High redshift ΛCDM cosmology: To bin or not to bin?},
journal = {Physics of the Dark Universe},
volume = {40},
pages = {101216},
year = {2023},
issn = {2212-6864},
doi = {https://doi.org/10.1016/j.dark.2023.101216},
url = {https://www.sciencedirect.com/science/article/pii/S221268642300050X},
author = {Eoin {Ó Colgáin} and M.M. Sheikh-Jabbari and Rance Solomon}
}

@article{Carrick_2015,
	doi = {10.1093/mnras/stv547},
  
	url = {https://doi.org/10.1093%2Fmnras%2Fstv547},
  
	year = 2015,
	month = {apr},
  
	publisher = {Oxford University Press ({OUP})},
  
	volume = {450},
  
	number = {1},
  
	pages = {317--332},
  
	author = {Jonathan Carrick and Stephen J. Turnbull and Guilhem Lavaux and Michael J. Hudson},
  
	title = {Cosmological parameters from the comparison of peculiar velocities with predictions from the 2M$\mathplus$$\mathplus$ density field},
  
	journal = {Monthly Notices of the Royal Astronomical Society}
}

@INPROCEEDINGS{Ellis1971,
       author = {{Ellis}, G. F. R.},
        title = "{Relativistic Cosmology}",
    booktitle = {General Relativity and Cosmology},
       editor = "{R.~K. Sachs}",
    publisher = "{Academic Press, New York}", 
         year = 1971,
        month = jan,
        pages = {104-182},
       adsurl = {https://ui.adsabs.harvard.edu/abs/1971grc..conf..104E}
}

@ARTICLE{Ellis1990,
       author = {{Ellis}, G. F. R.},
        title = "{The evolution of inhomogeneities in expanding Newtonian cosmologies}",
      journal = {Mon. Not. R. Astron. Soc.},
         year = 1990,
        month = apr,
       volume = {243},
        pages = {509-516},
       adsurl = {https://ui.adsabs.harvard.edu/abs/1990MNRAS.243..509E}
}

@article{Said_20,
    author = {Said, Khaled and Colless, Matthew and Magoulas, Christina and Lucey, John R and Hudson, Michael J},
    title = "{Joint analysis of 6dFGS and SDSS peculiar velocities for the growth rate of cosmic structure and tests of gravity}",
    journal = {Monthly Notices of the Royal Astronomical Society},
    volume = {497},
    number = {1},
    pages = {1275-1293},
    year = {2020},
    month = {07},
    issn = {0035-8711},
    doi = {10.1093/mnras/staa2032},
    url = {https://doi.org/10.1093/mnras/staa2032},
    eprint = {https://academic.oup.com/mnras/article-pdf/497/1/1275/33549975/staa2032.pdf},
}

@article{Whitbourn2014,
    author = {Whitbourn, J. R. and Shanks, T.},
    title = "{The local hole revealed by galaxy counts and redshifts}",
    journal = {Monthly Notices of the Royal Astronomical Society},
    volume = {437},
    number = {3},
    pages = {2146-2162},
    year = {2013},
    month = {11},
    issn = {0035-8711},
    doi = {10.1093/mnras/stt2024},
    url = {https://doi.org/10.1093/mnras/stt2024},
    eprint = {https://academic.oup.com/mnras/article-pdf/437/3/2146/18455902/stt2024.pdf},
}

@article{Bohringer2014,
    author = {B\"ohringer, Hans and Chon, Gayoung and Bristow, Martyn and Collins, Chris A.},
    title = "{The extended ROSAT-ESO Flux-Limited X-ray Galaxy Cluster Survey (REFLEX II)}: {V. Exploring a local underdensity in the Southern Sky}",
    eprint = "1410.2172",
    archivePrefix = "arXiv",
    primaryClass = "astro-ph.CO",
    doi = "10.1051/0004-6361/201424817",
    journal = "Astron. Astrophys.",
    volume = "574",
    pages = "A26",
    year = "2015"
}

@article{Krishnan_2020,
	doi = {10.1103/physrevd.102.103525},
  
	url = {https://doi.org/10.1103%2Fphysrevd.102.103525},
  
	year = 2020,
	month = {nov},
  
	publisher = {American Physical Society ({APS})},
  
	volume = {102},
  
	number = {10},
  
	author = {C. Krishnan and E.{\hspace{0.167em}
}{\'{O}} Colg{\'{a}}in and  Ruchika and A.{\hspace{0.167em}}A. Sen and M.{\hspace{0.167em}}M. Sheikh-Jabbari and T. Yang},
  
	title = {Is there an early Universe solution to Hubble tension?},
  
	journal = {Physical Review D}
}

@article{Sorrenti:2024ztg,
    author = "Sorrenti, Francesco and Durrer, Ruth and Kunz, Martin",
    title = "{The low multipoles in the Pantheon+SH0ES data}",
    eprint = "2403.17741",
    archivePrefix = "arXiv",
    primaryClass = "astro-ph.CO",
    month = "3",
    year = "2024",
    journal = {arXiv e-prints}
}

@article{Labb__2023,
   title={A population of red candidate massive galaxies ~600 Myr after the Big Bang},
   volume={616},
   ISSN={1476-4687},
   url={http://dx.doi.org/10.1038/s41586-023-05786-2},
   DOI={10.1038/s41586-023-05786-2},
   number={7956},
   journal={Nature},
   publisher={Springer Science and Business Media LLC},
   author={Labbé, Ivo and van Dokkum, Pieter and Nelson, Erica and Bezanson, Rachel and Suess, Katherine A. and Leja, Joel and Brammer, Gabriel and Whitaker, Katherine and Mathews, Elijah and Stefanon, Mauro and Wang, Bingjie},
   year={2023},
   month=feb, pages={266–269} }

@article{Boylan_Kolchin_2023,
   title={Stress testing ΛCDM with high-redshift galaxy candidates},
   volume={7},
   ISSN={2397-3366},
   url={http://dx.doi.org/10.1038/s41550-023-01937-7},
   DOI={10.1038/s41550-023-01937-7},
   number={6},
   journal={Nature Astronomy},
   publisher={Springer Science and Business Media LLC},
   author={Boylan-Kolchin, Michael},
   year={2023},
   month=apr, pages={731–735} }

@ARTICLE{Miliou2024,
       author = {{Miliou}, Eleftheria P. and {Tsagas}, Christos G.},
        title = "{Peculiar velocities in Friedmann universes with nonzero spatial curvature}",
      journal = {\prd},
         year = 2024,
        month = sep,
       volume = {110},
       number = {6},
          eid = {063540},
        pages = {063540},
          doi = {10.1103/PhysRevD.110.063540},
archivePrefix = {arXiv},
       eprint = {2404.19046},
 primaryClass = {gr-qc},
       adsurl = {https://ui.adsabs.harvard.edu/abs/2024PhRvD.110f3540M}
}

@BOOK{Peebles1980,
       author = {{Peebles}, P.~J.~E.},
        title = "{The large-scale structure of the universe}",
         year = 1980,
       adsurl = {https://ui.adsabs.harvard.edu/abs/1980lssu.book.....P}
}

@ARTICLE{Peebles1976,
       author = {{Peebles}, P.~J.~E.},
        title = "{The Peculiar Velocity Field in the Local Supercluster}",
      journal = {\apj},
         year = 1976,
        month = apr,
       volume = {205},
        pages = {318-328},
          doi = {10.1086/154280},
       adsurl = {https://ui.adsabs.harvard.edu/abs/1976ApJ...205..318P}
}

@ARTICLE{Nusser1991,
       author = {{Nusser}, Adi and {Dekel}, Avishai and {Bertschinger}, Edmund and {Blumenthal}, George R.},
        title = "{Cosmological Velocity-Density Relation in the Quasi-linear Regime}",
      journal = {\apj},
         year = 1991,
        month = sep,
       volume = {379},
        pages = {6},
          doi = {10.1086/170480},
       adsurl = {https://ui.adsabs.harvard.edu/abs/1991ApJ...379....6N}
}

@article{Clarkson_2011,
	doi = {10.1088/0034-4885/74/11/112901},
  
	url = {https://doi.org/10.1088%2F0034-4885%2F74%2F11%2F112901},
  
	year = 2011,
	month = {oct},
  
	publisher = {{IOP} Publishing},
  
	volume = {74},
  
	number = {11},
  
	pages = {112901},
  
	author = {Chris Clarkson and George Ellis and Julien Larena and Obinna Umeh},
  
	title = {Does the growth of structure affect our dynamical models of the Universe? The averaging, backreaction, and fitting problems in cosmology},
  
	journal = {Reports on Progress in Physics}
}

@misc{Pasten_Fractal,
  doi = {10.48550/ARXIV.2212.04529},
  
  url = {https://arxiv.org/abs/2212.04529},
  
  author = {Pastén, Erick and Cárdenas, Víctor H.},
  
  title = {A fractal LTB model cannot explain Dark Energy},
  
  publisher = {arXiv},
  
  year = {2022},
  
  copyright = {Creative Commons Zero v1.0 Universal}
}

@article{Zehavi:1998gz,
    author = "Zehavi, Idit and Riess, Adam G. and Kirshner, Robert P. and Dekel, Avishai",
    title = "{A Local hubble bubble from SNe Ia?}",
    eprint = "astro-ph/9802252",
    archivePrefix = "arXiv",
    doi = "10.1086/306015",
    journal = "Astrophys. J.",
    volume = "503",
    pages = "483",
    year = "1998"
}

@article{Davis_2019,
	doi = {10.1093/mnras/stz2652},
  
	url = {https://doi.org/10.1093%2Fmnras%2Fstz2652},
  
	year = 2019,
	month = {sep},
  
	publisher = {Oxford University Press ({OUP})},
  
	volume = {490},
  
	number = {2},
  
	pages = {2948--2957},
  
	author = {Tamara M Davis and Samuel R Hinton and Cullan Howlett and Josh Calcino},
  
	title = {Can redshift errors bias measurements of the Hubble Constant?},
  
	journal = {Monthly Notices of the Royal Astronomical Society}
}

@article{Conley_2010,
	doi = {10.1088/0067-0049/192/1/1},
  
	url = {https://doi.org/10.1088%2F0067-0049%2F192%2F1%2F1},
  
	year = 2010,
	month = {dec},
  
	publisher = {American Astronomical Society},
  
	volume = {192},
  
	number = {1},
  
	pages = {1},
  
	author = {A. Conley and J. Guy and M. Sullivan and N. Regnault and P. Astier and C. Balland and S. Basa and R. G. Carlberg and D. Fouchez and D. Hardin and I. M. Hook and D. A. Howell and R. Pain and N. Palanque-Delabrouille and K. M. Perrett and C. J. Pritchet and J. Rich and V. Ruhlmann-Kleider and D. Balam and S. Baumont and R. S. Ellis and S. Fabbro and H. K. Fakhouri and N. Fourmanoit and S. Gonz{\'{a}
}lez-Gait{\'{a}}n and M. L. Graham and M. J. Hudson and E. Hsiao and T. Kronborg and C. Lidman and A. M. Mourao and J. D. Neill and S. Perlmutter and P. Ripoche and N. Suzuki and E. S. Walker},
  
	title = {{SUPERNOVA} {CONSTRAINTS} {AND} {SYSTEMATIC} {UNCERTAINTIES} {FROM} {THE} {FIRST} {THREE} {YEARS} {OF} {THE} {SUPERNOVA} {LEGACY} {SURVEY}},
  
	journal = {The Astrophysical Journal Supplement Series}
}

@article{Colgain_2019,
	doi = {10.1088/1475-7516/2019/09/006},
  
	url = {https://doi.org/10.1088%2F1475-7516%2F2019%2F09%2F006},
  
	year = 2019,
	month = {sep},
  
	publisher = {{IOP} Publishing},
  
	volume = {2019},
  
	number = {09},
  
	pages = {006--006},
  
	author = {Eoin O
 Colgain},
  
	title = {A hint of matter underdensity at low $\less$i$\greater$z$\less$/i$\greater$?},
  
	journal = {Journal of Cosmology and Astroparticle Physics}
}

@article{Kazantzidis2020,
    author = "Kazantzidis, L. and Perivolaropoulos, L.",
    title = "{Hints of a Local Matter Underdensity or Modified Gravity in the Low $z$ Pantheon data}",
    eprint = "2004.02155",
    archivePrefix = "arXiv",
    primaryClass = "astro-ph.CO",
    doi = "10.1103/PhysRevD.102.023520",
    journal = "Phys. Rev. D",
    volume = "102",
    number = "2",
    pages = "023520",
    year = "2020"
}

@Article{Dainotti_22,
AUTHOR = {Dainotti, Maria Giovanna and De Simone, Biagio and Schiavone, Tiziano and Montani, Giovanni and Rinaldi, Enrico and Lambiase, Gaetano and Bogdan, Malgorzata and Ugale, Sahil},
TITLE = {On the Evolution of the Hubble Constant with the SNe Ia Pantheon Sample and Baryon Acoustic Oscillations: A Feasibility Study for GRB-Cosmology in 2030},
JOURNAL = {Galaxies},
VOLUME = {10},
YEAR = {2022},
NUMBER = {1},
ARTICLE-NUMBER = {24},
URL = {https://www.mdpi.com/2075-4434/10/1/24},
ISSN = {2075-4434},
DOI = {10.3390/galaxies10010024}
}

@inproceedings{seabold2010statsmodels,
title={Statsmodels: Econometric and Statistical Modeling with Python},
author={Seabold, J.S. and Perktold, J.},
booktitle={Proceedings of the 9th Python in Science Conference},
year={2010}
}

@misc{Aluri22,
  doi = {10.48550/ARXIV.2207.05765},
  year = 2022,
  url = {https://arxiv.org/abs/2207.05765},
  
  author = {Aluri, Pavan Kumar and Cea, Paolo and Chingangbam, Pravabati and Chu, Ming-Chung and Clowes, Roger G. and Hutsemékers, Damien and Kochappan, Joby P. and Krasiński, Andrzej and Lopez, Alexia M. and Liu, Lang and Martens, Niels C. M. and Martins, C. J. A. P. and Migkas, Konstantinos and Colgáin, Eoin Ó and Pranav, Pratyush and Shamir, Lior and Singal, Ashok K. and Sheikh-Jabbari, M. M. and Wagner, Jenny and Wang, Shao-Jiang and Wiltshire, David L. and Yeung, Shek and Yin, Lu and Zhao, Wen},
}

@article{Riess_2022,
	doi = {10.3847/2041-8213/ac5c5b},
  
	url = {https://doi.org/10.3847%2F2041-8213%2Fac5c5b},
  
	year = 2022,
	month = {jul},
  
	publisher = {American Astronomical Society},
  
	volume = {934},
  
	number = {1},
  
	pages = {L7},
  
	author = {Adam G. Riess and Wenlong Yuan and Lucas M. Macri and Dan Scolnic and Dillon Brout and Stefano Casertano and David O. Jones and Yukei Murakami and Gagandeep S. Anand and Louise Breuval and Thomas G. Brink and Alexei V. Filippenko and Samantha Hoffmann and Saurabh W. Jha and W. D'arcy Kenworthy and John Mackenty and Benjamin E. Stahl and WeiKang Zheng},
  
	title = {A Comprehensive Measurement of the Local Value of the Hubble Constant with 1 km s$\less$sup$\greater$-1$\less$/sup$\greater$ Mpc$\less$sup$\greater$-1$\less$/sup$\greater$ Uncertainty from the Hubble Space Telescope and the {SH}0ES Team},
  
	journal = {The Astrophysical Journal Letters}
}

@article{Wong_2019,
	doi = {10.1093/mnras/stz3094},
  
	url = {https://doi.org/10.1093%2Fmnras%2Fstz3094},
  
	year = 2019,
	month = {sep},
  
	publisher = {Oxford University Press ({OUP})},
  
	volume = {498},
  
	number = {1},
  
	pages = {1420--1439},
  
	author = {Kenneth C Wong and Sherry H Suyu and Geoff C-F Chen and Cristian E Rusu and Martin Millon and Dominique Sluse and Vivien Bonvin and Christopher D Fassnacht and Stefan Taubenberger and Matthew W Auger and Simon Birrer and James H H Chan and Frederic Courbin and Stefan Hilbert and Olga Tihhonova and Tommaso Treu and Adriano Agnello and Xuheng Ding and Inh Jee and Eiichiro Komatsu and Anowar J Shajib and Alessandro Sonnenfeld and Roger D Blandford and L{\'{e}
}on V E Koopmans and Philip J Marshall and Georges Meylan},
  
	title = {H0LiCOW {\textendash} {XIII}. A 2.4 per cent measurement of H0 from lensed quasars: 5.3$\upsigma$ tension between early- and late-Universe probes},
  
	journal = {Monthly Notices of the Royal Astronomical Society}
}

@article{Millon_2020,
	doi = {10.1051/0004-6361/201937351},
  
	url = {https://doi.org/10.1051%2F0004-6361%2F201937351},
  
	year = 2020,
	month = {jul},
  
	publisher = {{EDP} Sciences},
  
	volume = {639},
  
	pages = {A101},
  
	author = {M. Millon and A. Galan and F. Courbin and T. Treu and S. H. Suyu and X. Ding and S. Birrer and G. C.-F. Chen and A. J. Shajib and D. Sluse and K. C. Wong and A. Agnello and M. W. Auger and E. J. Buckley-Geer and J. H. H. Chan and T. Collett and C. D. Fassnacht and S. Hilbert and L. V. E. Koopmans and V. Motta and S. Mukherjee and C. E. Rusu and A. Sonnenfeld and C. Spiniello and L. Van de Vyvere},
  
	title = {{TDCOSMO}
},
  
	journal = {Astronomy \& Astrophysics}
}

@article{Wang_1998,
   title={Cluster Abundance Constraints for Cosmological Models with a Time‐varying, Spatially Inhomogeneous Energy Component with Negative Pressure},
   volume={508},
   ISSN={1538-4357},
   url={http://dx.doi.org/10.1086/306436},
   DOI={10.1086/306436},
   number={2},
   journal={The Astrophysical Journal},
   publisher={American Astronomical Society},
   author={Wang, Limin and Steinhardt, Paul J.},
   year={1998},
   month=dec, pages={483–490} }

@article{DiValentino_2021,
doi = {10.1088/1361-6382/ac086d},
url = {https://dx.doi.org/10.1088/1361-6382/ac086d},
year = {2021},
month = {jul},
publisher = {IOP Publishing},
volume = {38},
number = {15},
pages = {153001},
author = {Eleonora Di Valentino and Olga Mena and Supriya Pan and Luca Visinelli and Weiqiang Yang and Alessandro Melchiorri and David F Mota and Adam G Riess and Joseph Silk},
title = {In the realm of the Hubble tension—a review of solutions*},
journal = {Classical and Quantum Gravity}
}

@article{Sorrenti:2024ugq,
doi = {10.1088/1475-7516/2024/12/003},
url = {https://dx.doi.org/10.1088/1475-7516/2024/12/003},
year = {2024},
month = {dec},
publisher = {IOP Publishing},
volume = {2024},
number = {12},
pages = {003},
author = {Sorrenti, Francesco and Durrer, Ruth and Kunz, Martin},
title = {A local infall from a cosmographic analysis of Pantheon+},
journal = {Journal of Cosmology and Astroparticle Physics}
}

@article{KHALILIGOLMANKHANEH2024105081,
title = {Einstein field equations extended to fractal manifolds: A fractal perspective},
journal = {Journal of Geometry and Physics},
volume = {196},
pages = {105081},
year = {2024},
issn = {0393-0440},
doi = {https://doi.org/10.1016/j.geomphys.2023.105081},
url = {https://www.sciencedirect.com/science/article/pii/S0393044023003339},
author = {Alireza {Khalili Golmankhaneh} and Palle E.T. Jørgensen and Agnieszka Matylda Schlichtinger}
}

@ARTICLE{Perlmutter99,
       author = {{Perlmutter}, S. and {Aldering}, G. and {Goldhaber}, G. and {Knop}, R.~A. and {Nugent}, P. and {Castro}, P.~G. and {Deustua}, S. and {Fabbro}, S. and {Goobar}, A. and {Groom}, D.~E. and {Hook}, I.~M. and {Kim}, A.~G. and {Kim}, M.~Y. and {Lee}, J.~C. and {Nunes}, N.~J. and {Pain}, R. and {Pennypacker}, C.~R. and {Quimby}, R. and {Lidman}, C. and {Ellis}, R.~S. and {Irwin}, M. and {McMahon}, R.~G. and {Ruiz-Lapuente}, P. and {Walton}, N. and {Schaefer}, B. and {Boyle}, B.~J. and {Filippenko}, A.~V. and {Matheson}, T. and {Fruchter}, A.~S. and {Panagia}, N. and {Newberg}, H.~J.~M. and {Couch}, W.~J. and {Project}, The Supernova Cosmology},
        title = "{Measurements of {\ensuremath{\Omega}} and {\ensuremath{\Lambda}} from 42 High-Redshift Supernovae}",
      journal = {\apj},
         year = 1999,
        month = jun,
       volume = {517},
       number = {2},
        pages = {565-586},
          doi = {10.1086/307221},
archivePrefix = {arXiv},
       eprint = {astro-ph/9812133},
 primaryClass = {astro-ph},
       adsurl = {https://ui.adsabs.harvard.edu/abs/1999ApJ...517..565P}
}

@article{Seifert2024,
    author = {Seifert, Antonia and Lane, Zachary G and Galoppo, Marco and Ridden-Harper, Ryan and Wiltshire, David L},
    title = {Supernovae evidence for foundational change to cosmological models},
    journal = {Monthly Notices of the Royal Astronomical Society: Letters},
    volume = {537},
    number = {1},
    pages = {L55-L60},
    year = {2024},
    month = {12},
    issn = {1745-3925},
    doi = {10.1093/mnrasl/slae112},
    url = {https://doi.org/10.1093/mnrasl/slae112},
    eprint = {https://academic.oup.com/mnrasl/article-pdf/537/1/L55/61215381/slae112.pdf},
}

@article{Davis11,
   title={THE EFFECT OF PECULIAR VELOCITIES ON SUPERNOVA COSMOLOGY},
   volume={741},
   ISSN={1538-4357},
   url={http://dx.doi.org/10.1088/0004-637X/741/1/67},
   DOI={10.1088/0004-637x/741/1/67},
   number={1},
   journal={The Astrophysical Journal},
   publisher={American Astronomical Society},
   author={Davis, Tamara M. and Hui, Lam and Frieman, Joshua A. and Haugbølle, Troels and Kessler, Richard and Sinclair, Benjamin and Sollerman, Jesper and Bassett, Bruce and Marriner, John and Mörtsell, Edvard and Nichol, Robert C. and Richmond, Michael W. and Sako, Masao and Schneider, Donald P. and Smith, Mathew},
   year={2011},
   month=oct, pages={67} }

@ARTICLE{Ma2014,
       author = {{Ma}, Yin-Zhe and {Pan}, Jun},
        title = "{An estimation of local bulk flow with the maximum-likelihood method}",
      journal = {\mnras},
         year = 2014,
        month = jan,
       volume = {437},
       number = {2},
        pages = {1996-2004},
          doi = {10.1093/mnras/stt2038},
archivePrefix = {arXiv},
       eprint = {[arXiv:1311.6888]},
 primaryClass = {astro-ph.CO},
       adsurl = {https://ui.adsabs.harvard.edu/abs/2014MNRAS.437.1996M}
}

@ARTICLE{Scrimgeour2016,
       author = {{Scrimgeour}, Morag I. and {Davis}, Tamara M. and {Blake}, Chris and {Staveley-Smith}, Lister and {Magoulas}, Christina and {Springob}, Christopher M. and {Beutler}, Florian and {Colless}, Matthew and {Johnson}, Andrew and {Jones}, D. Heath and {Koda}, Jun and {Lucey}, John R. and {Ma}, Yin-Zhe and {Mould}, Jeremy and {Poole}, Gregory B.},
        title = "{The 6dF Galaxy Survey: bulk flows on 50-70 h$^{-1}$ Mpc scales}",
      journal = {\mnras},
         year = 2016,
        month = jan,
       volume = {455},
       number = {1},
        pages = {386-401},
          doi = {10.1093/mnras/stv2146},
archivePrefix = {arXiv},
       eprint = {[arXiv:1511.06930]},
 primaryClass = {astro-ph.CO},
       adsurl = {https://ui.adsabs.harvard.edu/abs/2016MNRAS.455..386S}
}

@ARTICLE{Watkins2009,
       author = {{Watkins}, Richard and {Feldman}, Hume A. and {Hudson}, Michael J.},
        title = "{Consistently large cosmic flows on scales of 100h$^{-1}$Mpc: a challenge for the standard {\ensuremath{\Lambda}}CDM cosmology}",
      journal = {\mnras},
         year = 2009,
        month = jan,
       volume = {392},
       number = {2},
        pages = {743-756},
          doi = {10.1111/j.1365-2966.2008.14089.x},
archivePrefix = {arXiv},
       eprint = {[arXiv:0809.4041]},
 primaryClass = {astro-ph},
       adsurl = {https://ui.adsabs.harvard.edu/abs/2009MNRAS.392..743W}
}

@ARTICLE{Colin2011,
       author = {{Colin}, Jacques and {Mohayaee}, Roya and {Sarkar}, Subir and {Shafieloo}, Arman},
        title = "{Probing the anisotropic local Universe and beyond with SNe Ia data}",
      journal = {\mnras},
         year = 2011,
        month = jun,
       volume = {414},
       number = {1},
        pages = {264-271},
          doi = {10.1111/j.1365-2966.2011.18402.x},
archivePrefix = {arXiv},
       eprint = {[arXiv:1011.6292]},
 primaryClass = {astro-ph.CO},
       adsurl = {https://ui.adsabs.harvard.edu/abs/2011MNRAS.414..264C}
}

@ARTICLE{Salehi2021,
       author = {{Salehi}, A. and {Yarahmadi}, M. and {Fathi}, S. and {Bamba}, Kazuharu},
        title = "{Cosmological bulk flow in the QCDM model: (in)consistency with {\ensuremath{\Lambda}}CDM}",
      journal = {\mnras},
         year = 2021,
        month = jun,
       volume = {504},
       number = {1},
        pages = {1304-1319},
          doi = {10.1093/mnras/stab909},
archivePrefix = {arXiv},
       eprint = {arXiv:2001.01743]},
 primaryClass = {astro-ph.CO},
       adsurl = {https://ui.adsabs.harvard.edu/abs/2021MNRAS.504.1304S}
}

@ARTICLE{Kashlinsky2008,
       author = {{Kashlinsky}, A. and {Atrio-Barandela}, F. and {Kocevski}, D. and {Ebeling}, H.},
        title = "{A Measurement of Large-Scale Peculiar Velocities of Clusters of Galaxies: Results and Cosmological Implications}",
      journal = {\apjl},
         year = 2008,
        month = oct,
       volume = {686},
       number = {2},
        pages = {L49},
          doi = {10.1086/592947},
archivePrefix = {arXiv},
       eprint = {[arXiv:0809.3734]},
 primaryClass = {astro-ph},
       adsurl = {https://ui.adsabs.harvard.edu/abs/2008ApJ...686L..49K}
}

@ARTICLE{Ellis2001,
       author = {{Ellis}, George F.~R. and {van Elst}, Henk and {Maartens}, Roy},
        title = "{General relativistic analysis of peculiar velocities}",
      journal = {Classical and Quantum Gravity},
         year = 2001,
        month = dec,
       volume = {18},
       number = {23},
        pages = {5115-5123},
          doi = {10.1088/0264-9381/18/23/308},
archivePrefix = {arXiv},
       eprint = {[arXiv:gr-qc/0105083]},
 primaryClass = {gr-qc},
       adsurl = {https://ui.adsabs.harvard.edu/abs/2001CQGra..18.5115E}
}

@ARTICLE{Maglara2022,
       author = {{Maglara}, Myrto and {Tsagas}, Christos G.},
        title = "{Peculiar velocities in the early Universe}",
      journal = {\prd},
         year = 2022,
        month = oct,
       volume = {106},
       number = {8},
          eid = {083505},
        pages = {083505},
          doi = {10.1103/PhysRevD.106.083505},
archivePrefix = {arXiv},
       eprint = {[arXiv:2207.09824]},
 primaryClass = {astro-ph.CO},
       adsurl = {https://ui.adsabs.harvard.edu/abs/2022PhRvD.106h3505M}
}

@misc{Tsagas2024,
      title={Bulk flows, general relativity and the fundamental role of the "peculiar" flux}, 
      author={Christos G. Tsagas},
      year={2024},
      eprint={2404.16719},
      archivePrefix={arXiv},
      primaryClass={gr-qc},
      url={https://arxiv.org/abs/2404.16719}, 
}

@article{Freedman_2020,
   title={Calibration of the Tip of the Red Giant Branch},
   volume={891},
   ISSN={1538-4357},
   url={http://dx.doi.org/10.3847/1538-4357/ab7339},
   DOI={10.3847/1538-4357/ab7339},
   number={1},
   journal={The Astrophysical Journal},
   publisher={American Astronomical Society},
   author={Freedman, Wendy L. and Madore, Barry F. and Hoyt, Taylor and Jang, In Sung and Beaton, Rachael and Lee, Myung Gyoon and Monson, Andrew and Neeley, Jill and Rich, Jeffrey},
   year={2020},
   month=mar, pages={57} }

@article{Madore_2020,
   title={Astrophysical Distance Scale: The AGB J-band Method. I. Calibration and a First Application},
   volume={899},
   ISSN={1538-4357},
   url={http://dx.doi.org/10.3847/1538-4357/aba045},
   DOI={10.3847/1538-4357/aba045},
   number={1},
   journal={The Astrophysical Journal},
   publisher={American Astronomical Society},
   author={Madore, Barry F. and Freedman, Wendy L.},
   year={2020},
   month=aug, pages={66} }

@article{Borghi_2022,
   title={Toward a Better Understanding of Cosmic Chronometers: A New Measurement of H(z) at z ∼ 0.7},
   volume={928},
   ISSN={2041-8213},
   url={http://dx.doi.org/10.3847/2041-8213/ac3fb2},
   DOI={10.3847/2041-8213/ac3fb2},
   number={1},
   journal={The Astrophysical Journal Letters},
   publisher={American Astronomical Society},
   author={Borghi, Nicola and Moresco, Michele and Cimatti, Andrea},
   year={2022},
   month=mar, pages={L4} }

@misc{Shajib_2024,
      title={Strong Lensing by Galaxies}, 
      author={A. J. Shajib and G. Vernardos and T. E. Collett and V. Motta and D. Sluse and L. L. R. Williams and P. Saha and S. Birrer and C. Spiniello and T. Treu},
      year={2024},
      eprint={2210.10790},
      archivePrefix={arXiv},
      primaryClass={astro-ph.GA},
      url={https://arxiv.org/abs/2210.10790}, 
}

@article{Agazie_2023,
   title={The NANOGrav 15 yr Data Set: Observations and Timing of 68 Millisecond Pulsars},
   volume={951},
   ISSN={2041-8213},
   url={http://dx.doi.org/10.3847/2041-8213/acda9a},
   DOI={10.3847/2041-8213/acda9a},
   number={1},
   journal={The Astrophysical Journal Letters},
   publisher={American Astronomical Society},
   author={Agazie, Gabriella and Alam, Md Faisal and Anumarlapudi, Akash and Archibald, Anne M. and Arzoumanian, Zaven and Baker, Paul T. and Blecha, Laura and Bonidie, Victoria and Brazier, Adam and Brook, Paul R. and Burke-Spolaor, Sarah and Bécsy, Bence and Chapman, Christopher and Charisi, Maria and Chatterjee, Shami and Cohen, Tyler and Cordes, James M. and Cornish, Neil J. and Crawford, Fronefield and Cromartie, H. Thankful and Crowter, Kathryn and DeCesar, Megan E. and Demorest, Paul B. and Dolch, Timothy and Drachler, Brendan and Ferrara, Elizabeth C. and Fiore, William and Fonseca, Emmanuel and Freedman, Gabriel E. and Garver-Daniels, Nate and Gentile, Peter A. and Glaser, Joseph and Good, Deborah C. and Gültekin, Kayhan and Hazboun, Jeffrey S. and Jennings, Ross J. and Jessup, Cody and Johnson, Aaron D. and Jones, Megan L. and Kaiser, Andrew R. and Kaplan, David L. and Kelley, Luke Zoltan and Kerr, Matthew and Key, Joey S. and Kuske, Anastasia and Laal, Nima and Lam, Michael T. and Lamb, William G. and W. Lazio, T. Joseph and Lewandowska, Natalia and Lin, Ye and Liu, Tingting and Lorimer, Duncan R. and Luo, Jing and Lynch, Ryan S. and Ma, Chung-Pei and Madison, Dustin R. and Maraccini, Kaleb and McEwen, Alexander and McKee, James W. and McLaughlin, Maura A. and McMann, Natasha and Meyers, Bradley W. and Mingarelli, Chiara M. F. and Mitridate, Andrea and Ng, Cherry and Nice, David J. and Ocker, Stella Koch and Olum, Ken D. and Panciu, Elisa and Pennucci, Timothy T. and Perera, Benetge B. P. and Pol, Nihan S. and Radovan, Henri A. and Ransom, Scott M. and Ray, Paul S. and Romano, Joseph D. and Salo, Laura and Sardesai, Shashwat C. and Schmiedekamp, Carl and Schmiedekamp, Ann and Schmitz, Kai and Shapiro-Albert, Brent J. and Siemens, Xavier and Simon, Joseph and Siwek, Magdalena S. and Stairs, Ingrid H. and Stinebring, Daniel R. and Stovall, Kevin and Susobhanan, Abhimanyu and Swiggum, Joseph K. and Taylor, Stephen R. and Turner, Jacob E. and Unal, Caner and Vallisneri, Michele and Vigeland, Sarah J. and Wahl, Haley M. and Wang, Qiaohong and Witt, Caitlin A. and Young, Olivia},
   year={2023},
   month=jun, pages={L9} }

@misc{Scott_2024,
      title={The Cosmic Neutrino Background}, 
      author={Douglas Scott},
      year={2024},
      eprint={2402.16243},
      archivePrefix={arXiv},
      primaryClass={astro-ph.CO},
      url={https://arxiv.org/abs/2402.16243}, 
}

@article{Yang_2014,
  title={Searching for a preferred direction with Union2. 1 data},
  author={Yang, Xiaofeng and Wang, FY and Chu, Zhe},
  journal={Monthly Notices of the Royal Astronomical Society},
  volume={437},
  number={2},
  pages={1840--1846},
  year={2014},
  publisher={Oxford University Press}
}

@article{Javanmardi_2015,
  title={Probing the isotropy of cosmic acceleration traced by Type Ia supernovae},
  author={Javanmardi, Behnam and Porciani, Cristiano and Kroupa, Pavel and Pflamm-Altenburg, Jan},
  journal={The Astrophysical Journal},
  volume={810},
  number={1},
  pages={47},
  year={2015},
  publisher={IOP Publishing}
}

@article{Sun_2018,
  title={Testing the anisotropy of cosmic acceleration from Pantheon supernovae sample},
  author={Sun, ZQ and Wang, FY},
  journal={Monthly Notices of the Royal Astronomical Society},
  volume={478},
  number={4},
  pages={5153--5158},
  year={2018},
  publisher={Oxford University Press}
}

@article{Tang_2023,
  title={Consistency of Pantheon+ supernovae with a large-scale isotropic universe},
  author={Tang, Li and Lin, Hai-Nan and Liu, Liang and Li, Xin},
  journal={Chinese Physics C},
  volume={47},
  number={12},
  pages={125101},
  year={2023},
  publisher={IOP Publishing}
}

@article{Marziani_2014,
   title={Highly accreting quasars: sample definition and possible cosmological implications},
   volume={442},
   ISSN={0035-8711},
   url={http://dx.doi.org/10.1093/mnras/stu951},
   DOI={10.1093/mnras/stu951},
   number={2},
   journal={Monthly Notices of the Royal Astronomical Society},
   publisher={Oxford University Press (OUP)},
   author={Marziani, P. and Sulentic, J. W.},
   year={2014},
   month=jun, pages={1211–1229} }

@article{Watson_2011,
doi = {10.1088/2041-8205/740/2/L49},
url = {https://dx.doi.org/10.1088/2041-8205/740/2/L49},
year = {2011},
month = {oct},
publisher = {The American Astronomical Society},
volume = {740},
number = {2},
pages = {L49},
author = {Watson, D. and Denney, K. D. and Vestergaard, M. and Davis, T. M.},
title = {A NEW COSMOLOGICAL DISTANCE MEASURE USING ACTIVE GALACTIC NUCLEI},
journal = {The Astrophysical Journal Letters}
}

@misc{DESC,
      title={The LSST Dark Energy Science Collaboration (DESC) Science Requirements Document}, 
      author={The LSST Dark Energy Science Collaboration and Rachel Mandelbaum and Tim Eifler and Renée Hložek and Thomas Collett and Eric Gawiser and Daniel Scolnic and David Alonso and Humna Awan and Rahul Biswas and Jonathan Blazek and Patricia Burchat and Nora Elisa Chisari and Ian Dell'Antonio and Seth Digel and Josh Frieman and Daniel A. Goldstein and Isobel Hook and Željko Ivezić and Steven M. Kahn and Sowmya Kamath and David Kirkby and Thomas Kitching and Elisabeth Krause and Pierre-François Leget and Philip J. Marshall and Joshua Meyers and Hironao Miyatake and Jeffrey A. Newman and Robert Nichol and Eli Rykoff and F. Javier Sanchez and Anže Slosar and Mark Sullivan and M. A. Troxel},
      year={2021},
      eprint={1809.01669},
      archivePrefix={arXiv},
      primaryClass={astro-ph.CO},
      url={https://arxiv.org/abs/1809.01669}, 
}

@article{Bardeen_1980,
  title = {Gauge-invariant cosmological perturbations},
  author = {Bardeen, James M.},
  journal = {Phys. Rev. D},
  volume = {22},
  issue = {8},
  pages = {1882--1905},
  numpages = {0},
  year = {1980},
  month = {Oct},
  publisher = {American Physical Society},
  doi = {10.1103/PhysRevD.22.1882},
  url = {https://link.aps.org/doi/10.1103/PhysRevD.22.1882}
}

@book{Wainwright_Ellis_1997, 
author = {Wainwright, Ellis},
place={Cambridge}, title={Dynamical Systems in Cosmology}, publisher={Cambridge University Press}, year={1997}}

@article{Barrow_2004,
doi = {10.1088/0264-9381/21/7/005},
url = {https://dx.doi.org/10.1088/0264-9381/21/7/005},
year = {2004},
month = {mar},
publisher = {},
volume = {21},
number = {7},
pages = {1773},
author = {John D Barrow and Christos G Tsagas},
title = {Dynamics and stability of the Gödel universe},
journal = {Classical and Quantum Gravity}
}

@misc{tzartinoglou_2024,
      title={The deceleration parameter in perturbed Bianchi universes with a peculiar-velocity "tilt"}, 
      author={Amalia Tzartinoglou and Christos G. Tsagas},
      year={2024},
      eprint={2405.17592},
      archivePrefix={arXiv},
      primaryClass={gr-qc},
      url={https://arxiv.org/abs/2405.17592}, 
}

@article{Watkins_2023,
   title={Analysing the large-scale bulk flow using cosmicflows4: increasing tension with the standard cosmological model},
   volume={524},
   ISSN={1365-2966},
   url={http://dx.doi.org/10.1093/mnras/stad1984},
   DOI={10.1093/mnras/stad1984},
   number={2},
   journal={Monthly Notices of the Royal Astronomical Society},
   publisher={Oxford University Press (OUP)},
   author={Watkins, Richard and Allen, Trey and Bradford, Collin James and Ramon, Albert and Walker, Alexandra and Feldman, Hume A and Cionitti, Rachel and Al-Shorman, Yara and Kourkchi, Ehsan and Tully, R Brent},
   year={2023},
   month=jul, pages={1885–1892} }

@article{Kashlinsky_2011,
   title={MEASURING THE DARK FLOW WITH PUBLIC X-RAY CLUSTER DATA},
   volume={732},
   ISSN={1538-4357},
   url={http://dx.doi.org/10.1088/0004-637X/732/1/1},
   DOI={10.1088/0004-637x/732/1/1},
   number={1},
   journal={The Astrophysical Journal},
   publisher={American Astronomical Society},
   author={Kashlinsky, A. and Atrio-Barandela, F. and Ebeling, H.},
   year={2011},
   month=apr, pages={1} }

@BOOK{Binney_2008,
       author = {{Binney}, James and {Tremaine}, Scott},
        title = "{Galactic Dynamics: Second Edition}",
         year = 2008,
       adsurl = {https://ui.adsabs.harvard.edu/abs/2008gady.book.....B}
}


\end{document}